\documentclass[12pt,lot, lof]{puthesis}
\newcommand{\proquestmode}{}

\title{Density Matrix Embedding Theory and Strongly Correlated Lattice Systems}

\submitted{September 2017}  
\copyrightyear{2017}  
\author{Bo-Xiao Zheng}
\adviser{Garnet K.-L. Chan}  
\department{Chemistry}

\usepackage[]{amsmath}
\usepackage{amsfonts}
\usepackage[]{color}
\usepackage{overpic}

\usepackage{graphicx}
\usepackage{subfigure}
\usepackage{tablefootnote}
\usepackage{footnote}
\usepackage{appendix}
\usepackage{verbatim}
\usepackage{bbm}

\usepackage{multirow}
\usepackage{longtable}

\usepackage{booktabs}
\graphicspath{{ch-intro/pics/}{ch-dmet/pics/}{ch-scaling/pics/}{ch-diagram/pics/}{ch-stripe/pics/}{ch-threeband/pics/}}

\ifdefined\printmode

\usepackage{url}

\else

\ifdefined\proquestmode

\usepackage{hyperref}
\hypersetup{bookmarksnumbered}

\makeatletter
\hypersetup{pdftitle=\@title,pdfauthor=\@author}
\makeatother

\else

\usepackage{hyperref}
\hypersetup{colorlinks,bookmarksnumbered}

\makeatletter
\hypersetup{pdftitle=\@title,pdfauthor=\@author}
\makeatother

\fi 
\fi 

\ifodd 0
\renewcommand{\maketitlepage}{}
\renewcommand*{\makecopyrightpage}{}

\abstract{
This thesis describes the development of the density matrix embedding theory (DMET) and
its applications to lattice strongly correlated electron problems. We introduced
a broken spin and particle-number symmetry DMET formulation to study the
high-temperature superconductivity and other low-energy
competing states in models of the cuprate superconductors.
These applications also relied on
(i) the development and adaptation of
approximate impurity solvers beyond exact diagonalization, including the density
matrix renormalization group, auxiliary-field quantum Monte Carlo and
active-space based quantum chemistry techniques, which expanded the sizes
of fragments treated in DMET; and
(ii) the theoretical development and numerical investigations for the finite size scaling behavior of DMET.

Using these numerical tools, we computed a comprehensive ground state phase diagram of
the standard and frustrated Hubbard models on the square lattice with well-controlled
numerical uncertainties, which confirms the existence of the $d$-wave superconductivity
and various inhomogeneous orders in the Hubbard model.
We also investigated the long-sought strong coupling, underdoped regime of the Hubbard
model in great detail, using various numerical techniques including DMET, and determined
the ground state being a highly-compressible, filled vertical stripe at $1/8$ doping in
the coupling range commonly considered relevant to cuprates.
The findings show both the relevance and limitations of
the one-band Hubbard model in studying the cuprate superconductivity.

Therefore, we further explored the three-band Hubbard model and downfolded cuprate Hamiltonians
from first principles, in an attempt to understand the physics beyond the one-band
model. We also extended the DMET formulation to finite temperature using the superoperator
representation of the density operators, which is potentially a powerful tool to investigate
finite-temperature properties of cuprates and other strongly correlated electronic systems.

}

\else

\abstract{

}

\acknowledgements{
First and foremost, I cannot thank my advisor Garnet Chan enough for his excellent
advising and mentoring. Garnet taught me how to be creative and curious, identify and solve
problems, care about details and work on research with passion.
He has always been supportive and helpful outside science as well. No doubt, the five years as
a graduate student working with Garnet was the happiest time I have lived.

It is challenging to work in the field of computational studies of high-temperature superconductivity.
I doubt I can achieve anything in this thesis without my collaborators.
I am especially grateful to Steven White and Shiwei Zhang, our long-time collaborators.
Steve is a profound and knowledgeable scholar. Every time in our discussion, he can hit the
point and give me an ``aha'' moment. Shiwei is a pioneer and expert in the field
of numerical simulations, and has given me a lot of help along the way.
Their students and postdocs, Hao Shi, Chia-Min Chung and Mingpu Qin, have provided a lot
of help for my work on the Hubbard model as well, and our collaborations were presented in Chapter
~\ref{chpt:scaling} and \ref{chpt:stripe}. I am also thankful to Philippe Corboz, Georg Ehlers
and Reinhard Noack for their irreplaceable contributions that resulted in the work presented
in Chapter~\ref{chpt:stripe}. I must thank Lucas Wagner and Hitesh Changlani for insightful
discussions on the three-band model, and Andreas Gr\"uneis and George Booth for their help
with the downfolded cuprate Hamiltonian, which are included in Chapter~\ref{chpt:threeband}.
I also thank Sandeep Sharma for helping me implement
DMRG for BCS calculations and Joshua Kretchmer for the work on the dynamical cluster formulation
of DMET. I must thank Qiming Sun as well for his continuous help on quantum chemistry theories and
programming.

I am especially grateful to Andrew Millis for organizing the Simons Collaboration on the
Many Electron Problem,
an excellent platform for junior scientists like me to connect and communicate with, and learn from
peers and experts in the field. Andy also had a lot of interest in my research, and
invited me to share my results at the APS March meeting.
I thank Emanuel Gull and David Huse, who guided me when I first entered the condensed matter field.
Emanuel and his postdoc James LeBlanc led a Simons Collaboration project on benchmarking
numerical methods, from which I learned a lot. I also thank my early collaborator Bryan
Clark, the discussions with whom helped me figure out how to break particle-number symmetry
in DMET.
I am also grateful to Andy and David Reichman for allowing me to stay with their groups at Columbia
when Garnet moved to Caltech. It was a lot of fun to hang out and to
talk about science with their group members, in particular Soumyo Mukherjee and Ara Go.
I thank Dominika Zgid, Gustavo Scuseria and Toru Shiozaki as well for their insightful suggestions
and comments on my research work.

I thank David Huse and Roberto Car for serving on my graduate committee, who taught me
critical, rigorous and thoroughly thinking. I also thank Annabella Selloni for agreeing to join
my thesis committee and read my thesis.

I feel so lucky for meeting so many talented and friendly people with diverse background
in the Chan group. I am especially thankful to Qiming Sun, Sandeep Sharma, George Booth, Gerald
Knizia, Qiaoni Chen, Barbara Sandhoefer,
Joshua Kretchmer, Sheng Guo and Carlos Jim\'enez-Hoyos, during the discussions and
collaborations with whom I gained understandings of theories, programming and research in general.
I am also grateful to Mark Watson, Jun Yang, Weifeng Hu, Naoki Nakatani, Roberto Olivares-Amaya
and James McClain who were in the group and welcomed me when I started. People who joined
later and left, including Tom Watson, Brecht Verstichel, Sebastian Wouters, Michael Roemelt, Helen van
Aggelen and Rahul Maitra were great colleagues.
Current members and visitors, including Chong Sun, Zhengdong Li, Narbe Mardirossian,
Elvira Sayfutyarova, Alexander Sokolov, Enrico Ronca, Ushnish Ray, Jiajun Ren and Denghui Xing
are nice to me both inside and outside the lab. I especially thank Chong and Denghui for going
to the gym with me during my stay at Caltech. I am sure the friendship with many of my colleagues
will last. I would also like to thank Tim Berkelbach and David Limmer for their kind help when
I applied for research positions last year.

I must also thank Princeton Reseach Computing and National Energy Research Scientific Computing
Center (NERSC) for providing wonderful computational resources and support over the five years.
In particular, Bill Wichser helped me solve many problems when using the Princeton computer clusters.
I would also thank U.S. Department of Energy (DOE) and the Simons Foundation for the financial support
of my research. The Collaboration on the Many Electron Problem was also supported by the Simons Foundation,
where I learned a lot from its summer schools, meetings and the informal discussions.

Finally, I would like to thank my family. I am grateful to my parents and grandparents, who raised me
and encouraged me to pursue my dream. I thank my husband and soulmate Yu
for so many things in my life and career, in particular, the support he gave me when I had setbacks.

}

\dedication{
To my beloved husband, Yu.}

\fi  

\begin{document}

\makefrontmatter


\chapter{Introduction} \label{chpt:intro}
Solving the quantum many-body problem is one of the greatest scientific challenges nowadays.
The fact that quantum mechanics is responsible for ``a large part of physics and the whole of
chemistry'' ~\cite{dirac1929quantum} makes it an appealing tool to understand and predict material
properties and chemical processes from pure mathematical calculations. The advances
in solving the many-electron Schr\"odinger equation harvest the power of quantum
mechanics to deliver faster and cheaper materials discovery, which is a key driving force
in advancing human civilization both historically and today.

Superposition and entanglement are the two most important properties of quantum mechanics that results
in the exotic phenomena. It is not coincidental that they are also the source of 
the extreme complexity in obtaining the numerical solutions of quantum many-body problems.
Superposition means we are dealing with probabilities
over the entire phase space; and entanglement means that the probabilities are not independent.
Thus, with a growing number of quantum objects, the solutions we seek require keeping track of the
outer product space of all the probabilities --- an exponentially growing space that becomes
intractable for any real world applications. One may argue that it has long been known
-- before the discovery of quantum
mechanics -- that high-dimension integrals can be evaluated stochastically via Monte Carlo,
but the way superposition in quantum mechanics works, via the probability amplitudes,
(along with the Fermi statistics)
causes the negative sign problem in all forms of quantum Monte Carlo, leading to exponentially
slow convergence~\cite{troyer2005computational}.

Recent progress in quantum computing~\cite{feynman1982simulating,mooij1999josephson,monroe2013scaling,devoret2013superconducting},
especially the algorithm development and experimental realizations of quantum simulation
~\cite{georgescu2014quantum,wecker2014gate,barends2015digital,o2016scalable,babbush2017low},
seems to provide an alternative path to overcome the fundamental exponential barrier.
However, despite the decades-long engineering effort yet to devote, universal, exact solutions
of general interacting fermion problems still requires a stunning $O(n^9)$ computational
complexity in quantum simulations~\cite{wecker2014gate}, despite a large prefactor over
classical computers.

Inevitably, approximations based on the observations of physical systems have to
be applied to simplify the problem. Perhaps the most popular approximation so far is the mean-field
theory, which replaces the electron interaction with its average effects. A mean-field
solution is usually correct about $\sim 99\%$ of the total energy, and, if one does not look
at the fine details, most of the electron density. Usually, only the electrons
near the Fermi surface are affected when electron interactions are re-introduced.
Depending on many factors including the mean-field energy gap and the strength of the electron
interaction (strictly speaking the ``remaining'' interaction \textit{not} described by the
mean field, or termed \textit{electron correlation}), the system can either be slightly
affect with the qualitative nature unchanged,
or go through phase transitions and behave entirely different from the mean-field picture.
In the first scenario where the mean-field theory is qualitatively correct, quantitative
accuracy can be achieved by introducing perturbative corrections.
Methods such as the GW
approximation~\cite{hedin1965new,Aryasetiawan1998,hedin1999correlation} and
random phase approximation~\cite{altick1964correlation} from the condensed matter field,
and many-body perturbation theory~\cite{moller1934note,kelly1964many} and coupled clusters~\cite{bartlett1978many,bartlett1981many,purvis1982full}
from quantum chemistry, are examples of improving on top of mean-field theory.
In the other scenario where mean-field theory gives qualitatively wrong pictures, such as the
wrong phase of matter, going from the mean-field solution to the real ground state requires
crossing a phase transition point, which is generally impossible for perturbative approaches, and
one may have to start with other limits.

Another valuable concept is the \textit{locality}. Although quantum mechanics itself allows
entanglement between quantum objects from any distance, the interactions between electrons decay
with distance, and thus for low energy states the correlation functions eventually vanish
at long distance.~\footnote{In some cases the correlation function converges to a constant
different from zero in the infinite distance limit, which represents a long-range order.
One can break the associated symmetry in the wavefunction, and the redefined
correlation function
       $\lim_{r\rightarrow \infty}\langle (O(0)-\bar{O})(O(r) - \bar{O})\rangle =0$.}
The principle
of locality thus allows ignoring certain long-range couplings in calculations, and have been
applied to reduce the computational cost of many mean-field based methods
(such as the linear-scaling density functional theory
~\cite{yang1995density,white1996linear,kudin2000linear},
local correlation techniques~\cite{schutz2001low,neese2009efficient,yang2012orbital} and
many-body expansion~\cite{brueckner1955many,dahlke2007electrostatically}),
as well as develop non-perturbative electronic structure methods that does
not rely on the mean-field theory (such as density matrix renormalization group and other
tensor network methods~\cite{white1992density,white1993density,schollwock2005density,
Schollwok2011,verstraete2008matrix,kraus2010fermionic,corboz2009fermionic}).

When it comes to solving large or open systems, another idea that naturally arises is
\textit{embedding}. Starting with an approximate description of the entire system, we can
refine the description of a small piece of the system which we care most, by solving
that piece using higher level methods while coupled to its environment.
To go one step further, if we improve the description on every piece of the entire
system, or, in the lattice settings, use the translational invariance to obtain a better
description for every piece of the system, we can somehow combine the information to
get a better description of the entire system; if the new description of the whole
system has the same form as the original one, we can embed the fragment into the updated
environment again, until reaching a fixed point. The self-consistent
version of embedding turns out to be very powerful in tackling many electron problems.

We would like to do a deeper analysis of the embedding approach.
When solving the coupled fragment-environment problem, it is necessary to approximate the
environment by removing irrelevant degrees of freedom, or we still face the intractable entire
Hilbert space. The simplest implementation of the idea is to include one or a few layers of neighboring
atoms in the fragment calculations. This essentially uses the locality principle, while ignores
the fact that only the electrons near the Fermi surface are the most important, therefore is
inefficient (requires many layers to converge); It does not allow self-consistent improvement
either.

The first and most popular embedding approach for lattice strong correlation problems
(where the electron correlation changes the qualitative phyiscal picture)
is the dynamical mean-field theory (DMFT)~\cite{Georges1996,Kotliar2006}. DMFT uses the
mean-field Green's function of the lattice problem to compute the ``hybridization'', the
frequency-dependent quantity required to make the fragment's standalone, non-interacting
Green's function look like its local Green's function in the context of the lattice. The hybridization is then used to approximate the environment when solving the impurity model using semi-exact methods
such as truncated configuration interaction~\cite{zgid2012truncated}
or continuous-time quantum Monte Carlo (CT-QMC)~\cite{rubtsov2005continuous,werner2006continuous,gull2008continuous}.
\footnote{Usually the
hybridization cannot be directly applied in the solvers, and an additional step of \textit{bath
discretization} is introduced, which uses a set of non-interacting orbitals
to reproduce the spectrum of the hybridization.
The setup thus includes an interacting fragment and a set of non-interacting bath orbitals,
therefore is called the (Anderson) \textit{impurity model}.
The fragment is usually called the impurity in this context.
It usually requires an infinite number of bath orbitals to exactly reproduce the hybridization.
One has to truncate the number of bath orbitals in practice,
causing the \textit{bath discretization error}.}
From the impurity model solution one extracts the \textit{self energy}, which is essentially
a Green's function kernel resulted from electron correlation. The self energy is then
used to update the lattice single-particle Green's function to include contributions from
electron correlation. The new lattice Green's function is then used to compute the
hybridization again and start a new iteration.

DMFT demonstrated the feasibility of embedding with strong quantum mechanical coupling,
and have been successfully applied to lattice fermion models and real materials
with strong correlation, such as the metal-insulator transition of transition metal oxides
and high-temperature superconductivity
~\cite{Kotliar2006,lichtenstein2001finite,wang2012covalency,lichtenstein2000antiferromagnetism,gull2013superconductivity}.
The reason that embedding methods like DMFT can give qualitatively correct pictures even when
starting from the non-interacting solution is that the feedback from the exact impurity model
solution pushes the lattice solution to have the correct physical picture, eg., to break the correct
symmetry. However, the applications of DMFT
 are limited due to its high computational cost and numerical instability, particularly at
low temperature, where CT-QMC encounters severe sign problems. Because of the potentially infinite
number of bath orbitals, and other numerical issues in fitting the hybridization,
DMFT is only able to access small fragments at low temperature or ground state,
whose results are still far from the thermodynamic limit.

\begin{figure}[htpb]
	\centering
	\includegraphics[width=0.9\textwidth]{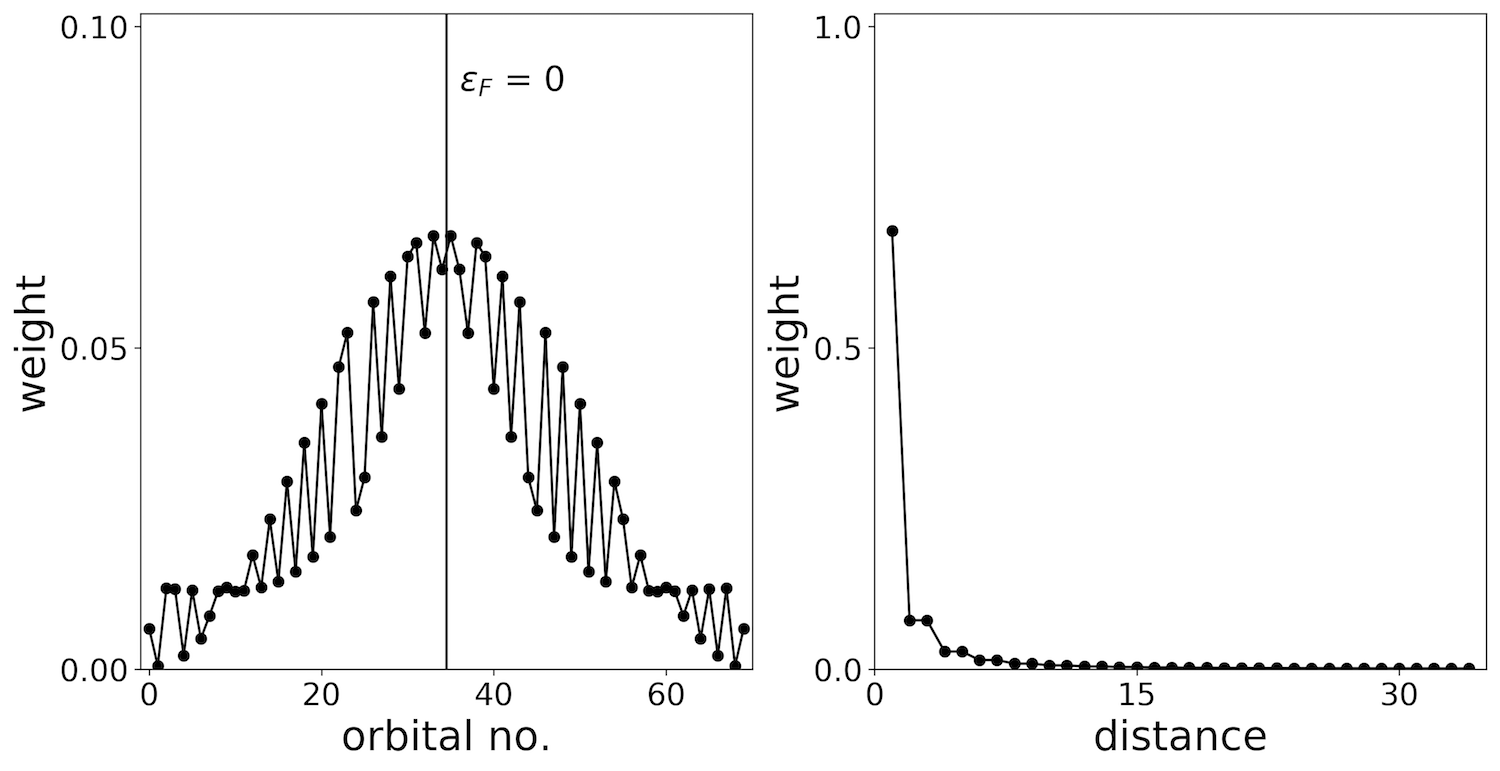}
	\caption{The characterization of bath subspace in DMET. The numerical experiment is
		performed on a half-filled
		70-site tight-binding Hamiltonian with a 2-site fragment.
	Left panel: The squared norm of projections of canonical orbitals
	to the bath space, ordered by orbital energy.
	Right panel: The squared norm of projections of lattice sites to the bath space,
	ordered by the distance to the fragment (only half of the environmental sites are
	shown because of the inversion symmetry).
}
	\label{fig:intro:bath}
\end{figure}

The density matrix embedding theory (DMET)~\cite{Knizia2012,Knizia2013} is developed
to directly target the ground state
where many interesting physical phenomena emerge. DMET is a wavefunction based embedding scheme
which shares many similar concepts with DMFT, but constructed differently. Unlike in DMFT,
the bath orbitals in DMET have physical correspondence --- linear combinations of environmental
orbitals that have the strongest entanglement with the
fragment of interest. Because of the way the bath orbitals are formed, they naturally have
higher weights on sites near the fragment, and on canonical orbitals closer to
the Fermi surface (Fig.~\ref{fig:intro:bath}). In other words, the bath is
selected by a tradeoff between the locality and the proximity to the Fermi surface.
Another advantage of the DMET bath is that the number of bath orbitals is no more
than the number of fragment orbitals. Thus, there is no truncation in the bath
space, and one can treat much larger fragments in DMET than in DMFT, and does not have to worry
about the numerical difficulties in fitting the hybridization function.

In the thesis work, we use DMET to study the cuprate high-temperature superconductivity (HTSC).
HTSC has wide applications in the high-technology industries and scientific reseach, although
these materials are still hard to make and require low temperatures to stay in the superconducting
phase. The ultimate goal in the scientific studies of HTSC is to find systematic ways to increase
its transition temperature. Despite the potential economic value, HTSC is also of much
theoretical interest because of its mysterious pairing mechanism, as well as the rich phases
other than superconductivity that emerge in the material~\cite{machida1989magnetism,
ding1996spectroscopic,vershinin2004local,fauque2006magnetic,sachdev2003colloquium}.

Although the details are far from settled, we actually understand quite a lot qualitative
understandings of cuprate HTSC.
The key of HTSC in cuprates lies in the CuO$_2$ plane, which is shared by all the materials
in this family. The essential physics of the CuO$_2$ plane is well approximated by the one-
and three-band Hubbard model, and the even simpler $t$-$J$ model~\cite{Dagotto1994},
in the sense that the most relevant phases in cuprates, such as antiferromagnetism,
$d$-wave superconductivity, and various inhomogeneous charge, spin and pairing orders
arise in theoretical analysis and/or numerical solutions of these models~\cite{emery1990phase,dagotto1992superconductivity,white1989numerical,halboth2000d,schulz1990incommensurate,white2003stripes,scalettar1991antiferromagnetic,Moshchalkov2001,gull2013superconductivity,Miyazaki2002}. The locations
of these orders in the phase diagram are roughly known for both cuprates and
these derived models. The most intriguing pseudogap phase, which may correspond to inhomogeneous
orders in the ground state, has been studied both experimentally and theoretically, and have
produced many theoretical hypotheses
~\cite{Fleck2001,Chen2004a,Li2006,valla2006ground,Sedrakyan2010,Lee2014}.

However, the problem in the studies of
cuprate HTSC is that, although we understand what phases might appear in cuprates (and
the Hubbard and $t$-$J$ models), and what mechanisms may be behind the phases,
we do not know exactly what actually happens, because many
of the candidate orders are packed in a small energy scale, and can be stabilized by
small changes in the parameters. The low energy scale associated with the various
competing orders makes it difficult to address the problem with pure theory or
simple calculations; numerical studies without enough energy accuracy, although can produce
various relevant orders in roughly correct regions, do not give a definitive answer either;
while accurate, quasi-exact numerical studies usually can only be applied to finite clusters
that are not large enough to support long-range orders. All it requires to solve the issue
is concrete, numerically precise simulation to resolve all the competing states and mechanisms
that appear in the material. Thus, DMET seems to have its
unique advantage as it gives accurate estimates of ground state energy while supporting
long-range order even with small fragment sizes (although to determine the energies and
orders more accurately, larger fragment calculations are necessary to enable extrapolation).

Thus, the goal of my thesis work is to provide well-controlled numerical studies of
HTSC that has enough energy resolution to distinguish the competing orders. We primarily work with
the one-band Hubbard model because it seems a good balance between conciseness and validity,
not to mention the historical importance of the model itself. Experimental realization of
the Hubbard model ground state is also within reach in the next decades
~\cite{hart2015observation,mazurenko2017cold}. To reach this
goal, we have (i) developed broken spin and particle-number symmetry DMET, which can
bring the lattice mean-field solution to the correct phase; (ii) developed
efficient impurity solvers that scales polynomial with the fragment size, that enables
performing DMET on large fragments where rich physics can arise; (iii) determined the
finite-size scaling that translates energies and observables from finite fragment calculations
into quantities in the thermodynamic limit; and (iv) calibrated error estimators for energies
and order parameters that allows us to draw concrete conclusions. As we will demonstrate,
while the one-band Hubbard model is relevant to cuprates, it misses various important
interactions in the real materials and has limitations regarding resolving the detailed
low-energy physics of cuprates. Thus, we go beyond the one-band model to study the three-band
Hubbard model and downfolded \textit{ab initio} cuprate Hamiltonians, which take into account
the missing interactions and can give material specific predictions.

The thesis is organized as follows. In Chapter~\ref{chpt:dmet}, we discuss the theoretical formulation of
DMET, including the basic idea, the broken-symmetry formulation, various impurity solvers
and useful theories and techniques in implementation. Chapter~\ref{chpt:scaling} to Chapter
\ref{chpt:stripe} will focus on the one-band Hubbard model. Chapter~\ref{chpt:scaling} introduces
a joint work
to study the finite-size scaling of two forms of DMET algorithms. In Chapter~\ref{chpt:diagram},
we present
a calibrated ground-state phase diagram of the standard and frustrated 2D Hubbard model with
a wide range of coupling strengths and dopings relevant to cuprates and the $d$-wave
superconductivity. With DMET, the energy accuracy achieved is one to two orders of
magnitude higher than previous studies. In Chapter~\ref{chpt:stripe}, we introduce a
joint work using various numerical methods to study the $1/8$-doping point of the 2D
Hubbard model in depth, where we are able to definitively determine the ground
state and various low-energy states. Chapter~\ref{chpt:threeband} presents a review
of DMET studies of the three-band model and downfolded cuprate Hamiltonians. Finally,
in Chapter~\ref{chpt:finiteT}, we lay out a finite-temperature DMET formulation which
can be applied to study temperature-dependent properties of cuprates and other strongly
correlated materials.
\chapter{Density Matrix Embedding Theory} \label{chpt:dmet}

\section{Introduction} \label{sec:dmet:intro}
Density matrix embedding theory (DMET)~\cite{Knizia2012,Knizia2013,wouters2016five}
aims at accurately describing small fragments strongly coupled to
an extended system. To achieve this goal, it focuses on correctly treating the quantum
entanglement between fragments and their environment.

There are many conceptual similarities between DMET and dynamical mean-field theory (DMFT),
a Green's function based embedding method that treats strongly correlated fermion systems. Both of
them obtain a mean-field-like solution for the entire system and build a set of bath states
to represent the environmental degrees of freedom coupled to the fragment. They also both
create an impurity model to compute the exact solution of the fragment, in the presence of
the bath. Then that information is used to improve the mean-field solution of the entire
system. The process is self-consistent in both methods.

Unlike DMFT, DMET uses wavefunctions
as its primary variable. This choice gives DMET various advantages, such as better suited for ground
state calculations, and that the number of bath states is finite, compared to
the potentially infinite number of bath orbitals in DMFT,
which allows DMET to have significantly smaller computational cost the DMFT.
Thus, one can use DMET to treat systems or properties which were computationally
intractable with DMFT.

This chapter will focus on the formulation of DMET for ground state calculations in lattice
systems. We will introduce a finite-temperature formulation in Chapter~\ref{chpt:finiteT}.
Other spectral, phononic and molecular extensions of DMET are out of the scope of
this thesis work, but can be found in
Refs.~\cite{Booth2013,sandhoefer2016density,Wouters2016,Tsuchimochi2015,welborn2016bootstrap,
bulik2014electron}.
Sec.~\ref{sec:dmet:theory}
introduces the formulation of DMET for normal state calculations. Sec.~\ref{sec:dmet:bcs}
extends the formulation to superconducting states. Sec.~\ref{sec:dmet:solver}
reviews impurity solvers developed or used as part of the thesis work.
In Sec.~\ref{sec:dmet:practice}, we will discuss issues in implementing DMET and strategies
to deal with them.

\section{Theoretical Framework} \label{sec:dmet:theory}
A large part of any embedding theory is determined by how the environment is represented in
the embedding calculations.
In DMET, this is done by the Schmidt decomposition
of the approximate solution of the entire system. In Sections ~\ref{sec:dmet:theory:exact} and
~\ref{sec:dmet:theory:slater}, we introduce the construction of the bath and the impurity model
in detail. In Sec.~\ref{sec:dmet:theory:potential}, we introduce the correlation potential and
how it is optimized. 
In Sec.~\ref{sec:dmet:theory:mu}, we discuss the necessity and ways
to fit chemical potential in DMET.
In Sec.~\ref{sec:dmet:theory:expectation},
we introduce the computation of expectations values in DMET.
Solving the impurity model is an important, but
relatively separate part in DMET algorithms, so we introduce the impurity solvers separately
in Sec.~\ref{sec:dmet:solver}.

Throughout this section we will use a spinless notation to convey only the essential idea of DMET.
Switching to the spinful representation is straightforward.

\subsection{The Exact Embedding} \label{sec:dmet:theory:exact}
The Fock space of electrons (the second-quantized representation of many-electron states)
can be naturally partitioned into two subsystems. We simply divide the orbitals (single-particle
states) into two sets (with the number of orbitals $n_A$ and $n_B$, respectively) and name them subsystem A and B. The Hilbert space of the entire system is thus
the direct product of the subsystem Hilbert spaces,
$\mathcal{H}=\mathcal{H}_A\otimes\mathcal{H}_B$. And the orthonormal
basis of $\mathcal{H}$ is $\{|\psi_i^A\rangle|\psi_j^B\rangle\}$, where $\{|\psi_i^A\rangle\}$
and $\{|\psi_j^B\rangle\}$ are the orthonormal basis of $\mathcal{H}_A$ and
$\mathcal{H}_B$, respectively.

Any state in $\mathcal{H}$ can then be written as
\begin{equation}
	|\psi\rangle=\Psi_{ij}|\psi_i^A\rangle|\psi_j^B\rangle.
	\label{eq:dmet:theory:bipartite}
\end{equation}
Note that we use the Einstein notation for implicit summation here.
The coefficient $\Psi$ is a $2^{n_A}\times2^{n_B}$ matrix.

We can perform the singular value decomposition (SVD) on this matrix, which gives $\Psi = U\Sigma V^{\dagger}$, where
$U$ and $V$ are unitary matrices and $\Sigma$ is a diagonal matrix of dimension $2^{n_A}\times2^{n_B}$.
The number of non-zero elements in $\Sigma$ is $\min\{2^{n_A}, 2^{n_B}\}$.
Without losing the generality, we assume $n_A \le n_B$, and let
the diagonal elements of $\Sigma$ be $\{\sigma_k\} (k = 1,\cdots, 2^{n_A})$, then
\begin{equation}
	\Psi_{ij}=\sigma_k U_{ik} V_{jk}^{*}.
	\label{eq:dmet:theory:svd}
\end{equation}
Let $|\alpha_k\rangle=U_{ik}|\psi_i^A\rangle$ and $|\beta_k\rangle=V_{jk}^*|\psi_j^B\rangle$, we have
\begin{equation}
	|\psi\rangle=\sigma_k|\alpha_k\rangle|\beta_k\rangle.
	\label{eq:dmet:theory:schmidt}
\end{equation}
$\{|\alpha_k\rangle\}$ and $\{|\beta_k\rangle\}$ are a special pair of biorthogonal basis
in A and B, which (a) give a diagonal expansion of the wavefunction
$|\psi\rangle$; and (b) the sizes of which equal to
the size of the smaller subspace $\mathcal{H}_A$, which is $2^{n_A}$, 
even for the bigger subspace $\mathcal{H}_B$.
(One can, of course, complete the basis of the larger subspace $\mathcal{H}_B$ with
the complement of $|\beta_k\rangle$.)

This is called the \textit{Schmidt decomposition}~\cite{ekert1995entangled} of
state $|\psi\rangle$. The coefficients $\sigma_k$ and basis $\alpha_k$, $\beta_k$ are Schmidt
coefficients and Schmidt basis, respectively. The math of SVD guarantees that the Schmidt decomposition
is unique (up to a phase), i.e., invariant of $\{|\psi_i^A\rangle\}$ and $\{|\psi_i^B\rangle\}$
chosen to perform the decomposition. It also implies that any biorthogonal basis pair satisfying
Eq.~\ref{eq:dmet:theory:schmidt} is the same Schmidt decomposition of $|\psi\rangle$, up to
a phase.

The Schmidt decomposition naturally defines a set of bath for the fragment problem:
if we let subsystem A be the fragment and the subsystem B be the rest, i.e., the environment,
$\{|\alpha_k\rangle\}$ is a complete basis for the fragment, while $\{|\beta_k\rangle\}$ spans
a small subspace in the environment with the following properties: (a) it has direct entanglement with the fragment; (b) its size equals to the size of the fragment, and thus independent of the
size of the environment. We can thus define $\{|\beta_k\rangle\}$ as the bath, in which we
solve the coupled fragment and bath problem by projecting the Hamiltonian to this subspace.

Thus, given the partition of the fragment and the environment, any wavefunction representation of the
entire system defines a set of bath, with which one can embed the fragment in the environment
quantum mechanically. In particular, in the limit where $|\psi\rangle$ is the exact ground-state
wavefunction of the entire system, the
embedding is exact. By exact embedding, we mean any observables in the fragment, obtained from the impurity
model, equals to the same observables obtained by solving the entire lattice system exactly.

Suppose in Eq.~\ref{eq:dmet:theory:schmidt}, $|\psi\rangle=|\Psi\rangle$ is the ground state of the entire
system, and the ground-state energy $E_0=\langle \Psi|\hat{H}|\Psi\rangle $.
The bath Hilbert space is the space spanned by environmental Schmidt basis
$\mathcal{H}_\beta=span\{|\beta_k\rangle\}$ while the fragment space is
$\mathcal{H}_\alpha=span\{|\alpha_k\rangle\}$.
We define the impurity model Hamiltonian in the space
$\mathcal{H}_\alpha\otimes\mathcal{H}_\beta$ by
projecting the entire system Hamiltonian $\hat{H}$ to the subspace
\begin{equation}
	\hat{H}_{\text{emb}} = P\hat{H}P
	\label{eq:dmet:theory:projection}
\end{equation}
where the projector
\begin{equation}
	P=\sum_{k,l}|\alpha_k\rangle|\beta_l\rangle\langle\beta_l|\langle\alpha_k|.
	\label{eq:dmet:theory:projector}
\end{equation}

Since $\langle\Psi|\hat{H}_{\text{emb}}|\Psi\rangle=\langle\Psi|P\hat{H}P|\Psi\rangle
=\langle\Psi|\hat{H}|\Psi\rangle=E_0$, the ground state solution of the impurity model
is exactly $|\Psi\rangle$, the same as the ground state of the entire system. It obvious
matches all the observables with the exact ground state.

\subsection{Embedding with Slater Determinants} \label{sec:dmet:theory:slater}
The exact embedding is an ideal scenario but impossible to realize. After all, there is no point to
do any embedding when we can obtain the exact solution for the entire system. Another subtle,
but a much more serious issue is that the Schmidt decomposition of any general many-body
wavefunction scales exponentially with the size of the entire lattice. It is thus desirable
to find a class of approximate wavefunctions, which
\begin{itemize}
	\item allows computationally tractable Schmidt decomposition; and
	\item can be systematically improved to approach the exact wavefunction.
\end{itemize}

It turns out that there is no perfect candidate that satisfies both properties.
One can, however, loose the second condition to only require observables in the correlated
fragment solution to be systematically improvable. In this case, Slater
determinants can be used to approximate the lattice wavefunction.

A Slater determinant consists of a set of $N$ occupied orbitals (canonical orbitals)
as linear combinations of $n$ local orbitals (i.e., atomic orbitals or lattice sites).
\begin{equation}
	|\psi\rangle=c_1^{\dagger}\cdots c_N^{\dagger}|0\rangle.
	\label{eq:dmet:theory:slater}
\end{equation}
The occupied orbitals $c_p^{\dagger}=C_{ip}a_i^{\dagger}$, where $a_i^{\dagger}$ creates an electron
on site (orbital) $i$ and $C_{n\times N}$ is the orbital coefficient matrix. For simplicity, we assume
both the cannonical orbitals and the local orbitals are orthonormal sets.
The one-body density matrix is defined as
\begin{equation}
	\rho_{ij}=\langle \psi|a_i^{\dagger}a_j|\psi\rangle=C_{ip}C_{jp}^*.
	\label{eq:dmet:theory:rho}
\end{equation}
In matrix form, $\rho=CC^{\dagger}$. All higher-order density matrices and observables of the
Slater determinants can be computed through the one-body density matrix. Also, the
one-body density matrix is invariant under the rotations of occupied orbitals.

The Schmidt basis of Slater determinants can be constructed by the rotations of the occupied orbitals. Consider
a bipartite of the system, where the first $n_A$ sites belong to subsystem A, and the other $n_B=n-n_A$ sites belong
to subsystem B. We also assume $n_A < n_B$ and number of electrons $N > n_A$. In this setting, A is the fragment of
interest, while B is the environment.
There exists a unitary transformation of the coefficient matrix $C$, such that
\begin{equation}
	\tilde{C}=CR=
	\begin{bmatrix}
		P& \mathbf{0}\\
		Q& E
	\end{bmatrix}
	\label{eq:dmet:theory:rotation}
\end{equation}
where $R$ is a unitary matrix, $P$, $Q$ and $E$ are of dimensions $n_A\times n_A$, $n_B\times n_A$ and
$n_B\times(n-n_A)$, respectively. The upper right corner of the transformed coefficient matrix is zero, meaning
all but $n_A$ occupied orbitals are restricted to subsystem B. As we have seen before, the transformation leaves
the one-body density matrix, and thus the Slater determinant invariant, therefore
\begin{equation}
	\begin{split}
		|\psi\rangle&=\prod_{k=1}^{n_A}\tilde{c}_k^{\dagger}\prod_{l=n_A+1}^N\tilde{c}_l^{\dagger}|0\rangle\\
		&=\prod_{k=1}^{n_A}(p_k c_{A,k}^{\dagger}+q_k c_{B,k}^{\dagger})
		\prod_{l=n_A+1}^N c_{B,l}^{\dagger}|0\rangle\\
		&=\sum_{i_1,\cdots,i_{n_A}\in\{0,1\}}
		\prod_{k=1}^np_k^{i_k}q_k^{1-i_k} |i_1,\cdots,i_{n_A}\rangle_A
		\otimes |1-i_1,\cdots,1-i_{n_A};\Psi_c\rangle_B
	\end{split}
	\label{eq:dmet:theory:slater_rotated}
\end{equation}
where
\begin{equation}
	c_{A,k}^{\dagger}=\frac{1}{p_k}\sum_{i=1}^{n_A}P_{ik}a_i^{\dagger},
	c_{B,k}^{\dagger}=\frac{1}{q_k}\sum_{j=1}^{n_B}Q_{j, k}a_{j+n_A}^{\dagger},
	c_{B,l}^{\dagger}=C_l^{\dagger}=\sum_{j=1}^{n_B}E_{j, l}a_{j+n_A}^{\dagger}
\end{equation}
and $p_k = (\sum_{i=1}^{n_A}|p_{ik}|^2)^{1/2}$, $q_k = (\sum_{i=1}^{n_B}|q_{ik}|^2)^{1/2}$
are the normalization factors.

Note that the columns of $P, Q$ do not have to be orthogonal. But we will show later that there exists a set of
$P, Q$ where the columns are orthogonal. (The columns of $Q$ are always orthogonal with columns of $E$.) If this
is true, $c_{A, k}^{\dagger}, c_{B_k}^{\dagger}$ are orthogonal fermions operators, and in particular
$\{c_{A, k}^{\dagger}\}$ spans the Fock space of subsystem A. Under this condition, we used the following notation
in Eq.~\ref{eq:dmet:theory:slater_rotated}
\begin{equation}
	\begin{split}
		|i_1,\cdots,i_{n_A}\rangle_A&=\prod_{k=1}^{n_A}(c_{A,k}^{\dagger})^{i_k}|0\rangle_A\\
		|j_1,\cdots,j_{n_A};\Psi_c\rangle_B&
		=\prod_{k=1}^{n_A}(c_{B,k}^{\dagger})^{j_k} |\Psi_c\rangle_B
		=\prod_{k=1}^{n_A}(c_{B,k}^{\dagger})^{j_k}\prod_{l=n_A+1}^N c_{B,l}^{\dagger}|0\rangle_B.
	\end{split}
	\label{eq:dmet:theory:basis}
\end{equation}

Eq.~\ref{eq:dmet:theory:basis} defines a pair of biorthogonal basis for subsystem A and B, and, according to the
uniqueness of Schmidt decomposition, Eq.~\ref{eq:dmet:theory:slater_rotated} is actually the Schmidt decomposition
of the Slater determinant. The impurity model is thus equivalent to solving the complete active space (CAS) problem
with active orbitals $\{c_{A_k}^{\dagger}, c_{B_k}^{\dagger}\} (k\in[1, n_A])$ and core orbitals
$\{c_{B_l}\} (l\in[n_A+1, N])$. Since the rotation within the active space does not affect the solution, we can simply
use the site basis $\{a_i^{\dagger}\}(i\in[1, n_A])$ instead of $\{c_{A_k}^{\dagger}\}$. Thus, only the
bath and core orbitals $\{c_{B_k}^{\dagger}\}$ need to be specified for the impurity model.
(Note the bath orbitals are related to but different from the bath states $\{|\beta_k\rangle\}$ from
Sec.~\ref{sec:dmet:theory:exact} which are many-body states.)

There are a few different but equivalent approaches to obtain the bath orbitals using the one-body density matrix.
Use $\tilde{C}=CR$ to compute the one-body density matrix, we have
\begin{equation}
	\rho=\tilde{C}\tilde{C}^{\dagger}=
	\begin{bmatrix}
		PP^{\dagger}& PQ^{\dagger}\\
		QP^{\dagger}& QQ^{\dagger}+EE^{\dagger}
	\end{bmatrix}=
	\begin{bmatrix}
		\rho_A& \rho_{AB}\\
		\rho_{AB}^{\dagger}& \rho_B
	\end{bmatrix}.
	\label{eq:dmet:theory:rho_block}
\end{equation}
Let $\rho_A=U\Gamma U^{\dagger}$ be the eigendecomposition, then we have $P=U\Gamma^{1/2}$, and
$Q=(P^{-1}\rho_{AB})^{\dagger}$. The bath orbitals are thus defined by normalizing each column of $Q$.
Note that here the columns of $P$ are orthogonal, because $U$ is unitary and $\Gamma^{1/2}$ is diagonal.
Since $PP^{\dagger}+QQ^{\dagger}=I$, it follows that the columns of $Q$ are also orthogonal. One can
perform a simple transformation, for instance, $P^\prime=PT$ and $Q^\prime=Q(T^{-1})^{\dagger}$ where $T$ is any
non-singular square matrix, to make columns of $P^\prime$ and $Q^\prime$ non-orthogonal. This verifies
our claim that the representation in Eq.~\ref{eq:dmet:theory:rotation} is not unique, and there exist
solutions where columns of $P$ and $Q$ are orthogonal.

An equivalent way to obtain bath orbitals is to diagonalize the environmental part of the density matrix $\rho_B$.
The eigenstates with eigenvalues between 0 and 1 are the bath orbitals, while those with eigenvalue 1 are core
orbitals. One can as well perform SVD on $\rho_{AB}=U\Sigma V^{\dagger}$, where the matrix $V$ gives the
orthonormalized coefficients of bath orbitals.

In summary, given a Slater determinant, we can formulate the embedding calculation by obtaining the bath and
core orbitals of the environment, and the impurity model becomes a CASCI problem.

\subsection{Correlation Potential} \label{sec:dmet:theory:potential}
In Sec.~\ref{sec:dmet:theory:slater}, we present in detail how to perform an embedding calculation given a
Slater determinant wavefunction of the lattice. We now discuss the parameterization and optimization
of the determinant.

In DMET, the determinant is parameterized as the ground state of a non-interacting Hamiltonian
\begin{equation}
	H_{\text{mf}}=h+u
	\label{eq:dmet:theory:corr_pot}
\end{equation}
where the core Hamiltonian $h$ is either the one-body part of the full Hamiltonian
$H_1$, or the Fock matrix $F$, depending on the nature of the problem.
For instance, in lattice problems where the interactions are usually local,
the bare one-body Hamiltonian $H_1$ is preferred; while in molecular calculations, it is better
to use the Fock matrix $F$. The additional one-body term $u$, usually called
the \textit{correlation potential}, mimics the effective two-body interaction, is to be determined.

Compared to using the orbital coefficients $C$ as primary variable, an auxiliary potential
$u$ is favored in various ways: the constraints on the $u$ (Hermitian) is much simpler than the
constraints on $C$ (unitary), and there is strong physical interpretations for $u$ as the effective
potential due to electron correlation.

To approximate the behavior of local correlation, we restrict the correlation potential to be local
to each fragment.
In lattice systems, it means $u$ is block-diagonal on the supercells chosen as fragments,
and each diagonal block is identical because of the translational invariance.
\begin{equation}
	u=
	\begin{bmatrix}
		u_C&&&\\
		&\ddots&&\\
		&&u_C&\\
		&&&u_C
	\end{bmatrix}.
	\label{eq:dmet:theory:corr_pot_block}
\end{equation}

Given the correlation potential $u$, one obtains the lattice Slater determinant $|\psi\rangle$ and the
correlated wavefunction $|\psi_{\text{emb}}\rangle=|\Psi\rangle\otimes|\Psi_c\rangle$. In the notation,
$|\Psi\rangle$ is the correlated wavefunction of the impurity model (fragment + bath), while
$|\Psi_c\rangle$ is the core environmental wavefunction defined in Sec.~\ref{sec:dmet:theory:slater}. It becomes
apparent that using Slater determinant to construct the impurity model does not change any expectation values
in the core space. Therefore, one only expects the observables in the fragment to improve upon
the mean-field results. Because one has no access to the exact values of these observables,
to evaluate the quality of the embedding calculation, and thus $u$, we use the similarity between $|\psi\rangle$
and $|\psi_{\text{emb}}\rangle$.

There are many metrics to choose from, but with DMET, the most common choice is the one-body density
matrix. The cost function is thus
\begin{equation}
	f(u)=||\rho_{\Psi}(u)-\rho_{\psi}(u)||^2
	\label{eq:dmet:theory:cost}
\end{equation}
where $\rho_{\Psi,ij}$ is the one-body density matrix of the correlated wavefunction and $\rho_{\psi}$
is the that of the Slater determinant, both in the basis of the impurity model. Minimizing the cost
thus minimizes the difference between the mean-filed level and correlated solutions of the impurity
model at one-particle level. There are various implications of this cost function:
\begin{itemize}
	\item Both the fragment and bath parts of the density matrix were used, indicating that we try
		to maintain a balance between the accuracy of the fragment itself and its coupling
		to the environment.
	\item Since the mean-field density matrix $\rho_{\psi}$ is idempotent (even when projected to
		the impurity model), generally the cost function cannot be reduced to zero.
\end{itemize}

Other cost functions, such as the density matrix difference on the fragment, or even simpler, the difference
in occupation numbers of the fragment orbitals
, were proposed~\cite{Bulik2014}. Empirically, we find the full density matrix
formulation works best for lattice models, probably because of the balance it achieves between the fragment
itself and the coupling to the environment.

Eq.~\ref{eq:dmet:theory:cost} defines an unconstrained optimization problem, which one can solve with
standard minimization procedure. However, the gradient of the cost function is
\begin{equation}
	f^\prime(u)=\sum_{ij}(\rho_{\Psi,ij}(u)-\rho_{\psi,ij}(u))
	(\frac{d\rho_{\Psi,ij}}{du}-\frac{d\rho_{\psi,ij}}{du})
	\label{eq:dmet:theory:cost_grad}
\end{equation}
where the response of $\rho_{\Psi}$ with respect to changes in $u$ cannot be computed analytically and the
numerical gradient involves solving the impurity model multiple times. Therefore, we use a self-consistent
procedure to perform the minimization
\begin{enumerate}
	\item Compute $\rho_{\Psi}^*=\rho_{\Psi}(u)$.
	\item $\min_u f(u)=||\rho_{\Psi}^*-\rho_{\psi}(u)||^2$ with $\rho_{\Psi}^*$ fixed;
		If $||\Delta u|| > \varepsilon_u$, go back to step 1.
\end{enumerate}
Here $\varepsilon_u$ is the convergence threshold for the correlation potential $u$.

One last issue related to the correlation potential is the choice of impurity model Hamiltonian.
In exact embedding, we simply project the entire lattice Hamiltonian to the impurity model
(Eq.~\ref{eq:dmet:theory:projection}). We can continue using this approach when embedding
with Slater determinants; because there are electron interactions between bath
orbitals, this approach is often referred as the \textit{interacting bath}. Alternatively,
one can use the correlation potential to replace the electron interactions on bath orbitals, and include
the interactions only between the fragment orbitals, called the \textit{non-interacting bath}
\begin{equation}
	\hat{H}_{\text{emb}}^{\text{NI}}=P[H_1+u+P_F(H_2-u)P_F]P
	\label{eq:dmet:theory:ni}
\end{equation}
where $H_2$ is the two-body part of the original Hamiltonian, and $P_F$ is the projection to the
fragment.
The non-interacting bath has an origin analogous to those used in the DMFT impurity model, which
are restricted to be non-interacting and varied to match the hybridization function.

It is often preferable to use the non-interacting bath formulation in lattice DMET calculations,
although the interacting bath seems more elegant. There are various theoretical arguments for
that. One argument is that embeddings from
Slater determinants may not have enough flexibility to obtain a good enough approximation for the
ground state Schmidt basis, while the non-interacting approach, by directly encoding $u$ in the impurity
Hamiltonian, can access a larger space to approximate the ground state properties
of the fragment efficiently.
Another more physical argument is that, since the bath orbitals in lattice systems often
extend many unit cells, the screening plays a role and it is well-known that direct downfolding of the
electron interactions works poorly in this context. The use of $u$ to replace these long-range
interactions in the non-interacting bath, to some extent, renormalizes the electron interaction
and could improve the results. Computationally, the non-interacting bath is also favored as it
avoids the task of sometimes formidable integral transformation which scales $O(n^5)$ with the size of
the entire lattice.

The detailed equations and algorithms for this section are presented in Appendix~\ref{sec:algo:corr_fit}.

\subsection{Chemical Potential Optimization} \label{sec:dmet:theory:mu}
A problem in the embedding calculations is how to control the number of electrons in the
fragment. Following the embedding formulation in Sec.~\ref{sec:dmet:theory:slater} and ~\ref{sec:dmet:theory:potential},
there is no guarantee what number of electrons in the fragment we get from impurity model calculations.
This does not only affect the values
of the observables, but is itself problematic in a lattice problem, where
the number of electrons per fragment is usually well defined. It is thus desirable to have the number
of electrons exactly match what we expect for the lattice model.

To do so, we introduce the chemical potential term $\mu$. We recognize that there is a gauge
freedom between $\mu$ and the diagonal terms of correlation potential $u$ in the mean field, i.e.,
\begin{equation}
	\mu^\prime=\mu+\phi,\thinspace u^\prime=u+\phi\sum_ia_i^{\dagger}a_i.
	\label{eq:dmet:theory:gauge}
\end{equation}
This gauge freedom is lost in the impurity model calculation with non-interacting bath, as one can see in
Eq.~\ref{eq:dmet:theory:ni} that the correlation potential is only added to the bath orbitals. Therefore,
one can vary $\mu$ and the diagonal of $u$ together following Eq.~\ref{eq:dmet:theory:gauge} so that the
mean-field solution (thus the bath orbitals) does not change, while the relative energy levels of the
fragment and the bath orbitals are shifted. Thus, we can perform chemical potential optimization while
solving the impurity model. The chemical potential obtained in this procedure is usually a good approximation
of the chemical potential in the physical sense. The details of a self-adaptive chemical potential fitting
algorithm, and how it is
built into the DMET iterations are explained in Appendix~\ref{sec:algo:chem_fit}.


\subsection{Expectation Values} \label{sec:dmet:theory:expectation}
In general, there are two types of expectation values of interest in DMET. Local observables,
such as occupations and local order parameters can be directly computed using the correlated
wavefunction $|\Psi\rangle$. These expectation values should be computed only if they are fully
within the fragment. Reciprocal space
observables, such as band structure and spin structure factors, cannot be formally defined for DMET
correlated wavefunction $|\Psi\rangle\otimes|\Psi_c\rangle$ because of the lack of translational
invariance. One could, however, obtain rough estimates from the associated lattice Slater determinants.
Other nonlocal observables, such as long-range correlation functions,
can be computed by taking expectation values of the locally correlated entire system wavefunction
$|\Psi\rangle\otimes|\Psi_c\rangle$. However, these observables will have a smooth transition from
the full many-body results to the mean-field values, and are seldom of practical use.

Another important expectation value is the energy per supercell (fragment). Note the DMET energy is
different from the impurity model energy
$E_\text{emb}=\langle \Psi|\hat{H}^{\text{(NI)}}_\text{emb}|\Psi\rangle $. We consider
the bipartite decomposition of the lattice Hamiltonian
\begin{equation}
	\hat{H}=H_{\text{frag}}+H_{\text{env}}+H_{\text{frag-env}}.
	\label{eq:dmet:theory:energy}
\end{equation}
Obviously, the fragment part $H_{\text{frag}}$ should be included and $H_{\text{env}}$ should not.
The coupling term $H_{\text{frag-env}}$ is split between the fragment and environment, by taking
a partial trace of the second-quantized terms, i.e.
\begin{equation}
	e_{\text{frag}}=\sum_{i\in\text{fragment},j}h_{ij}\rho_{ij}
	+\frac{1}{2}\sum_{i\in\text{fragment},jkl}(ik||jl)\Gamma_{ik,jl}
	\label{eq:dmet:theory:e_frag}
\end{equation}
where $\rho_{ij}$ and $\Gamma_{ik,jl}=
\langle \Psi|\langle\Psi_c|a_{i}^{\dagger}a_{j}^{\dagger}a_la_k|\Psi\rangle|\Psi_c\rangle$
are one- and two-body density matrices of the correlated wavefunction
$|\Psi\rangle\otimes|\Psi_c\rangle$. This is equivalent to a full trace,
with scaling factors equal to the fraction of the indices in the fragment. (For example,
for term $(ik||jl)\Gamma_{ik,jl}$ where indices $i,k,l$ are in the fragment, the scaling
factor is $3/4$.) This is an intuitive way to understand the decomposition of the coupling energy.

\section{The Broken Particle-Number Symmetry Formalism} \label{sec:dmet:bcs}
In this section, we extend the generic DMET formulation to broken particle-number
symmetry problem. They correspond to systems with superconductivity, where the
pairing order parameters $\langle a_{i\alpha}^{\dagger}a_{j\beta}^{\dagger}\rangle $
are non-zero (The $\alpha$ and $\beta$ are spin labels. We consider
singlet pairing only.)

Historically, superconducting wavefunctions with broken particle-number symmetry
were first obtained in the mean-field solution of
fermion systems with effective attractive interactions, induced by electron-phonon coupling
~\cite{bardeen1957microscopic,bardeen1957theory}.
The size of the broken particle-number symmetry, measured by the magnitudes of pairing
terms, describes the existence and robustness of the superconductivity.
In pure electronic systems, such as the high-temperature superconductors, there are no
attractive terms between electrons, and mean-field solutions never spontaneously
break the particle number symmetry, although the exact solution in the thermodynamic limit
has long-range pairing correlations. In DMET, however, we can enable pairing
in the correlation potential, which is determined self-consistently,
and it turns out the superconducting solutions can be stabilized spontaneously.

In Sec.~\ref{sec:dmet:bcs:mf} and ~\ref{sec:dmet:bcs:embedding}, we introduce the
mean-field and embedding aspects of DMET in the presence of pairing. In our derivation,
the spin-unrestricted formulation is used.
In Sec.~\ref{sec:dmet:bcs:symm}, we discuss why this formulation can give
correct results of superconductivity.

\subsection{The BCS Wavefunction and BdG Equation} \label{sec:dmet:bcs:mf}
The spin-unrestricted Bogoliubov transformation is defined as
\begin{equation}
	\begin{split}
		c_{p\alpha}^{\dagger}&=u^\alpha_{ip}a_{i\alpha}^{\dagger}+v^\beta_{iq}a_{i\beta}\\
		c_{p\beta}^{\dagger}&=u^{\beta}_{ip}a_{i\beta}^{\dagger}+v^{\alpha}_{iq}a_{i\alpha}
	\end{split}
	\label{eq:dmet:bcs:bogoliubov}
\end{equation}
where $c_{p\sigma}^{\dagger}$ creates a \textit{quasiparticle} of spin $\sigma$,
$a_{i\sigma}^{\dagger}$ creates an electron of spin $\sigma$ in orbital $i$. For simplicity, we assume the coefficients are all real numbers. The quasiparticles
defined here have good $S_z$ quantum numbers, but the particle number symmetry is broken.
This is the so-called \textit{singlet pairing} that preserves $S_z$ symmetry, and is usually the case in
both conventional and high-temperature superconductors.

The transformation can also be written in matrix form
\begin{equation}
	\begin{bmatrix}
		c_\alpha^{\dagger} & c_\beta
	\end{bmatrix}=
	\begin{bmatrix}
		a_\alpha^{\dagger} & a_\beta
	\end{bmatrix}
	\begin{bmatrix}
		U_\alpha& V_\alpha\\
		V_\beta & U_\beta
	\end{bmatrix}=	\begin{bmatrix}
		a_\alpha^{\dagger} & a_\beta
	\end{bmatrix}S.
	\label{eq:dmet:bcs:bogoliubov_matrix}
\end{equation}
A complete set of Bogoliubov transformation has $n$ orthonormal
quasiparticles for each spin. In this case, the matrix $S$ is unitary
(one can verify it by checking the anticommutators), therefore
\begin{equation}
	U_\sigma U_\sigma^{\dagger}+V_\sigma V_\sigma^{\dagger}=I
	,\thinspace
	V_\sigma U_{\bar\sigma}^{\dagger}+U_\sigma V_{\bar\sigma}^{\dagger}=0
	\label{eq:dmet:bcs:uv_condition}
\end{equation}
and the inverse Bogoliubov transformation
\begin{equation}
	\begin{bmatrix}
		a_\alpha^{\dagger} & a_\beta
	\end{bmatrix}=
	\begin{bmatrix}
		c_\alpha^{\dagger} & c_\beta
	\end{bmatrix}S^{\dagger}=
	\begin{bmatrix}
		c_\alpha^{\dagger} & c_\beta
	\end{bmatrix}
	\begin{bmatrix}
		U_\alpha^{\dagger}&V_\beta^{\dagger}\\
		V_\alpha^{\dagger}&U_\beta^{\dagger}
	\end{bmatrix}.
	\label{eq:dmet:bcs:bogoliubov_inverse}
\end{equation}

A BCS wavefunction $|\psi\rangle$ are defined as the vacuum of a complete set of Bogiliubov quasiparticles, i.e.,
\begin{equation}
	c_{p\sigma}|\psi\rangle=0.
	\label{eq:dmet:bcs:bcs_wfn}
\end{equation}
Like the Slater determinants, the observables of BCS wavefunctions are also determined entirely
by its one-body density matrices. Besides the standard one-body density matrices, BCS wavefunctions
have non-zero \textit{pairing density matrices}
$\kappa^{\alpha\beta}_{ij}=\langle \psi|a_{i\alpha}a_{j\beta}|\psi\rangle $. The density matrices
of BCS wavefunctions can be evaluated using Eq.~\ref{eq:dmet:bcs:bogoliubov_inverse} and
~\ref{eq:dmet:bcs:bcs_wfn}, thus
\begin{equation}
	\begin{split}
	G=&
	\begin{bmatrix}
		\rho_\alpha&\kappa_{\beta\alpha}^{\dagger}\\
		\kappa_{\beta\alpha}&I-\rho_\beta^T
	\end{bmatrix}
	=\langle\psi|
	\begin{bmatrix}
		a_\alpha^{\dagger} \\ a_\beta
	\end{bmatrix}
	\begin{bmatrix}
		a_\alpha&a_\beta^{\dagger}
	\end{bmatrix}|\psi\rangle\\
	=&
	\langle \psi|S
	\begin{bmatrix}
		c_\alpha^{\dagger} \\ c_\beta
	\end{bmatrix}
	\begin{bmatrix}
		c_\alpha&c_\beta^{\dagger}
	\end{bmatrix}
	S^{\dagger}|\psi\rangle
	=
	S
	\begin{bmatrix}
		0&0\\
		0&1
	\end{bmatrix}S^{\dagger}=
	\begin{bmatrix}
		V_\alpha V_\alpha^{\dagger} & V_\alpha U_\beta^{\dagger}\\
		U_\beta V_\alpha^{\dagger}  & U_\beta U_\beta^{\dagger}
	\end{bmatrix}
	\end{split}
	\label{eq:dmet:bcs:generalized_rdm}
\end{equation}
i.e.,
\begin{equation}
	\rho_\sigma=V_\sigma V_\sigma^{\dagger},
	\thinspace\kappa_{\beta\alpha}=-\kappa_{\alpha\beta}^T=U_\beta V_\alpha^{\dagger}.
	\label{eq:dmet:bcs:rdms}
\end{equation}
$G$ is the \textit{generalized one-body density matrix} of $|\psi\rangle$, as
it contains the information of all the channels of the one-body density matrices.


The BCS wavefunctions are the ground state solutions of non-interacting Hamiltonians with pairing
terms
\begin{equation}
	\hat{H}=h_{ij\sigma}a_{i\sigma}^{\dagger}a_{j\sigma}+
	\Delta_{ij}a_{i\alpha}^{\dagger}a_{j\beta}^{\dagger}+h.c..
	\label{eq:dmet:bcs:bcs_mf}
\end{equation}
It is also the mean-field solution of an interacting Hamiltonian, with the assumption
that $\langle a_{i\alpha}a_{j\beta}\rangle \neq 0$. In this case, the two-body terms
are approximated as
\begin{equation}
	\langle a_{i\mu}^{\dagger}a_{j\nu}^{\dagger}a_{l\nu}a_{k\mu}\rangle\approx
	\langle a_{i\mu}^{\dagger}a_{k\mu}\rangle \langle a_{j\nu}^{\dagger}a_{l\nu}\rangle
	-\delta_{\mu\nu}\langle a_{i\mu}^{\dagger}a_{l\mu}\rangle
	\langle a_{j\mu}^{\dagger}a_{k\mu}\rangle
	+\delta_{\mu\bar\nu}\langle a_{i\mu}^{\dagger}a_{j\nu}^{\dagger}\rangle
	\langle a_{l\nu}a_{k\mu}\rangle
\end{equation}
where the first two terms are Coulomb and exchange terms in Hartree-Fock theory, while the last
term comes from pairing. This formulation is thus called Hartree-Fock-Bogoliubov (HFB) theory.
By forming the Fock matrix, HFB essentially solves for the ground state of an effective Hamiltonian
similar to that in Eq.~\ref{eq:dmet:bcs:bcs_mf} as well. The pairing density matrix becomes
non-zero for certain values of the two-body interactions.

In both cases, the problem reduces to finding the quasiparticles by solving the
Bogoliubov-de Gennes (BdG) equation~\cite{Gennes1966}, in analogous to the Hartree-Fock-Roothaan equation
in Hartree-Fock theory. Given the chemical potential $\mu$, the BdG equation is~\cite{Yamaki2004}
\begin{equation}
	\begin{bmatrix}
		h_\alpha-\mu I& \Delta\\
		\Delta^{\dagger}& -h_\beta+\mu I
	\end{bmatrix}
	\begin{bmatrix}
		U_\alpha& V_\alpha\\
		V_\beta&U_\beta
	\end{bmatrix}=
	\begin{bmatrix}
		U_\alpha& V_\alpha\\
		V_\beta&U_\beta
	\end{bmatrix}
	\begin{bmatrix}
		\varepsilon_\alpha&\\
		&-\varepsilon_\beta
	\end{bmatrix}
	\label{eq:dmet:bcs:bdg}
\end{equation}
where $\varepsilon_\alpha$ and $\varepsilon_\beta$ are all positive. The number of
$\{\varepsilon_\alpha\}$ equals to the number of $\{\varepsilon_\beta\}$ for a $S_z=0$ system.
One can also see from here why the ground state of Eq.~\ref{eq:dmet:bcs:bcs_mf}
is the quasiparticle vacuum, since the Hamiltonian is transformed to
\begin{equation}
	\hat{H}=E_0+\sum_{p\sigma}\varepsilon_{p\sigma}c_{p\sigma}^{\dagger}c_{p\sigma}
	\label{eq:dmet:bcs:diagonal}
\end{equation}
where $E_0=\langle \psi|\hat{H}|\psi\rangle $ is the ground state energy.

To apply the BCS formulation to DMET, we allow the correlation potential to
include pairing terms, i.e.
\begin{equation}
	u_C=\sum_{ij\sigma}h_{ij\sigma}a_{i\sigma}^{\dagger}a_{j\sigma}
	+\sum_{ij}\Delta_{ij}a_{i\alpha}^{\dagger}a_{j\beta}^{\dagger}+h.c.
	\label{eq:dmet:bcs:corr_pot}
\end{equation}
where $i,j$ go over all the sites in a fragment. The mean-field solution $|\psi\rangle$ now
becomes a BCS wavefunction. To obtain the correct mean-field solution, one needs to adjust the chemical potential
$\mu$ so that $|\psi\rangle$ gives the pre-defined number of electrons.

\subsection{Embedding the Quasiparticles} \label{sec:dmet:bcs:embedding}
In this section, we take the mean-field BCS wavefunction and discuss the formation of the bath
orbitals. At first look, the BCS wavefunction, defined as the quasiparticle vacuum, is not a
product state and thus not possible to perform the Schmidt decomposition analogous to Slater
determinants. As shown in Ref.~\cite{Zheng2016}, the Schmidt decomposition of a BCS wavefunction is equivalent to
finding an embedded quasiparticle active space. Here, we introduce a new approach
that converts the BCS wavefunction to a product state~\cite{Datta1999}, which results in similar mathematical
operations to the case of Slater determinants.

We start by defining a simple vacuum (background) $|\mathrm{vac}\rangle$ as a ferromagnetic state
where all the $n$ lattice sites are occupied by spin-down ($\beta$) electrons, and let the quasiparticles
\begin{equation}
	d_{p\alpha}^{\dagger}=c_{p\beta}.
\end{equation}
We have $d_{p\alpha}|\mathrm{vac}\rangle=0$ and $c_{p\alpha}|\mathrm{vac}\rangle=0$
since $d_{p\alpha}$ and $c_{p\alpha}$ bring down $S_z$ by $1/2$,
but $|{\mathrm{vac}}\rangle$ is already the lowest eigenstate of $S_z$.

We can thus write the BCS wavefunction as
\begin{equation}
	|\psi\rangle = \prod_pd_{p\alpha}^{\dagger}|\mathrm{vac}\rangle
	\label{eq:dmet:bcs:product}
\end{equation}
up to a phase, because the state on the right-hand side is the quasiparticle vacuum of $\{c_{p\sigma}\}$:
\begin{equation}
	\begin{split}
	c_{p\alpha}(\prod_qd_{q\alpha}^{\dagger}|\mathrm{vac}\rangle)
	&=(-)^n\prod_qc_{q\beta}c_{p\alpha}|\mathrm{vac}\rangle=0\\
	c_{p\beta}(\prod_qd_{q\alpha}^{\dagger}|\mathrm{vac}\rangle)
	&=c_{p\beta}\prod_qc_{q\beta}|\mathrm{vac}\rangle=0.
	\end{split}
	\label{eq:dmet:bcs:product_equiv}
\end{equation}

Since Eq.~\ref{eq:dmet:bcs:product} rewrites the BCS state into a product state
(the vacuum $|\mathrm{vac}\rangle$ is also a product state), we view it as a Slater determinant
with the occupied orbitals $\{d_{p\alpha}^{\dagger}\}$, and its one-body density matrix can be obtained
using the orbital coefficients of $\{d_{p\alpha}^{\dagger}\}$
\begin{equation}
	d_\alpha^{\dagger}=c_\beta=
	\begin{bmatrix}
		a_\alpha^{\dagger}&a_\beta
	\end{bmatrix}
	\begin{bmatrix}
		V_\alpha\\
		U_\beta
	\end{bmatrix}.
	\label{eq:dmet:bcs:d_rep}
\end{equation}
It thus turns out the density matrix of $\{d_{p\alpha}^{\dagger}\}$ is acutally
the generalized density matrix (Eq.~\ref{eq:dmet:bcs:generalized_rdm}) of $|\psi\rangle$.
The Schmidt decomposition of $|\psi\rangle$ is then
\begin{equation}
	|\psi\rangle=\prod_{k=1}^{2n_A}(p_kd_{A,k}^{\dagger}+q_kd_{B,k}^{\dagger})|\mathrm{vac}\rangle_A
	\otimes\prod_{l=2n_A+1}^nd_{B,l}^{\dagger}|\mathrm{vac}\rangle_B
  \label{eq:dmet:bcs:schmidt}
\end{equation}
where the notations are similar to those in Eq.~\ref{eq:dmet:theory:slater_rotated}. The $|\mathrm{vac}\rangle_A$
and $|\mathrm{vac}\rangle_B$ states are the ferromagnetic states of the fragment and the environment, respectively.
The bath and core orbitals $d_{B}^{\dagger}$ can be obtained by diagonalizing the environmental part of
the generalized density matrix $G_B$, similar to normal state DMET.

Note that the fragment and bath orbitals obtained
in this procedure are all of spin $\alpha$, and the core state has spin $-n_{A}$. (the factor $1/2$ from electron
spin is taken into account.) The impurity model is thus a spinless fermionic system of $4n_A$ sites
in which $2n_A$ are occupied. Alternatively, we can perform a partial particle-hole transformation on half
of the $4n_A$ modes, and make the core $S_z=0$. The impurity model then becomes a fermion problem of $2n_A$ (spatial)
orbitals without particle-number symmetry, while $S_z=0$. The latter is used in the thesis work, because
the pairing is usually a small effect in the problems of interest, thus the partial particle-hole transformation
results in a better physical picture, where all the quasiparticles become electron-like (i.e. $||v||>||u||$), and
fits better into available impurity solvers.

After the partial particle-hole transformation, the impurity model consists of $n_A$ fragment orbitals
$\{a_{i\sigma}^{\dagger}\}$ and $n_A$ bath orbitals for each spin
\begin{equation}
	d_{B,\alpha}^{\dagger}=a_\alpha^{\dagger}V_{B,\alpha}+a_\beta U_{B,\beta},\thinspace
	d_{B,\beta}^{\dagger}=a_\beta^{\dagger}V_{B,\beta}+a_\alpha U_{B,\alpha}.
	\label{eq:dmet:bcs:bath}
\end{equation}
And the core wavefunction becomes
\begin{equation}
	|\Psi_c\rangle=\prod_{k=1}^{n_A}d_{B,k\beta}\prod_{l=2n_A+1}^nd_{B,l}^{\dagger}|\mathrm{vac}\rangle_B.
	\label{eq:dmet:bcs:core}
\end{equation}
We discuss how to project the lattice Hamiltonian to the embedding basis in Appendix~\ref{sec:formula:bcs_integral}.

\subsection{Broken Symmetry DMET} \label{sec:dmet:bcs:symm}
We have discussed how to construct the impurity model with broken particle-number symmetry.
The correlation potential and chemical potential optimizations are similar to the case of the normal state,
although the math becomes more complicated (detailed in the Appendix). Here, we discuss the physical
side of broken symmetry DMET calculations, i.e.,
\begin{itemize}
	\item When and how does DMET give a broken symmetry solution?
	\item What controls the quality of order parameters obtained in DMET?
	\item How do the DMET orders connect to the exact solutions in the thermodynamic limit?
\end{itemize}

We use the example of pairing order to address these questions. By including pairing terms in the
correlation potential, we obtain a BCS mean-field solution. That gives an impurity model Hamiltonian
with non-zero pairing terms in the environmental degrees of freedom. This can be viewed as a
pinning field similar to those used in finite lattice calculations, although more complicated and structured.
The pinning field thus induces non-zero pairing orders in the fragment, the magnitude of which depends
on the susceptibility of the fragment.

In the correlation potential fitting
stage, we try to match the lattice wavefunctions to the correlated solutions, thus forcing the mean-field
order parameters to increase or decrease adaptively, and changing the magnitudes of the pinning field
and the order parameters in the correlated calculations. At convergence, the order parameters obtained in
the impurity model calculations thus become a good approximation of the actual long-range order.
The order parameters become exact in the limit of infinitely large fragments.

If the true ground state
of the entire lattice has a long-range order, the susceptibility is infinite and the symmetry is spontaneously
broken. In this case, DMET will also give a broken symmetry solution, which does not disappear as the size
of the fragment grows.
If, however, the order is short-ranged, the order parameters ultimately decay to zero as the
distance to the fragment boundaries grows. In practice, these order parameters are often zero even with
minimal fragment sizes thanks to DMET self-consistency.

In the case where long-range order does exist, one can obtain the exact order parameter by growing the fragment size
to infinity. Alternatively, we can compute the order parameters with several fragment sizes and extrapolate to the
thermodynamic limit. (It usually extrapolates to zero if the order is short-ranged.)
Thus, the accuracy of these order parameters are determined by the largest sizes we
can achieve in DMET calculations, as well as the function forms used for the extrapolation.

In summary, DMET coupled with finite size extrapolation is possible to exract order parameters
in the thermodynamic limit. In Sec.~\ref{sec:dmet:practice:extrapolation} and
Chapter~\ref{chpt:scaling}, we will introduce the theory and empirical studies of DMET fragment size
extrapolation.

\section{Impurity Solvers} \label{sec:dmet:solver}
Impurity solvers are used to solve the quantum many-body problem the impurity models,
either exactly or approximately. An advantage of DMET is that one can choose from
a wide range of solvers already developed in the quantum chemistry or computational
condensed matter communities, depending on the problems at hand. In this section, we
introduce the common impurity solvers, both for normal state calculations and
BCS calculations. We will focus on the principles and the appropriate contexts of
these solvers.

\subsection{Exact Diagonalization} \label{sec:dmet:solver:ed}
The most direct way is to diagonalize the many-body Hamiltonian and obtain the full
Fock space representation of the ground-state wavefunction.
We write the wavefunction as an expansion of the occupation strings (subject to
symmetries)
\begin{equation}
	|\Psi\rangle=\sum_{n_1,n_2,\cdots,n_K}C_{n_1,n_2,\cdots,n_K}
	|n_1,n_2,\cdots,n_K\rangle
	\label{eq:dmet:solver:exact}
\end{equation}
where $n_k\in\{|0\rangle, |\alpha\rangle, |\beta\rangle, |\alpha\beta\rangle\}$
is the occupation of orbital $k$. In exact diagonalization (ED),
We minimize the total energy and find the optimal coefficients $C_{n_1,n_2,\cdots,n_K}$.
The method is also called \textit{full configurational interaction} (FCI) because of the Fock space expansion.
The cost of ED scales exponentially with the system size, as we can see, the number of parameters
in the giant tensor $C_{n_1,n_2,\cdots,n_K}$ scales $O(4^K)$.
In addition, naive eigendecomposition requires writing down the Hamiltonian matrix in this basis,
and then perform the diagonalization routine, which takes $O(N^2)$ space and $O(N^3)$ time. The computational
cost soon becomes formidable with about merely 8 to 10 orbitals.

We can save some memory and computational cost using the Lanczos or
Davidson algorithms~\cite{lanczos1950iteration,davidson1975iterative} to compute only the lowest one
or several eigenstates without explicitly writing down the Hamiltonian matrix. This
is often referred as the \textit{direct CI}~\cite{saunders1983direct,knowles1984new}.
The Davidson algorithm is described in Appendix ~\ref{sec:algo:davidson} in detail. The most
time-consuming step in direct CI is to compute $\hat{H}|\Psi\rangle$, which is achieved by
taking the each individual term in $\hat{H}$ and acting on each occupation string in
the expansion of $|\Psi\rangle$. This, in fact, uses the sparsity of the Hamiltonian matrix
and results in roughly $O(K^4 4^K)$ time complexity (terms in $\hat{H}$ $\times$ basis
in $|\Psi\rangle$).

Due to the exponential complexity, even with Davidson algorithm, ED (or FCI) is able to treat
only up to $\sim20$ orbitals with current computational power. It is of limited uses as a DMET
impurity solver, but can serve as the benchmark for other approximate impurity solvers.

\subsection{Density Matrix Renormalization Group}\label{sec:dmet:solver:dmrg}
The density matrix renormalization group (DMRG)~\cite{white1992density,white1993density,Schollwok2011} approximates the exact wavefunction coefficients
Eq.~\ref{eq:dmet:solver:exact} as the a tensor product
\begin{equation}
	C_{n_1,n_2,\dots,n_K}=\sum_{i_1,\dots,i_K}A_{i_1}^{n_1}A_{i_1,i_2}^{n_2}\dots A_{i_{K-1}}^{n_K}.
	\label{eq:dmet:solver:mps}
\end{equation}
The representation is called the \textit{matrix product states} (MPS) because each $A^{n_k}$
is a matrix. The uncontracted indices $\{n_k\}$ are called the \textit{physical indices}, while $\{i_k\}$
are called the \textit{auxiliary indices}.

Eq.~\ref{eq:dmet:solver:mps} is exact if we allow the dimensions of auxiliary indices $i_k$ to reach
$4^{[K/2]}$, though in this case, the number of parameters scales exponentially. In practice, the
dimension is restricted to have the maximum \textit{bond dimension} $M$. The expression thus
becomes a variational ansatz, whose accuracy is a monotonically increasing function of $M$.

DMRG is an efficient algorithm to optimize the MPS using the sweep algorithm, i.e., the tenors
$\{A^{n_k}\}$ are updated one at a time, from left to right (forward) and then from right to left
(backward), until the energy is converged.
In each step during the sweep, one projects the Hamiltonian to a renormalized subspace around orbital $k$
and solve for the ground state. The subspace wavefunction is then compressed to keep the bond dimension $M$
unchanged during the calculation.

Specifically, when optimizing tensor $A^{n_k}$ in the forward sweep, the Hamiltonian is projected onto the basis
\begin{equation}
	|\psi_{i_{k-1}, n_k, n_{k+1}, i_{k+1}}\rangle=
	|L_{i_{k-1}}\rangle\otimes|n_k n_{k+1}\rangle\otimes|R_{i_{k+1}}\rangle
	\label{eq:dmet:solver:dmrg_basis}
\end{equation}
where
\begin{equation}
	\begin{split}
		|L_{i_{k-1}}\rangle=&
		\sum_{i_1,\dots,i_{k-2}}A_{i_1}^{n_1}\dots A_{i_{k-2},i_{k-1}}^{n_{k-1}}|n_1,\dots,n_{k-1}\rangle\\
		|R_{i_{k+1}}\rangle=&
		\sum_{i_{k+2},\dots,i_{K-1}}A_{i_{k+1},i_{k+2}}^{n_{k+2}}\dots A_{i_{K-1}}^{n_k}|n_{k+1},\dots,n_{K}\rangle
	\end{split}
	\label{eq:dmet:solver:dmrg_lrbasis}
\end{equation}
are the left and right renormalized basis, respectively. Thus, the renormalized Hamiltonian is $H^{(k)}=P^k\hat{H}P^k$,
where the projector
	\begin{equation}
		P^k=
		\sum_{i_{k-1},n_k,n_{k+1},i_{k+1}}|\psi_{i_{k-1}, n_k, n_{k+1}, i_{k+1}}\rangle
		\langle\psi_{i_{k-1}, n_k, n_{k+1}, i_{k+1}}|.
	\end{equation}
The process of building the renormalized basis and the subspace Hamiltonian is called \textit{blocking}.
We then solve the Schr\"odinger equation in the renormalized subspace
\begin{equation}
	(H^{(k)})_{i_{k-1},i_k;i_{k-1}^\prime,i_k^\prime}^{n_k,n_{k+1};n_k^\prime,n_{k+1}^\prime}C_{i_{k-1}^\prime,i_k^\prime}^{n_k^\prime,n_{k+1}^\prime}=E^{(k)}C_{i_{k-1},i_k}^{n_k,n_{k+1}}.
	\label{eq:dmet:solver:dmrg_e}
\end{equation}
The corresponding wavefunction
$|\Psi^{(k)}\rangle=\sum_{i_{k-1},i_k,n_k,n_{k+1}}C_{i_{k-1},i_k}^{n_k,n_{k+1}}
|\psi_{i_{k-1}, n_k, n_{k+1}, i_{k+1}}\rangle$ is an approximation of the ground state wavefunction, while
$E^{(k)}$ gives an upper bound for the ground state energy.

To obtain $A^{n_k}$,we perform the singular value decomposition (SVD) on the subspace wavefunction
\begin{equation}
	C_{i_{k-1},i_k}^{n_k,n_{k+1}}=\lambda_sU_{i_{k-1},s}^{n_k}V_{i_k,s}^{n_{k+1}}.
	\label{eq:dmet:solver:dmrg_svd}
\end{equation}
The left tensor $U_{i_{k-1},s}^{n_k}$ thus corresponds to $A^{n_k}$. However, since the dimensions of $i_{k-1}$
and $i_k$ are $M$, and the dimensions of $n_k$ and $n_{k+1}$ are $4$, the dimension of $s$ is up to $4M$. To keep the
bond dimension unchanged, we need to discard all but the largest $M$ singular values.
\footnote{Note that the truncation
is meaningful only when the left and right renormalized basis are orthonormal. This is possible through the
\textit{canonicalization} of the MPS tensors. Readers can refer to, e.g., Ref.~\cite{Schollwok2011} for an in-depth
discussion.} The kept states, associated with the largest singular values,
have the strongest entanglement with the right part of the system.
The total norm of the discarded singular values, referred as the
\textit{discarded weight}, and is an important indicator to quantify the accuracy of DMRG calculations, and can be
used to extrapolate to $M=\infty$ limit. The process of renormalizing and truncating the left basis is called
\textit{decimation}. Note that in each blocking and decimation step, although two physical sites are explicitly
involved, only the tensor on one of them is updated.
After updating $A^{n_k}$, we continue to build the new left and right renormalized basis $\{|L_{i_{k}}\rangle\}$ and
$|R_{i_{k+2}}\rangle$, and optimize the next tensor $A^{n_{k+1}}$. The backward sweep is similar but the
direction is reversed.

The algorithm we discuss here is often referred as the \textit{two-dot} algorithm since there are two orbitals involved
in the block and decimation step. There exists the \textit{one-dot} algorithm in which the renormalized basis
$\{|R_{i_{k}}\rangle\}$ is used instead of $\{|n_{k+1} R_{i_{k+1}}\rangle\}$. However, there is no truncation
in the one-dot algorithm because the number of non-zero singular values is only up to $M$.
Thus the wavefunction is not optimized at all during the one-dot sweeps. In a DMRG calculations, one usually
use the two-dot algorithm to optimize the wavefunction and then perform the one-dot sweeps to compute observables
such as the density matrices.

When optimally implemented, the scaling of DMRG is $O(M^2K^4)+O(M^3K^3)$ per sweep for general Hamiltonians including
dense two-body interaction terms~\cite{chan2002highly}. Given $M$, the computational error depends on the nature of the system, and is
guaranteed to decrease with increasing $M$. Besides, when the discarded weight is small enough, the energy and
density matrices have a linear relationship with the discarded weight. One can thus perform
DMRG calculations for several different bond dimensions $M$, and extrapolate to the $M=\infty$ limit (i.e. the exact
ground state) where the discarded weight is zero.

We notice the MPS ansatz in Eq.~\ref{eq:dmet:solver:mps} requires ordering the orbitals.
As a result, the performance of DMRG is system dependent and orbital-order dependent.
In principle, the order can be arbitrary, but DMRG works best when the entanglement
between the bipartitions of the ordered orbitals is minimized. For instance, gapped one-dimensional
physical systems usually require the same bond dimension $M$ for any length, resulting in a
cheap $O(L^4)$ scaling for obtaining near-exact ground-state solutions.
In general, given the Hamiltonian and the orbital ordering, for constant accuracy,
$M$ scales exponentially with the entanglement entropy between the bipartitions of the ordered orbitals. For
ground-state wavefunctions, the entanglement entropy often follows the \textit{area law}, i.e., being proportional
to the contact area between the two parts. In the example of 1D systems, the entanglement entropy is
thus a constant, and results in constant $M$. For general quantum chemistry Hamiltonians and orbitals,
one can manually or use algorithms to find a reasonable order by looking at the interaction integrals between
orbitals, thus minimizing the computational cost of DMRG.

In DMET calculations, we work with general quantum chemistry Hamiltonians in the impurity model.
We perform the calculations in the so-call
\textit{split localized} orbitals~\cite{bytautas2003split}
(or the quasiparticles in the electron-like representation,
as discussed in Sec.~\ref{sec:dmet:bcs:embedding})
which are ordered using a genetic algorithm~\cite{olivares2015ab}.
The implementation is adapted from the BLOCK quantum chemistry DMRG package~\cite{Chan2011,sharma2012spin},
with the addition of broken particle-number symmetry and various performance improvements.
Because of the ability to treat up to $\sim60$ orbitals accurately within reasonable
computational effort, DMRG is the main impurity solver used in the thesis work.

\subsection{Auxliary Field Quantum Monte Carlo}\label{sec:dmet:solver:afqmc}
Auxiliary field quantum Monte Carlo (AFQMC)~\cite{Sugiyama1986,Zhang2013,Shi2015,Shi2016} is a stochastic method to obtain the ground state of a fermion
Hamiltonian~\cite{sorella1989novel}. It performs the imaginary time evolution of a trial wavefunction
\begin{equation}
	|\Psi_0\rangle \propto \lim_{\beta\rightarrow\infty}e^{-\beta \hat{H}}|\Psi_T\rangle.
  \label{eq:dmet:solver:afqmc_evolve}
\end{equation}
The time evolution is carried out using the second-order Trotter-Suzuki decomposition,
\begin{equation}
e^{-\beta \hat{H}}=(e^{-\tau H})^n=(e^{-\frac{\tau}{2} H_1}e^{-\tau H_2}e^{-\frac{\tau}{2} H_1})^n+O(\beta\tau^2)
\end{equation}
where $H_1$ and $H_2$ are the one- and two-body parts of the Hamiltonian.

Given any Slater determinant 
$|\Psi\rangle=|\phi_{1\alpha}\dots\phi_{N\alpha}\rangle\otimes|\phi_{1\beta}\dots\phi_{N\beta}\rangle$
and any one-body operator
\begin{equation}
	K=\sum_{ij\sigma}k_{ij\sigma}a_{i\sigma}^{\dagger}a_{j\sigma},
\end{equation}
the canonical transformation $e^{K}|\Psi\rangle$ can be carried out exactly, giving another Slater determinant
$|\Psi^\prime\rangle =e^{K}|\Psi\rangle=
|\phi_{1\alpha}^\prime\dots\phi_{N\alpha}^\prime\rangle\otimes
|\phi_{1\beta}^\prime\dots\phi_{N\beta}^\prime\rangle$ with 
the coefficient matrices
\begin{equation}
  \Phi_\sigma^\prime=\left(
  \phi_{1\sigma}^\prime,\dots,\phi_{N\sigma}^\prime
  \right)=e^{k_\sigma}\Phi_\sigma.
  \label{eq:dmet:solver:afqmc_boost}
\end{equation}
The matrix multiplication in Eq.~\ref{eq:dmet:solver:afqmc_boost} gives the $O(N^3)$
scaling of the AFQMC algorithm (where $N$ is the number of electrons per spin). Starting with a Slater determinant
trial wavefunction $|\Psi_T\rangle$, the propagation of the one-body Hamiltonian can be computed using
Eq.~\ref{eq:dmet:solver:afqmc_boost}, by letting $K=-\frac{\tau}{2}H_1$.

The propagation of the two-body part of the Hamiltonian is rewritten
as an multidimensional integral over one-body propagations using a Hubbard-Stratonovich (HS)
transformation~\cite{stratonovich1957method}
\begin{equation}
	e^{-\tau \lambda \hat{v}^2}=\int_{-\infty}^\infty dx
	\frac{e^{-x^2/2}}{\sqrt{2\pi}}e^{x\sqrt{-2\tau\lambda}\hat{v}}
	\label{eq:dmet:solver:afqmc_hs}
\end{equation}
by writing the two-body Hamiltonian as the summation of squares, using the Hermitian
symmetry of the two-body integrals.
We diagonalize the Hermitian matrix
$V_{ik,lj}=(ik||jl)=\sum_{\gamma}R_{ik,\gamma}\lambda_\gamma R_{lj,\gamma}^*$.
(For simplicity, we use the spinless notation.) Thus,
\begin{equation}
	\begin{split}
		H_2=&\frac{1}{2}\sum_{ijkl}(ik||jl)a_{i}^{\dagger}a_{j}^{\dagger}a_{l}a_{k}\\
		=&\frac{1}{2}\sum_{ijkl}(ik||jl)a_{i}^{\dagger}a_{k}a_{j}^{\dagger}a_{l}
		-\frac{1}{2}\sum_{ijkl}(ik||jl)c_i^{\dagger}c_l\delta_{jk}\\
		=&\frac{1}{2}\sum_{\gamma}\lambda_\gamma
		(\sum_{ik}R_{ik,\gamma}a_{i}^{\dagger}a_{k})
		(\sum_{jl}R_{lj,\gamma}^*a_{j}^{\dagger}a_{l})
		-\frac{1}{2}\sum_{il}(\sum_j(ij||jl))c_i^{\dagger}c_l.
	\end{split}
	\label{eq:dmet:solver:afqmc_h2}
\end{equation}
Since $H_2$ is Hermitian, we can symmetrize the expression as
\begin{equation}
	H_2
	=\frac{1}{4}\sum_\gamma \lambda_\gamma\{\hat\rho_\gamma, \hat\rho_\gamma^{\dagger}\}+\hat{\rho}_0
	\label{eq:dmet:solver:afqmc_h2_symm}
\end{equation}
where $\rho_\gamma=\sum_{ik}R_{ik,\gamma}a_{i}^{\dagger}a_{k}$ and
$\hat{\rho}_0=-\frac{1}{4}\sum_{il}(\sum_j(ij||jl)+(ji||lj))c_i^{\dagger}c_l$. We thus have
\begin{equation}
	\{\hat\rho_\gamma, \hat\rho_\gamma^{\dagger}\}
	=\frac{1}{2}[(\hat\rho_\gamma+\hat\rho_\gamma^{\dagger})^2-(\hat\rho_\gamma-\hat\rho_\gamma^{\dagger})^2]
	=\frac{1}{2}[\hat\rho_{\gamma+}^2-\hat\rho_{\gamma-}^2].
	\label{eq:dmet:solver:afqmc_h2_sqaure}
\end{equation}
And the time evolution of $H_2$ becomes
\begin{equation}
	\begin{split}
	e^{-\tau H_2}=&e^{-\tau \hat\rho_0}\prod_{\gamma}\int_{-\infty}^\infty 
	dx_{\gamma+}dx_{\gamma-}\frac{e^{-(x_{\gamma+}^2+x_{\gamma-}^2)/2}}{2\pi}
	e^{x_{\gamma+}\sqrt{-\frac{1}{4}\tau\lambda_\gamma}\hat\rho_{\gamma+}+
	x_{\gamma-}\sqrt{\frac{1}{4}\tau\lambda_\gamma}\hat\rho_{\gamma-}}\\
	=&e^{-V_0+\sum_\gamma [V(x_{\gamma+})+V(x_{\gamma-})]}.
	\end{split}
	\label{eq:dmet:solver:afqmc_h2_evolve}
\end{equation}
Unfortunately, this integral cannot be evaluated analytically.
The \textit{auxiliary field} $\{x_{\gamma\pm}\}$ is sampled at each time slice to obtain a stochastic
representation of the propagation,  and thus of the ground state wavefunction
$|\Psi_0\rangle$ as a sum of walkers.

In the thesis work, we use AFQMC as the impurity solver only for the half-filled
Hubbard model. In this case, a simplified  discrete form of HS transformation exists
\begin{equation}
\begin{split}
  e^{-\tau Un_{i\alpha}n_{i\beta}}&=e^{-\tau U(n_{i\alpha}+n_{i\beta})/2}
  \sum_{x_i=\pm1}\frac{1}{2}e^{\gamma x_i(n_{i\alpha}-n_{i\beta})}\notag\\
  &=\sum_{x_i=\pm1} e^{ V(x_i,\tau)}
\end{split}
  \label{eq:dmet:solver:afqmc_U}
\end{equation}
where $x_i$ is a binary auxiliary field, and $\cosh{\gamma}=\exp(-\tau U/2)$.
Eq.~\ref{eq:dmet:solver:afqmc_U} is often termed \textit{spin decomposition}, in contrast to another
possible formed called \textit{charge decomposition}. The choice of different transformations does affect
the accuracy and efficiency of AFQMC calculations~\cite{Shi2013}.

In the thesis work, observables are calculated from the pure estimators, where the summations are similarly sampled,

{
\tiny
  \begin{equation}
    \langle \hat{O}\rangle =\lim_{n\rightarrow \infty}\frac{\sum_{\vec{x}_1} \dots \sum_{\vec{x}_n} \sum_{\vec{x}_1^\prime}\dots \sum_{\vec{x}_n^\prime}
    \langle \Psi_T|\prod_{j^\prime=1}^n(e^{-\frac{\tau}{2} H_1}e^{-\hat{{V}}(\vec{x}_{j^\prime}^\prime,\tau)}e^{-\frac{\tau}{2} H_1})\hat{O}
    \prod_{j=1}^n(e^{-\frac{\tau}{2} H_1}e^{-\hat{{V}}(\vec{x}_j,\tau)}e^{-\frac{\tau}{2} H_1})|\Psi_T\rangle }
    {\sum_{\vec{x}_1} \dots \sum_{\vec{x}_n} \sum_{\vec{x}_1^\prime}\dots \sum_{\vec{x}_n^\prime}
    \langle \Psi_T|\prod_{j^\prime=1}^n(e^{-\frac{\tau}{2} H_1}e^{-\hat{{V}}(\vec{x}_{j^\prime}^\prime,\tau)}e^{-\frac{\tau}{2} H_1})
    \prod_{j=1}^n(e^{-\frac{\tau}{2} H_1}e^{-\hat{{V}}(\vec{x}_j,\tau)}e^{-\frac{\tau}{2} H_1})|\Psi_T\rangle }
  \label{eq:dmet:solver:afqmc_ops}
\end{equation}
}

\noindent where, in the Hubbard case, $\hat{{V}}(\vec{x}, \tau)=\sum_{i=1}^N V(x_{i}, \tau)$.
The energy may be computed using a simpler estimator (the mixed estimator) where the
propagation of the bra is omitted.

The fermion sign problem arises because the individual terms in the denominator of
Eq.~\ref{eq:dmet:solver:afqmc_ops} can be both positive and negative (or complex in the
case of general two-body interaction) and
lead to a vanishing average with infinite variance. When there is a sign problem,
a constrained path approximation can be invoked in the calculation which removes
the problem with a gauge condition using a trial wavefunction
~\cite{PhysRevLett.90.136401,PhysRevB.55.7464,Zhang1995}.
In certain models, however, such as the half-filled repulsive Hubbard model
on a bipartite lattice, the sign problem
does not arise because the overlap between every walker and the trial
wavefunction is guaranteed non-negative.
It turns out that, in these models, the DMET impurity model Hamiltonian
is also sign-problem free as long as  certain constraints are enforced on the correlation potential.
For the half-filled Hubbard model on a bipartite lattice, the condition is
\begin{equation}
  u_{ij,\alpha}+(-)^{i+j}u_{ij,\beta}=\delta_{ij}U
  \label{eq:dmet:solver:afqmc_phsymm}
\end{equation}
where the parity term $(-)^{i+j}$ takes opposite signs for the two sublattices.
The derivation of this constraint is given in Appendix~\ref{sec:formula:sign}.

In cases where the sign problem is absent, AFQMC is an excellent impurity solver which obtains
the exact ground state (within statistical error bounds) with $O(N^3)$ computational cost. The
algorithm can also be massively parallelized, leading to fast solutions even for large fragments.
When the sign problem does exist, AFQMC is still a powerful solver, although the results
suffer from the uncontrolled constrained phase (CP) error.

\subsection{Complete Active Space Methods}\label{sec:dmet:solver:cas}
The complete active space (CAS) methods are widely used in quantum chemistry to study systems with
static (strong) correlation. The idea is to choose a subset of strongly correlated orbitals as the active space,
and perform ED in the active space, while other orbitals are treated at the mean-field level.
The wavefunction ansatz in CAS calculations is
\begin{equation}
	|\Psi\rangle=\sum_{n_1,\dots,n_K\in\{0,1\}, n_1+\dots+n_K=N_a}
	C_{n_1,\dots,n_K} c_{1}^{\dagger}\dots c_{N-N_a}^{\dagger}
	(c_{N-N_a+1}^{\dagger})^{n_1}(c_{N-N_a+K}^{\dagger})^{n_K}|0\rangle
	\label{eq:dmet:solver:cas:ansatz}
\end{equation}
where $N_a$ is the number of electrons in the active space, $i = 1,\cdots, N-N_a)$ are core (occupied) orbitals label,
$i = N-N_a+1,\cdots, N-N_a+K)$ are active orbitals labels and $i = N-N_a+K, \cdots, N$ are virtual (unoccupied)
orbitals.

One can perform a one-shot CAS calculation to obtain the active space coefficients $C_{n_1,\dots,n_K}$, using
orbital coefficients from, e.g., Hartree-Fock calculations. This is called the \textit{complete active
space configurational interaction} (CASCI). To further improve the accuracy, one can optimize the orbitals
as well as the CI coefficients, which is called \textit{complete active
space self-consistent field} (CASSCF)~\cite{roos1980complete,roos1987complete}.

In DMET calculations where not all the orbitals in the impurity model are strongly correlated, such as
when dealing with \textit{ab initio} Hamiltonians, one can
choose to use CASCI or CASSCF to reduce the computational cost. In addition to ED, CASCI/CASSCF can
also use DMRG or AFQMC to solve the active space problem. In the thesis work, an extension of CASCI and
CASSCF to BCS quasiparticle active space are implemented and applied. We follow the optimization scheme
outlined in Ref.~\cite{sun2017general}, and adapted from the CASSCF routine in
PySCF~\footnote{https://github.com/sunqm/pyscf}. The details of the formulation
are described in Appendix~\ref{sec:formula:casscf_bcs}.

\section{Practical Issues} \label{sec:dmet:practice}
In this section, we introduce several topics encountered in practice.
Sec.~\ref{sec:dmet:practice:convergence} explains the various numerical tricks used in DMET
to accelerate the convergence of correlation potential.
Sec.~\ref{sec:dmet:practice:dca} introduces the dynamical cluster formulation
of DMET to deal with the problem of broken translational symmetry, which is undesirable
in various applications. In Sec.~\ref{sec:dmet:practice:extrapolation} we analyze the fragment
size convergence of cluster DMET and the dynamical cluster formulation.

\subsection{Correlation Potential Convergence} \label{sec:dmet:practice:convergence}
\subsubsection{Direct Inversion in the Iterative Subspace}\label{sec:dmet:practice:convergence:diis}
Because we do not update the impurity model solutions when fitting the correlation potential,
the convergence can be rather slow in the late stage of DMET iterations. One scenario is,
while the correlation potential $u$ is optimized according to Eq.~\ref{eq:dmet:theory:cost_grad},
the correlated density matrix $\rho_\Psi(u)$ and the mean-field density matrix $\rho_\psi(u)$
move in roughly the same direction, but since $\rho_\Psi(u)$ is not updated until next DMET cycle,
the step sizes we take are much smaller than the optimal.
Another scenario is that we take too large step sizes in changing
$\rho_\psi(u)$ when $\rho_\Psi(u)$ and $\rho_\psi(u)$ move in opposite directions.
Of course many situations are not so well defined, but from these examples one sees how the
convergence problem arises in DMET correlation potential optimization.

These self-consistent field problems in have been known to quantum chemists for decades. One
classical method to accelerate convergence is the
\textit{direct inversion in the iterative subspace} (DIIS) algorithm
~\cite{pulay1980convergence,pulay1982improved,csaszar1984geometry}, which tries to predict
the optimum using information from the last few iterations.

In the context of DMET correlation potential fitting, after obtaining a set of trial correlation potentials
$\{u^{(i)}\}$ in previous cycles, we can define the \textit{residual vector} associated with
$u^{(i)}$ as
\begin{equation}
	\Delta u^{(i)} = u^{(i+1)} - u^{(i)}.
	\label{eq:dmet:practical:diis_residual}
\end{equation}
We then approximate the optimal $u^*$ as a linear combination $\{u^{(i)}\}$,
i.e., $u=\sum_i c_iu^{(i)}$, where the coefficients minimize the norm of the residual
vector correspond to $u$, subject to the normalization condition
\begin{equation}
	c_i=\text{argmin}_{c_i} ||\Delta u||^2,\thinspace s.t.\thinspace \sum_i c_i=1
	\label{eq:dmet:practical:diis_min}
\end{equation}
where  $\Delta u= \sum_i c_i\Delta u^{(i)}$.
The minimization problem in Eq.~\ref{eq:dmet:practical:diis_min} can be solved using the Lagrangian multiplier
\begin{equation}
	\mathcal{L}=c^TBc - 2\lambda(\sum_i c_i - 1)
	\label{eq:dmet:practical:diis_l}
\end{equation}
where $B_{ij}=\langle \Delta u^{(i)},\Delta u^{(j)}\rangle $. Thus
\begin{equation*}
	\frac{\partial L}{\partial c_i}=2\sum_j B_{ij}c_j - 2\lambda=0,\thinspace
	\frac{\partial L}{\partial \lambda}=\sum_i c_i - 1 = 0
\end{equation*}
or in matrix form
\begin{equation}
	\begin{bmatrix}
		B& -\mathbbm{1}\\
		-\mathbbm{1}^T & 0
	\end{bmatrix}
	\begin{bmatrix}
		c\\
		\lambda
	\end{bmatrix}=
	\begin{bmatrix}
		0\\
		-1
	\end{bmatrix}
	\label{eq:dmet:practical:diis_eq}
\end{equation}
where $\mathbbm{1}$ is a $n\times 1$ vector filled with $1$. In practice, we turn on DIIS
when $\Delta u^{(i)}$ is smaller than a certain threshold, and limit the maximum dimension
of vector $c$, by kicking out the old trial vectors. When
chemical potential is also optimized, we use the joint vector $[u, \mu]$ in DIIS.

\subsubsection{Zero-Trace Condition}\label{sec:dmet:practice:convergence:zero_trace}
In broken particle-number symmetry calculations, another issue can arise if the number of
electrons in the lattice mean-field solution $|\psi\rangle$ that minimizes the cost function is different
from the target value.
When we are sufficiently close to converging the correlation potential, suppose
the correlation potential is $u$ and $u^\prime$ before and after the correlation potential
fitting, while the chemical potential is $\mu$.
Because the correlation potential is close to convergence, its elements barely change except
for those on the diagonal, which control the number of electrons. Thus we assume
$u^\prime\approx u+\delta I$, i.e., the diagonal been shifted by a constant.
Then in the next DMET iteration, we need to increase (decrease) $\mu$ to $\mu+\delta$ so that
the number of electrons of $|\psi\rangle$ equals to the target value
(note that before the fitting in the last DMET iteration, $|\psi\rangle$ has the correct
electron number, and that we have the gauge invariance Eq.~\ref{eq:dmet:theory:gauge}).
Thus, although $u$ and $\mu$ are updated in the last iteration, the mean-field wavefunction
stays the same; after solving the impurity model with chemical potential fitting, the correlation
potential and chemical potential will go back to $u$ and $\mu$!
This cycle will then go on forever, while the DMET solution is not improving at all.
Note that this is not a problem in normal state DMET calculations, because shifting the diagonal elements
of $u$ does not change the lattice mean-field solution.

In broken particle-number symmetry calculations,
one can simply add a constraint on the mean-field number of electrons in the correlation
potential fitting step. However, this constraint is highly non-linear in the parameter space,
making the optimization much harder. Instead, we simply require
$\text{Tr}(\Delta u)=0$ in the optimization (by projecting out the changes in this direction at
each step of the optimization), or even simpler, use
$\Delta u^\prime= \Delta u - \frac{1}{n}\text{Tr}(\Delta u)I$
(where $n$ is the dimension of the $u$) after converging the unconstrained optimization.
The zero-trace condition is turned on after a few initial iterations, which removes
redundant the degree of freedom between the diagonal of $u$ and $\mu$. This degree of freedom is
optimized only during the chemical potential fitting. This trick solves the problem above at an
extremely low cost.

\subsection[Intracluster Translational Symmetry]{Intracluster Translational Symmetry~\footnote{Based on work published in Phys. Rev. B \textbf{95}, 045103 (2017). Copyright 2017, American Physical Society.~\cite{zheng2017cluster}}} \label{sec:dmet:practice:dca}

\begin{figure}[htpb]
  \centering
  \subfigure[]{
    \includegraphics[width=\columnwidth]{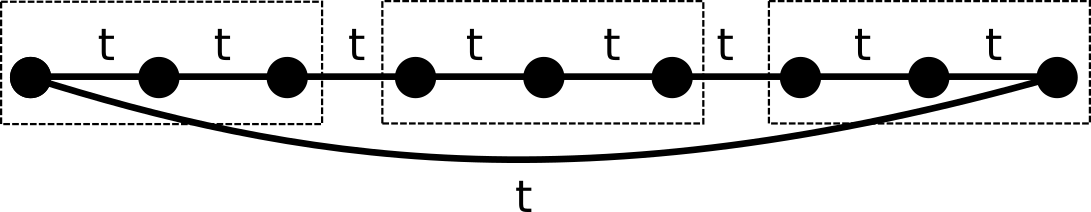}
    \label{fig:dmet:practical:lattice}
  }
  \subfigure[]{
    \includegraphics[width=0.4\columnwidth]{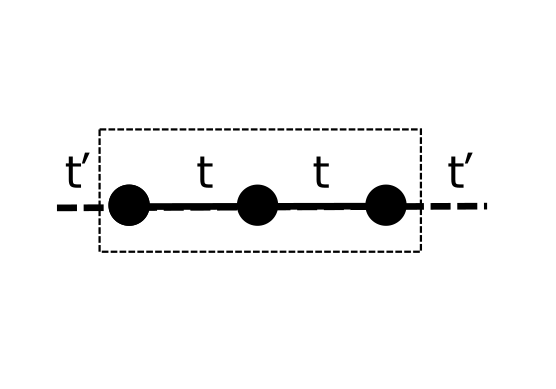}
    \label{fig:dmet:practical:cdmet}
  }
  \subfigure[]{
    \includegraphics[width=0.4\columnwidth]{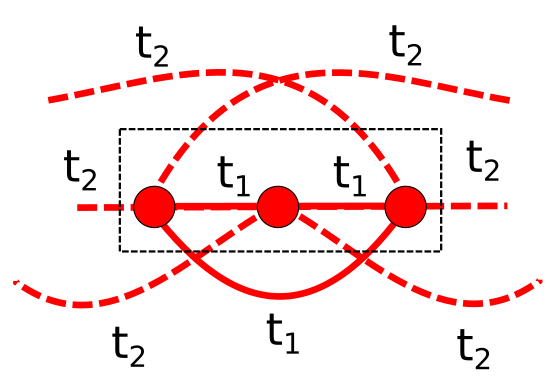}
    \label{fig::dmet:practical:dca}
  }
  \caption{Translational symmetry in DMET. (a) The original lattice with translational symmetry,
	  divided into 3 supercells. (b) The DMET fragment with broken intracluster
	  translational symmetry, between the central site and the edge sites.
    	(c) The DCA-DMET fragment restores the intracluster translational
	symmetry through a basis transformation and interaction coarse-graining.}
  \label{fig::dmet:practical:translational}
\end{figure}

Lattice DMET calculations with more than one orbitals in the fragment (referred as cluster DMET, or CDMET)
suffers from broken intracluster translational symmetry due to the boundary effects
(Fig.~\ref{fig:dmet:practical:cdmet}).
The violence of translational symmetry causes conceptual and practical problems, such as
the somewhat arbitrary measurement of observables in CDMET calculations.
The dynamical cluster approximation (DCA)~\cite{PhysRevB.58.R7475,PhysRevB.61.12739,fotso2012dynamical}, originated from the DMFT community, defines a transformation
of the lattice Hamiltonian such that the restriction to a finite fragment retains
the periodic boundary within the fragment, thus restoring the intracluster
translational symmetry (Fig.~\ref{fig::dmet:practical:translational}).

\begin{figure}[htpb]
  \centering
  \includegraphics[width=\columnwidth]{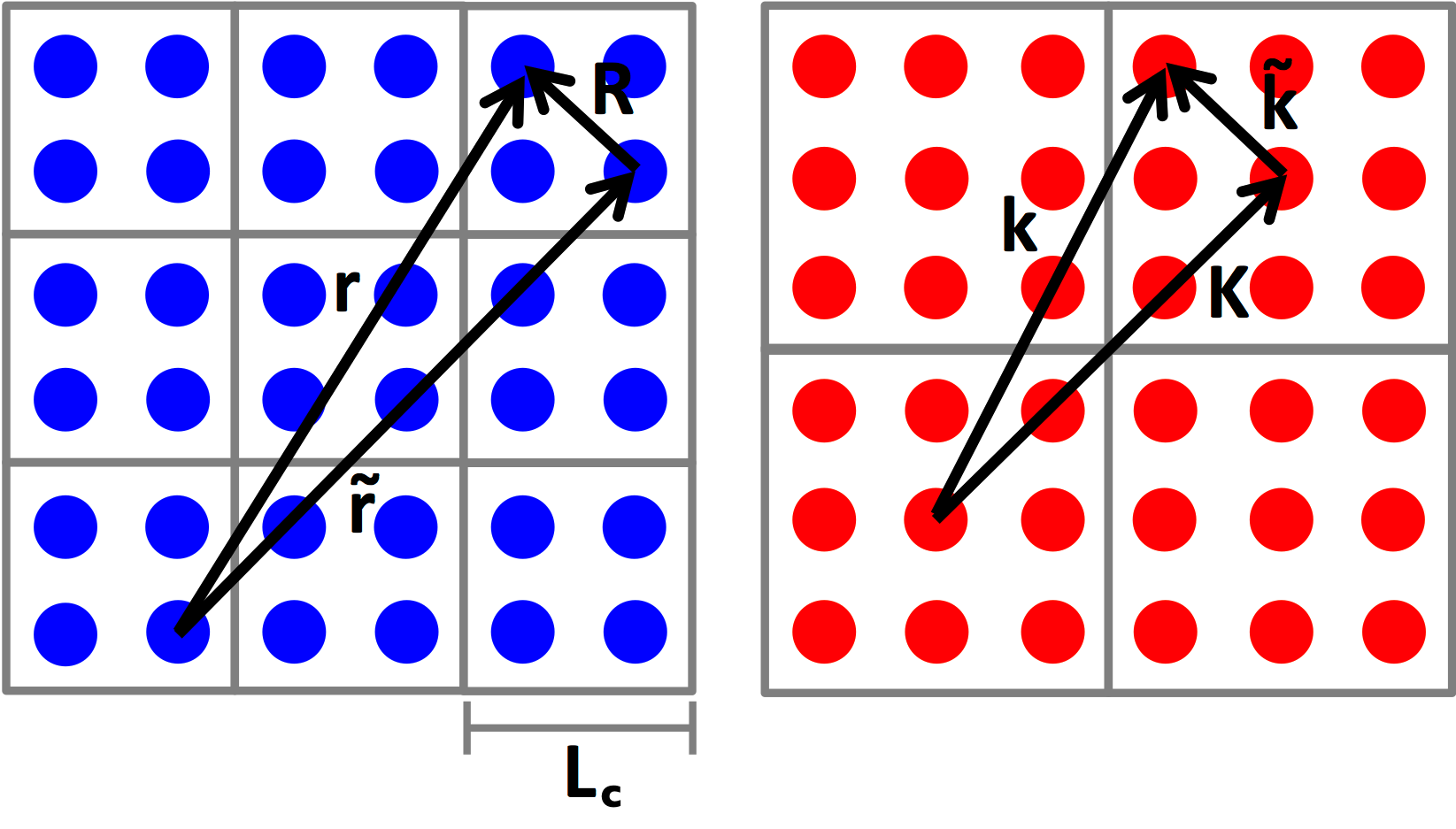}
  \caption{Definition of the real (left) and reciprocal (right) lattice vectors for the DCA transformation for a ``hypercubic" fragment with $L_c=2$. The intercluster component of the real lattice vector, $\tilde{\mathbf{r}}$, labels the origin of the fragment, and the intracluster component, $\mathbf{R}$, labels the site within the fragment. The reciprocal space of $\tilde{\mathbf{r}}$ and $\mathbf{R}$ are labeled by $\tilde{\mathbf{k}}$ and $\mathbf{K}$, respectively.}
  \label{fig:dmet:practical:dca_vec}
\end{figure}

The DCA transformation involves two steps: a basis rotation which redefines
the lattice one-body Hamiltonian, and  a coarse graining of
the two-body interaction~\cite{PhysRevB.58.R7475,PhysRevB.61.12739,Maier2005a,Potthoff2007}.
To introduce the DCA transformation, we first define  the intra- and intercluster components
of the real and reciprocal lattice vectors (Fig.~\ref{fig:dmet:practical:dca_vec}),

\begin{equation}
  \mathbf{r} = \mathbf{R}+\tilde{\mathbf{r}}, \ \ \mathbf{k} = \mathbf{K}+ \tilde{\mathbf{k}}.
  \label{eq:dmet:practical:dca_decomp}
\end{equation}

For simplicity we assume ``hypercubic'' lattices (in arbitrary dimension)
with orthogonal unit lattice vectors with linear dimension $L$, and
``hypercubic'' fragment with linear dimension $L_c$. The corresponding supercell lattice (of fragments)
then has orthogonal lattice vectors of magnitude $L_c$, and the
total number of supercells along each linear dimension is $L/L_c$.
The intracluster lattice vector,  $\mathbf{R} = (R_1, R_2, \cdots)$
and reciprocal lattice vector $\mathbf{K} = 2\pi/L_c ( N_1,  N_2, \cdots)$
where $0\le  R_i, N_i < L_c; \ R_i, N_i \in \mathbb{Z}$,
and intercluster components $\tilde{\mathbf{r}}=L_c (\tilde{r}_1, \tilde{r}_2 \cdots)$,
$\tilde{\mathbf{k}}=2\pi/L ( \tilde{n}_1,  \tilde{n}_2, \cdots)$,
with $0 \le \tilde{r}_i,\tilde{n}_i < L/L_c; \ \tilde{r}, \tilde{n} \in \mathbb{Z}$,
are uniquely defined for any $\mathbf{r}$ and $\mathbf{k}$.

Our goal is to obtain a Hamiltonian which is {\it jointly} periodic in the
intracluster and intercluster lattice vectors, $\mathbf{R}$ and $\tilde{\mathbf{r}}$.
Such a jointly periodic  basis  is provided by the product functions
$e^{-i \tilde{\mathbf{k}} \cdot \tilde{\mathbf{r}}} e^{-i \mathbf{K} \cdot{\mathbf{R}}}$. 
From one-body Hamiltonian $H_1$ defined in reciprocal space,
$H_1=\sum_\mathbf{k} H_1 (\mathbf{k}) a_\mathbf{k}^{\dagger}a_\mathbf{k}$,
  and with the mapping in Eq.~\ref{eq:dmet:practical:dca_decomp},
  we identify the diagonal DCA Hamiltonian matrix elements in the jointly periodic basis as
  \begin{equation}
    H_1 (\mathbf{k}) \to  H_1^{\mathrm{DCA}} (\mathbf{\tilde{k}, \mathbf{K}}).
  \end{equation}
  The inverse Fourier transformation then gives the DCA matrix elements on the real-space lattice. The Fourier transforms between
 the
   different single particle Hamiltonians  are summarized as:
\begin{equation}
  \begin{split}
  H_1(\mathbf{r})&\xrightarrow[]{e^{-i\mathbf{k}\cdot \mathbf{r}}}H_1(\mathbf{k})\xrightarrow[]{\mathbf{k}=\tilde{\mathbf{k}}+\mathbf{K}} H_1^\text{DCA}(\tilde{\mathbf{k}},\mathbf{K})\\
  &  \xrightarrow[]{e^{i\tilde{\mathbf{k}}\tilde{\mathbf{r}}}}
  \xrightarrow[]{e^{i\mathbf{K}\cdot\mathbf{R}}}H_1^{\text{DCA}}(\tilde{\mathbf{r}},\mathbf{R}).
  \end{split}
  \label{eq:dmet:practical:dca_flow}
\end{equation}
The resultant real-space matrix elements, $H_1^{\text{DCA}}(\tilde{\mathbf{r}},\mathbf{R})$, thus only depend on the inter- and intracluster
separation between sites.
The transformation from $h(\mathbf{r}) \to h_\text{DCA}(\tilde{\mathbf{r}}, \mathbf{R})$ is simply a basis transformation of $h$, with the rotation matrix defined as~\cite{Potthoff2007} 
\begin{equation}
  C_{\mathbf{R}+\tilde{\mathbf{r}}, \mathbf{R}^\prime+\tilde{\mathbf{r}}^\prime}=\sum_{\mathbf{K},\tilde{\mathbf{k}}}e^{-i[\mathbf{K} \cdot (\mathbf{R}^\prime-\mathbf{R})+\tilde{\mathbf{k}}(\tilde{\mathbf{r}}^\prime-\tilde{\mathbf{r}})+\tilde{\mathbf{k}} \cdot \mathbf{R}^\prime]}.
  \label{eq:dmet:practical:dca_unitary}
\end{equation}

Viewing the DCA transformation as a basis rotation suggests that the same transformation
should be extended to the interaction terms as well, generating nonlocal interactions.
However, in DCA one uses a ``coarse-grained" interaction in momentum space, which reduces
the effect of nonlocal interactions to within the fragment~\cite{Maier2005a}.
The coarse-grained interaction is obtained by averaging the Fourier transformed
interaction term over the intercluster reciprocal vectors for a given intracluster
reciprocal vector.

A special case is the Hubbard model, where such coarse-graining leaves
the local $Un_{i\alpha}n_{i\beta}$ term unchanged in the transformed Hamiltonian.
Note that the coarse-grained Hubbard interaction is nonlocal if transformed back to the
original site basis using the rotation in Eq.~\ref{eq:dmet:practical:dca_unitary}.

We can thus perform DMET using the DCA transformed Hamiltonian. To preserve intracluster
translational symmetry of the DMET results, the correaltion potential $u_C$ is also
required to be translational invariant within a fragment (See Appendix~\ref{sec:formula:dca_pot}). We refer this formulation as
DCA-DMET. We present the numerical tests of DCA-DMET, as well as CDMET in Chapter~\ref{chpt:scaling}.

\subsection[Cluster Size Extrapolation]{Cluster Size Extrapolation~\footnote{Based on work published in Phys. Rev. B \textbf{95}, 045103 (2017). Copyright 2017, American Physical Society.~\cite{zheng2017cluster}}} \label{sec:dmet:practice:extrapolation}
As discussed in Sec.~\ref{sec:dmet:bcs:symm}, extrapolation with the fragment size is an important
tool to improve upon finite fragment DMET results. In this section, we present the theories of the
cluster size scaling for energy and intensive observables. We analyze both CDMET and DCA-DMET in
the Hubbard model on a $d$-dimensional hypercubic lattice, although most of the conclusions we
derive here apply to other Hamiltonians as well. For the energy, we use a perturbation
argument to obtain the leading term of the finite-size scaling; for the more complicated case of
intensive observables, we suggest a plausible scaling form.

We consider the following factors to derive the DMET finite-size scaling for the Hubbard model:
(a) the open boundary in CDMET; (b) the gapless spin excitations of quantum antiferromagnets;
(c) the coupling between the impurity and bath; (d) the modification of the hoppings of the
in DCA-DMET.

\begin{figure}[htpb]
  \centering
  \includegraphics[width=\columnwidth]{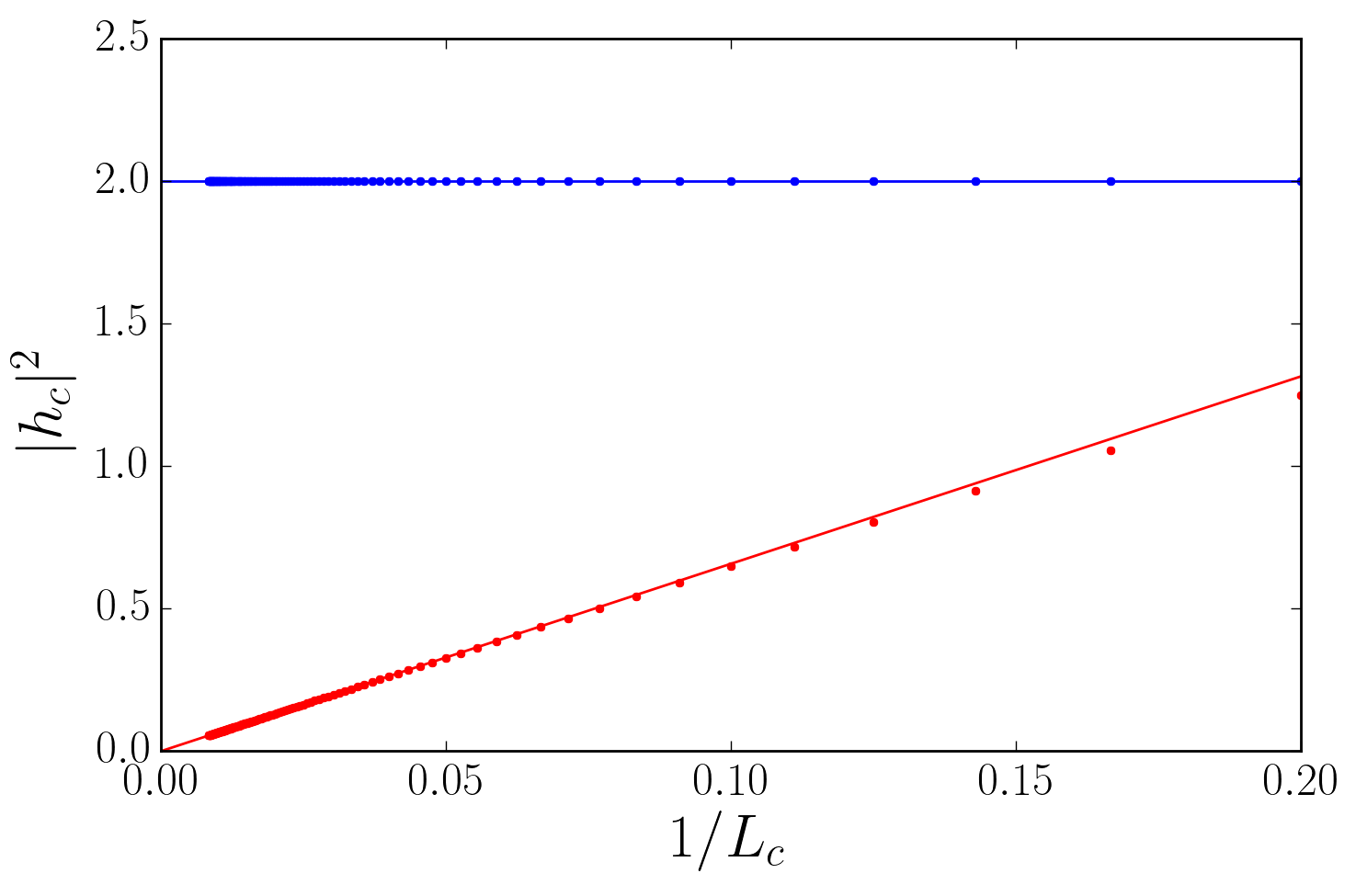}
  \caption{Sum-of-square of the one-body fragment-environment coupling Hamiltonian
	  $|h_c|^2=\sum_{i\in C=0,j\in C^\prime\neq0}|h_{ij}|^2$
    for the CDMET and DCA-DMET formulations, in one-dimension. The fittings follow
    constant (CDMET) and $1/L_c$ (DCA) scalings, respectively.}
  \label{fig:dmet:practical:coupling}
\end{figure}

We start with the CDMET energy. We first consider the {\it bare} fragment in CDMET
(i.e. without the bath) which is just the finite size truncation of the infinite
system. For a gapped system, we expect an open boundary to lead to a finite-size
energy error (per site) proportional to the surface area to volume
ratio~\cite{Fisher1972}, i.e.
\begin{equation}
  e(L_c) = {e(\infty)} + \frac{a_0}{L_c} + \cdots
  \label{eq:dmet:practical:open_FS}
\end{equation}
where $e(L_c)$ is the energy per site for an $L_c^d$ site fragment and $e(\infty)$ is the
energy per site in the thermodynamic limit (TDL). If, in the TDL, there are gapless modes,
a more careful analysis is required. The Hubbard model studied here has gapless spin excitations.
These yield a finite size error of $O(1/L_c^{d+1})$ in a fragment with periodic boundary
condition (PBC)~\cite{Fisher1989,Hasenfratz1993,Huse1988,Sandvik1997}, which is subleading
to the surface finite size error introduced by the open boundary in
Eq.~\ref{eq:dmet:practical:open_FS} for $d>0$.

We next incorporate the CDMET bath coupling. Each
site on the fragment boundary couples to the bath, yielding a total Hamiltonian coupling of
$O(1)$ per boundary site (see Fig.~\ref{fig:dmet:practical:coupling}).
The total ``perturbation'' to the
bare fragment Hamiltonian is then $O(L_c^{d-1})$, which leads to a first order
energy correction per site. Thus, in CDMET, the open boundary and the error in the bath
orbitals together cause the leading contribution to the finite size error
\begin{equation}
{e(L_c)_\text{CDMET}} =  {e(\infty)} + \frac{a_0'}{L_c}  + \cdots
\end{equation}
in any dimension.

For DCA-DMET, the above argument must be modified in two ways: first, the fragment
uses PBC, and second, the formulation modifies intercluster and
intracluster hoppings. Similarly, we start with
the bare periodic fragment (without any modification of the intracluster hoppings).

In the TDL, for a gapped state with short-range interactions, all correlation
functions decay exponentially (e.g. Wannier functions are exponentially localized)
and we expect an exponential convergence of the energy with respect to fragment size.

However, in the Hubbard model, as previously mentioned, the gapless spin excitations
give a finite-size energy error (per site) of $O(1/L_c^{d+1})$.
The leading order finite-size scaling for the bare periodic fragment
is thus expected to be
\begin{equation}
	e(L_c) = e(\infty) + \frac{a_0}{L_c^{d+1}} + \cdots.
\end{equation}

The DCA-DMET  Hamiltonian  modifies the periodic fragment
Hamiltonian by changing both the intracluster and intercluster hopping terms.
The intracluster hopping terms are modified by a term of order
$O(1/L_c^2)$, and the intercluster hopping terms are modified so as to generate
a coupling between each site in the fragment and the bath with a total
interaction strength of $O(1/L_c^2)$ (see Fig.~\ref{fig:dmet:practical:coupling}).
Since there are $L_c^d$ sites in the fragment,  the total magnitude of the DCA-DMET
perturbation (including the contributions of both intracluster and intercluster terms)
is $O(L_c^{d-2})$.

For one dimension, the perturbation and fragment-bath coupling give a contribution
 with the same scaling as the contribution of the gapless modes,
while in two and higher dimensions, they give the leading term in the finite-size error.
 Thus combining the three sources of finite-size error we expect in any dimension
the energy per site scaling of DCA-DMET to be
\begin{align}
	e(L_c)_\text{DCA-DMET} &= {e(\infty)} + \frac{a_0'}{L_c^2} + \cdots.
  \end{align}
Note that the scaling of the CDMET and DCA-DMET energies is the same as is
found for CDMFT and DCA.

The finite size scaling of intensive quantities is more tricky to
analyze~\cite{Maier2002}.
For an observable $Q$ we have the relation
$\langle Q \rangle = \lim_{r\to\infty} \langle Q(0) Q(r) \rangle^{1/2} $,
where $\langle Q(0) Q(r)\rangle$ is the correlation function.
It is often argued that the error in $\langle Q \rangle$ in
a large finite fragment behaves like
\begin{equation}
  \Delta Q  \sim  [\langle Q(0) Q(R)\rangle^{1/2} -
  \langle Q(0) Q(\infty)\rangle^{1/2}]
  \label{eq:dmet:practical:dQ}
\end{equation}
where $R$ is the largest length in the fragment~\cite{Huse1988} $\sim L_c/2$.
For CDMET, where the cluster is only coupled to the symmetry-broken
bath at the boundary, we assume the form in Eq.~\ref{eq:dmet:practical:dQ} holds,
with additional corrections from the
fragment size, expanded as a Taylor series
\begin{align}
	\Delta Q(L_c) =  \left(a + \frac{b}{L_c} + \cdots\right)
	[\langle Q(0) Q(R)\rangle^{1/2} - \langle Q(0) Q(\infty)\rangle^{1/2}].
	\label{eq:dmet:practical:intensive}
\end{align}
Eq.~\ref{eq:dmet:practical:intensive} is a heuristic form and its
correctness will be assessed in our numerical results in Chapter~\ref{chpt:scaling}.
Taking the local magnetic moment $m = \langle S_z\rangle$ as an example,
the correlation function $\langle S_z(0) S_z(r)\rangle$ at large $r$
behaves like $a \sqrt{\ln r} /r$  in the 1D Hubbard model, and $a+b/r$ in
the 2D square-lattice Hubbard model at half-filling.
Consequently, we can assume a scaling form in 1D of
 \begin{equation}
   m(L_c)_\text{CDMET} = \sqrt{\frac{\sqrt{\ln L_c/2}}{L_c/2}}
   \left(a + \frac{b}{L_c} + \cdots\right)
   \label{eq:dmet:practical:1dmscaling}
 \end{equation}
and in 2D of
\begin{equation}
  m(L_c)_\text{CDMET} = a + \frac{b}{L_c} + \frac{c}{L_c^2} + \cdots.
  \label{eq:dmet:practical:2dmscaling}
\end{equation}

For DCA-DMET, however, every fragment site, not just those at the
boundary, is coupled to a set of bath orbitals, which provides a
symmetry-breaking pinning field. This means that there is no simple
connection to the scaling of correlation function (with respect to distance)
of the system. Thus, there is no obvious theoretical argument for any
form of scaling for the observables in DCA-DMET, and we can, at best,
assume the Taylor expansion in both one- and two-dimensions,
\begin{equation}
  m(L_c)_\text{DCA-DMET} = a + \frac{b}{L_c} + \frac{c}{L_c^2} + \cdots.
  \label{eq:dmet:practical:dcamscaling}
\end{equation}

\chapter[Cluster Size Convergence of the Density Matrix Embedding
Theory and Its Dynamical Cluster Formulation]
{Cluster Size Convergence of the Density Matrix Embedding
Theory and Its Dynamical Cluster Formulation~\footnote{Based on work published in Phys. Rev. B \textbf{95}, 045103 (2017). Copyright 2017, American Physical Society.~\cite{zheng2017cluster}}} \label{chpt:scaling}

We present in this chapter the numerical studies of cluster size convergence of the
energy and observables using two forms
of DMET: the original CDMET and the DCA-DMET, motivated by the dynamical cluster
approximation (see Sec.~\ref{sec:dmet:practice:dca}).
Both methods are applied to the half-filled one- and two-dimensional Hubbard models
using a sign-problem free AFQMC impurity solver
(see Sec.~\ref{sec:dmet:solver:afqmc}), which allows for the treatment of large
impurity clusters of up to 100 sites. While
CDMET is more accurate at smaller fragment sizes, DCA-DMET exhibits
faster asymptotic convergence towards the TDL. In addition to investigating
the cluster size convergence scaling, these calculations produce accurate estimates
for the energy and local moment of the two-dimensional Hubbard model for
$U/t=2, 4, 6$. The results compare favorably with the best data available
in the literature, and help resolve earlier uncertainties in the moment for $U/t=2$.

In Sec.~\ref{sec:scaling:intro}, we introduce the background and motivation
of the study. In Sec.~\ref{sec:scaling:method}, we briefly describe the methods
and parameters used in the calculations. We present the calculation results
in Sec.~\ref{sec:scaling:result} and discuss the implications of the results
in Sec.~\ref{sec:scaling:conclusion}.

\section{Introduction} \label{sec:scaling:intro}
An critical dimension in numerical studies of lattice models is the ability
to study the physical properties in TDL. To do so, one typically considers
finite sized clusters of increasing sizes under some choice of boundary
conditions, followed by a finite size scaling of the observables.
Embedding methods accelerate the finite size convergence, by mapping the
bulk problem onto an auxiliary impurity model, where a small fragmen of
the physical interacting sites is coupled to special ``bath sites'' that
mimic the effects of the neglected environment.

As discussed in Sec.~\ref{sec:dmet:bcs:symm}, the cluster size extrapolation is
a great tool to extend the scope of DMET studies. A detailed analysis of the
finite-size scaling of DMET not only is theoretically interesting, but helps
obtaining accurate TDL estimates in all future DMET studies.

In this chapter, we perform such a detailed investigation in the context of the
half-filled 1D and 2D square lattice Hubbard models using the AFQMC impurity
solver, implemented by Zhang, et. al.~\cite{Sugiyama1986,Zhang2013}. Because
of the absence of the sign problem, these models serve as an excellent testing
bed for the purpose, as we are able to study DMET fragments with up
to 100 fragment sites. Using this solver further facilitates direct comparisons
to bare (i.e. not embedded) AFQMC calculations in the literature that used
very large clusters (with up to 1058 sites) with periodic (PBC), anti-periodic
(APBC), modified (MBC), and twisted boundary (TBC) conditions
~\cite{Sorella2015,Qin2016}. The comparison provides a direct demonstration
of the benefits of embedding, versus simply modifying the boundary conditions.

Similar to previously established scalings for CDMFT and DCA
~\cite{Fisher1972,Maier2002,Biroli2002,Aryanpour2005,Biroli2005}, the CDMET
and DCA-DMET converge $O(1/L_c)$ and $O(1/L_c^2)$ from the theoretical analysis
(Sec.~\ref{sec:dmet:practice:extrapolation}). In this chapter, we perform
CDMET and DCA-DMET calculations on 1D and 2D Hubbard model at half-filling
for $U/t=4, 8$ and $U/t=2, 4, 6$, respectively, to confirm the scalings
of the TDL energies and to investigate the convergence of the TDL spin-moments.

For the energies, our results provide high accuracy benchmarks with small error bars.
Converging finite-size effects for spin-moment has well-known pitfalls, and 
existing data in the literature do not always agree
~\cite{Varney2009,Wang2014,LeBlanc2015,Sorella2015,Qin2016}.
Where an agreement is observed, our new
estimates confirm the existing data with comparable or improved error bars. In the
case of $U/t=2$ where severe finite size effects are found, our data resolves
between the earlier incompatible estimates in the literature.

\section{Methods} \label{sec:scaling:method}
The Hubbard Hamiltonian is defined as 
\begin{equation}
	H=-t\sum_{\langle ij\rangle\sigma}a_{i\sigma}^\dag a_{j\sigma}+\sum_iUn_{i\alpha}n_{i\beta}
	\label{eq:scaling:hubbard}
\end{equation}
where $a_{i\sigma}^\dag$ ($a_{i\sigma}$) creates (destroys) a particle
of spin $\sigma$ at site $i$, $\langle ij\rangle$ denotes nearest neighbors,
and $n_{i\sigma}=a_{i\sigma}^\dag a_{i\sigma}$.

We apply the standard spin-unrestricted, normal state CDMET algorithm detailed
in Appendix~\ref{sec:algo:normal} to the Hubbard models,
with the following modifications: (i) the correlation potential is restricted to
preserve the particle-hole symmetry so that the AFQMC impurity solver does suffer
from the sign problem (Eq.~\ref{eq:dmet:solver:afqmc_phsymm});
and (ii) the chemical potential fitting is
skipped since $\mu\equiv U/2$ at half-filling.

The DCA-DMET calculations are performed similarly with the modified Hamiltonian
described in Sec.~\ref{sec:dmet:practice:dca}, which restores intracluster
translational symmetry. There is also stronger restrictions on the correlation
potential in DCA-DMET, as it preserves both the translational and the particle-hole
symmetry.

In the calculations of this chapter, we use the AFQMC impurity solver,
whose implementation described in Refs.~\cite{Zhang2013,Shi2015,Shi2016}
with small modifications to treat Hamiltonians with broken $S^2$ symmetry
~\footnote{The implementation was done by Hao Shi and Shiwei Zhang.}.
Both the energy and the one-body density matrix (required for the DMET
self-consistency) are computed by the pure estimator,
Eq.~\ref{eq:dmet:solver:afqmc_ops}. We converge the standard deviation of
all elements in the one-body density matrix to be less than 0.001, to make
the AFQMC statistical errors (and thus DMET statistical convergence errors)
orders of magnitude smaller than the finite cluster size error. This results
in considerably higher statistical accuracy for extensive quantities than
typically obtained in the AFQMC literature.

The finite fragment results are extrapolated to obtain TDL estimates following the
scalings suggested by the theoretical analysis in Sec.~\ref{sec:dmet:practice:extrapolation}.
The quality of these extrapolations is assessed.

\section{Results} \label{sec:scaling:result}
We present the CDMET and DCA-DMET calculations on the half-filled 1D and 2D
Hubbard models, focusing on the finite-size convergence of the energy and
local observables. As discussed in section~\ref{sec:scaling:method} the DMET
correlation potential preserves $S_z$ symmetry but is allowed to break $S^2$
symmetry. For the Hubbard models studied here, all the converged self-consistent
DMET solutions explicitly break $S^2$ symmetry.

In 1D, we compare our results against exact results from the Bethe Ansatz
(BA), while in 2D, we compare to literature benchmark data from AFQMC calculations
scaled to the TDL~\cite{Wang2014,Sorella2015,Qin2016}, DMRG calculations
scaled to the TDL~\cite{LeBlanc2015}, and iPEPS calculations
scaled to zero truncation error~\cite{Corboz2016}.

\subsection{1D Hubbard Model} \label{sec:scaling:result:1d}
We study fragment clusters with $N_\text{imp} = L_c \leq 24$ sites on a DMET
auxiliary lattice with $N=L=480$ (even $N_c$) or $N=L=480 + L_c$ (odd $N_c$)
sites. The auxiliary lattice uses PBC, and as the DCA-DMET impurity Hamiltonian
becomes complex for even $N_c$, we only use auxiliary lattices with an odd
$N_c$ in the DCA-DMET calculations. We study two couplings $U/t=4$
(moderate coupling) and $U/t=8$ (strong coupling). When starting from
uniform antiferromagnetic initial guesses for the correlation potential,
it usually takes 4 to 8 DMET iterations to converge the calculations.

\begin{figure}[thpb]
	\centering
  \subfigure[U/t = 4]{
	  \includegraphics[width=0.85\columnwidth]{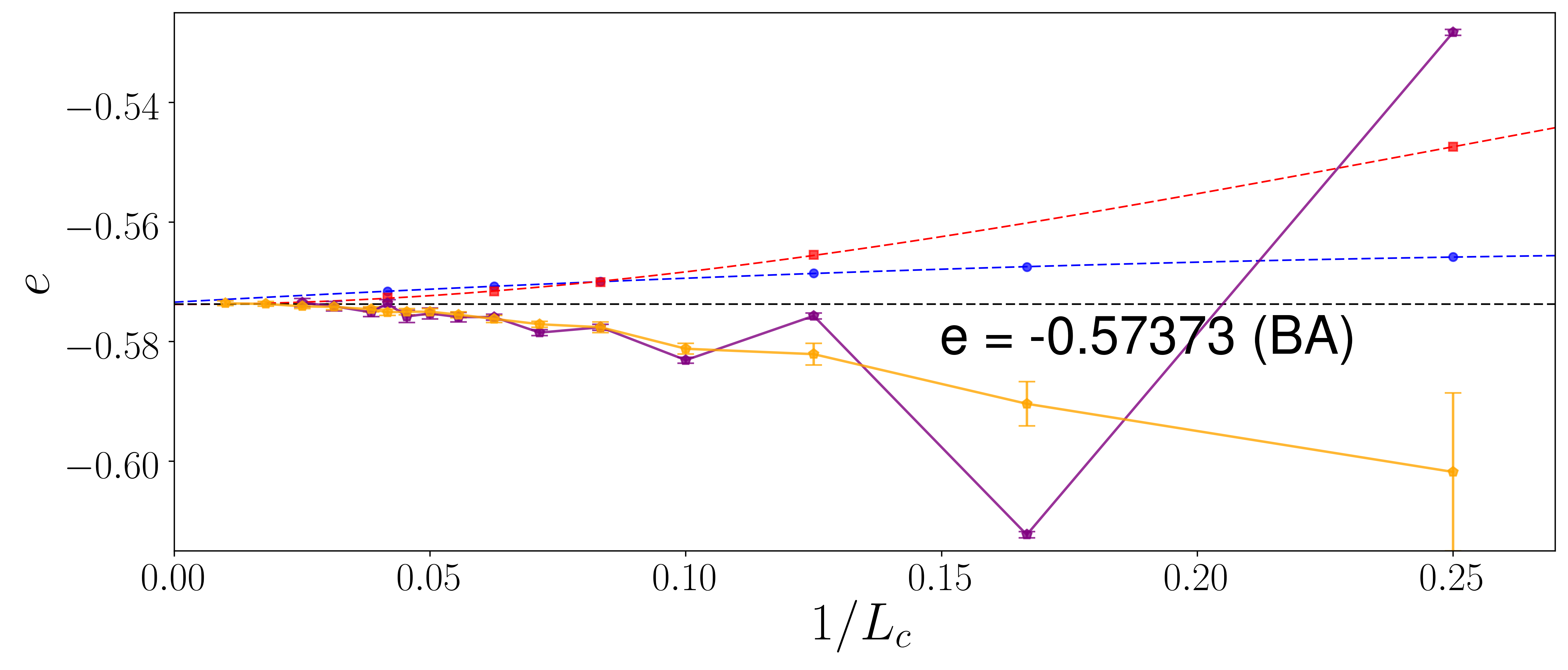}
	  \label{fig:scaling:energy1D_U4}
  }
  \subfigure[U/t = 8]{
	  \includegraphics[width=0.85\columnwidth]{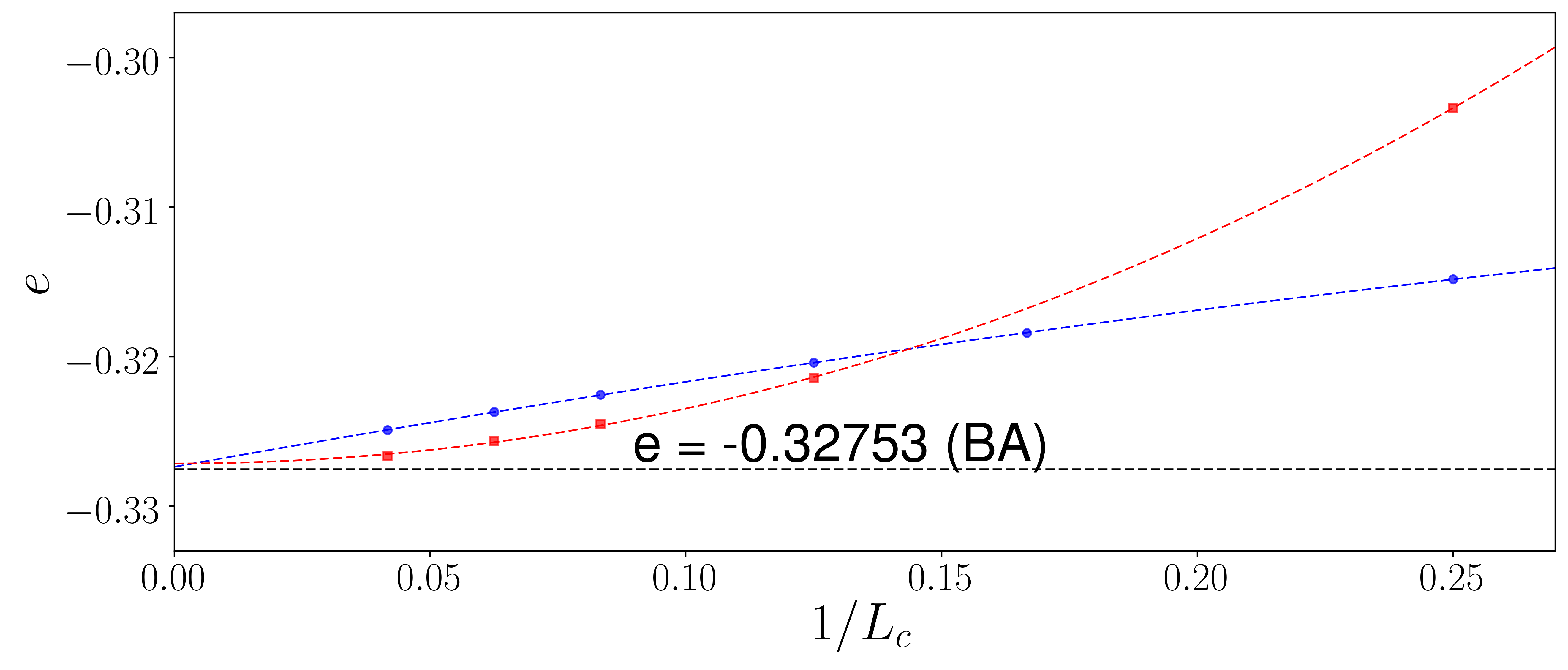}
	  \label{fig:scaling:energy1D_U8}
  }
  \caption{Energy per site, $e$, for the half-filled 1D Hubbard model
	  versus inverse impurity size, $1/L_c$, from CDMET (blue) and
	  DCA-DMET (red). For comparison, we also plot the same numbers from AFQMC with PBC
    (purple) and TABC (orange) for $U/t=4$.
    The extrapolations use $e=a+bL_c^{-1}+cL_c^{-2}$ for
    CDMET and $e=a+bL_c^{-2}+cL_c^{-3}$ for DCA-DMET.}
  \label{fig:scaling:energy1D}
\end{figure}

Fig.~\ref{fig:scaling:energy1D} shows the energy per site as a function of inverse
impurity size $1/L_c$.  Statistical error bars associated with the AFQMC solver are
not shown here as they are too small to be visible; this is true for all the CDMET
and DCA-DMET results presented in this chapter.
As shown in Table~\ref{tab:scaling:energy1D}, the extrapolated energies are in
generally good agreement with the exact Bethe ansatz TDL data, with a deviation of
less than $0.001t$.
To further improve the accuracy, we include the subleading terms
in the energy extrapolation, i.e.  $a + b/L_c + c/L_c^2$  for CDMET
and $a + b/L_c^2 + c/L_c^3$
for DCA-DMET (dashed lines in Fig.~\ref{fig:scaling:energy1D}). 
This improves the extrapolated TDL 
results significantly, with the single exception of DCA-DMET
at $U/t=8$, where the coefficient of the cubic term is not
statistically significant [$c=0.08(9)$] and the deviation is
already very small. The subleading terms are more important
at $U/t=4$ than at $U/t=8$. This is consistent with the smaller
gap at weaker coupling, that introduces stronger finite size effects.

\begin{table}[htpb]
\centering
\caption{CDMET and DCA-DMET cluster size extrapolation of the energy per site
(in units of $t$) for the 1D half-filled Hubbard model.}
\label{tab:scaling:energy1D}

\begin{tabular}{p{0.15\columnwidth} p{0.25\columnwidth} p{0.25\columnwidth} p{0.25\columnwidth}}
\toprule
\multicolumn{2}{c}{\textbf{Extrapolation}} & $\mathbf{U/t=4}$         & $\mathbf{U/t=8}$ \\      \midrule
\multirow{2}{*}{CDMET}    & $a + b/L_c$                    & -0.5724(3)  & -0.3267(2)  \\
                          & $a + b/L_c + c/L_c^2$       & -0.5734(1)  & -0.3274(1)  \\ \midrule
\multirow{2}{*}{DCA-DMET} & $a + b / L_c^2$                 & -0.5729(4)  & -0.3273(1)  \\
                          & $a + b/L_c^2 + c/L_c^3$        & -0.5738(1)  & -0.3272(1)  \\ \midrule
\multicolumn{2}{l}{Bethe Ansatz}                    & -0.57373    & -0.32753    \\
\bottomrule
\end{tabular}
\end{table}

To further numerically test the scaling form for the DCA-DMET
extrapolation, we include a linear $1/L_c$ term in the DCA-DMET
scaling form, i.e. $a + b/L_c + c/L_c^2$.
While the coefficient of the linear term is statistically
significant at $U/t=4$, the extrapolated TDL energy acquires a
larger uncertainty [-0.5749(6)], while for $U/t=8$, the
coefficient of $1/L_c$ term becomes statistically insignificant
[$b=0.003(5)$]. This supports the leading finite-size scaling
of the DCA-DMET energy per site as being $O(1/L_c^2)$.
The finite size scaling of the energy observed for CDMET and DCA-DMET is
consistent with similar data observed for CDMFT and DCA~\cite{Maier2002,Maier2005a}.

In Fig.~\ref{fig:scaling:energy1D_U4}, we plot the AFQMC results with
periodic (PBC) and twist-average (TABC) boundary conditions as well.
While the PBC energy oscillates strongly for all cluster sizes, the convergence
of TABC is much smoother. The finite-size scaling of bare cluster AFQMC (PBC and
TABC) appears to be quadratic in inverse size, which is consistent with
the spin-wave theory predictions in 1D~\cite{Huse1988}, and coincides with
the scaling of DCA-DMET. Therefore, with large clusters, the finite-size errors
of bare cluster AFQMC and DCA-DMET are comparable and smaller than those of
CDMET, while CDMET is much more accurate for small fragments.

\begin{figure}[htpb]
	\centering
	 \includegraphics[width=0.9\columnwidth]{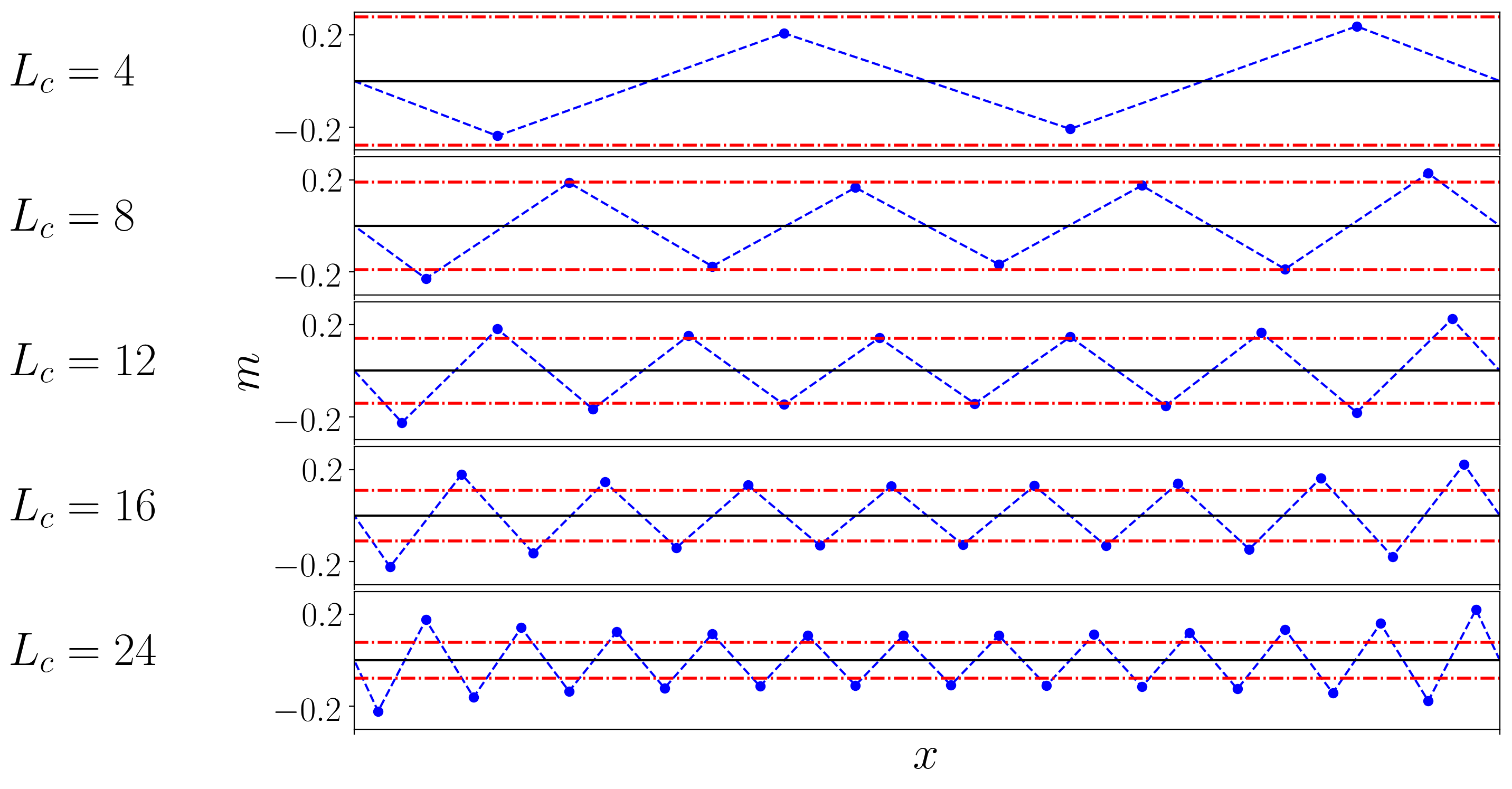}
  \caption{Local spin moments $m$ from  CDMET (blue) and DCA-DMET (red) in finite
  fragment calculations at $U/t=4$ in the 1D Hubbard model.
  $x$ is the site index scaled to the interval $[0, 1]$ for the CDMET results.
  }
  \label{fig:scaling:spin1d}
\end{figure}
\begin{figure}[htpb]
	\centering
	\includegraphics[width=0.9\columnwidth]{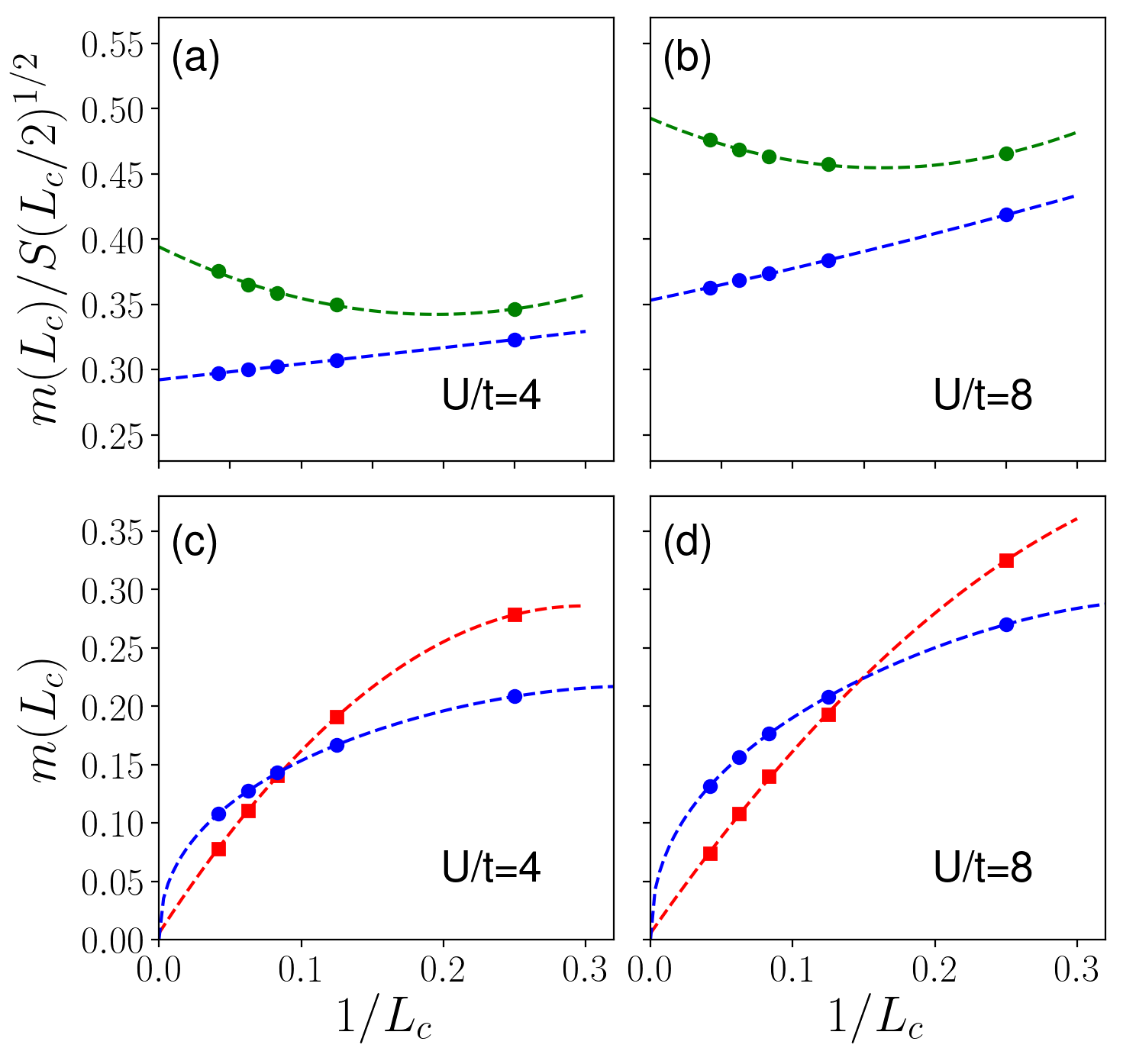}
	\caption{Cluster size extrapolation of the
		AF order parameters in the 1D Hubbard model.
	(a) (b) CDMET AF order parameters $m(L_c)$ divided by spin correlation function
  $S(L_c/2)^{1/2}$, versus inverse impurity cluster size $1/L_c$
  for $U/t=4$ and $U/t=8$ (blue: center average, green: entire cluster average).
  The extrapolation uses the form $m(L_c)/S(L_c/2)^{1/2}=a+bL_c^{-1}+cL_c^{-2}$,
  see Eq.~\ref{eq:dmet:practical:1dmscaling} for details.
  (c) (d) DCA-DMET and CDMET (center average) AF order parameters $m(L_c)$
  versus inverse impurity cluster size $1/L_c$ for $U/t=4$ and $U/t=8$.
  The extrapolation for DCA-DMET values uses the form
  $m(L_c)=a+bL_c^{-1}+cL_c^{-2}$, see Eq.~\ref{eq:dmet:practical:dcamscaling}
  for details.}
	\label{fig:scaling:spin1d_extrap}
\end{figure}

We now turn to the spin orders. Although there is no true long-range
antiferromagnetic (AF) order in 1D, the finite correlated fragment calculations yield
non-zero spin moments, which should extrapolate to zero in the TDL.
The local spin moments $m$ are plotted in Fig.~\ref{fig:scaling:spin1d}.
We see that the spin moments in the CDMET fragment are largest at the boundary
with the AF environment, and decay towards the center. We can understand this
because quantum fluctuations are incompletely treated in the bath orbitals,
and thus they are overmagnetized. This effect is propagated to the boundary
of the CDMET fragment.
Note that the fragment sites in a DCA-DMET cluster are all equivalent, and
are equally coupled to the environment, resulting in an equal spin magnitude
for all sites, to within the statistical error of the solver. In
Fig.~\ref{fig:scaling:spin1d} we use the two horizontal lines to
represent the spin magnitudes from the DCA-DMET calculations.

To determine the magnetic order parameter from CDMET, we consider two
possible definitions:
(a) the average $|m|$ for the central pair (or the plaquette in 2D);
(b) the average $|m|$ over the entire fragment. These definitions are
equivalent for DCA-DMET. In CDMET, they agree in the limit of small
clusters ($L_c=2$) and large clusters ($L_c\rightarrow\infty$), but differ
inbetween.

The AF order parameters for different fragment sizes are plotted in
Figs.~\ref{fig:scaling:spin1d_extrap} for different $U$.
For CDMET, we fit the order parameter to the scaling form in
Eq.~\ref{eq:dmet:practical:1dmscaling}, up to second order.
The fits are shown in Figs.~\ref{fig:scaling:spin1d_extrap}(a), (b), and
are quite good for both types of measurements. For the average $|m|$ of
the central pair, an almost straight line is observed at both couplings,
with the quadratic term close to vanishing ($c=0.00(4)$ for $U/t=4$ and
$c=0.12(7)$ for $U/t=8$).
The average $|m|$ over the entire fragment requires a larger $c$ for a
good fit. This is because $|m|$ is measured at different points which
corresponds to averaging over different effective lengths $L_c$ in
Eq.~\ref{eq:dmet:practical:1dmscaling}. Averaging over
Eq.~\ref{eq:dmet:practical:1dmscaling} yields the same leading scaling
but introduces larger subleading terms.
Overall, the error decreases much more rapidly by using the center
average, consistent with observations in CDMFT~\cite{Biroli2005}.

For DCA-DMET, the scaling form Eq.~\ref{eq:dmet:practical:dcamscaling} truncated
at second order works well. This correctly predicts the vanishing local moments
at the TDL ($a=0.005(1)$ at $U/t=4$ and $a=0.005(4)$ at $U/t=8$). The $O(1/L_c)$
scaling of DCA-DMET thus converges faster than CDMET, whose leading term is
$\left(\frac{\sqrt{\log(L_c/2)}}{L_c/2}\right)^{1/2}\sim L_c^{-1/2}$.

While the smallest fragments in CDMET report a smaller magnetization than
seen in DCA-DMET (and thus can be regarded as ``closer'' to the TDL) the
cross-over between the DCA-DMET and CDMET moments occurs at smaller clusters
than for the energy.

\subsection{2D Hubbard Model} \label{sec:scaling:result:2d}

We now show results from the half-filled 2D Hubbard model at $U/t=2,4,6$.
We use square fragments of size $N_{\textrm{imp}}=L_c\times L_c$, where
for CDMET $L_c=2,4,6,8,10$ and for DCA-DMET $L_c=4,6,8,10$.
The $2\times 2$ plaquette is not used in the finite-size scaling  of
DCA-DMET as it is known from DCA studies to exhibit anomalous
behavior~\cite{Maier2002}, which we also observe.
Also at $U/t=6$, we do not present results for $L_c=10$, as we are unable
to converge the statistical error to high accuracy in the AFQMC calculations
(within our computational time limits). The total lattices we used have
linear lengths of around $L=120$ ($N=L\times L$), adjusted to fit even (CDMET)
or odd (DCA-DMET) $N_c$, as in the 1D case.
As in 1D, we initialize the correlation potential as a diagonal matrix with
uniform AF terms. The 2D calculations thus take slightly more self-consistent
iterations (about 10) than in 1D to converge.

\begin{figure}[htpb]
  \subfigure[U/t = 2]{
	  \includegraphics[width=0.45\columnwidth]{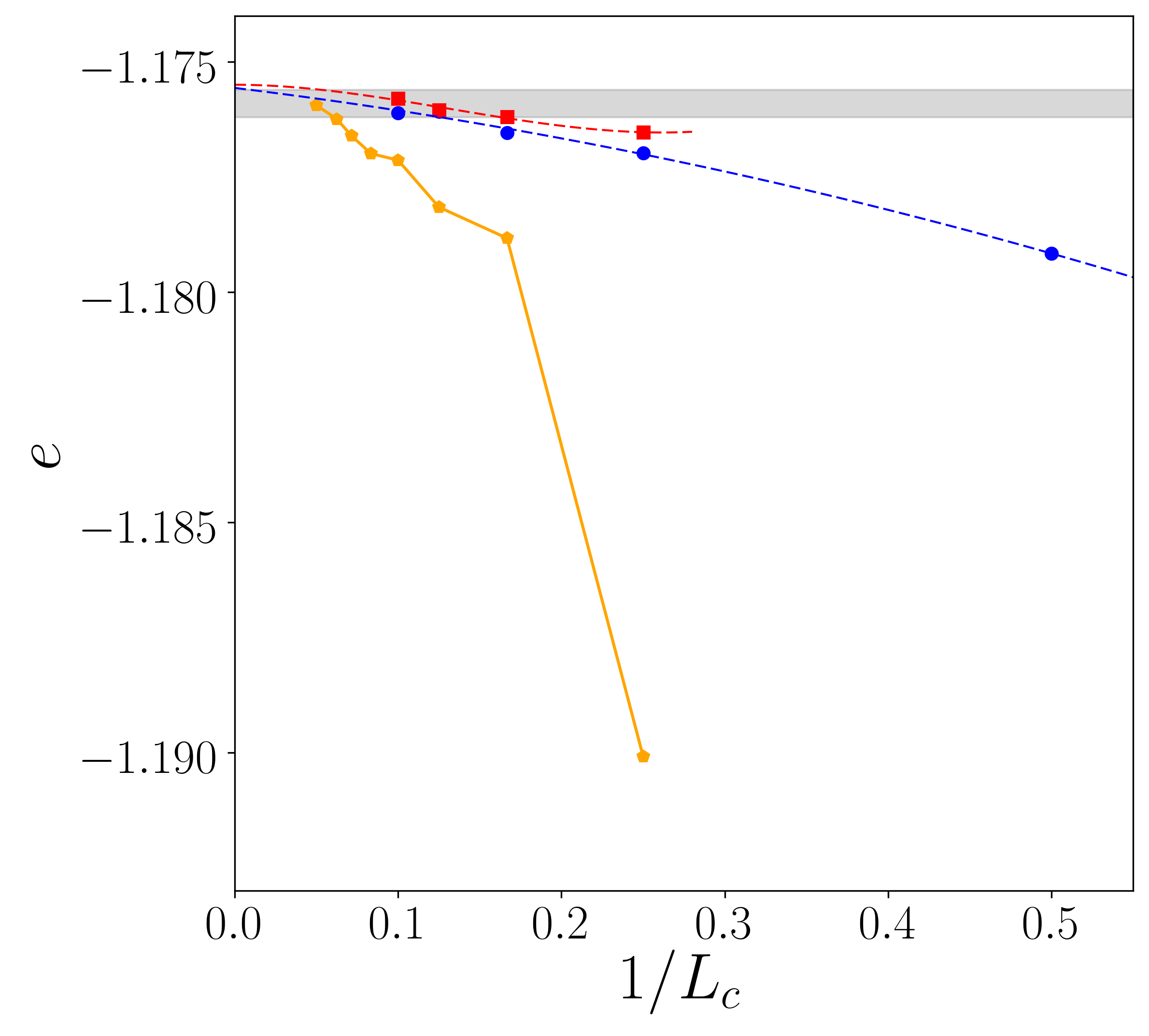}
	  \label{fig:scaling:energy2D_U2}
  }
  \subfigure[U/t = 4]{
	  \includegraphics[width=0.45\columnwidth]{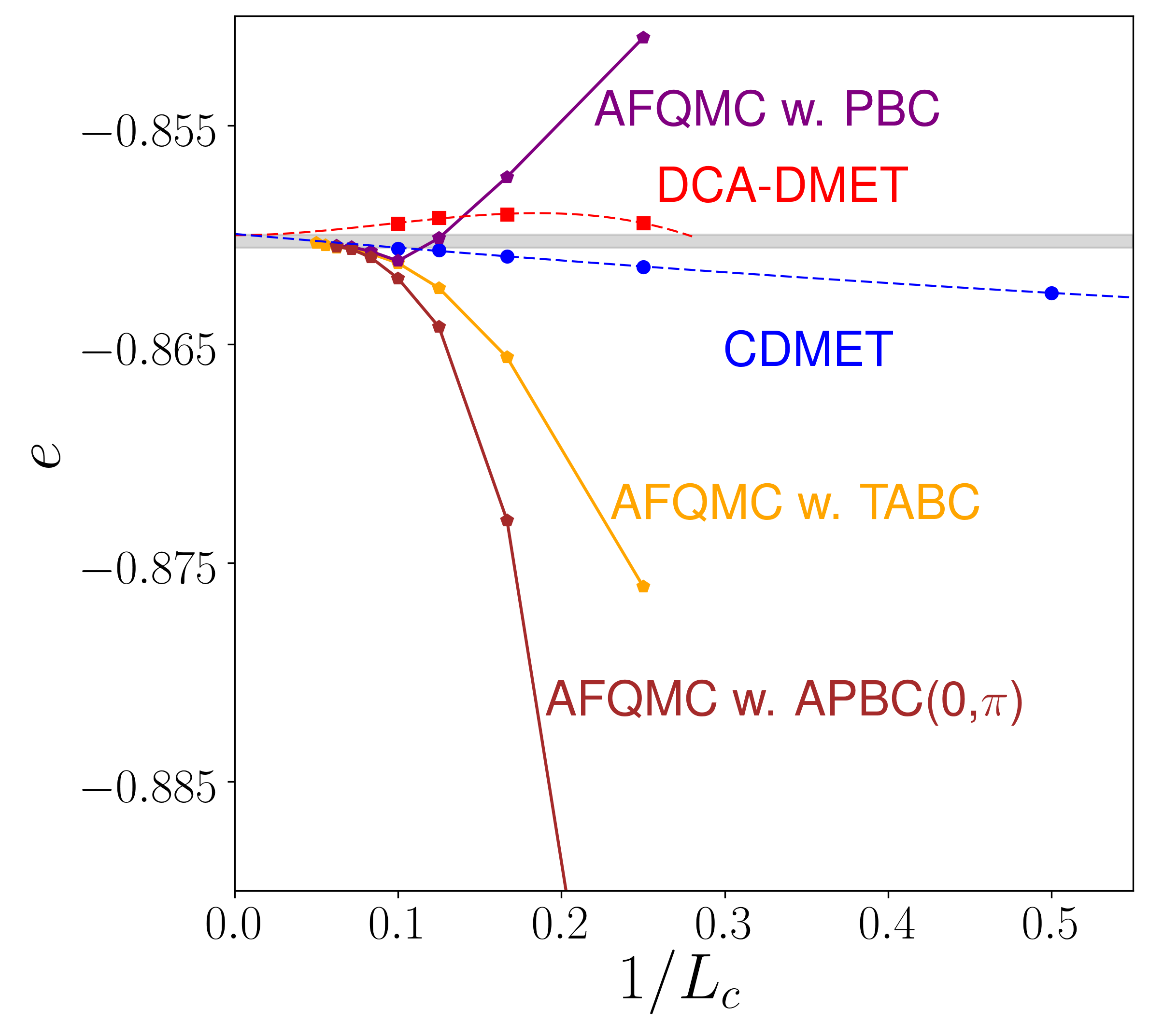}
	  \label{fig:scaling:energy2D_U4}
  }
  \subfigure[U/t = 6]{
	  \includegraphics[width=0.45\columnwidth]{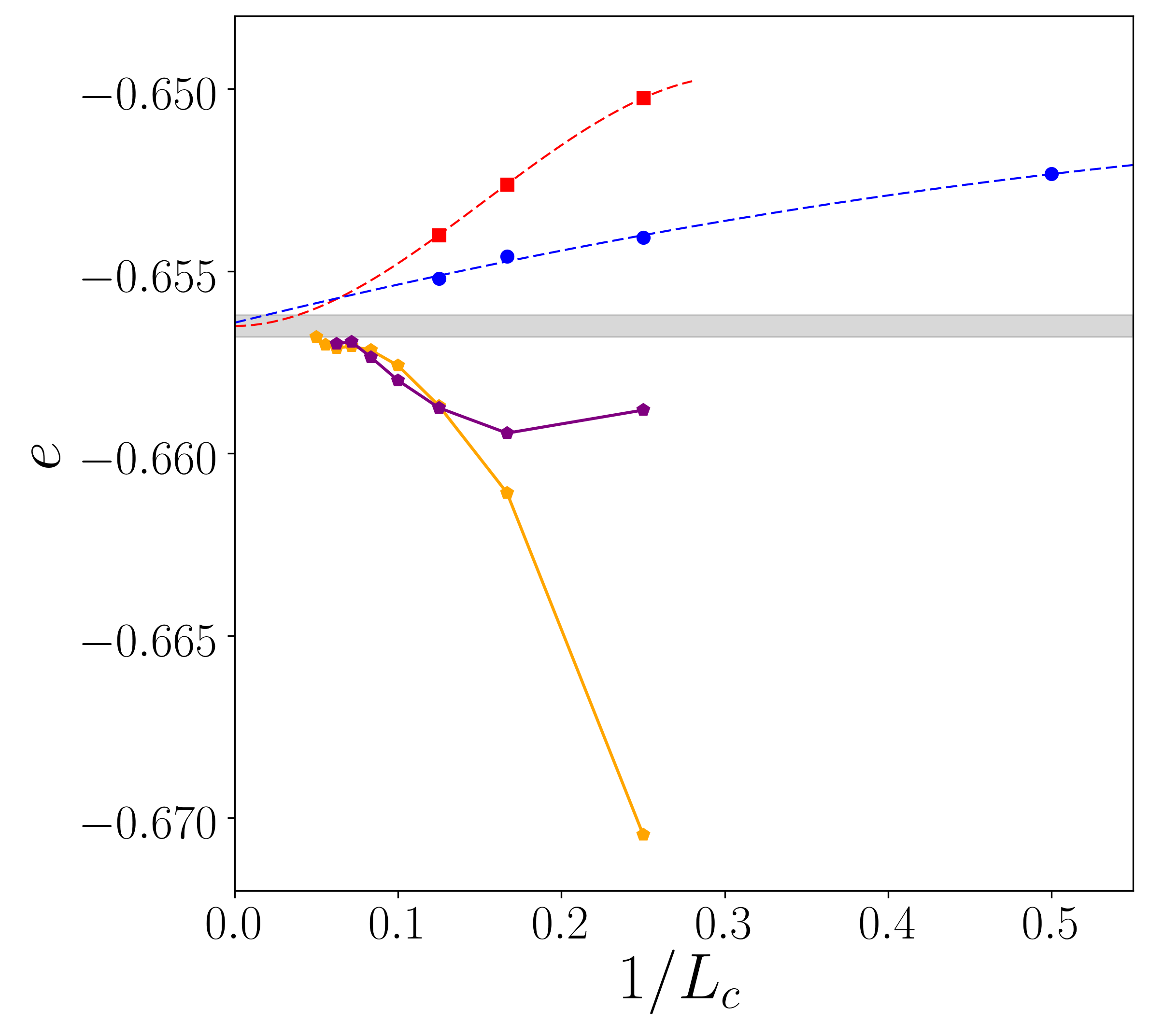}
	  \label{fig:scaling:energy2D_U6}
  }
  \caption{Energy per site $e$ versus $1/L_c$ in the 2D Hubbard model from
	  CDMET (blue),  DCA-DMET (red) and finite system AFQMC
    (orange: TABC, purple: PBC, brown: APBC for y-direction and PBC for x-direction) (from Ref.~\cite{Qin2016}).
    The consensus range illustrated by the grey-shaded region represents the TDL results of AFQMC, DMRG and iPEPS calculations in 
    Refs.~\cite{LeBlanc2015,Sorella2015,Corboz2016,Qin2016}.
  }
  \label{fig:scaling:Energy_2D}
\end{figure}

In Fig.~\ref{fig:scaling:Energy_2D}, we show the cluster size dependence of the
energy per site; the data is tabulated in Table~\ref{tab:scaling:Energy_2D}.
Because there are no exact TDL results for the 2D Hubbard model, we show gray
ribbons as ``consensus ranges'', obtained from the TDL estimates of several
methods including (i) AFQMC extrapolated to infinite size~\cite{Sorella2015,Qin2016},
(ii) DMRG extrapolated to infinite size~\cite{LeBlanc2015}, and (iii) iPEPS
extrapolated to zero truncation error~\cite{Corboz2016}. To show the effects
of embedding versus bare cluster AFQMC calculations we also plot the AFQMC
results of Ref.~\cite{Qin2016} on finite lattices with up to 400 sites, using
TABC for $U/t=2, 4, 6$, as well as periodic (PBC) and anti-periodic (APBC)
boundary conditions for $U/t=4$.

In 2D, both CDMET and DCA-DMET appear to display much higher accuracy for small
fragments, compared to in 1D. Although DMET is not exact in the infinite dimensional
limit, this is similar to the behavior of DMFT, which improves with increasing
coordination number~\cite{Georges1996}.
The DMET energies for each fragment size are, as expected, much closer to the TDL
estimates than the finite system AFQMC energies, even when twist averaging is employed
to reduce finite size effects. For example, the $2\times 2$ CDMET energy is competitive
with the $8 \times 8$ AFQMC cluster energy with twist averaging. Further, the convergence
behavior generally appears smoother in DMET than with the bare clusters, likely due to 
smaller shell filling effects. 
Combining these benefits, we find that using DMET gives several orders of magnitude
savings in computation time to achieve a given energy accuracy in the TDL estimate,
as compared to using bare cluster calculations alone.
This illustrates the benefits of using bath orbitals to approximately represent
the environment in an embedding.

\begin{table}[thpb]
\centering
\caption{Finite size extrapolation of the energy for the 2D half-filled Hubbard model. }
\label{tab:scaling:Energy_2D}

\begin{tabular}{p{0.15\columnwidth} p{0.21\columnwidth} p{0.18\columnwidth} 
	p{0.18\columnwidth} p{0.18\columnwidth}}
	\toprule
\multicolumn{2}{l}{\textbf{Methods}} & $\mathbf{U/t=2}$ & $\mathbf{U/t=4}$ & $\mathbf{U/t=6}$\\
\midrule
\multirow{2}{*}{CDMET} & $a + b/L_c$ & -1.1752(1) & -0.8601(1) & -0.6560(2) \\
		   & $a + b/L_c + c/L_c^2$ & -1.1756(3) & -0.8600(1) & -0.6564(6)\\
\midrule
\multirow{2}{*}{DCA-DMET} & $a + b/L_c^2$ &-1.1758(1)&-0.8593(2)&-0.6550(4)\\
& $a + b/L_c^2 + c/L_c^3$ &-1.1755(2)&-0.8600(2)&-0.6565$^a$
\\
\midrule
\multirow{2}{*}{AFQMC} & TABC~\cite{Qin2016} &-1.1760(2)&-0.8603(2)&−0.6567(3)\\
		& MBC~\cite{Sorella2015} & -1.17569(5)&-0.86037(6) & -\\
\midrule
\multicolumn{2}{l}{DMRG~\cite{LeBlanc2015}} &-1.176(1)&-0.8605(5)&-0.6565(1)\\
\midrule
\multicolumn{2}{l}{iPEPS~\cite{Corboz2016}} &-&-0.8603(5)&-\\
\midrule
\multicolumn{2}{l}{Consensus range} &-1.1758(3)&-0.8603(3)&-0.6565(3)\\
\bottomrule
\end{tabular}
\raggedright
{\footnotesize  $a$ Uncertainty cannot be computed due to insufficient data points in the fit.}
\end{table}

We now discuss our TDL estimates. As in the 1D Hubbard model, we use the scaling
forms proposed in Sec.~\ref{sec:dmet:practice:extrapolation}, i.e.
$a + b/L_c (+ c/L_c^2)$  for CDMET and $a + b/L_c^2 (+ c/L_c^3)$ for DCA-DMET.
The results are summarized in Table~\ref{tab:scaling:Energy_2D} and plotted in
Fig.~\ref{fig:scaling:Energy_2D}.
The TDL energy estimates fall within the TDL consensus range, with an error bar
competitive with the best large-scale ground state calculations.
The DMET estimates are also all in agreement (within 2$\sigma$) of our earlier
CDMET extrapolations that only used fragments of up to $4\times 4$ sites in
Refs.~\cite{Zheng2016,LeBlanc2015}. The largest deviation from our earlier small
fragment DMET extrapolations is for $U/t=2$ where finite size effects are strongest;
the current estimates of $-1.1756(3)$ (CDMET) and $-1.1755(2)$ (DCA-DMET)
can be compared with our small fragment estimate of $-1.1764(3)$, and
 the recent TDL estimate of Sorella of $-1.17569(5)$, obtained by extrapolating
 AFQMC energies from clusters as large as 1058 sites, using modified (periodic) boundary
 conditions~\cite{Sorella2015}.
Note that the subleading terms are more important for accurate extrapolations
in 2D than they are in 1D. This is simply because we do not reach as large linear
dimensions in 2D as in 1D, which means that we are not fully in the asymptotic regime.
For the same reason it is more difficult to see the crossover between the convergence
of DCA-DMET and CDMET. For $U/t=2$, it appears advantageous to use the DCA-DMET formulation
already for fragments of size $L_c\ge 4$, while at $U/t=4, 6$ it appears necessary
to go to fragments larger than the largest linear size used in this study, $L_c=10$.

\begin{table}
  \centering
  \caption{Estimated staggered magnetization for the 2D half-filled Hubbard model at TDL.}
  \label{tab:scaling:AF_2D}

 \begin{tabular}{p{0.36\columnwidth} p{0.18\columnwidth} 
	p{0.18\columnwidth} p{0.18\columnwidth}}
	\toprule
\textbf{Methods} & $\mathbf{U/t=2}$ & $\mathbf{U/t=4}$ & $\mathbf{U/t=6}$\\
\midrule
CDMET&0.115(2)&0.226(3)&0.275(8)\\
DCA-DMET&0.120(2)&0.227(2)&0.261$^a$\\
\midrule
DQMC~\cite{Varney2009}&0.096(4)&0.240(3)&0.283(5)\\
Pinning field QMC~\cite{Wang2014}&0.089(2)&0.215(10)&0.273(5)\\
AFQMC w. TABC~\cite{Qin2016}&0.119(4)&0.236(1)&0.280(5)\\
AFQMC w. MBC~\cite{Sorella2015}&0.120(5)&-&-\\
\bottomrule
  \end{tabular}
\raggedright
{\footnotesize  $a$ Uncertainty cannot be computed due to insufficient data points in the fit.}
\end{table}

\begin{figure}[thpb]
  \centering
  \subfigure[U/t = 2]{
	  \includegraphics[width=0.85\columnwidth]{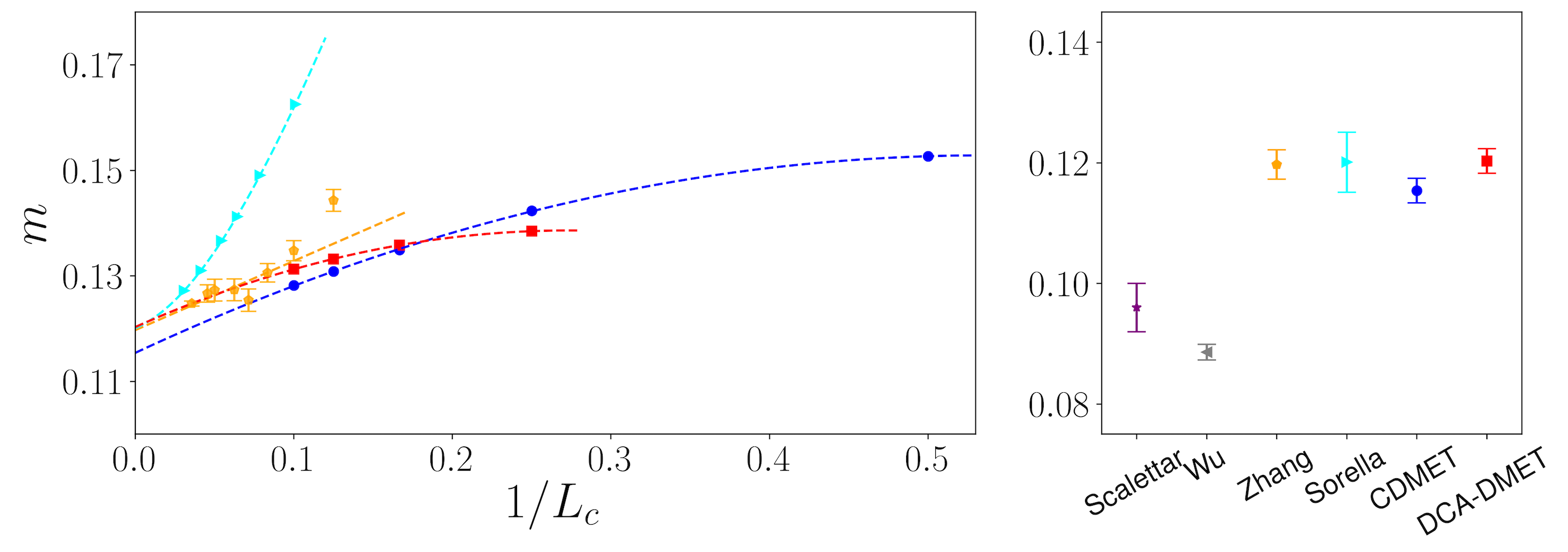}
	  \label{fig:scaling:AF2D_U2}
  }
  \subfigure[U/t = 4]{
	  \includegraphics[width=0.85\columnwidth]{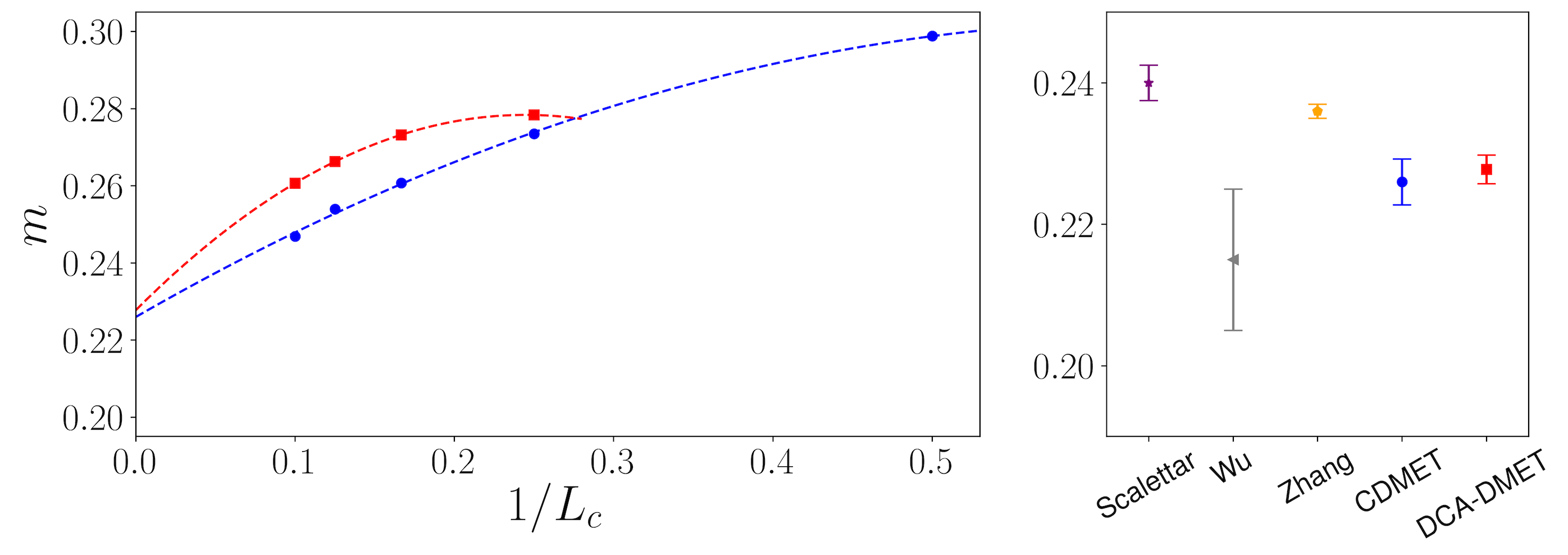}
	  \label{fig:scaling:AF2D_U4}
  }
  \subfigure[U/t = 6]{
	  \includegraphics[width=0.85\columnwidth]{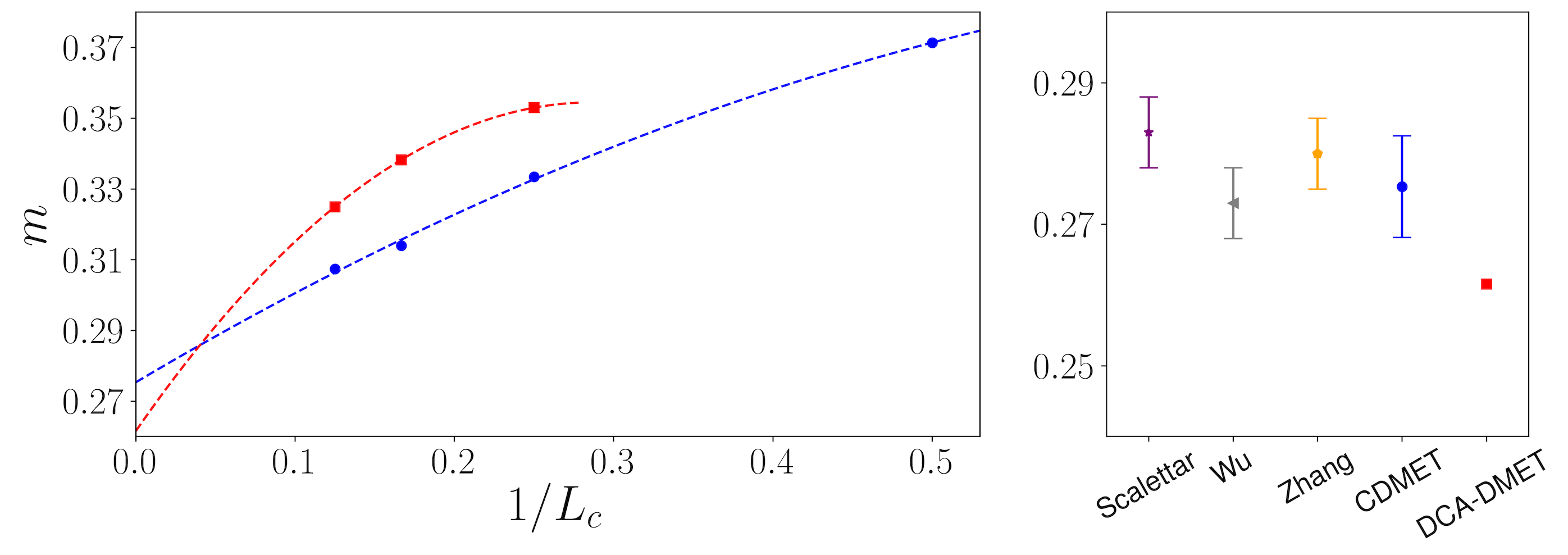}
	  \label{fig:scaling:AF2D_U6}
  }
  \caption{Left Panel: Antiferromagnetic order parameter $m$ versus $1/L_c$ in the 2D Hubbard
	  model from CDMET (blue), DCA-DMET (red) and finite system AFQMC using
	  TABC~\cite{Qin2016} (orange) and modified boundary conditions~\cite{Sorella2015}
	  (cyan).The DMET results extrapolate to the TDL uses the form
	  $m(L) = a + bL_c^{-1} + cL_c^{-2}$.
    Right Panel: CDMET and DCA-DMET TDL estimates with error bars including
    fitting and AFQMC statistical uncertainties, compared to the determinantal
    Monte Carlo simulations by Scalettar and coworkers~\cite{Varney2009},
    pinning field QMC simulations by Wu and coworkers~\cite{Wang2014},
    AFQMC with TABC by Qin et. al.~\cite{Qin2016} and the modified boundary
    conditions by Sorella~\cite{Sorella2015}.}
  \label{fig:scaling:AForder}
\end{figure}

The AF order in the half-filled 2D Hubbard model is long-ranged in the ground state.
In the left part of Fig.~\ref{fig:scaling:AForder}, the AF order parameters from DMET finite
fragment calculations are plotted and extrapolated, with the right panel showing comparisons
of TDL estimates with the other methods. In addition, we summarize the extrapolated TDL
estimates for the AF order parameters in Table.~\ref{tab:scaling:AF_2D}.
For CDMET, the order parameters are measured as the average magnitude of the central
plaquette. We fit the magnetization data to the form suggested in 
Section~\ref{sec:dmet:practice:extrapolation}, i.e.
$a+ b/L_c + c/L_c^2$ for both CDMET and DCA-DMET.
These fits lead to good agreement between the CDMET and DCA-DMET TDL estimates,
supporting the scaling form used. At $U/t=4$, the CDMET and DCA-DMET TDL moments are
in good agreement with the estimates from two different AFQMC calculations,
with competitive error bars. At $U/t=6$, the CDMET TDL moment is consistent with
the two AFQMC estimates although the DCA-DMET estimate is somewhat smaller than the
two AFQMC estimates. (We do not have errors bars for the $U/t=6$ DCA-DMET moment as
we are fitting 3 data points to a 3 parameter fit).

The TDL magnetic moment at $U/t=2$ is an example for which current literature
estimates are in disagreement. While earlier AFQMC calculations in 
Refs.~\cite{Varney2009,Wang2014,LeBlanc2015} appear to give an estimate close
to $m\sim 0.09$, the AFQMC estimates from recent work of Sorella~\cite{Sorella2015}
and Qin et al~\cite{Qin2016}\footnote{The AFQMC result of antiferromagnetic order
parameter at $U/t=2$ in Ref.~\cite{LeBlanc2015} has an error in the extrapolation
to the TDL, which was corrected in Ref.~\cite{Qin2016}.}
using larger clusters and modified and twist
average boundary conditions predict a moment of $m \sim 0.120(5)$ and $0.119(4)$,
respectively. This is much closer to our earlier DMET result of $m \sim 0.133(5)$
extrapolated from small clusters of up to $4\times 4$ in size. Revising this with the
larger CDMET and DCA-DMET clusters in this work we can now confirm the larger value
of the TDL magnetic moment, $m \sim 0.115(2)$ (CDMET) and $m \sim 0.120(2)$ (DCA-DMET)
with very small error bars.
The underestimate of the moment seen in earlier QMC work is likely due to the non-monotonic
convergence of the moment with cluster size when using PBC, as identified in
Sorella's work~\cite{Sorella2015}. In contrast to PBC calculations and the TABC
calculations shown here (orange) which display some scatter, the dependence on fragment
size is very mild once embedding is introduced. This once again highlights the ability
of the embedded approach to capture some of the relevant aspects even of long-wavelength
physics, leading to good convergence of local observables.

\section{Conclusions} \label{sec:scaling:conclusion}
We carried out a detailed study of the fragment size convergence of density matrix
embedding theory, using an auxiliary field quantum Monte Carlo solver (AFQMC) in
order to large cluster sizes inaccessible from other impurity solvers.
In addition to the cluster density matrix embedding formulation (CDMET),
we study the ``dynamical cluster'' variant (DCA-DMET) that restores translational
invariance in the fragment and accelerates finite size convergence.
Using the half-filled one- and two-dimensional Hubbard models where AFQMC has
no sign problem, as examples, we numerically explored the finite size convergence
of the energy and the magnetization. The energy convergence of CDMET and DCA-DMET
goes like $O(1/L_c)$ and $O(1/L_c^2)$ respectively, where $L_c$ is the linear dimension
of the fragment, similar to that observed in cellular dynamical mean-field theory
and the dynamical cluster approximation.
The convergence of the magnetization follows a scaling relation related to the
magnetic correlation function in CDMET, while the DCA-DMET converges more quickly
than CDMET.
In the case of the 2D Hubbard model, our thermodynamic limit extrapolations from
both CDMET and DCA-DMET are competitive with the most accurate estimates
in the literature, and in the case of $U/t=2$ where finite size effects are
particularly strong,  help to determine the previously uncertain magnetic moment.

In all the cases we studied in this chapter, the use of density matrix embedding,
as compared to computations using bare clusters with any form of boundary condition,
significantly decreased the computational cost required to obtain a given error in
the TDL estimate, sometimes by orders of magnitudes. Since the computational scaling
of the AFQMC solver employed here is quite modest with fragment size (cubic) this
improvement would only be larger when using other, more expensive solvers.

The DCA-DMET formulation appears superior for large fragments due to the faster
asymptotic convergence, however, it is typically less accurate for small fragments
than CDMET. When performed in conjunction, the consistency of TDL estimates from
CDMET and DCA-DMET serves as a strong check on the reliability of the DMET TDL
extrapolations.
\chapter[Ground-State Phase Diagram of the Square Lattice Hubbard Model]
{Ground-State Phase Diagram of the Square Lattice Hubbard Model~\footnote{Based on work published in Phys. Rev. B \textbf{93}, 035126 (2016). Copyright 2016, American Physical Society.~\cite{Zheng2016}}} \label{chpt:diagram}
In this chapter, we present the ground-state phase diagram of the Hubbard and
frustrated Hubbard models (with non-zero next-nearest-neighbor hopping) on the
square lattice with density matrix embedding theory using clusters of up to 16 sites.
We provide an error model to estimate the reliability  of the computations and
complexity of the physics at different points in the diagram.
We find superconductivity in the ground-state as well as competition between
 inhomogeneous charge, spin, and pairing states at low doping.
The estimated errors in the study are below $T_c$ in the cuprates and on the
scale of contributions in real materials that are neglected in the Hubbard model.

In Sec.~\ref{sec:diagram:intro}, we introduce the background of the study, focusing
on the physics of the 2D Hubbard model. In Sec.~\ref{sec:diagram:method}, we
present the detailed specifications of our DMET calculations and
in a separate Section~\ref{sec:diagram:error} we discuss the techniques
to quantify the errors in the TDL estimates.
We present the results and conclusions of our
calculations in Sec.~\ref{sec:diagram:result} and
Sec.~\ref{sec:diagram:conclusion}.

\section{Introduction} \label{sec:diagram:intro}

The Hubbard model~\cite{PhysRevLett.10.159,Kanamori01091963,hubbard1963electron}
is one of the simplest quantum lattice models of correlated electron materials.
Its one-band realization on the square lattice plays a central role in understanding
the essential physics of high temperature  superconductivity~\cite{Zhang1988,Dagotto1994}.
Rigorous, near exact results are available
in certain limits~\cite{scalapino2007numerical}:
at high temperatures from series expansions~\cite{PhysRevLett.97.187202,Khatami2011,Khatami2015,PhysRevB.31.4403},
in infinite dimensions from converged dynamical mean-field theory~\cite{georges1992hubbard,Georges1996,PhysRevB.77.033101,PhysRevB.86.125114}, and 
at weak coupling from perturbation theory~\cite{schweitzer1991weak} and
renormalization group analysis~\cite{halboth2000renormalization,raghu2010superconductivity}.
Further, at half-filling, the model has no
fermion sign problem, and unbiased determinantal quantum Monte Carlo simulations
can be converged~\cite{Varney2009}.
Away from these limits, however, approximations are necessary.
Many numerical methods have been applied to the model at both 
finite and zero temperature, including fixed-node, constrained-path,
determinantal, and variational quantum Monte Carlo (QMC)~\cite{PhysRevB.58.R14685,Becca2000,PhysRevLett.72.2442,Tocchio2008,PhysRevB.55.7464,PhysRevB.78.165101,Chang2010,yokoyama1987variational,PhysRevB.76.180504,yamaji1998variational,PhysRevB.43.12943},
density matrix renormalization group (DMRG)~\cite{white2000phase,scalapino2001numerical,white2003stripes},
and dynamical cluster (DCA)~\cite{PhysRevB.58.R7475,PhysRevB.61.12739},
(cluster) dynamical mean-field theories (CDMFT) ~\cite{Lichtenstein2000,PhysRevLett.87.186401}, and variational cluster approximations (VCA)~\cite{Potthoff2003,Dahnken2004}.
(We refer to DCA/CDMFT/VCA collectively as Green's function cluster theories).
These pioneering works have revealed rich phenomenology in the phase diagram
including metallic, antiferromagnetic,  $d$-wave (and other kinds of) superconducting
phases, a pseudogap regime, and inhomogeneous orders such as stripes, and charge,
spin, and pair-density waves, as well as phase separation~\cite{scalapino2007numerical,PhysRevB.43.12943,PhysRevB.58.R14685,
Lichtenstein2000,Becca2000,yamaji1998variational,white2003stripes,
PhysRevB.76.180504,PhysRevB.78.165101,Chang2010,Aichhorn2005,Senechal2005,
Aichhorn2006,Halboth2000,Schulz1990,White1989,Chubukov1995,Igoshev2010,Capone2006,
Gull2013,Kato1990,Moreo1990,Miyazaki2002,Mizusaki2006,Maier2006,Maier2005,
Macridin2004,Jarrell2001,Otsuki2014,Chen2012}. However, as different numerical
methods have yielded different pictures of the ground-state phase diagram, a
quantitative picture of the ground-state phase diagram has yet to emerge.

It is the goal of this numerical study to produce such a quantitative picture as
best as possible across the \textit{full} Hubbard model phase diagram
below $U/t = 8$. We perform the calculations on the Hubbard model
with the density matrix embedding theory (DMET), which is very accurate
in this regime~\cite{Knizia2012,Knizia2013,Chen2014,Bulik2014,
Fetter2003,sun2014exact,bulik2014electron,Booth2013},
together with clusters of up to 16 sites and thermodynamic extrapolation.
We carefully calibrate errors in our calculations, giving error bars to
quantify the remaining uncertainty in our phase diagram.
These error bars also serve, by proxy, to illustrate
the complexity of the underlying physics for different Hubbard parameters.
The accuracy we achieve is significantly higher than attained by earlier
comparable Green's function cluster calculations for the ground state.
We also carefully estimate the finite size effects, which we find to have
a crucial impact on the location of the phase boundaries of the antiferromagnetic
and $d$-wave superconducting (SC) orders, in contrast to some early ground-state
studies~\cite{Aichhorn2006}.

\section{Methods} \label{sec:diagram:method}
We study the one-band (frustrated) Hubbard model on the $L\times L$
square lattice
\begin{equation}
    H=-t\sum_{\langle ij\rangle \sigma}a_{i\sigma}^{\dagger}a_{j\sigma}-t^\prime\sum_{\langle\langle ij\rangle \rangle \sigma}a_{i\sigma}^{\dagger}a_{j\sigma}
    +U\sum_{i}n_{i\alpha}n_{i\beta}
  \label{eq:diagram:hubbard}
\end{equation}
where $\langle \ldots\rangle$ and $\langle \langle \ldots\rangle \rangle $ 
denote nearest and next-nearest neighbors, respectively, 
$a_{i\sigma}^{(\dagger)}$  destroys (creates) a particle on site $i$ with
spin $\sigma$, and $n_{i\sigma}=a_{i\sigma}^{\dagger}a_{i\sigma}$ is the
number operator.
The standard Hubbard model corresponds to $t'=0$ (we fix $t=1$).
We further study the frustrated model with  $t'=\pm 0.2$.

We use the broken particle-number symmetry, spin-unrestricted DMET formulation
in this chapter. The basic principles of the method are outlined in
Sec.~\ref{sec:dmet:bcs}, with detailed equations and algorithms presented
in Appendix~\ref{sec:algo:bcs}.
To obtain the ground-state phase diagram, we carry out
DMET calculations using 2$\times$2, 4$\times$2, 8$\times 2$,
and 4$\times$4 fragments, cut from a bulk square lattice with $L=72$.
We considered $t'=0, 0.2, -0.2$, and $U=2,4,6,8$, and various densities
between $n=0.6-1$.
The impurity model ground-state $|\Psi\rangle$ is determined using
a DMRG solver with a maximum number of renormalized states $M=2000$
(DMET self-consistency is performed up to $M=1200$),
and which allows for $U(1)$ and $SU(2)$ spin symmetry breaking (see
Sec.~\ref{sec:dmet:solver:dmrg} and Appendix~\ref{sec:formula:dmrg_bcs}).
The energy per site $e$, local spin moment
$m=\frac{1}{2}(n_{i\alpha}-n_{i\beta})$,
double occupancy $D=\langle n_{i\alpha} n_{i\beta}\rangle$,
and local $d$-wave pairing $d_{sc}=\frac{1}{\sqrt{2}}
(\langle a_{i\alpha}a_{j\beta}\rangle+\langle a_{j\alpha}a_{i\beta}\rangle)$
were measured from the fragment part of $|\Psi\rangle$.

Local observables either are close to uniform in the entire
fragment or exhibit tendencies to inhomogeneity.
In the former case, we average the local observables in the central
plaquette (so the average is consistent for all fragment sizes)
and extrapolated to the thermodynamic limit
(TDL) using a linear relationship with $N_c^{-1/2}$
as verified in Chapter~\ref{chpt:scaling}.
For the later case, reliable average and extrapolation are usually not
possible, and we report these case as inhomogeneous order where
a single order parameter cannot be extracted.

\section{Error Estimation} \label{sec:diagram:error}
The finite fragment DMET energies and measurements
contain 3 sources of error relative to the exact TDL. These are from
(i) errors in the DMET self-consistency, (ii) finite $M$
in the DMRG solver (This is only significant for the $8\times 2$
and $4\times 4$ clusters, corresponding to 32 fragment plus bath
sites in the impurity model.),which also induces error in the
correlation potential $u$, (iii) finite \textit{fragment} size.
The error from the use of a finite $72\times 72$ bulk lattice is so small
as to not affect any of the significant digits presented here.

To estimate the TDL result, we (i) estimate DMET self-consistency
quality by the convergence of expectation values in the last iterations,
(ii) extrapolate DMRG energies and expectation values
at finite $M$ to infinite $M$, using the linear relation with DMRG density
matrix truncation error~\cite{white2007neel}, (iii) estimate the error in
$u$ due to finite $M$, by extrapolating expectation values from
self-consistent $u(M)$ obtained with different solver accuracy,
(iv) extrapolate fragment size to infinite size, with
the  $N_c^{-1/2}$ scaling appropriate to a non-translationally-invariant
fragment (Sec.~\ref{sec:dmet:practice:extrapolation}).
Each of (i) to (iv) gives an estimate of an uncertainty component
(for linear extrapolations, we use the 1$\sigma$ standard deviation),
which we combine to obtain a single error bar on the DMET TDL estimates.

\begin{figure}[htpb]
  \begin{overpic}[width=0.65\columnwidth]{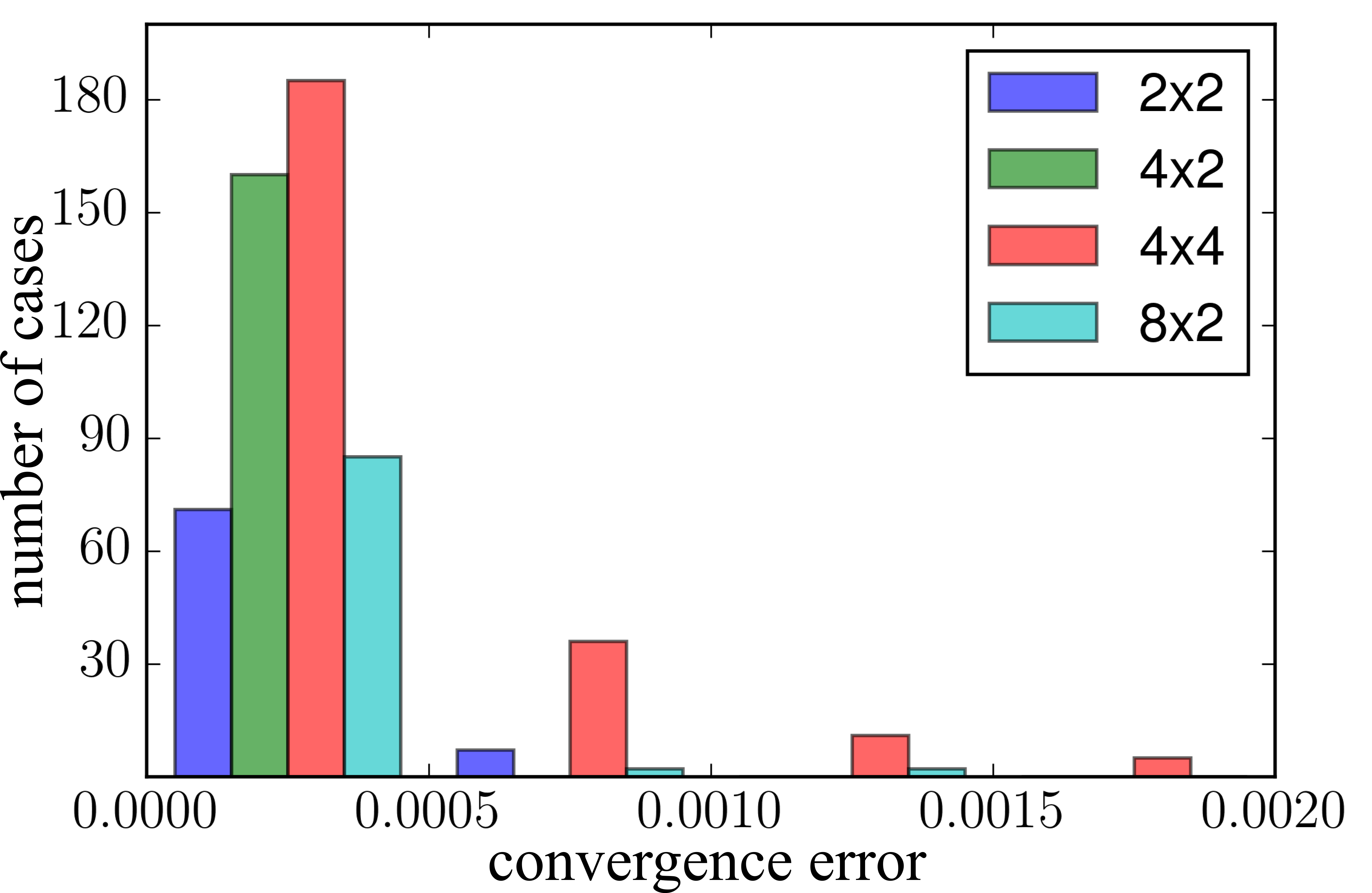}
    \put (100,37) {
        \begin{tabular}{ccc}
          \toprule
	  \textbf{Shape} & \textbf{Mean} $\mathbf{(\times10^{-4})}$&\\
	  \midrule
          2$\times$2&2.2\\
          4$\times$2&1.8\\
          4$\times$4&3.7\\
          8$\times$2&2.0\\
          \bottomrule
        \end{tabular}
    }
  \end{overpic}
  \caption{Distribution and average value (side table) of the DMET
  self-consistency error in the energy per site (units of $t$)
  for each fragment shape.}
  \label{fig:diagram:conv_err}
\end{figure}

We use the energy per site as an example to discuss the error estimation
in detail. The DMET self-consistency error is estimated as
$\frac{1}{2}|e^{(n-1)}-e^{(n)}|$, where $e^{(n)}$ and $e^{(n-1)}$ are
the energies per site of the last two DMET self-consistency iterations.
A typical DMET calculation oscillates between two slightly different
solutions with the magnitude of the oscillations decreasing with the
number of iterations. We thus use the range of oscillation as a
representation of the self-consistency error. The error distributions
across the range of calculations in this chapter are shown in
Fig.~\ref{fig:diagram:conv_err}, with the average values on the side.
For most points in the phase diagram, and for all fragment sizes,
the self-consistency error is less than $0.0005t$.
For $4\times4$ fragments DMET calculations are the harder to converge,
due to larger error in the impurity solvercalculations, giving a largest
error of up to $0.002t$, and an average self-consistency error approximately
twice as large as that for the other fragment shapes.

For fragments larger than the $2\times$2 fragment (where
our DMRG solver is not exact), there is the error due to using
finite $M$ in the DMRG impurity solver. The  error due to finite
$M$ has two components:
\begin{enumerate}
  \item  variational error in the DMRG calculation, which is
	  usually assumed proportional to the density matrix
	  truncation error $\delta_w$,
  \item the DMET correlation potential error $\delta_u$, as
	  $\delta_u$ is a function of the impurity density matrices,
	  and these have an error for finite $M$.
\end{enumerate}

For the $4\times2$ and $8\times2$ fragments, the second source
$\delta_u$ appears negligible. For these clusters, we carry out the DMET
self-consistency with smaller $M$ to obtain the converged correlation
potential $u$, then do a few DMRG calculations using large $M$ with
fixed correlation potential $u$ to extrapolate to the $M\to \infty$
exact solver limit, thus obtaining the first error component due to
the DMRG solver.

\begin{figure}[htpb]
  \subfigure[$U/t=4$  $e = -1.033(2)$]{
   \includegraphics[width=0.31\columnwidth]{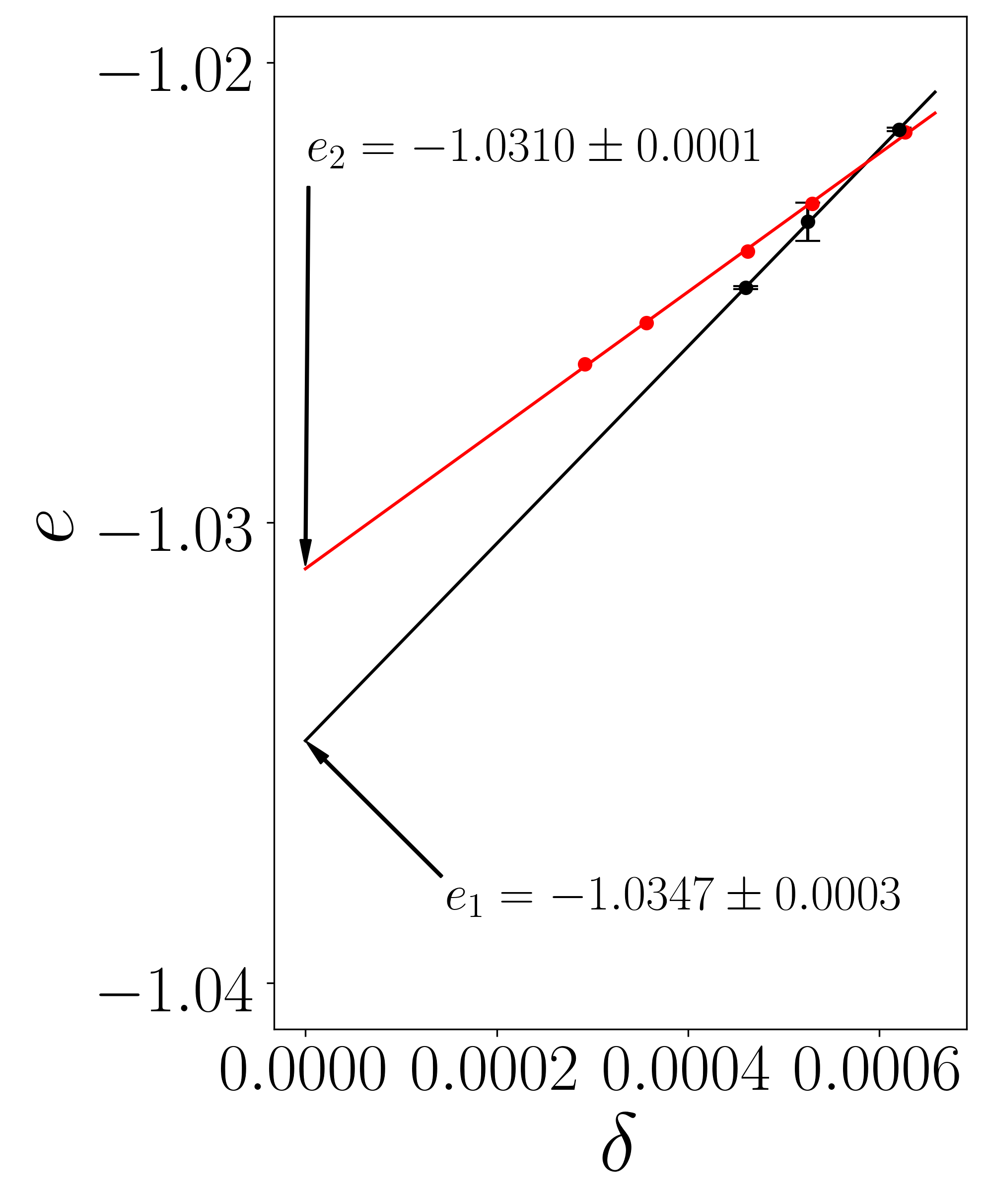}
    \label{fig:diagram:M_extrap_U4}
  }
  \subfigure[$U/t=6$  $e = -0.866(2)$]{
    \includegraphics[width=0.31\columnwidth]{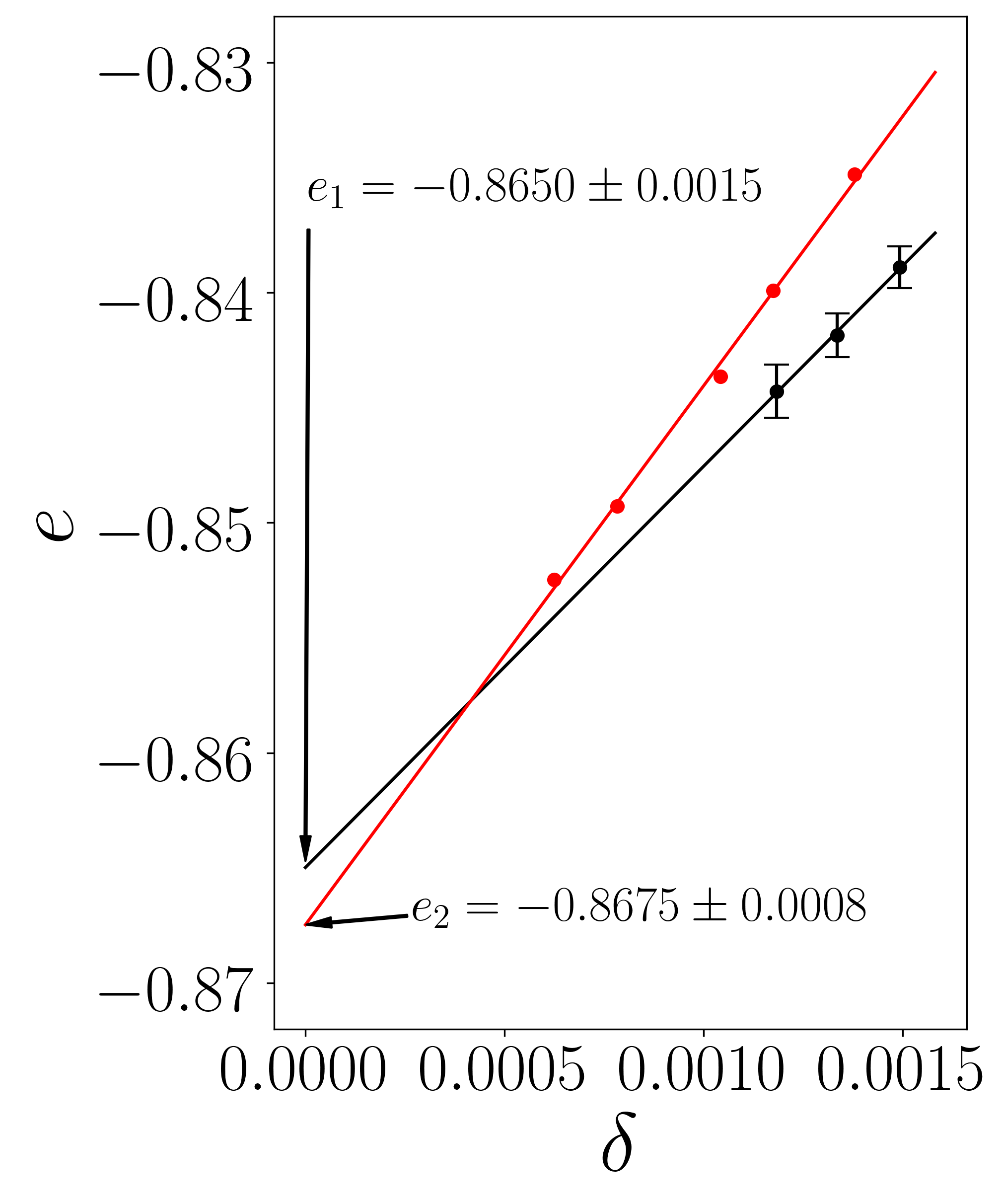}
    \label{fig:diagram:M_extrap_U6}
  }
  \subfigure[$U/t=8$  $e = -0.748(4)$]{
   \includegraphics[width=0.31\columnwidth]{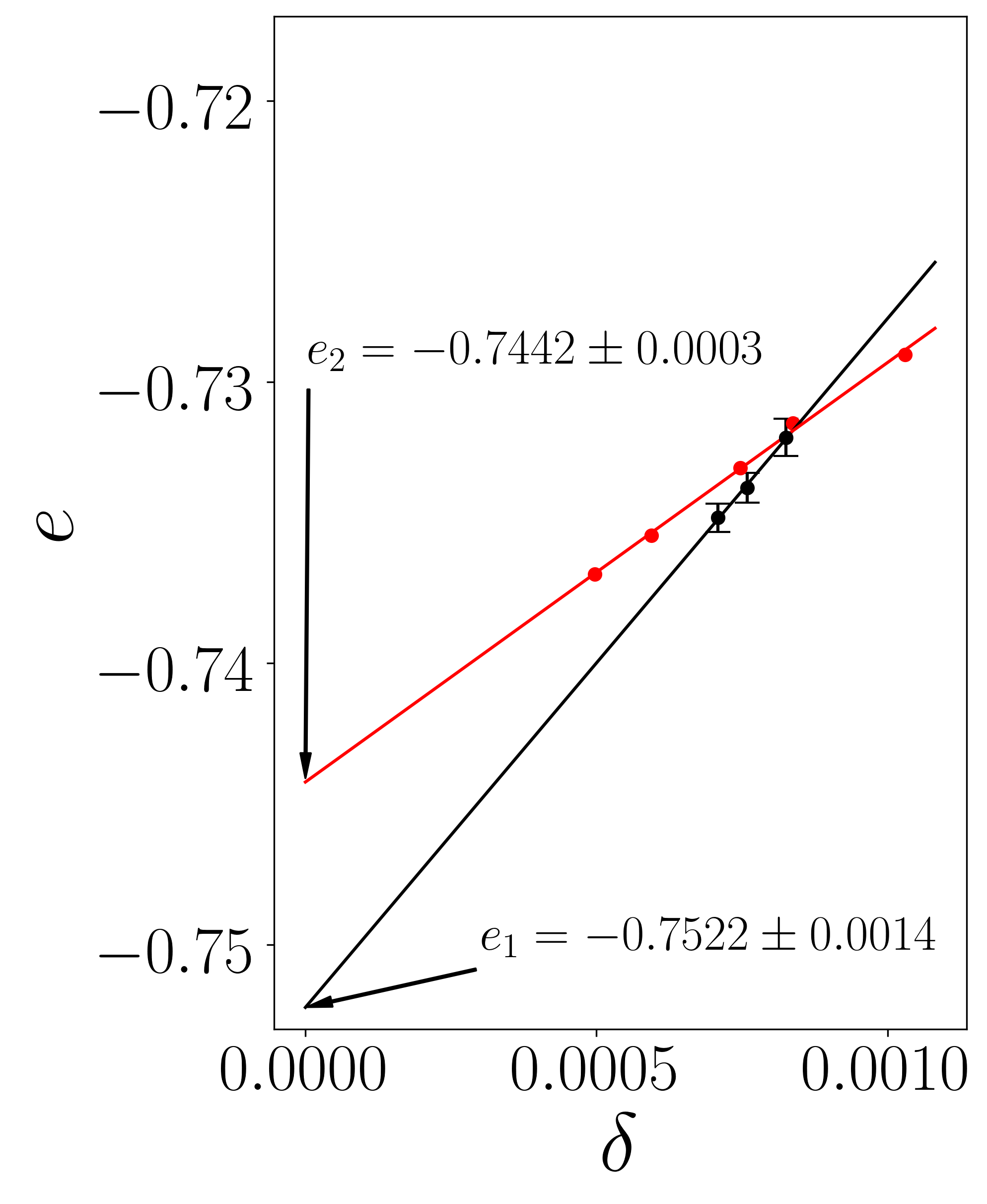}
    \label{fig:diagram:M_extrap_U8}
  }
  \caption{Estimation of the DMRG $M=\infty$ energy per site and the
	  associated error bars due to finite $M$, for $4\times4$ fragments.
  (See the text for detail.) The plots are shown for $t'=0,n=0.875$.}
  \label{fig:diagram:M_extrap}
\end{figure}

For $4\times4$ fragments, the $U/t=2$ data is processed in this way as well.
However, for other values of $U$ using the $4\times4$ fragments,
the DMRG truncation error can reach $10^{-3}$ for low to intermediate
doping with computational tractable $M$, making the contribution from
$\delta_u$ significant. To compensate for this, we first carry out the
DMET self-consistency cycles with a series of different $M$'s up to 1200
, and linearly extrapolate the energy to the $M=\infty$ limit, $e_1$.
This thus extrapolates errors from {\it both} source 1 and 2,
assuming $\delta_u\propto\delta_w$. Another further set of DMRG
calculations are done with $M$ up to 2000, using the converged
correlation potential from the DMET self-consistency with the
largest $M$. This second set of results are then extrapolated
again against the truncated error to obtain an energy $e_2$,
which only accounts for the error from source 1.
Although the linear relation between the source 2 error and
the truncation error need not hold in general, in practice,
we find that $\delta_u=\frac{1}{2}|e_1-e_2|$ gives a reasonable
estimate of $\delta_u$.
Therefore, we report the $4\times4$ fragment energy per site
as $e_{4\times4}=\frac{1}{2}(e_1+e_2)$, with a final uncertainty of
\begin{equation}
	\delta e_{4\times4}^2=\delta_u^2+\delta e_1^2+\delta e_2^2
	\label{eq:diagram:error_4x4}
\end{equation}
where $\delta e_1$ is a combination of the linear regression
uncertainty and the uncertainties of the original data points
(from DMET self-consistency error), while $e_2$ does not have
any self-consistency error.
Fig.~\ref{fig:diagram:M_extrap} illustrates the set of
computations and linear extrapolations performed with each
$4\times4$ cluster to obtain the $4\times 4$ cluster energy
and error estimate.

After obtaining the energy per site and other observables for each
fragment size, we extrapolate to the TDL using the relation
$\Delta e_{N_c}\propto N_c^{-1/2}$.
Since both the $4 \times 4$ and $2\times 8$ fragments are of 16 sites, 
we must choose which one to use in the extrapolation. We believe that
the $4\times4$ fragments have less finite size error than 
the $8 \times 2$ fragments, and thus we generally use these in
the  extrapolation. However, at certain points in the phase diagram
(e.g. at strong coupling, or negative $t^\prime$) 
there is a strong tendency towards inhomogeneity, and 
the new order the new order the $4\times4$ fragments,
resulting in a much higher energy than for the $8\times 2$ fragments.
In such cases, namely, when (a) the $4\times4$ and $8\times2$ fragments
show different orders, and (b) the $8\times2$ cluster is lower
in energy, we use the $8\times2$ cluster energy for the extrapolation.

\begin{figure}[htpb]
  \centering
  \includegraphics[width=\columnwidth]{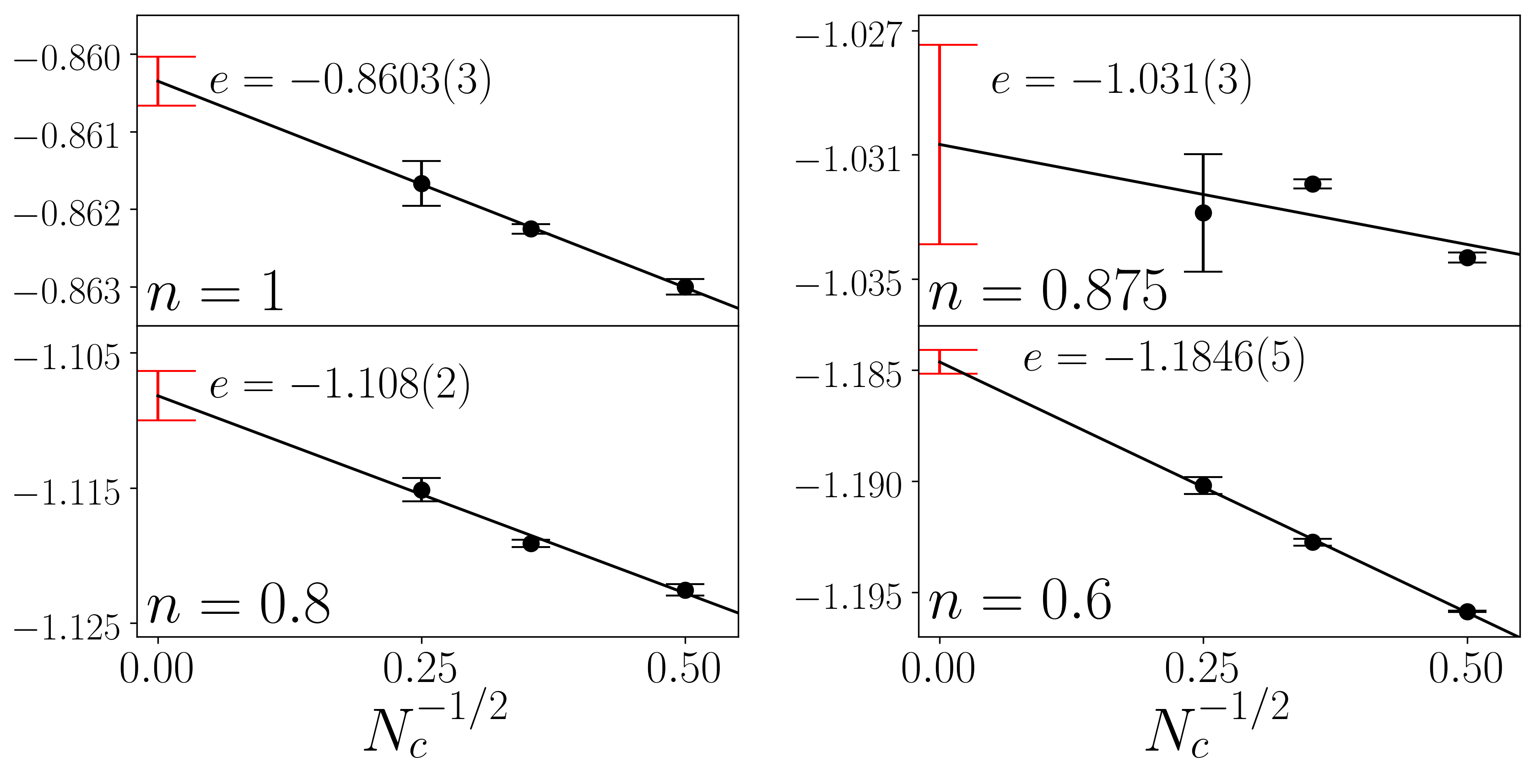}
  \caption{Fragment size extrapolation of the energies per site for $U=4, t'=0$
	  at various fillings. The black dots are finite size results. The red
	  error bars are the confidence intervals for the TDL estimates.}
  \label{fig:diagram:size_extrap}
\end{figure}

The fragment size extrapolation works considerably well given the limited
number of data points and small sizes of the clusters, although it contributes
the main source of error in the uncertainty of TDL estimates.
In Fig.~\ref{fig:diagram:size_extrap} we show examples of the finite size extrapolation
at $U/t=4$. The error bars shown for TDL estimates include both the error from
the linear extrapolation and the propagated uncertainties from the finite
fragment calculations.
At half-filling and in the overdoped region ($n<0.8$), the
linear relation appears quite good even with these small clusters. In the
underdoped region, however, the energy per site is more strongly dependent
on the cluster shape, often because the system has a strong tendency to
establish an inhomogeneous phase (see Sec.~\ref{sec:diagram:result} for
detail). Nontheless, even in the underdoped region, the error model appears
to give a reliable estimate of the energy per site at the TDL, albeit with
a large uncertainty.

\begin{figure}[htpb]
  \centering
  \subfigure[$t^\prime=0.2$]{
    \includegraphics[width=\columnwidth]{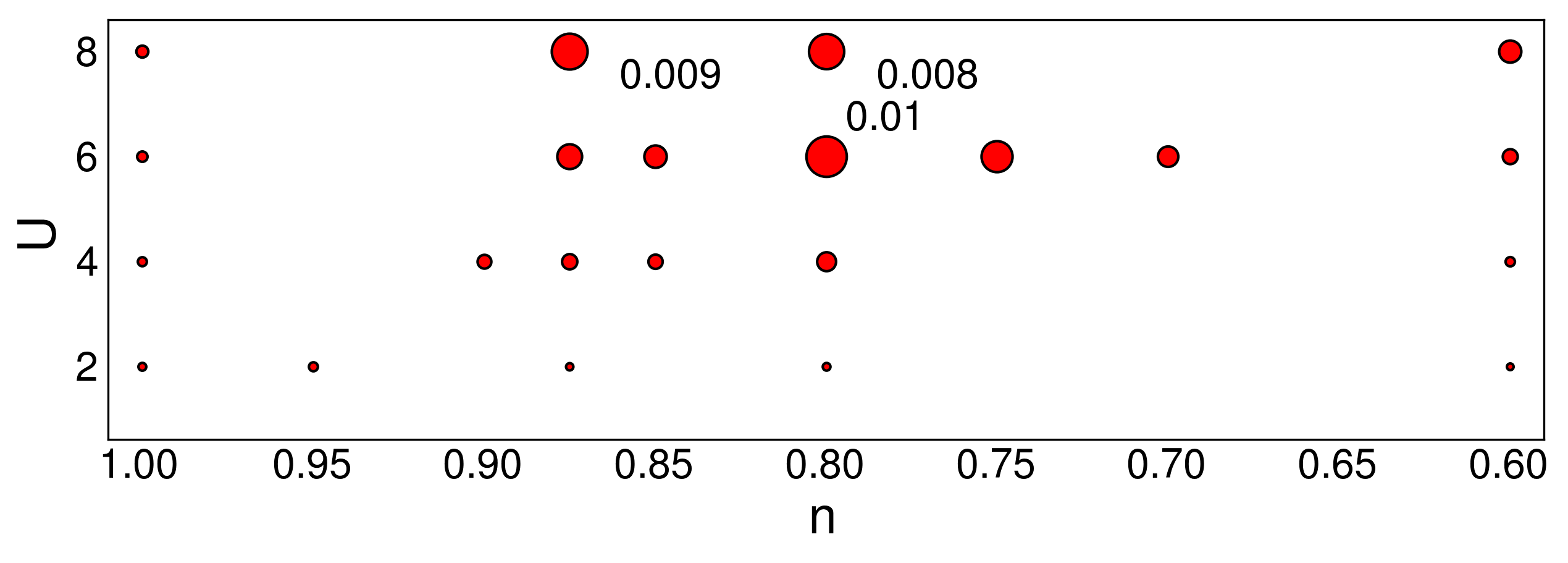}
    \label{fig:diagram:error_p}
  }
  \subfigure[$t^\prime=0$]{
    \includegraphics[width=\columnwidth]{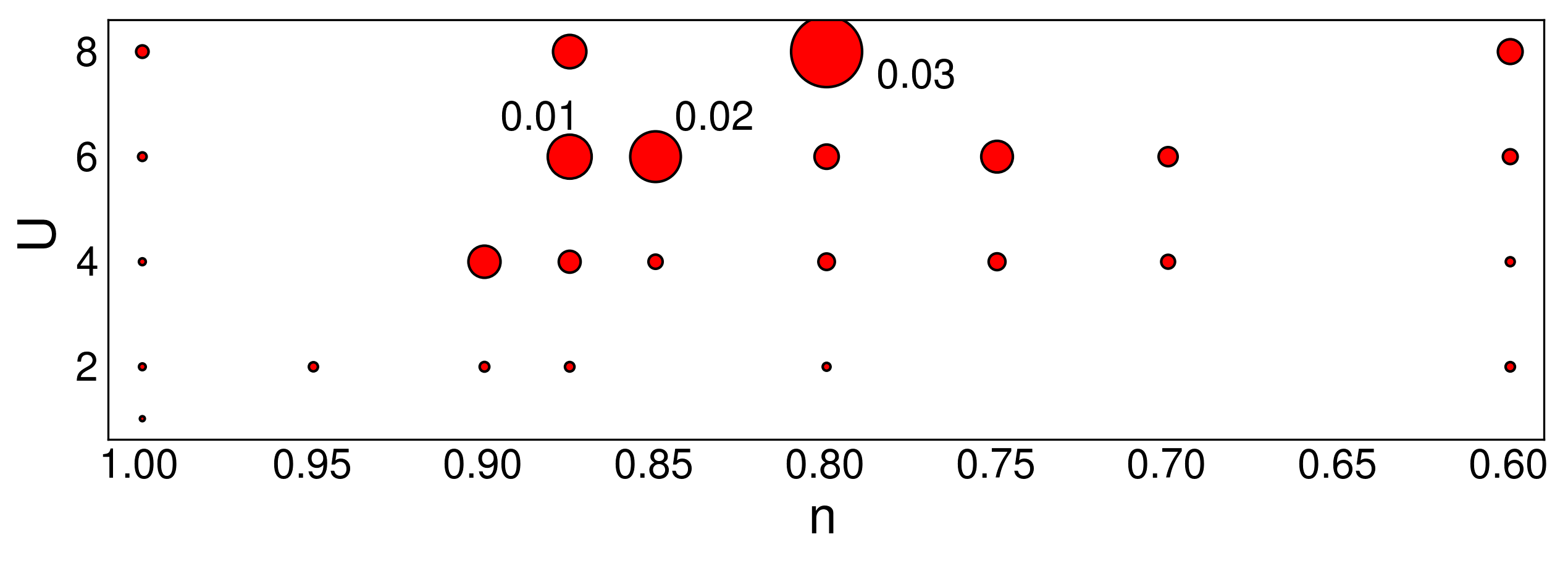}
    \label{fig:diagram:error_0}
  }
  \subfigure[$t^\prime=-0.2$]{
    \includegraphics[width=\columnwidth]{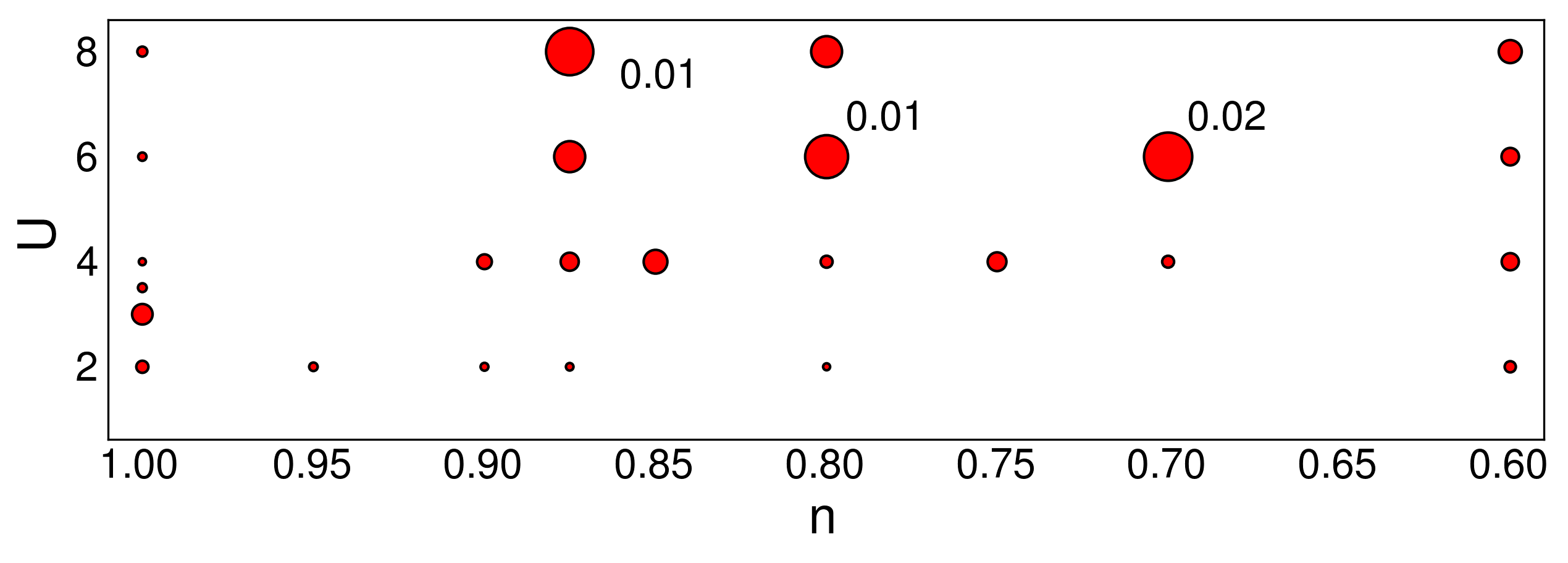}
    \label{fig:diagram:error_n}
  }
  \caption{Numerical uncertainty map of DMET energies per site for the (frustrated)
  Hubbard model with $t^\prime=0.2, 0$ and $-0.2$. The areas of the circles are
  proportional to the estimated uncertainties.}
  \label{fig:diagram:error}
\end{figure}

Fig.~\ref{fig:diagram:error} shows the uncertainties of the energy per site TDL
estimates for all the points in our calculations. As one would normally expect,
the accuracy away from half filling is significantly lower than at half
filling, with the largest errors found in the underdoped region of $n = 0.8\sim 0.9$,
where the solution is sensitive to cluster shapes because of phase boundaries
and/or the onset of competing inhomogeneous orders, thus introducing the largest
errors from fragment size extrapolation. We also see from the
Fig.~\ref{fig:diagram:error} that the maximum uncertainty for $t^\prime=0.2$
is smaller than that for $t^\prime=0$ and $t^\prime=-0.2$
($0.01t$ versus $0.03t$ and $0.02t$, respectively), implying the completing
order effect is weaker for positive $t^\prime$ than in the zero or
negative $t^\prime$ case.

We discussed energies per site in this section as an example. The order parameters
are extrapolated to the TDL and the uncertainties associated with them are
estimated, following the same procedure. Whenever necessary, we use the
$Z$-score (estimated value divided by its standard deviation) to describe
the robustness of the order.

\section{Results} \label{sec:diagram:result}
The quantitative analysis of the sources of error gives us confidence to
distinguish the physics from artifacts in the orders we see from DMET calculations.
We now look at the calculation results and try to interpret them, with the map
of uncertainties (Fig.~\ref{fig:diagram:error}) in mind.

In Sec.~\ref{sec:diagram:result:overview}, we present the phase diagram using
our TDL estimates of the all the parameter sets. In the subsequent sections
we analyze the results of different regions of the phase diagram in greater detail.

\subsection{Overview} \label{sec:diagram:result:overview}

\begin{figure}[htpb]
  \centering
  \includegraphics[width=\columnwidth]{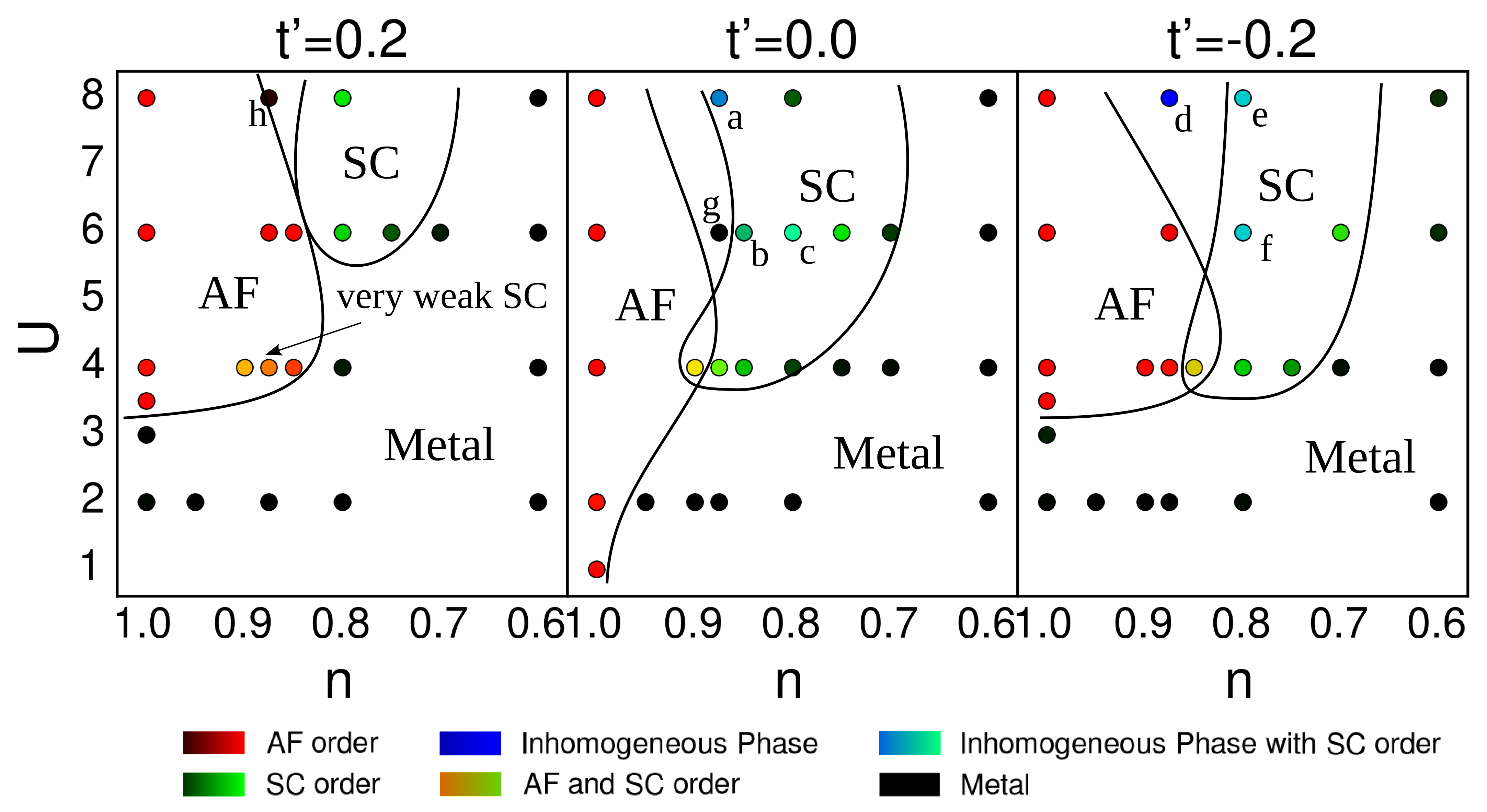}
  \caption{Phase diagrams of the standard and frustrated Hubbard models.
  Orders are represented with three primary colors:
  red (antiferromagnetism), green ($d$-wave superconductivity) and
  blue (inhomogeneity),
  with the brightness proportional to the robustness of the order.
  The points highlighted with letters: (a) local phase separation;
  (b) $d$-wave SC with a slight modulation in $(\pi,\pi)$ direction;
  (c) SC with a weak spin density wave (SDW);
  (d) a ``classic'' stripe phase; (e) stripe with pair-density wave (PDW)
  coexisting with SC;
  (f) CDW and spin $\pi$-phase shift;
  (g) and (h) intermediate points between AF and SC where both order
  parameters extrapolate to zero. Phase boundaries are guides only.
  \label{fig:diagram:phase}
}
\end{figure}

We present the DMET phase diagrams in Fig.~\ref{fig:diagram:phase}.
Interestingly, they feature many behaviors previously proposed in different studies.
In particular, we observe (i) an antiferromagnetic (AF) phase at half-filling;
(ii) a metallic phase at large dopings and at small $U$, enhanced by frustration;
(iii) a region of $d$-wave superconducting (SC) order at intermediate dopings and
sufficiently large $U$; (iv) a region of coexisting AF and SC order; (v) a region
rich with inhomogeneous charge, spin, and superconducting orders that are very
sensitive to the Hubbard parameters; (vi) points in between the AF and SC phase
where the AF and SC orders extrapolate to zero.
(The metallic phase is predicted to be unstable at  weak coupling and large
dopings from weak coupling expansions~\cite{Metzner1989,raghu2010superconductivity},
but the relevant parameter region is outside the scope of this study.)
At $t^\prime = 0$,for $U/t = 8,n = 0.875$,an SC state with strong inhomogeneity
appears which creates large uncertainties in the extrapolated order parameters;
thus, the precise location of the SC phase boundary at $U/t = 8$ is uncertain.
Overall, the existence of all the phases and their boundaries are consistent
with the general understandings of the 2D Hubbard model.

\begin{figure}[htbp]
  \centering  
  \subfigure[$U/t=2$]{
    \includegraphics[width=0.42\columnwidth]{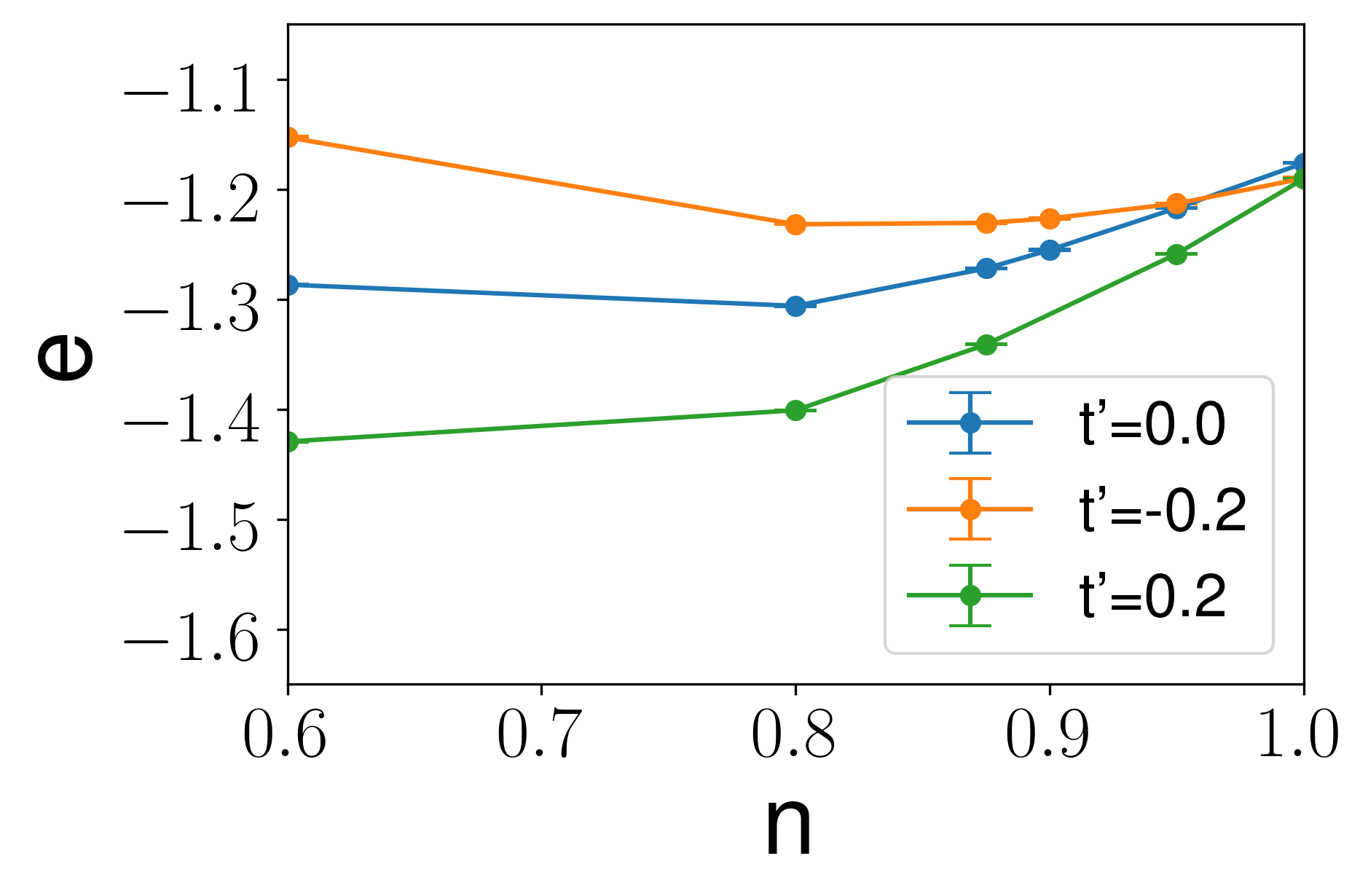}
    \label{fig:diagram:U2energy}
  }
  \subfigure[$U/t=4$]{
    \includegraphics[width=0.42\columnwidth]{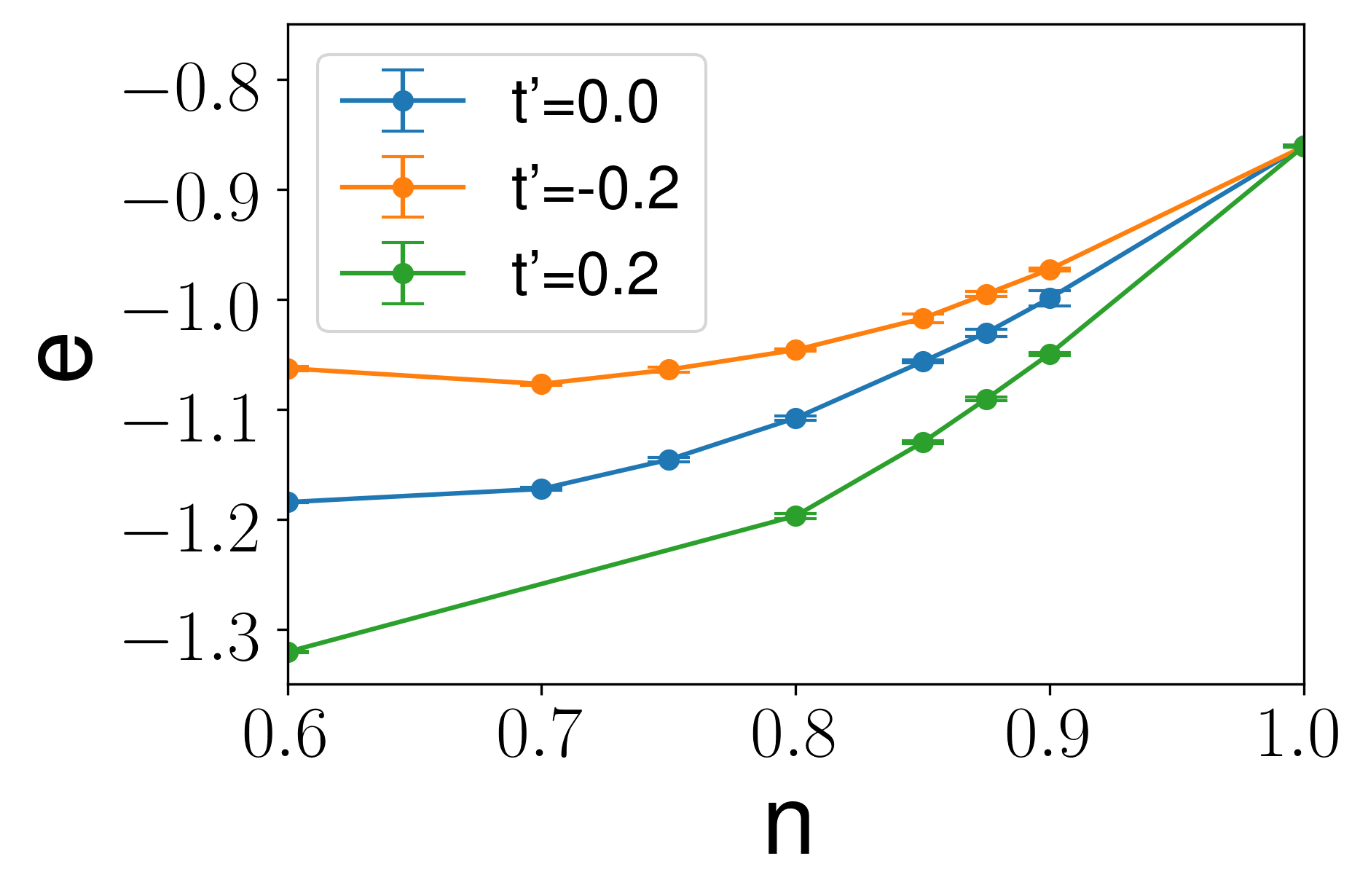}
    \label{fig:diagram:U4energy}
  }
  \subfigure[$U/t=6$]{
    \includegraphics[width=0.42\columnwidth]{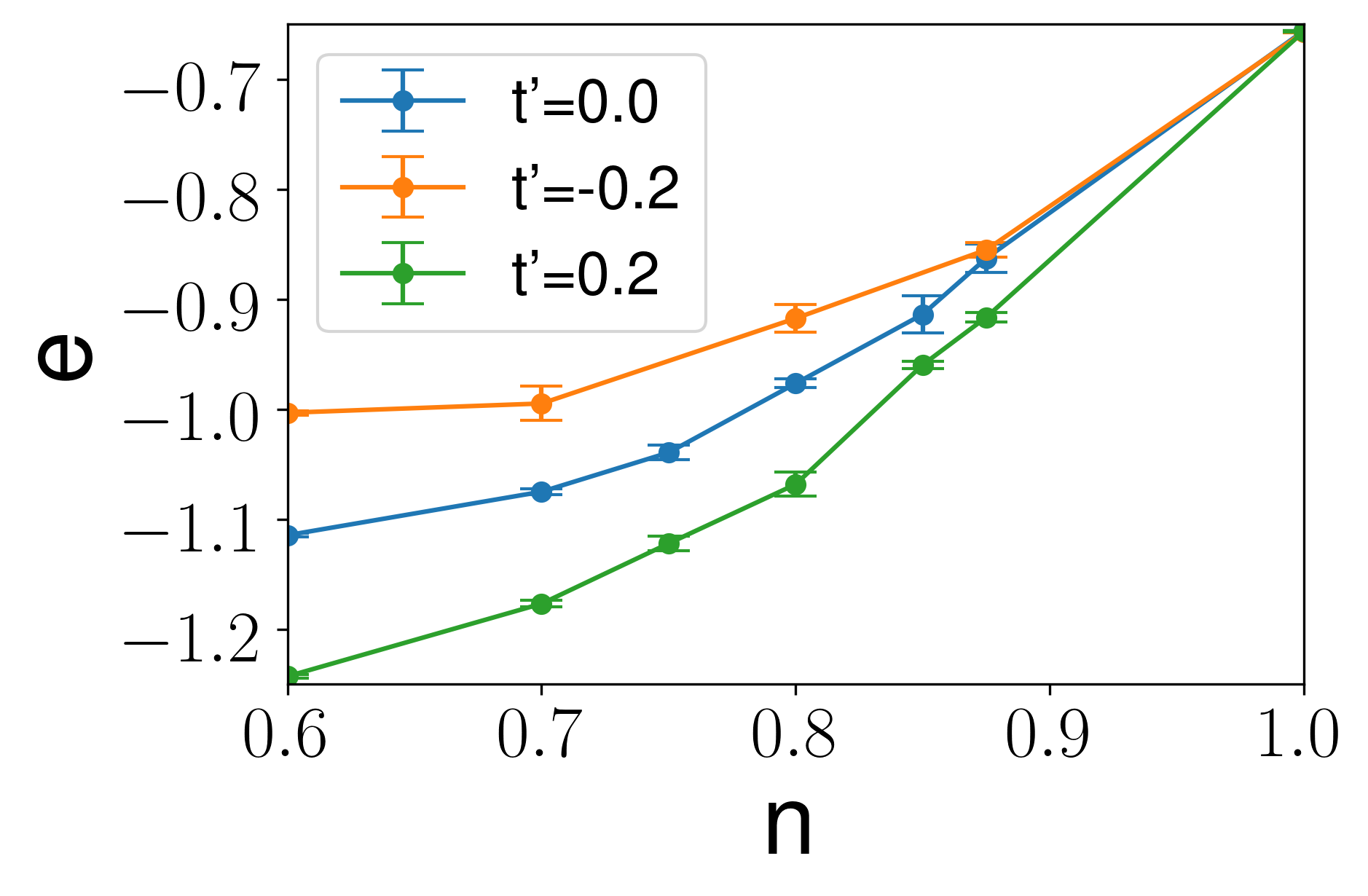}
    \label{fig:diagram:U6energy}
  }
  \subfigure[$U/t=8$]{
    \includegraphics[width=0.42\columnwidth]{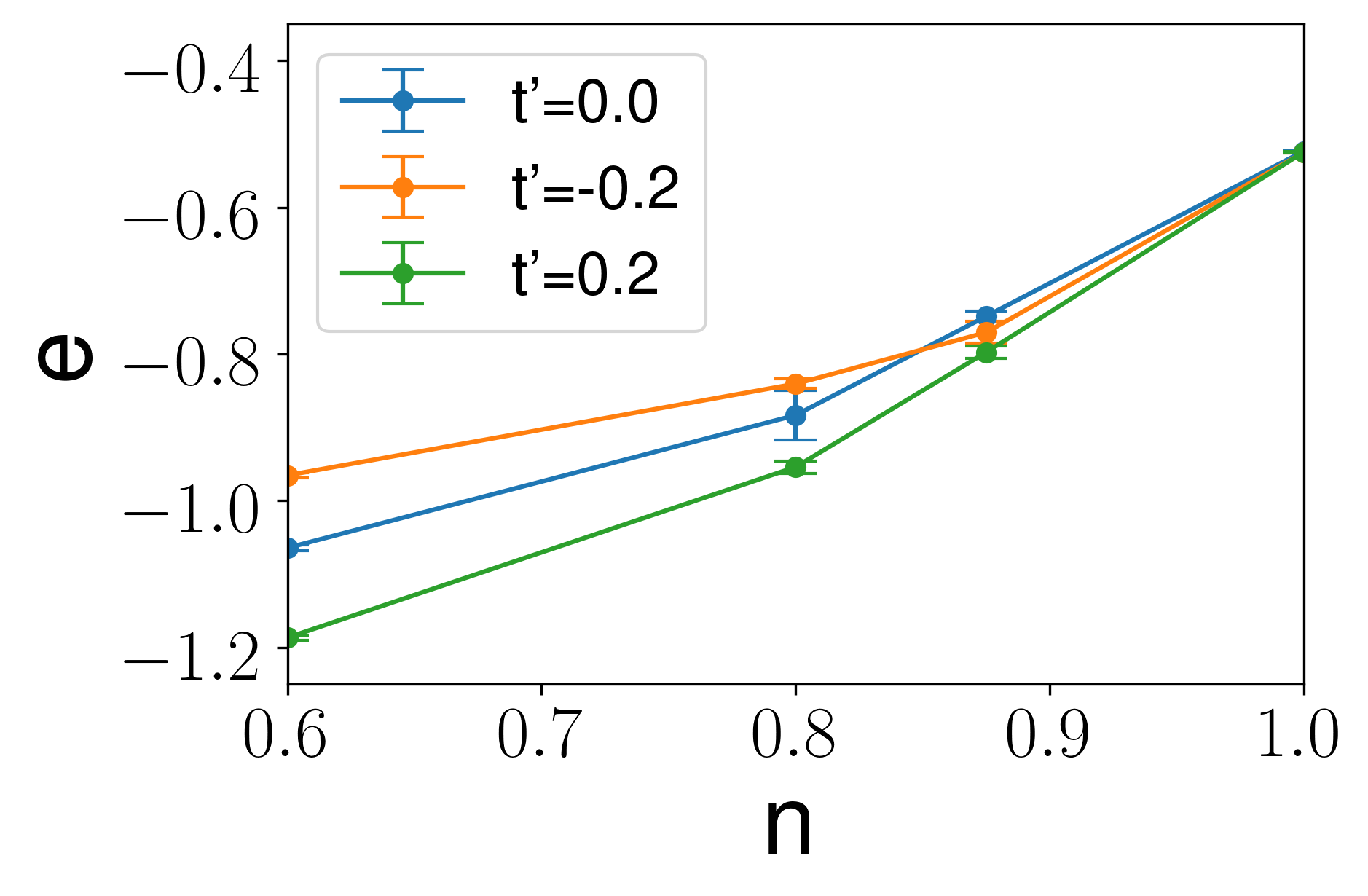}
    \label{fig:diagram:U8energy}
  }
  \caption{DMET estimates of the 2D Hubbard model energies per site in the TDL.}
  \label{fig:diagram:energy}
\end{figure}

In Fig.~\ref{fig:diagram:energy}, we show the estimated TDL energies
across the parameter space. At half-filling, the energy in the frustrated models
$t^\prime=\pm0.2$ are slight below those in the standard Hubbard model, due to
the kinetic energy relaxation. The effect is smaller for larger $U$ where the
electrons are more confined. At large doping, e.g., $n\le0.8$, the energy order
is dominated by the kinetic effects, i.e. $e_{t^\prime=-0.2}>e_{t^\prime=0}
>e_{t^\prime=0.2}$.
The energy curves show more complicated behavior in the underdoped region,
especially at large $U$ and for $t^\prime=0$ and $-0.2$, indicating complicated
behaviors in this region, as we will discuss later.

\subsection{Half-Filling Results} \label{sec:diagram:result:half}
\begin{figure}[htpb]
	\centering
	\subfigure[]{
		\includegraphics[width=0.46\textwidth]{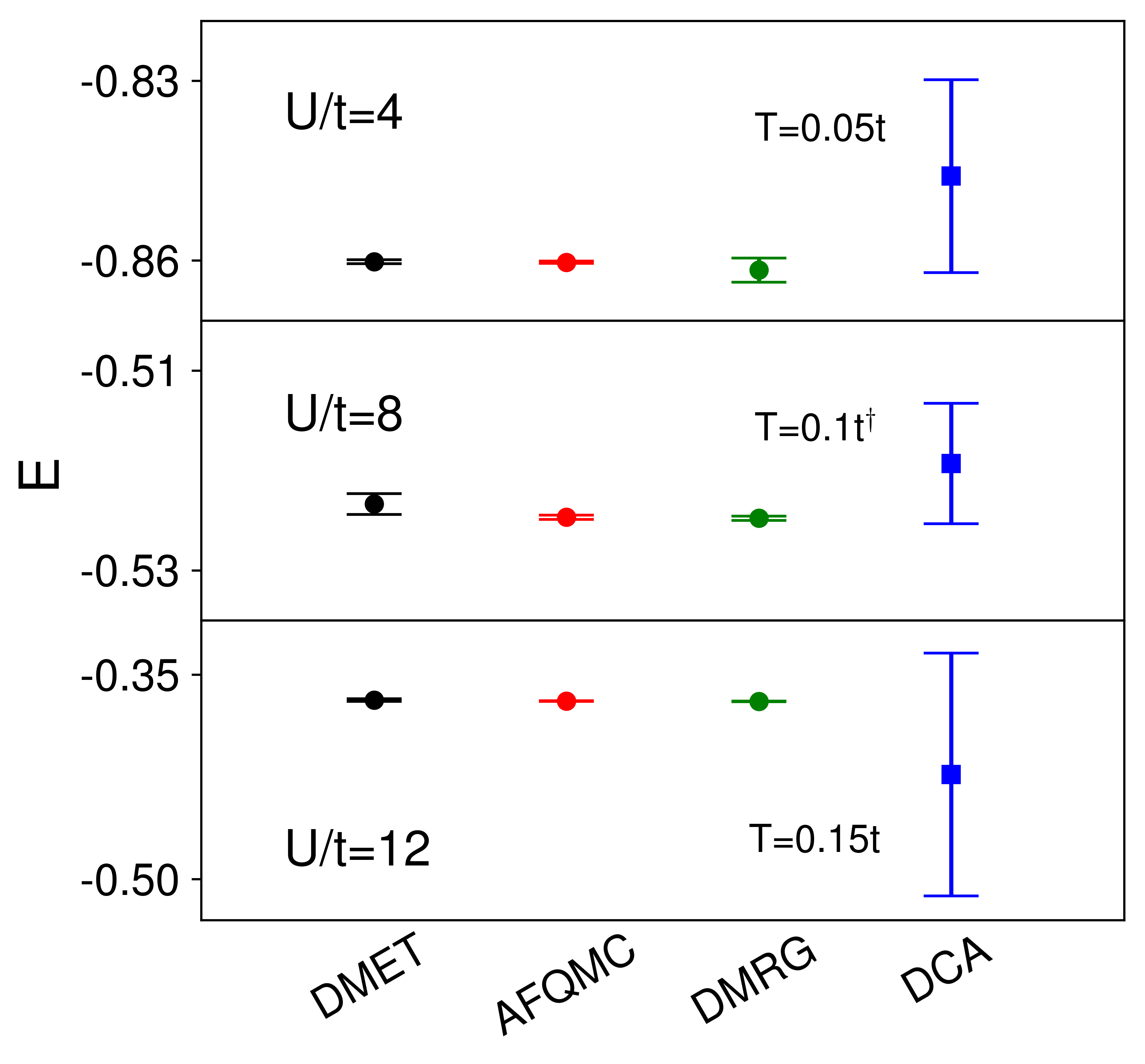}
		\label{fig:diagram:e_half}
	}
	\subfigure[]{
		\includegraphics[width=0.46\textwidth]{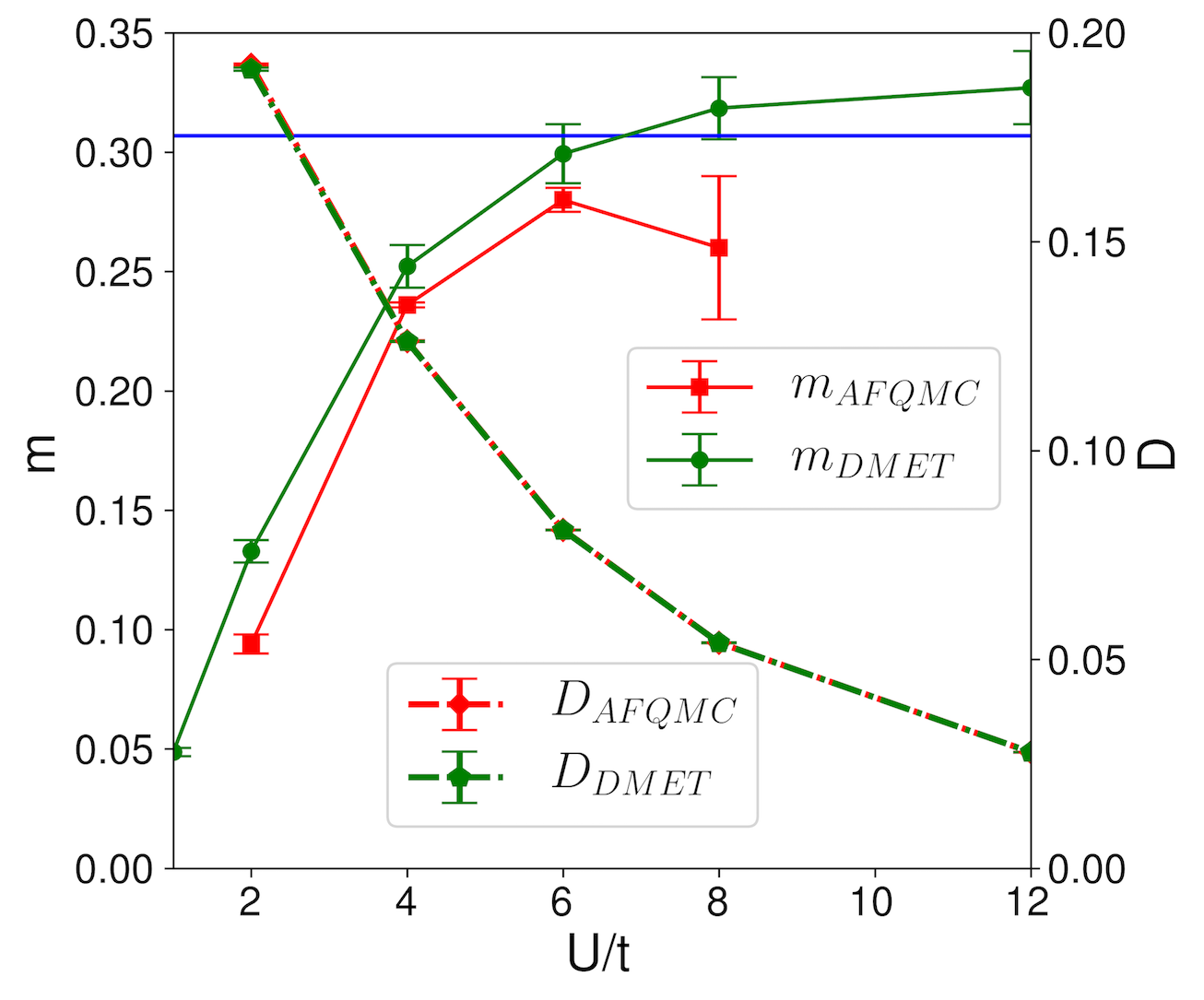}
		\label{fig:diagram:af_do_half}
	}
	\caption{Results for the half-filled ($t^\prime=0$) Hubbard model.
	(a) Energy per site. Ground state estimates from DMET, AFQMC and DMRG,
	compared to a recent DCA study~\cite{LeBlanc2013}.
	The temperatures are the lowest published values in the DCA study.
	$^{(\dagger)}$ DCA data at $U/t$=8 is from 50-site cluster calculations,
	and not extrapolated to the TDL. (b) Staggered magnetization ($m$)
	and double occupancy($D$) at half-filling. The  blue line is the
	spin-$\frac{1}{2}$ Heisenberg limit $m=0.3070(3)$~\cite{PhysRevB.56.11678}.
}
	\label{fig:diagram:half}
\end{figure}

We now look at the results for the half-filled case in detail. In 
Table~\ref{tab:diagram:energy} and Fig.~\ref{fig:diagram:half} we
compare the DMET energies per site, double occupancies,
and staggered magnetizations with exact estimates at half-filling, as obtained
from auxiliary field quantum Monte Carlo (AFQMC)
extrapolated to infinite size ~\cite{Qin2016}, and DMRG on long open
cylinders, extrapolated to infinite width and length~\cite{LeBlanc2015}.
For comparison, we also show recent DCA energies computed 
at the lowest published temperatures, $T=0.05-0.15$t~\cite{LeBlanc2013}.

\begin{table}[htpb]
  \centering
  \caption{Ground-state energy per site of the half-filled
	  ($t^\prime=0$) 2D Hubbard model.
    All the numbers are extrapolated to the TDL.
    AFQMC and DMRG results are from Refs.~\cite{Qin2016,LeBlanc2015}.}
  \label{tab:diagram:energy}
  \begin{tabular}{p{0.15\textwidth}p{0.25\textwidth}p{0.25\textwidth}p{0.25\textwidth}}
    \toprule
          $\mathbf{U/t}$&\textbf{DMET}&\textbf{AFQMC}&\textbf{DMRG}\\
    \midrule
    2	&-1.1764(3)& -1.1763(2)&-1.176(2)\\
    4	&-0.8604(3)& -0.8603(2)&-0.862(2)\\
    6	&-0.6561(5)& -0.6568(3)&-0.658(1)\\
    8	&-0.5234(10)& -0.5247(2)&-0.5248(2)\\
    12  &-0.3686(10)&-0.3693(2)&-0.3696(3)\\
    \bottomrule
  \end{tabular}
\end{table}

The data shows the high accuracy of the DMET energies at half filling.
The error bars from DMET, AFQMC, and DMRG are all consistent, 
with an accuracy better than 0.001$t$. Indeed, the DMET error bars are
competitive with the exact ``statistical'' error bars of AFQMC up to $U/t = 6$.
As a point of reference, the DMET uncertainty is one to two orders of magnitude
smaller than finite temperature contributions to recent low-temperature benchmark
DCA calculations (Fig.~\ref{fig:diagram:e_half}), and is similarly two to three
orders of magnitude smaller than energy errors in earlier zero-temperature Green's
function cluster calculations~\cite{Aichhorn2007}.

Fig.~\ref{fig:diagram:af_do_half} further gives the half-filling staggered
magnetizations and double occupancies computed with DMET as compared with AFQMC.
The DMET double occupancies are obtained with similar error bars to the AFQMC
estimates, suggesting the good agreement of the total energy is not from
an effect of error cancellation.
The staggered magnetization exhibits larger discrepancies at the smallest
$U/t=2$, but with later revised AFQMC estimates $m\sim0.12$ the agreement
is excellent~\cite{Qin2016}. For $U/t\ge4$, DMET gives similar, or in fact more accurate
staggered magnetization than AFQMC. At our largest available $U/t=12$, we
find $m=0.327(15)$, slightly above the exact Heisenberg value $m=0.3070(3)$
~\cite{PhysRevB.56.11678} (corresponding to $U/t=\infty$ limit).

\begin{figure}[htpb]
  \centering
  \includegraphics[width=0.8\columnwidth]{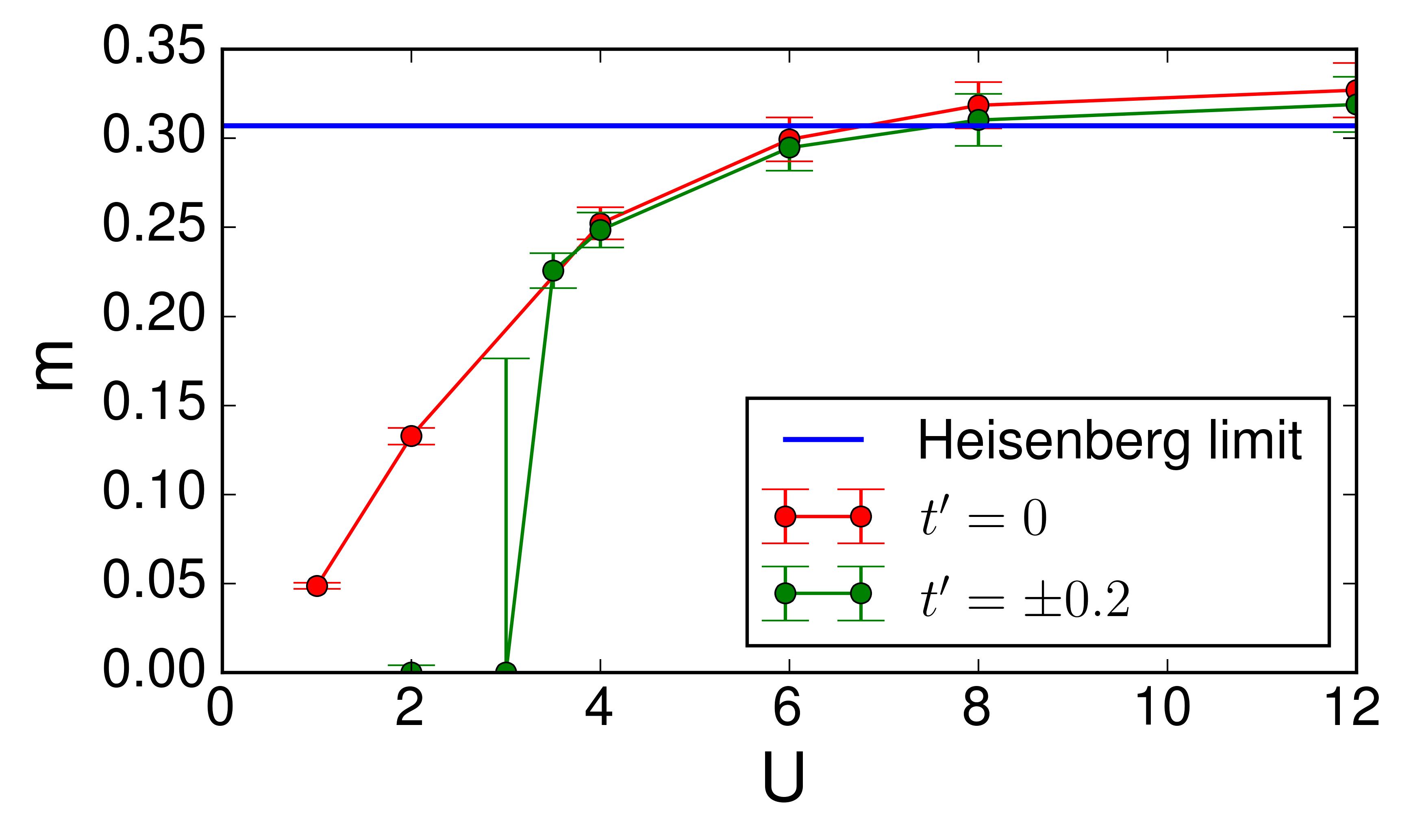}
  \caption{Staggered magnetization ($m$) of the half-filled
  Hubbard model for $t^\prime=\pm0.2$, compared to $t'=0$.}
  \label{fig:diagram:af_frustrated}
\end{figure}

The staggered magnetization for the frustrated Hubbard models at half filling
(compared to the  $t'=0$ standard Hubbard model and the Heisenberg limit)
are shown in Fig.~\ref{fig:diagram:af_frustrated}. Because of the particle-hole
parity, the magnetization curve is identical for $t^\prime=\pm0.2$.
The onset of antiferromagnetism is at finite $U$ in the frustrated model, somewhere
between $U/t=2$ and $3.5$. This is consistent with previous quantum Monte Carlo
simulations~\cite{Lin1987}.

The large error bar at $U/t=3$ reflects the increasing quantum fluctuations near
the phase boundary, leading to multiple possible solutions in self-consistent
embedding methods. The results are thus sensitive to initial conditions and
fragment shape, etc., resulting in a huge uncertainty in the TDL results.

\subsection{Doped Hubbard Model in Weak to Moderate Coupling}
\label{sec:diagram:result:smallU}

The impressing accuracy of half-filling results lend confidence to the DMET
TDL estimates of the energy per site and observables and their associated error bars.
The same procedure is used to compute the quantities and error bars for the
doped Hubbard model.

\begin{table}[htpb]
  \centering
  \caption{Ground-state energy per site of the ($t^\prime=0$) 2D Hubbard model
	  at $U/t=4$. All the numbers are extrapolated to the TDL.
    CP-AFQMC and DMRG results are from Refs.~\cite{Qin2016,LeBlanc2015}.}
  \label{tab:diagram:energy_U4}
  \begin{tabular}{p{0.15\textwidth}p{0.25\textwidth}p{0.25\textwidth}p{0.25\textwidth}}
    \toprule
    $\mathbf{n}$&\textbf{DMET}&\textbf{CP-AFQMC}&\textbf{DMRG}\\
    \midrule
    0.875&-1.031(3)&-1.026(1)&-1.028(2)\\
    0.8&-1.108(2)&-1.110(3)&-1.1040(14)\\
    0.6&-1.1846(5)&-1.185(1)&-\\
    0.3&-0.8800(3)&-0.879(1)&-\\
    \bottomrule
  \end{tabular}
\end{table}

\begin{figure}[htpb]
  \centering
  \includegraphics[width=\columnwidth]{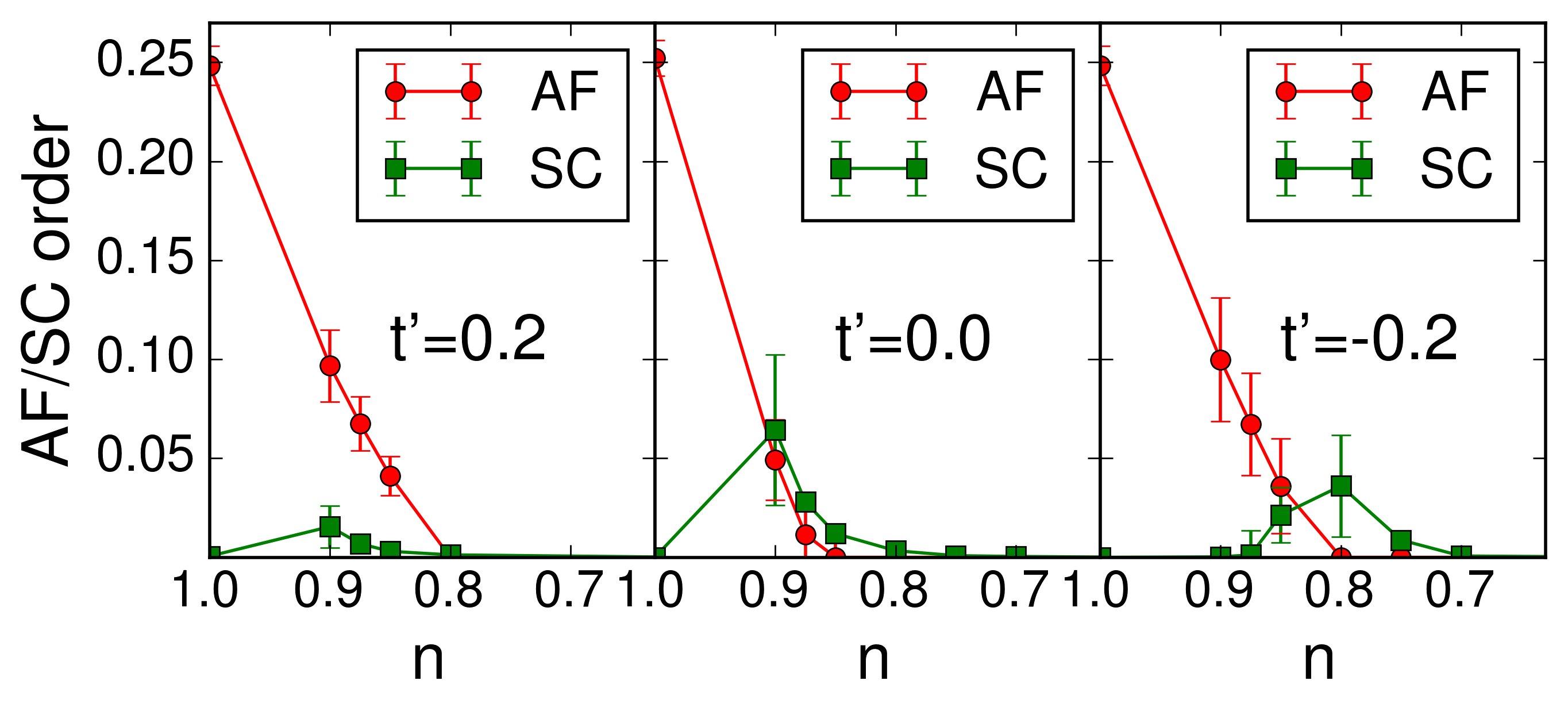}
  \caption{Antiferromagnetic (red circle) and ($d$-wave) superconducting
  (green square) order parameters for the (frustrated) 2D Hubbard models at $U/t=4$.}
  \label{fig:diagram:U4order}
\end{figure}

We start with $U/t=4$. For benchmark purpose, we
compare the energies per site from DMET, DMRG and the constrained path (CP) AFQMC,
a sign-free QMC with a bias that disappears at low density and small
$U$~\cite{PhysRevB.55.7464,PhysRevB.78.165101} in Table.~\ref{tab:diagram:energy_U4}.
For $n\le0.6$, a parameter regime where CP-AFQMC is very accurate,
the DMET and CP-AFQMC energies agree to $0.001t$, while the error bars from
DMET, CP-AFQMC and DMRG are comparable in the underdoped region.

Fig.~\ref{fig:diagram:U4order} shows the averaged (over the central plaquette)
AF and $d$-wave SC order parameters as a function of filling for $U/t=4$.
As expected, the SC order lives at the proximity of the AF order and is
a small effect compared to the magnetization.
We find that for $t^\prime=0$, the peak in SC order is around
$\langle n\rangle=0.9$ and SC extends to $\langle n\rangle\sim$0.8.
The figures also show that next-nearest-neighbor hopping $t^\prime=0.2$
stabilizes AF versus SC, and the reverse is true for $t^\prime=-0.2$.
The suppression (enhancement) of SC order with positive (negative)
$t^\prime$ is consistent with the stronger superconductivity found
in hole-doped materials~\cite{Pavarini2001,Huang2001,Eberlein2014}.

The presence of SC in the Hubbard model ground-state has previously
been much discussed. The strongest SC order found in DMET roughly
occurs in the same region as seen in earlier Green's function cluster
calculations~\cite{Aichhorn2006,Capone2006}.
However, this region is not typically found to be superconducting
in ground-state wavefunction calculations using DMRG and AFQMC on
finite lattices, even though such calculations achieve significantly
higher energy accuracies than the Green's function cluster studies
~\cite{Chang2010,white2003stripes,Zhang1997,LeBlanc2015}.
The significance of the DMET result is that the energy error bar in
this region (e.g., $0.001t$ per site for $U/t = 4, n = 0.8, t^\prime = −0.2$)
is comparable to or better than the accurate ground-state wavefunction
calculations, yet SC order is still seen. This strongly suggests
that SC is, in fact, the ground-state order.

\begin{table}[t]
  \centering
  \caption{Energies per site for various 16-site fragments at $U/t=4$.}
  \label{tab:diagram:U4_compare}
  \begin{tabular}{p{0.15\textwidth}p{0.25\textwidth}p{0.25\textwidth}p{0.25\textwidth}}
\toprule
	  $\mathbf{t^\prime}$&$\mathbf{n}$&$\mathbf{e_{8\times2}}$&$\mathbf{e_{4\times4}}$\\
		     \midrule
		     \multirow{2}{*}{0.2}        &0.8  & -1.2036(2) & -1.204(2)\\
    		     &0.875&-1.0944(1)&-1.095(1)\\
		     \midrule
		     \multirow{2}{*}{0}         &0.8  &  -1.1164(2)&-1.1151(8)\\
		     &0.875 & -1.0284(1)$^*$&-1.033(2)\\
		     \midrule
		     \multirow{3}{*}{-0.2}        &0.8&-1.10483(6)$^*$&-1.0507(4)\\
    			&0.85&-1.0162(1)$^*$&-1.020(2)\\
    			&0.875&-0.9966(1)$^*$&-0.9989(7)\\
    \bottomrule
  \end{tabular}
 
  {\raggedright\footnotesize
	  $^*$ These $8\times2$ fragment results show significant inhomogeneity.
  }

\end{table}

\begin{figure}[htbp]
  \centering
  \subfigure[$t^\prime=0.2$]{
    \includegraphics[width=0.7\columnwidth]{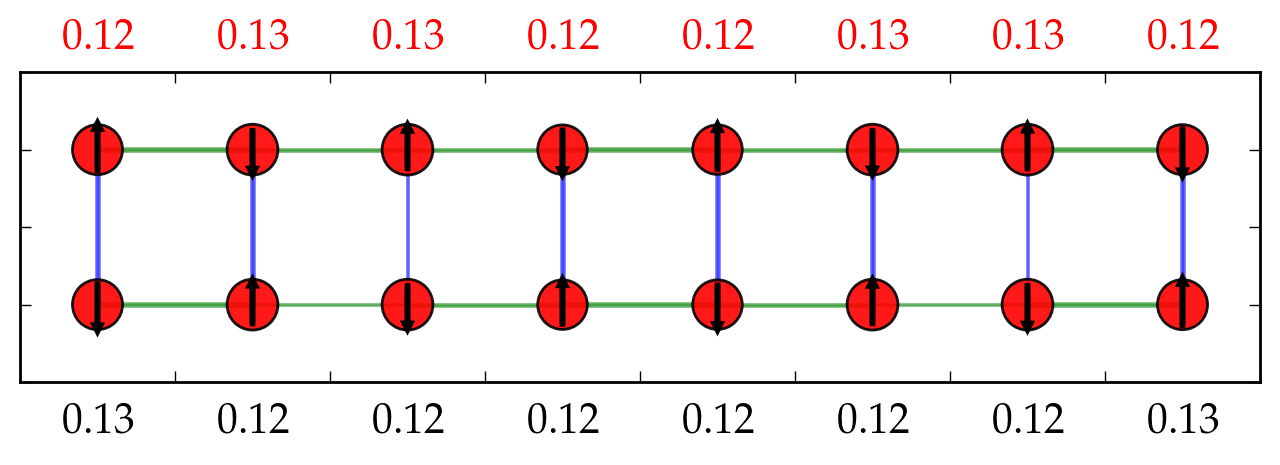}
    \label{fig:diagram:U4_82_p}
  }
  \subfigure[$t^\prime=0$]{
    \includegraphics[width=0.7\columnwidth]{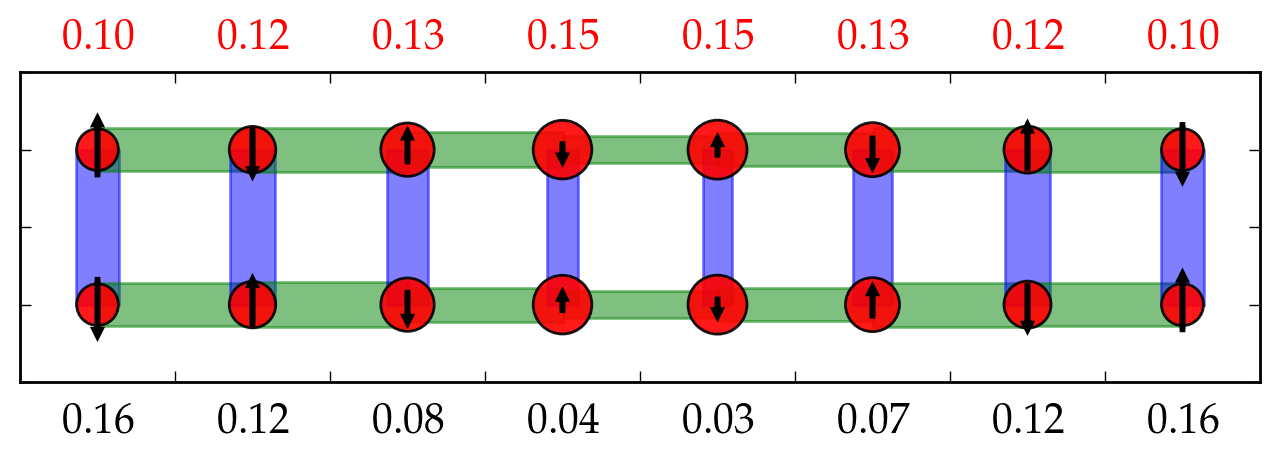}
    \label{fig:diagram:U4_82_0}
  }
  \subfigure[$t^\prime=-0.2$]{
   \includegraphics[width=0.7\columnwidth]{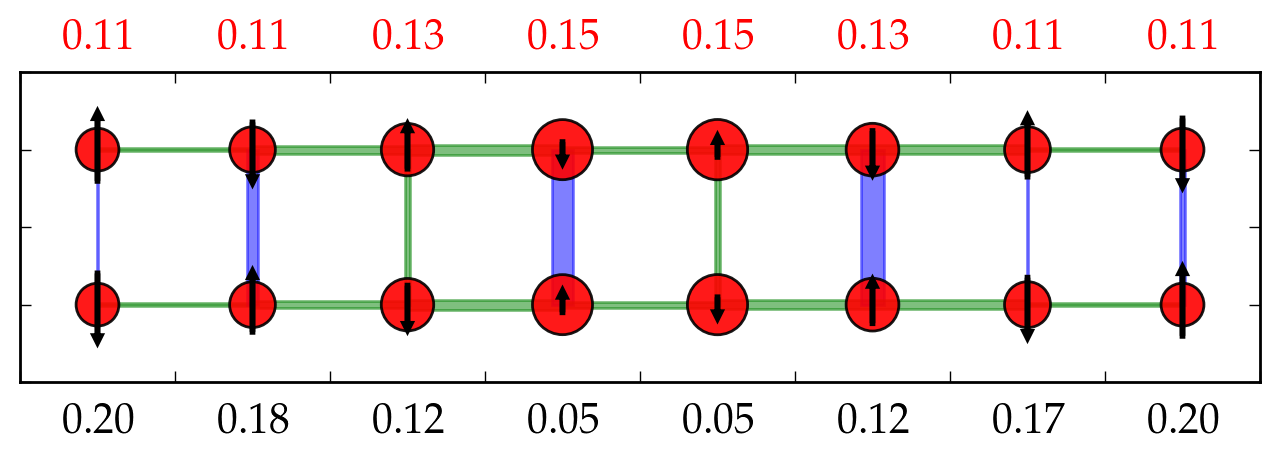}
    \label{fig:diagram:U4_82_n}
  }
  \caption{Charge, spin and pairing orders for the $8\times2$ fragment
  calculations at $U/t=4, n=0.875$.}
  \label{fig:diagram:U4_82}
\end{figure}

Even at moderate $U/t=4$, there is already tendency towards inhomogeneity.
The $8\times2$ fragment calculations result in an inhomogeneous state at
dopings $n=0.8$ to $0.875$, although the energies are significantly
higher than obtained with the $4\times4$ fragments at the same fillings.
In Fig.~\ref{fig:diagram:U4_82_0} and ~\ref{fig:diagram:U4_82_n}, we
show two such examples with spin and pairing modulation. As a comparison, we show the $t^\prime=0.2$
result (Fig.~\ref{fig:diagram:U4_82_p}) with the same doping to verify
that this is not an artifact. Table~\ref{tab:diagram:U4_compare} further
shows the energies per site in the underdoped region at $U/t=4$. All the
$4\times4$ fragment calculations are homogeneous, while some of the $8\times2$
calculations give an inhomogeneous solution, at a higher energy, suggesting that
the ground state at $U=4$ is homogeneous, or inhomogeneous with a very long
wavelength that does not fit in our cluster shapes.

We then briefly discuss results at weak coupling $U/t=2$. We find the
antiferromagnetism (in the standard Hubbard model) is destroyed already at
small doping $x=0.05$ away from half-filling. (For the frustrated
Hubbard model there is no AF order even at half-filling) While the
extrapolated AF order is zero, the uncertainty $\delta m$ is still large
($\sim 0.05$ at $x=0.05$), reflecting short-range AF fluctuations still
exist. $\delta m$ decays exponentially as we increase doping. At $U/t=2$
we do not find $d$-wave superconductivity to within in numerical precision.

\subsection{Doped Hubbard Model in Stronger Coupling} \label{sec:diagram:result:largeU}

The stronger coupling region of the 2D Hubbard model has a more direct connection with
the cuprate physics. We are most interested in the underdoped region between
the AF and the SC phases. In this region, a variety of spin-density
~\cite{Scalapino1986,Schulz1990,Kato1990,Chang2010,PhysRevB.89.155134,leprevost2015intertwined,Schulz1990,Chubukov1995,Igoshev2010,Kato1990},
charge-density~\cite{Poilblanc1989,Vojta1999,Melikyan2005,Chang2010}, 
pair-density wave~\cite{Chen2004a,Melikyan2005,Lee2014,berg2009charge}, and 
stripe orders~\cite{white1998density,Hellberg1999,white2000phase,white2003stripes,Hager2005,Corboz2011,leprevost2015intertwined,Miyazaki2002,Mizusaki2006},
have been posited in both the Hubbard model and the simpler $t$-$J$ model.
These inhomogeneous phases are proposed to be relevant in the pseudogap physics
~\cite{Moshchalkov2001,Fleck2001,Chen2004a,
valla2006ground,Li2006,Sedrakyan2010,Lee2014}.
Recent infinite projected entangled pair state (iPEPS) studies of the
$t$-$J$ model and Hubbard model at large $U/t=8$ suggest that inhomogeneous
and homogeneous states are near degenerate at low doping and can be stabilized
with small changes in the model parameters~\cite{Corboz2011,Corboz2016}.

\begin{figure}[htpb]
	\centering
    \subfigure[$n=0.875$ $t^\prime=0$]{
      \includegraphics[width=0.75\columnwidth]{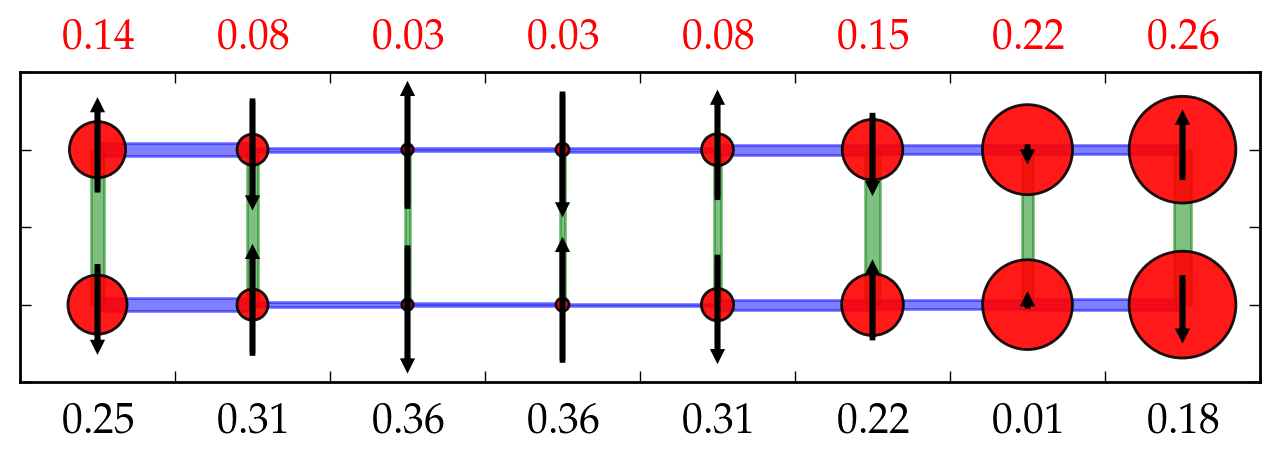}
      \label{fig:diagram:U8n875t0Imp82}
    }
    \subfigure[$n=0.875$ $t^\prime=-0.2$]{
      \includegraphics[width=0.75\columnwidth]{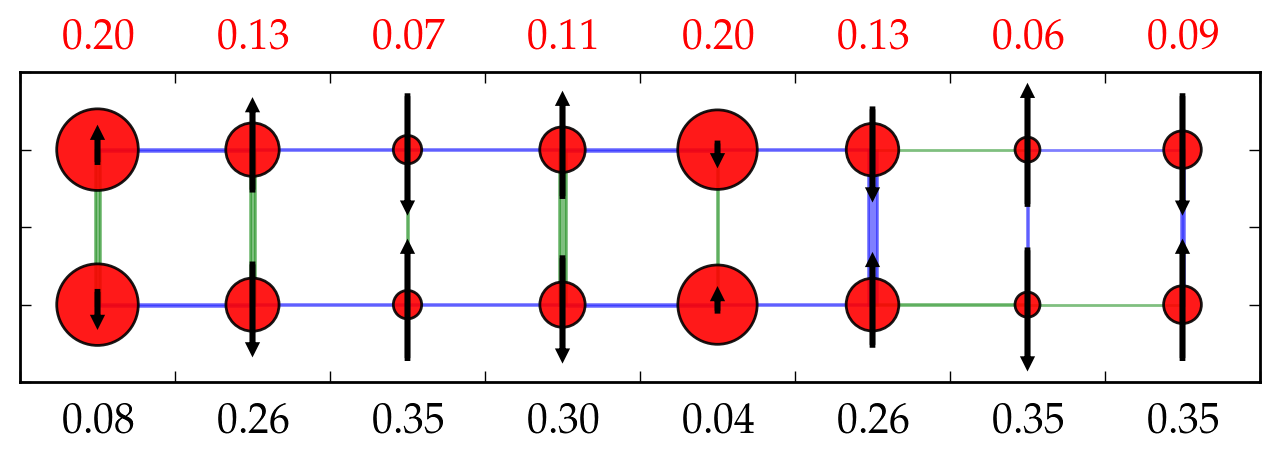}
      \label{fig:diagram:U8n875tnImp82}
    }
    \subfigure[$n=0.875$ $t^\prime=0.2$]{
      \includegraphics[width=0.75\columnwidth]{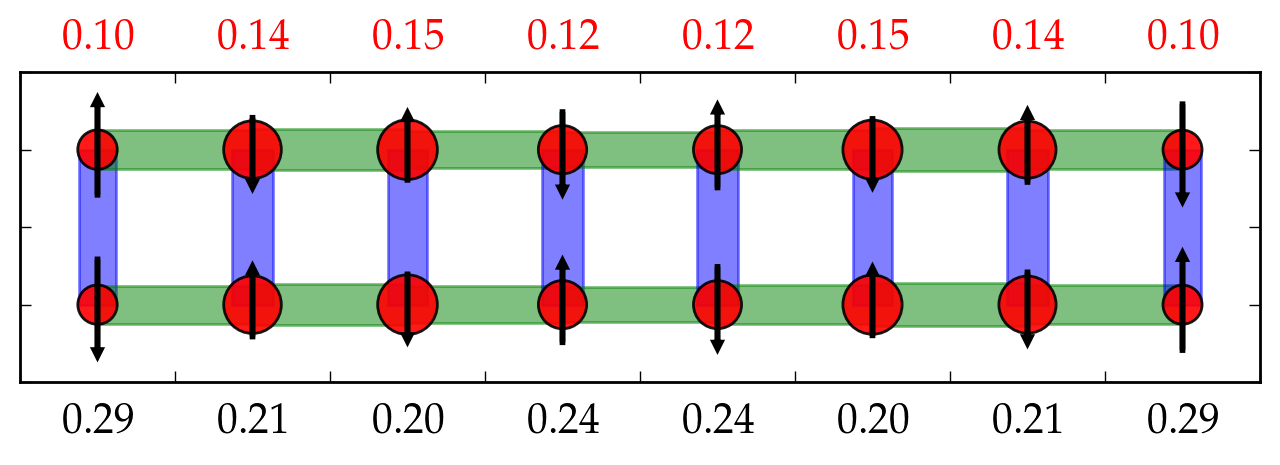}
      \label{fig:diagram:U8n875tpImp82}
    }
    \subfigure{
      \includegraphics[width=0.85\columnwidth]{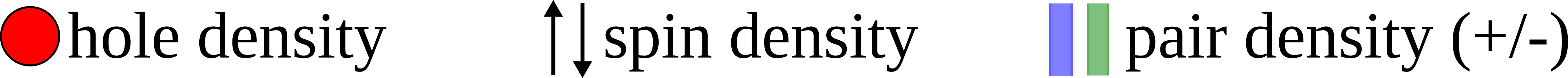}
    }
    \caption{Local charge, spin and pairing orders in the (frustrated)
    Hubbard model at various points in the strong coupling regime ($U/t$=8).}
  \label{fig:diagram:stripes}
\end{figure}

\begin{figure}[htpb]
  \centering
  \subfigure[$U/t=4$ $e_{\text{diff}}=0.002$]{
    \includegraphics[width=0.75\columnwidth]{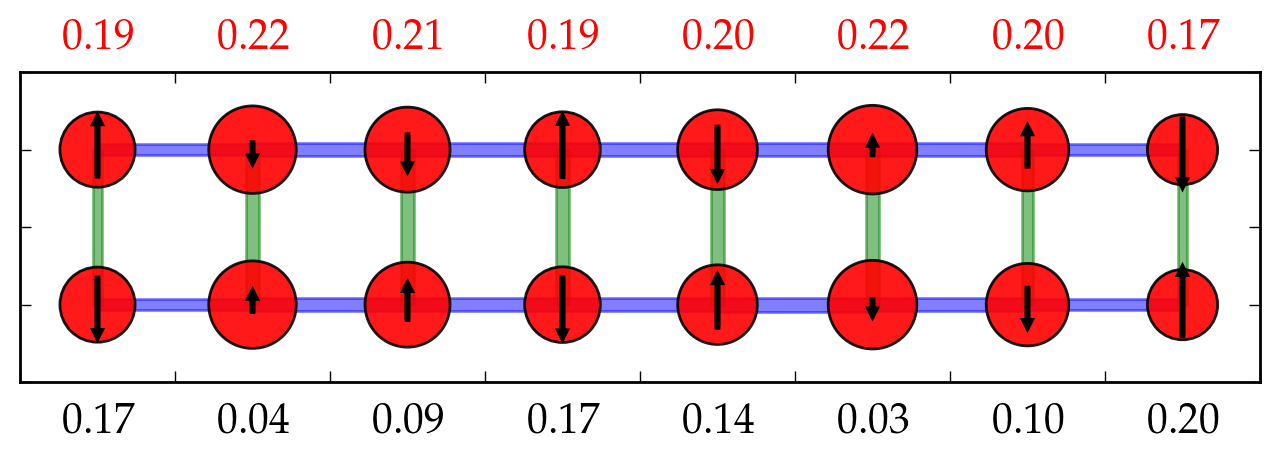}
    \label{fig:diagram:U4n800tnImp82}
  }
  \subfigure[$U/t=6$ $e_{\text{diff}}=-0.001$]{
    \includegraphics[width=0.75\columnwidth]{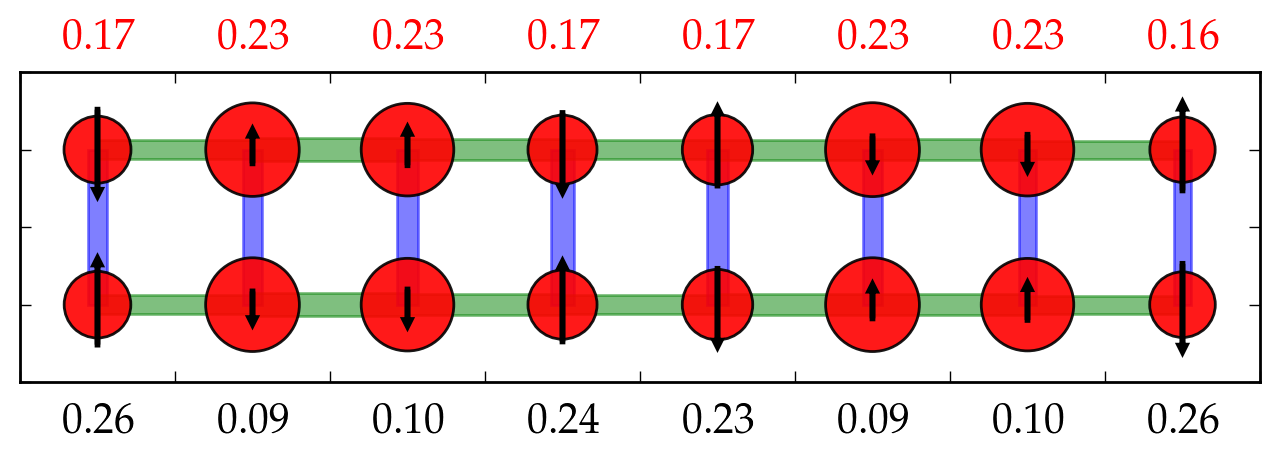}
    \label{fig:diagram:U6n800tnImp82}
  }
  \subfigure[$U/t=8$ $e_{\text{diff}}=-0.003$]{
    \includegraphics[width=0.75\columnwidth]{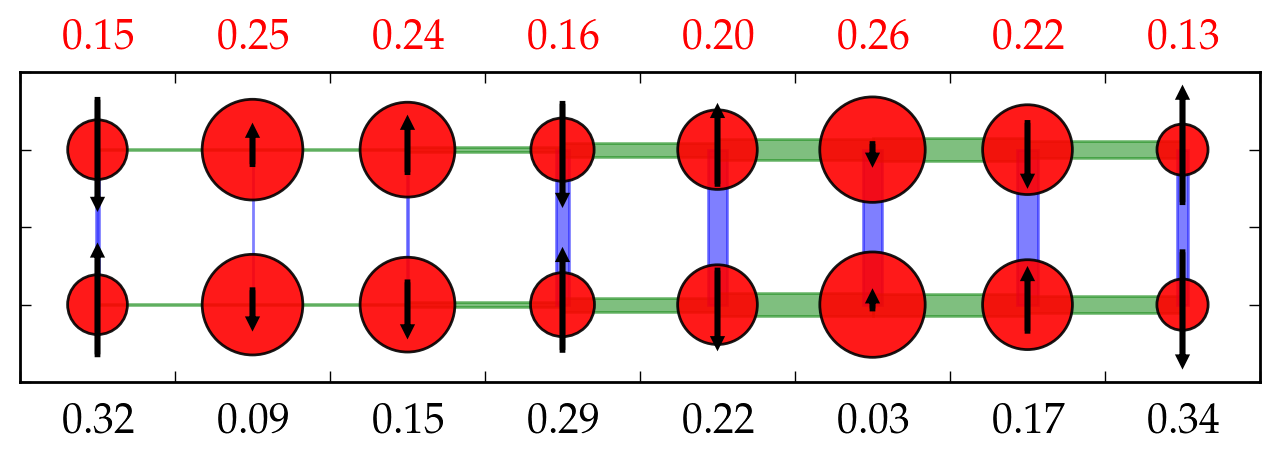}
    \label{fig:diagram:U8n800tnImp82}
  }
  \caption{Evolution of the inhomogeneous patterns and stabilities for
	  $n=0.8, t^\prime=-0.2$ with respect to coupling strength.
	  $e_{\text{diff}} = e_{8\times2} - e_{4\times4}$ for each point.
  	At $U/t=8$, both 16-site fragments show inhomogeneous orders.}
  \label{fig:diagram:evolve}
\end{figure}

Our work indicates similar behavior in the Hubbard model. For large $U$ and
low doping  $n=0.875-0.8$ we find many points with inhomogeneous orders.
Interestingly, the kinds of inhomogeneity we observe are extremely rich,
and some representative examples are shown in Fig.~\ref{fig:diagram:stripes}.
These correspond to (i) a local phase separation between a half- filled,
antiferromagnetic phase and a superconducting ribbon
[Fig.~\ref{fig:diagram:U8n875t0Imp82}];
(ii) a classic stripe phase order [Fig.~\ref{fig:diagram:U8n800tnImp82}]
very similar to that seen in earlier DMRG ladder studies~\cite{white2003stripes}
(there is also a coexisting weak PDW, exhibiting a sign change across the cell,
consistent with earlier stripe proposals~\cite{berg2009charge});
(iii) inhomogeneities in the pairing order coexisting with the
charge and spin orders in, e.g., Fig.~\ref{fig:diagram:U8n875tpImp82},
similar to a recent theoretical proposal (see, e.g., Ref.~\cite{Lee2014}).
Fig.~\ref{fig:diagram:U8n875tpImp82} shows an example at 1/8 doping with
positive $t^\prime$, where the inhomogeneity is much weaker.
Again the next-nearest-neighbor hopping $t^\prime$ plays an important role
in stablishing or destroying the inhomogeneous orders, providing an
explanation for the particle-hole asymmetry in cuprates.

We also see from the results how the inhomogeneity develops
with increasing interaction $U$. Fig.~\ref{fig:diagram:evolve} shows the
inhomogeneous patterns in $8\times2$ fragments and their energies per site
relative to those of $4\times4$ fragments at the same doping ($x=0.2$) for
$U/t$ from $4$ to $8$.
The change in coupling strength results in stronger spin modulation and
charge localization, and, at $U/t=8$ the pairing order also start to show
inhomogeneity. At the same time, the $e_{\text{diff}}$ decreases from positive
to negative values, indicating the inhomogeneous order is favored at stronger
coupling.

It is important to note that the $8\times2$ fragment geometry does not
always lead to inhomogeneity, nor are all the $4\times 4$ fragments
are homogeneous. (For instance, the same parameter as in
Fig.~\ref{fig:diagram:U8n800tnImp82} leads to an incommensurate inhomogeneous
order in the $4\times4$ fragment, while many $t^\prime=0.2$ calculations see
homogeneous $8\times2$ fragments.)
Instead, at points where the tendency towards inhomogeneity
is strong, we find a significant lowering of the energy associated with the
inhomogeneous order, reflected either in a much lower energy of an inhomogeneous
inhomogeneous $8\times2$ fragment relative to the $4\times4$ fragment,
or inhomogeneity in {\it both} $8\times2$ and $4\times4$ fragments.
Thus while it is not possible with our fragment sizes to extrapolate details
of the inhomogeneities in the TDL (for example, the particular wavelengths
of the spin, charge, and pairing instabilities, or diagonal versus vertical
stripe patterns), the evidence points strongly to some forms of inhomogeneity
surviving in the TDL at the indicated parts of the phase diagram.

Overall, in the underdoped regions of $U/t=8$ and $t^\prime=0$ and $-0.2$,
various inhomogeneous orders are well established in $8\times2$ and even $4\times4$
fragment calculations, providing strong evidence to the inhomogeneity in the TDL;
While at $U/t=6$, the calculations seem near the transition between the homogeneous
and the inhomogeneous regimes, and give more ambiguous results, thus it
is hard to determine what happens in the TDL ground state.

\section{Conclusions} \label{sec:diagram:conclusion}
We have computed a ground-state phase diagram for the Hubbard model on
the square lattice using cluster DMET. The accuracy achieved by DMET appears
competitive with the exact ground-state benchmarks available at half-filling,
while away from half-filling our error model suggests that the calculations
remain very accurate. We observe AF and metallic phases and robust $d$-wave
pairing. Further in parts of the phase space, our calculations strongly suggest
that inhomogeneous phases are a feature of the thermodynamic limit, although the
precise inhomogeneous patterns require larger clusters to resolve and reflect
competition between different orders at very low energy scales.

However, for real materials such as the cuprates ($t\approx$3000K), the energy
resolution achieved here for most of the phase diagram is already below the
superconducting $T_c$, suggesting that the near degeneracy of competing orders
will be lifted by terms beyond those in the Hubbard model, such as long-range
charge and hopping terms, multi-orbital effects, and interlayer coupling. Moving
beyond the  Hubbard model to more realistic systems thus now appears of
principal relevance.

\chapter[Stripe Order in the Underdoped Region of the Two-Dimensional Hubbard Model]{Stripe Order in the Underdoped Region of the Two-Dimensional Hubbard Model~\footnote{Based on work posted in arXiv:1701.00054.~\cite{zheng2016stripe}}} \label{chpt:stripe}

Competing inhomogeneous orders are a central feature of correlated electron
materials including the high-temperature superconductors. The two-dimensional
Hubbard model serves as the canonical microscopic physical model. Multiple orders
have been proposed in the underdoped part of the phase diagram, which corresponds
to a regime of maximum numerical difficulty.

In Chapter~\ref{chpt:diagram}, we investigated the broad picture of the 2D Hubbard
model phase diagram over a wide range of parameter space; In this chapter, we turn
our focus on the strong-coupling, underdoped region of the phase diagram. We
introduce a collaborative work with experts in various state-of-the-art numerical
methods to perform exhaustive simulations to provide a definitive resolution of
the order in the underdoped ground state of the strong coupling regime, which was
left as an open question in Chapter~\ref{chpt:diagram}.
As we have shown before,
the possible inhomogeneous orders are very rich and sensitive to slight
changes in the parameters. Since it is a huge effort to deliver a detailed map
for the ground-state orders, we focus on the $1/8$-doped point to demonstrate
that it is possible to complete such a map with the latest development of numerical
method.

At this point, We find a stripe order that has a highly
compressible wavelength on an energy scale of a few Kelvin, with wavelength
fluctuations coupled to pairing order. The favored filled stripe order is
different from that seen in real materials, indicating the possibilities of
other interactions beyond the Hubbard model at play. Our results demonstrate
the power of modern numerical methods to solve microscopic models even in the
most challenging settings.

In Sec.~\ref{sec:stripe:intro}, we introduce the existing evidence and
characterizations of the stripe orders. In Sec.~\ref{sec:stripe:method},
we discuss the numerical methods and data analysis techniques employed
in the study. Sec.~\ref{sec:stripe:result} presents the computational results,
focusing on the competing orders in the $1/8$-doped Hubbard model at $U/t=8$,
while in Sec.~\ref{sec:stripe:conclusion}, we summarize the findings and
discuss future directions.

\section{Introduction} \label{sec:stripe:intro}
Competing inhomogeneous orders are a common feature in many strongly correlated
materials~\cite{Dagotto2005}. A famous example is found in the underdoped region
of the phase diagram of the high-temperature cuprate superconductors (HTSC).
Here, multiple probes, including neutron scattering, scanning tunneling
microscopy, resonant X-ray scattering, and nuclear magnetic resonance
spectroscopy all lend support to various proposed inhomogeneous orders, such as
charge, spin, and pair density waves, with suggested patterns ranging from
unidirectional stripes to checkerboards~\cite{comin2016resonant,julien2015}. Recent
experiments on cuprates  indicate that the observed inhomogeneous
orders are distinct from, and compete with, pseudogap
physics~\cite{parker2010fluctuating,gerber2015three}.

Much theoretical effort has been directed to explain the origin of the
inhomogeneities~\cite{Fradkin2015}.  Numerical calculations on microscopic
lattice models have provided illuminating examples of the possible orders.
The prototypical lattice model to understand HTSC is the 2D Hubbard model on
the square lattice, with the Hamiltonian
\begin{equation}
  \hat{H} = -\sum_{\langle ij \rangle,\sigma \in \{\alpha,\beta\}} t
  a^\dag_{i\sigma} a_{j\sigma} + U \sum_i n_{i\alpha}
  n_{i\beta}
\end{equation}
where $a^{(\dagger)}$ and $n$ denote the usual fermion creation, annihilation,
and number operators, and $t$ and $U$ are the kinetic and on-site repulsion
energies.

A large number of numerical techniques have been applied to compute
the low-temperature and ground-state phase diagram of this model.
Early evidence for unidirectional stripe ordering in the Hubbard model
came from Hartree-Fock calculations~\cite{Poilblanc1989,Zaanen1989,machida89,schulz89},
while non-convex energy versus filling curves in exact
diagonalization of small clusters of the related $t$-$J$ model
were interpreted as signs of macroscopic phase separation
~\cite{Emery1990,Emery1990a}.
Since then, inhomogeneous orders have been observed both in the Hubbard
and $t$-$J$ models with density matrix renormalization group (DMRG)
~\cite{white1998density,white2003stripes,hager2005stripe},
variational quantum Monte Carlo~\cite{himeda02} and constrained path auxiliary field
quantum Monte Carlo (AFQMC)~\cite{chang2010spin}, (iPEPS)~\cite{Corboz2014}, 
density matrix embedding theory (DMET)~(see Chapter~\ref{chpt:diagram} and
Ref.~\cite{Zheng2016}), and functional renormalization group
~\cite{yamase2016coexistence} calculations amongst others, although not necessarily
the same kind of inhomogeneity is observed in each case.

However, there are other sophisticated simulations, for example, with 
variational and projector quantum Monte Carlo~\cite{Sorella2002,hu12}, and
cluster dynamical mean-field theory, which do not see, or are unable to resolve,
the inhomogeneous order~\cite{macridin2006phase,LeBlanc2015}.
The most recent studies with iPEPS~\cite{Corboz2014} and
DMET~\cite{Zheng2016}, as well as some earlier variational calculations
~\cite{himeda02,raczkowski07,chou08,chou10}, further show that both homogeneous
and inhomogeneous states can be observed and stabilized within the
same numerical methodology, with a small energy difference between
homogeneous and inhomogeneous states, on the order of $\sim 0.01t$ per site,
corresponding to a smaller energy scale than the common cuprate superconductivity
transition temperatures.

The small energy differences between orders means that very small biases in ground
state simulations, such as from an incomplete treatment of fluctuations, using
insufficiently accurate constraints to control the sign problem, bias towards low
entanglement states, or from finite size effects,  can easily stabilize one order
over the other. Similarly, the low temperatures needed to resolve between orders
is a challenge for finite temperature numerical methods
~\cite{white1989numerical,wu2016controlling}.

However, in this work we will demonstrate that, with the latest numerical techniques,
obtaining a {\it definitive} characterization of the ground state order in the
underdoped region of the 2D Hubbard model is now an achievable goal. As a representative
point in the phase diagram, we choose the iconic $1/8$ doping point at strong
coupling ($U/t=8$). Experimentally, this doping corresponds to a region of maximal
inhomogeneity in many HTSC's, and in the strong coupling regime it is
recognized as a point of maximum numerical difficulty and uncertainty in
simulations~\cite{LeBlanc2015}.

Using state-of-the-art computations with detailed cross checks and validation,
including newer methodologies such as infinite projected-entangled pair states
(iPEPS) and density matrix embedding theory (DMET) as well as recent developments in
established methodologies such as constrained-path auxiliary field quantum Monte Carlo
(AFQMC) and density matrix renormalization group (DMRG), and with exhaustive accounting
for finite size effects combined with calculations directly in the thermodynamic
limit, we are able to achieve unprecedented accuracy in this challenging region of the
ground-state phase diagram. In so doing, we can finally answer the question:
what is the order and physics found in the underdoped ground state of the 2D
Hubbard model?

\section{Methods} \label{sec:stripe:method}
\subsection{Overview} \label{sec:stripe:method:overview}
An important new strategy we bring to bear on this part of the Hubbard model
phase diagram is to combine the insights of multiple numerical tools with
complementary strengths and weaknesses. This approach, pioneered in an earlier
work on the Hubbard model~\cite{LeBlanc2015}, greatly increases the confidence
of the numerical characterization. To understand what each method contributes,
we briefly summarize the theoretical background and corresponding sources of error,
and then discuss the detailed settings for each method in this work.

\noindent {\bf Auxiliary field quantum Monte Carlo}.~\footnote{AFQMC calculations are performed by Mingpu Qin, Hao Shi and Shiwei Zhang.} AFQMC expresses the ground
state of a finite system through  imaginary time evolution (Eq.~\ref{eq:dmet:solver:afqmc_evolve}).
The projection is Trotterized, and the evolution reduces to a stochastic single-particle evolution in the presence of
auxiliary fields generated by the Hubbard-Stratonovich decoupling of the Hubbard repulsion.
Away from half-filling, this decoupling has a sign problem.
We use the constrained path (CP) approximation, to eliminate the sign problem at the cost of a bias
dependent on the quality of the trial state~\cite{Zhang1995,chang2008prb}.
In this work, the Trotter error is well converged and we report the statistical error bar.
To minimize the constrained path bias, we use several different trial states, including  self-consistent optimization of the trial state~\cite{qin2016coupling}.
The calculations are carried out on finite cylinders with open, periodic, and twist-averaged boundary conditions,
with widths of up to 12 sites, and lengths of up to 72 sites.
This method can reach large sizes and finite size effects are minimized. The uncontrolled
error is from the CP approximation.

\noindent {\bf Density matrix renormalization group}.~\footnote{The lattice basis DMRG calculations are performed by Chia-Min Chung and Steven R. White. The hybrid basis
	DMRG calculations are performed by Georg Ehlers and Reinhard M. Noack.} DMRG is a variational wavefunction
approximation using matrix product states (MPS), which are low-entanglement states with a
1D entanglement structure. The quality of the approximation is determined by the bond
dimension ($M$) of the MPS (Sec.~\ref{sec:dmet:solver:dmrg}). The calculations are carried
out on finite cylinders with widths of up to 7 sites, and lengths of up to 64 sites, 
with periodic boundary conditions in the short direction and open boundaries in the long direction.
Two different DMRG algorithms were used: one working in a pure (real-space) lattice basis, and
another in a mixed momentum/lattice (hybrid) basis, with the momentum representation
used along the short periodic direction~\cite{Motruk2016}. We remove the bond dimension error
and finite size error in the long direction by well-known extrapolation procedures, and report
the associated error bar~\cite{stoudenmire2012studying}. Consistency between the lattice and
hybrid DMRG algorithms provides a strong validation of this error bar. The remaining uncontrolled
error is the finite width error in the periodic direction.

\noindent {\bf Density matrix embedding}. DMET is a quantum embedding method which 
works directly at the thermodynamic limit, although interactions are only accurately treated
within  a fragment (see Sec.~\ref{sec:dmet:bcs})~\cite{Knizia2012}.
To solve for the ground state of the impurity model, consisting of
a supercell of the original lattice (the fragment) coupled to a set of self-consistently determined
bath orbitals, we use a DMRG solver~\cite{Chan2011}. We treat a series of fragment
supercells with up to 18 sites ($9\times2$) to target different low-lying states.
With the narrow shapes of the fragments used in this work, we are able to
perform DMRG calculations with negligible truncation errors; and we do not extrapolate the observables
to the thermodynamic limit (with respect to the fragment sizes). Thus, the error bar
reported in DMET only corresponds to the estimated error from incomplete self-consistency of the
impurity problem~(see Sec.~\ref{sec:diagram:error}). The remaining uncontrolled error is
the finite fragment size error.

\noindent {\bf Infinite projected entangled pair states}.~\footnote{iPEPS calculations are performed by Philippe Corboz.}
iPEPS is a variational approach using a low-entanglement tensor network ansatz
natural to 2D systems~\cite{verstraete2004,nishio2004,jordan2008classical}. The
calculations are carried out directly in the thermodynamic limit where
different supercell sizes including up to 16 sites
(independent tensors that can break symmetry) are used to target
different low-energy states. As in DMRG, the accuracy of the ansatz is
systematically controlled by the bond dimension $D$ of the
tensors. Estimates of quantities in the exact $D$ limit are obtained
using an empirical extrapolation technique which is a potential source
of uncontrolled error.

\noindent {\bf Cross-checks: systematic errors, finite size biases}. The use of multiple techniques allows
us to ameliorate the uncontrolled errors from one technique using information from another. For example,
by carrying out simulations on the same finite clusters in the AFQMC and DMRG calculations, we can estimate
the constrained path bias in AFQMC. Similarly, in the AFQMC calculations we can treat larger width cylinders
than in the DMRG simulations; thus we can estimate the finite width error in DMRG.
In other examples, DMRG may miss low-lying states with higher entanglement in finite bond dimension calculations.
If, however, these states show up in other methods, we can control the initial boundary conditions or trial states
of the DMRG calculations and target the missing states. These states may later show lower energy when extrapolated
to infinite $M$.

In all of the methods, there is a bias towards orders commensurate with the shape of the simulation cell, be 
it the finite lattice and boundary conditions in AFQMC/DMRG, or the fragment cluster in DMET,  or the supercell in iPEPS.
Using this bias, together with different boundary conditions and pinning fields, we
can stabilize different {\it meta-stable} orders. For example, by setting up clusters commensurate 
with multiple inhomogeneous orders and observing the order that survives, we can determine the relative energetics of 
the candidate states. We can fit the orders along the short axis or the long axis of the cluster to obtain 
two independent estimates of the energy. We have carried out exhaustive studies of about 100
different combinations of clusters, cells, and boundary conditions, 
to fully investigate the low-energy landscape of states. 

To characterize the orders, we use the local hole density $h_i=1 - (\langle n_{i\alpha}+n_{i\beta}\rangle)$,  
magnetic moment $m_i=\frac{1}{2} \langle n_{i\alpha}- n_{i\beta}\rangle$,  and pairing order 
$\kappa_{ij}=\frac{1}{\sqrt{2}}(a_{i\alpha}^{\dagger}a_{j\beta}^{\dagger}+a_{j\alpha}^{\dagger}a_{i\beta}^{\dagger})$
($i$ adjacent to $j$).

\subsection{Detailed Specifications} \label{sec:stripe:method:detail}
In this section, We discuss the process, scope and computational parameters
used in each of the computational methods in detail.

\subsubsection{Auxiliary field quantum Monte Carlo} \label{sec:stripe:method:detail:afqmc}
Two sets of AFQMC calculations are performed using the constrained-path AFQMC method
with self-consistently optimized, unrestricted Hartree-Fock trial wavefunctions~\cite{qin2016coupling}.

The first set is on cylinders of dimension $L_x \times L_y$ ($L_x > L_y$) with
open boundary conditions (OBC) along the $x$-direction and periodic boundary conditions (PBC) along the $y$-direction.
This allows the AFQMC calculations to be directly compared with DMRG results. We also run calculations with pinning fields to
fix the desired spin structures. Several types of antiferromagnetic (AF) pinning fields are applied along the open edges
of the cylinders depending on the target state. The types of pinning fields include: the in- and anti-phases pinning fields
along the two boundaries, and the pinning fields applied to only one edge.

With in- (anti-) phase pinning fields, we target an odd (even) number of nodes (where the $\pi$-phase shift happens) in
the system ($L_x$ is always even in the calculations); while pinning fields on only one edge are able to accommodate states
with different wavelengths so we can learn which wavelength survives in the competition.
The strength of the pinning fields is $|h| = 0.5t$ for all calculations.

In the second set of calculations, we use PBC or twist averaged boundary conditions (TABC) along both directions.
The twist averaging allows us to reduce the finite size errors in the total energy, thus giving a more unbiased estimation
of the energy order for various low-lying states.

For $U/t=8$, the wavelength 5, 6, 8, 10 and 12 stripes are explicitly studied in AFQMC using cylinders, while the energy
estimates from PBC and TABC calculations are obtained for wavelengths 5 to 10. Similar calculations are also performed
at $U/t=6$ and $U/t=12$.

\subsubsection{Density matrix renormalization group in the lattice basis} \label{sec:stripe:method:detail:lattice_dmrg}
Lattice-basis DMRG calculations are carried out on cylinders
of size $L_{x}\times L_{y}$ ($L_x > L_y$) as well. Both fixed particle
number (at $1/8$ doping) and broken particle-number symmetry calculations are
conducted.

To stabilize the states with particular wavelengths, we use as initial states product
states with holes in the desired locations, and apply temporary fields
and (site-dependent) chemical potentials on the whole cylinder in the first few sweeps.
A temporary chemical potential $\mu=2.0$ is applied on the sites where the holes
are supposed to be in the final striped state, and a temporary magnetic field of
strength $|h|=0.5$ is applied to fit the antiferromagnetic domains between holes.
After the first few sweeps (typically up to bond dimension $M=600$) the temporary
field and chemical potential are turned off, and the state is called ``stable'' in
the DMRG simulation if it keeps the same wavelength during subsequent sweeps.
In some cases, an AF pinning field of strength $|h|=0.5$ at the open edges is kept in the
full simulation to further stabilize the state. By extrapolating to the infinite length,
the pinning field does not affect the energy per site at the thermodynamic limit.

We next consider the pairing order. To measure the pairing order, we simulate the Hubbard
cylinder in the grand canonical ensemble, i.e., without particle number conservation.
The chemical potentials are tuned so that the expectation value of
the particle number is close to the desired value ($7/8$ of the number
of sites). We apply pairing pinning fields at the edges, to induce broken
particle-number symmetry, and observe how the pairing order decays into the bulk.

The lattice-basis DMRG simulations are carried out at $U/t=8$ and $U/t=12$,
the width 4, 6 and 7 cylinders to simulate stripes of various wavelengths.

\subsubsection{Density matrix renormalization group in the hybrid basis} \label{sec:stripe:method:detail:hybrid_dmrg}
Again, the hybrid-basis DMRG calculations are carried out on cylinders of
size $L_x\times L_y$, with OBC in the longitudinal $x$-axis and
PBC or APBC in the transverse $y$-direction.

The DMRG algorithm in a mixed--real-momentum-space (hybrid)
representation~\cite{Motruk2016} is used.
The hybrid-basis DMRG algorithm uses a real-space representation
in the longitudinal cylinder direction
and a momentum-space representation in the transverse direction.
The additional transverse-momentum quantum number
grants us a speedup over lattice-basis DMRG
whose cost grows with the width of the cylinder.
For width-6 Hubbard cylinders, the hybrid-space algorithm
is approximately 20 times faster than its real-space
counterpart.

To obtain the ground-state energy for fixed $L_y$ (while $L_x$ is infinite),
we perform consecutive extrapolations  first in the DMRG truncation error
$\Delta w$ and then in the inverse cylinder length $1/L_x$.

In order to directly target and stabilize different stripe configurations on
width-6 cylinders with PBC, we use a sine-shaped pinning field coupled to the
local charge density
\begin{equation}
	P = \sum_{x\,y\,\sigma} A
	\cos(k_x \, x \, + \,  \phi_0 )
	n^{\vphantom{\dagger}}_{x\,y\,\sigma}
\end{equation}
with suitable amplitude $A$, wave vector $k_x$, and phase $\phi_0$. Note that
the pinning field only acts on the charge density, while does not explicitly
break the spin or particle-number symmetries. The contribution to the ground-state
energy, $\langle \Psi_0 | P | \Psi_0 \rangle$, is subtracted after the DMRG
calculation. We found a field amplitude of $A=0.01$ to be sufficient
to stabilize the different stripe patterns and to improve the convergence of
the DMRG algorithm.
After the calculations are converged, the charge order can be measured
directly from the one-body density matrix, while the spin structure factor
and pairing correlation functions are measured to obtain the spin and pairing
order.

The calculations are carried out on width 4 to 8 cylinders of various lengths.
However, for width 8, we do not achieve sufficient convergence in terms of DMRG
density matrix truncation, despite using up to $M=35000$ states.

\subsubsection{Density matrix embedding theory} \label{sec:stripe:method:detail:dmet}
\begin{figure}[htpb]
  \centering
  \includegraphics[width=0.7\textwidth]{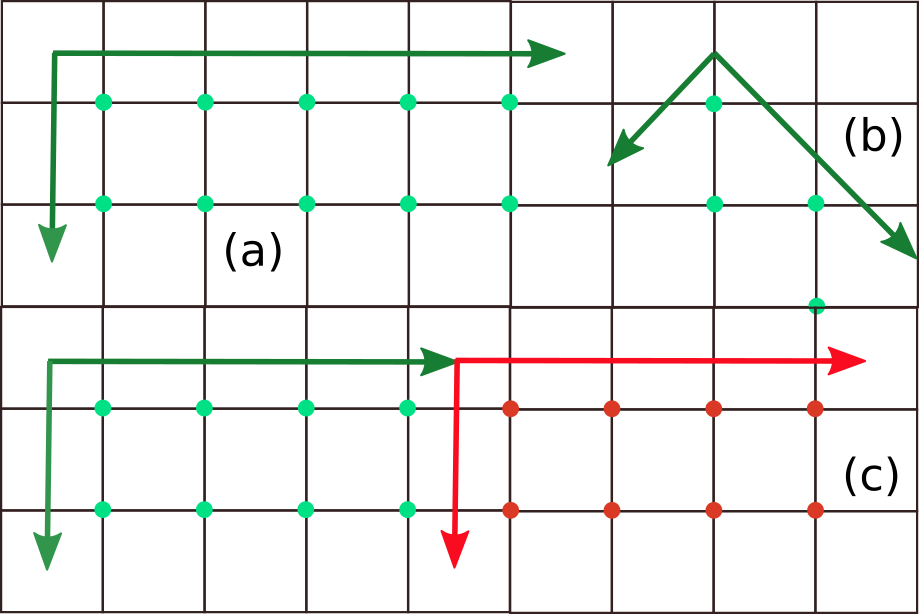}
  \caption{Fragment shapes used in the DMET calculations:
	  (a) $\lambda\times2$ fragments.
	  (b) $\lambda\sqrt{2}\times\sqrt{2}$ fragments.
          (c) $\lambda\times2$ fragments with spin inversions on neighboring
  supercells.}
  \label{fig:stripe:dmet_cells}
\end{figure}

We perform the DMET calculations on an auxiliary square lattice of dimension
$L\times L$, where $L=160$ for most fragment sizes ($L=168$ for $7\times2$
fragments and $L=162$ for $9\times2$ fragments). The correlation potential
is allowed to break spin and particle-number symmetries, and so do the
impurity model wavefunctions. The calculations are carried out as
described in Appendix~\ref{sec:algo:bcs}, and similar to those in Chapter
~\ref{chpt:diagram}, but a larger number of fragment sizes, shapes and
boundary conditions are explored. Various shapes of fragments are used
to accommodate uniform $d$-wave order, vertical and diagonal stripes.
The shapes of the fragments are summarized in Fig.~\ref{fig:stripe:dmet_cells}.
We do not attempt to do extrapolations of fragment size in this work
but compare the energies of different fragments directly.
This is making an implicit  assumption that the finite size errors are about the
same for different fragment shapes employed here, and this lack of TDL
extrapolation is the main systematic error. In our experience, however,
it is reasonable to directly compare energies of fragments of the same
orientation and family (e.g. $L \times 2$), which is confirmed in this
work by comparison with the other techniques.

$\lambda\times2$ cells [Fig.~\ref{fig:stripe:dmet_cells}(a)] are used
to study vertical stripes with odd wavelengths. As the AF order has a
$\pi$-phase shift at the domain wall, the AF order is commensurate with the
cell size. For even-wavelength stripes, the setup is similar, however,
to support a single domain wall, it is necessary for the spin wavelength
to be twice that of the charge wavelength. To allow this, rather than
using a large cluster of size $2\lambda\times2$, we modify the way the
correlation potential is added to the lattice wavefunction, i.e. by
swapping the spin channels between neighboring cells in the longitudinal
direction [Fig.~\ref{fig:stripe:dmet_cells}(c)], so translation by a unit
cell gives a time reversal, $n_{i\alpha} \to n_{i\beta}$ transformation.
Specifically, the local correlation potential in this case is written as
\begin{equation}
  u = \sum_{C_1}\sum_{i,j\in C_1}(\sum_\sigma u_{ij\sigma}a_{i\sigma}^{\dagger}a_{j\sigma} + \Delta_{ij}a_{i\alpha}^{\dagger}a_{j\beta}^{\dagger}+c.c.)
  + \sum_{C_2}\sum_{i,j\in C_2}(\sum_\sigma u_{ij\bar\sigma}a_{i\sigma}^{\dagger}a_{j\sigma} + \Delta_{ij}^\prime a_{i\alpha}^{\dagger}a_{j\beta}^{\dagger}+c.c.)
\end{equation}
where $C_1$ and $C_2$ label even and odd cells along the longitudinal $x$ direction.
Both $\Delta^\prime=\pm \Delta$ possibilities are tested in our calculations,
because we cannot determine the phase factor associated with the transformation
\begin{displaymath}
  a_{i\alpha}\rightarrow a_{i+R,\beta}, a_{i\beta}\rightarrow \pm a_{i+R,\alpha}.
\end{displaymath}
where the $R$ denotes translation by a unit cell. As shown in the results,
neither parameterization results in finite pairing order in the ground-state
of even wavelength stripes. This trick not only greatly reduces the computational
cost it would have taken to simulate the even-wavelength stripes, but also
makes the energies of odd and even wavelengths more comparable.

We also use the traditional cells in Fig.~\ref{fig:stripe:dmet_cells}(a) for
even wavelengths to simulate possible spin density wave states. One of
these states turns out to have lower energy than the stripe of the same
wavelength at $U/t=6$.

We also use the tilted clusters in Fig.~\ref{fig:stripe:dmet_cells}(b) to
accommodate diagonal stripes. As the finite-size effects are different
in regular and tilted clusters, we use both
$2\times2$ and $2\sqrt{2}\times\sqrt{2}$ clusters to obtain the
uniform $d$-wave state, to estimate the relative energies of the states
on regular and tilted lattices.

In all the calculations reported in this work, the DMRG solution of the
impurity model is converged in terms of truncation error, and the DMET
uncertainty comes solely from the convergence of the correlation potential.
We report the energy and its uncertainty as the average and half of the
difference of the last two DMET cycles, respectively.

\subsubsection{Infinite projected entangled-pair state} \label{sec:stripe:method:detail:ipeps}
An infinite projected entangled-pair state (iPEPS)~\cite{verstraete2004,Verstraete08,jordan2008} (also called a tensor
product state~\cite{nishino01,nishio2004}) is an efficient variational
tensor network ansatz for two-dimensional states in the thermodynamic limit
which obeys an area law of the entanglement entropy~\cite{eisert2010}.
The ansatz consists of a supercell of tensors which is periodically repeated
on a lattice, with one tensor per lattice site. Each tensor has a physical
index which carries the local Hilbert space of a lattice site and four
auxiliary indices which connect to the nearest-neighboring tensors on
a square lattice. Each auxiliary index has a certain dimension $D$, called
the bond dimension, with which the accuracy of the ansatz (the number of
variational parameters) can be controlled systematically. An iPEPS
with $D=1$ corresponds to a product state, and by increasing $D$ entanglement
can be systematically added. In this work we use bond dimensions up
to $D=16$ corresponding to highly-entangled states.

For translationally invariant states a supercell with only a single tensor
is needed. If the translational symmetry is spontaneously broken a supercell
compatible with the symmetry breaking pattern is required. For example,
a N\'eel ordered state requires a supercell with two different tensors
$A$ and $B$ (one for each sublattice), or a stripe state with period 5
requires a $5\times 2$ supercell with 10 independent tensors. A diagonal
stripe state with period $L/\sqrt{2}$ can be obtained in a
$L\times L$ rectangular supercell, or more efficiently by using a
$L\times 1$ supercell with $L$ different tensors and translation
vectors $v_1=(L,0)$, $v_2=(1,1)$.

By running simulations with different supercell sizes one obtains different
competing low-energy states. In order to determine which of these competing
low-energy states corresponds to the true ground state a systematic analysis
of the energy as a function of $D$ is required. Here we used the extrapolation
technique from Ref.~\cite{Corboz2016} in which the energy is plotted
as a function of the so-called truncation error $w$ in the simulation,
and then the extrapolation to the $w\rightarrow 0$ limit is taken to
determine the energy of each of the competing states. While in 2D it
is theoretically unknown how the energy depends on $w$, several
benchmarks~\cite{Corboz2016} have empirically shown that an accurate
estimate can be used using a polynomial fit.

In this work the optimization of the tensors has been done using an
imaginary time evolution based on the so-called full update~\cite{corboz2010}
(or fast-full update~\cite{phien15}), which is more accurate than the simple
update approach~\cite{jiang2008}.  Observables are evaluated by contracting
the two-dimensional tensor network in a controlled, approximate way, using
a variant~\cite{Corboz2011,Corboz2014} of the corner-transfer matrix (CTM)
method~\cite{nishino1996, orus2009-1}. The accuracy of the  contraction is
controlled by the ``boundary'' dimension $\chi$, which is chosen large enough
such that the resulting error is small (compared to the effect of the
finite~$D$). To increase the efficiency we make use of Abelian
symmetries~\cite{singh2010,bauer2011}. Fermionic statistics are taken
into account following the formalism explained in Refs
~\cite{Corboz09_fmera, corboz2010}.

\subsection{Estimating Long-range Coulomb Interaction} \label{sec:stripe:method:coulomb}
We estimate the long-range Coulomb interactions of various low-lying states to
understand how these interactions may change the relative order of the states
in real cuprates. We use charge order from DMET calculations, and compute
\begin{equation}
  e_{\text{Coul}} = \frac{1}{N_c}\sum_{i\in \text{frag}, j, i\neq j}
  (h_i-\bar{h})(h_j-\bar{h})/4\pi\varepsilon_0\varepsilon r_{ij}
\end{equation}
where $N_c$ is the size of the fragment, $h_i$ is the hole density on site $i$,
and $\bar{h}$ is the average hole density ($1/8$).

In atomic units, i.e. if we express the energy in Hartrees, and
distance in Bohr, $1/4\pi\varepsilon_0=1$. The appropriate dielectric constant
to use in a statically screened Coulomb interaction in the CuO$_2$ plane has
been estimated to lie between about $4$ and $27$
~\cite{schuttler2001screening,arrigoni2002stripes}.
We use a dielectric constant of $\varepsilon=15.5$, and a lattice constant
$a=3.78$\AA$=7.14 \text{ Bohr}$ corresponding to the lattice constant of
La$_2$CuO$_4$. We transform the computed Coulomb energy (per site) to units of
$t$, using $t\sim3000K\sim0.01 \text{ Hartree}$.

In 2D, the Coulomb summation converges reasonably fast. We choose a cutoff radius
as 300 lattice spacings and converge the Coulomb energy to the fourth digit in
units of $t$. This is of course just a crude, order of magnitude estimate,
because we neglect the effect of relaxation in the presence of the Coulomb
interaction, and there is significant uncertainty in the dielectric.

\section{Results} \label{sec:stripe:result}
Using the above methods, we carry out calculations for the ground
state of the 2D Hubbard model at $1/8$ doping at $U/t=8$.
The first check of reliability is the independent convergence of the
methods for the energy per site. While the quality of the ground-state
energy may be a poor proxy for the quality of the corresponding state
when the overall accuracy is low (as there are always many degenerate states
far above the ground state), well-converged energies are a much tighter
constraint on the ground state order, as any degeneracies must be below
the energy convergence threshold.

\begin{figure}[htpb]
  \centering
  \includegraphics[width=0.8\columnwidth]{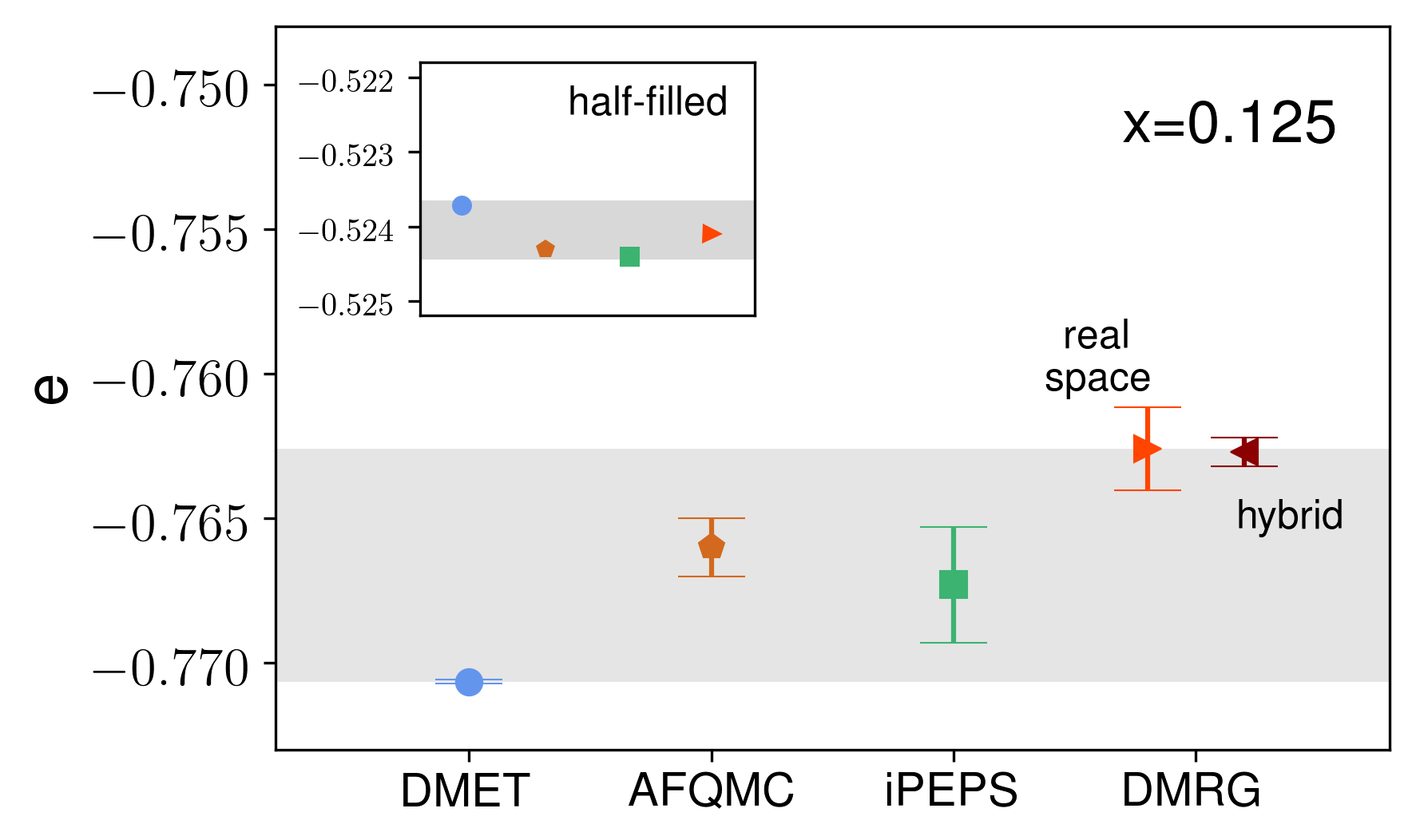}
  \caption{Best estimates of ground state energies per site
	  for the $1/8$-doped 2D Hubbard model at $U/t=8$ from
	  DMET, AFQMC, iPEPS and DMRG.
    	Inset: Best estimates of ground state energy for the
	half-filled 2D Hubbard model at $U/t=8$. Here and elsewhere
	in this chapter, error bars indicate only the estimable
	numerical errors of each method; uncontrolled systematic
errors are not included.} 
  \label{fig:stripe:gs_energy}
\end{figure}

\begin{figure}[htpb]
	\centering
	\includegraphics[width=0.65\textwidth]{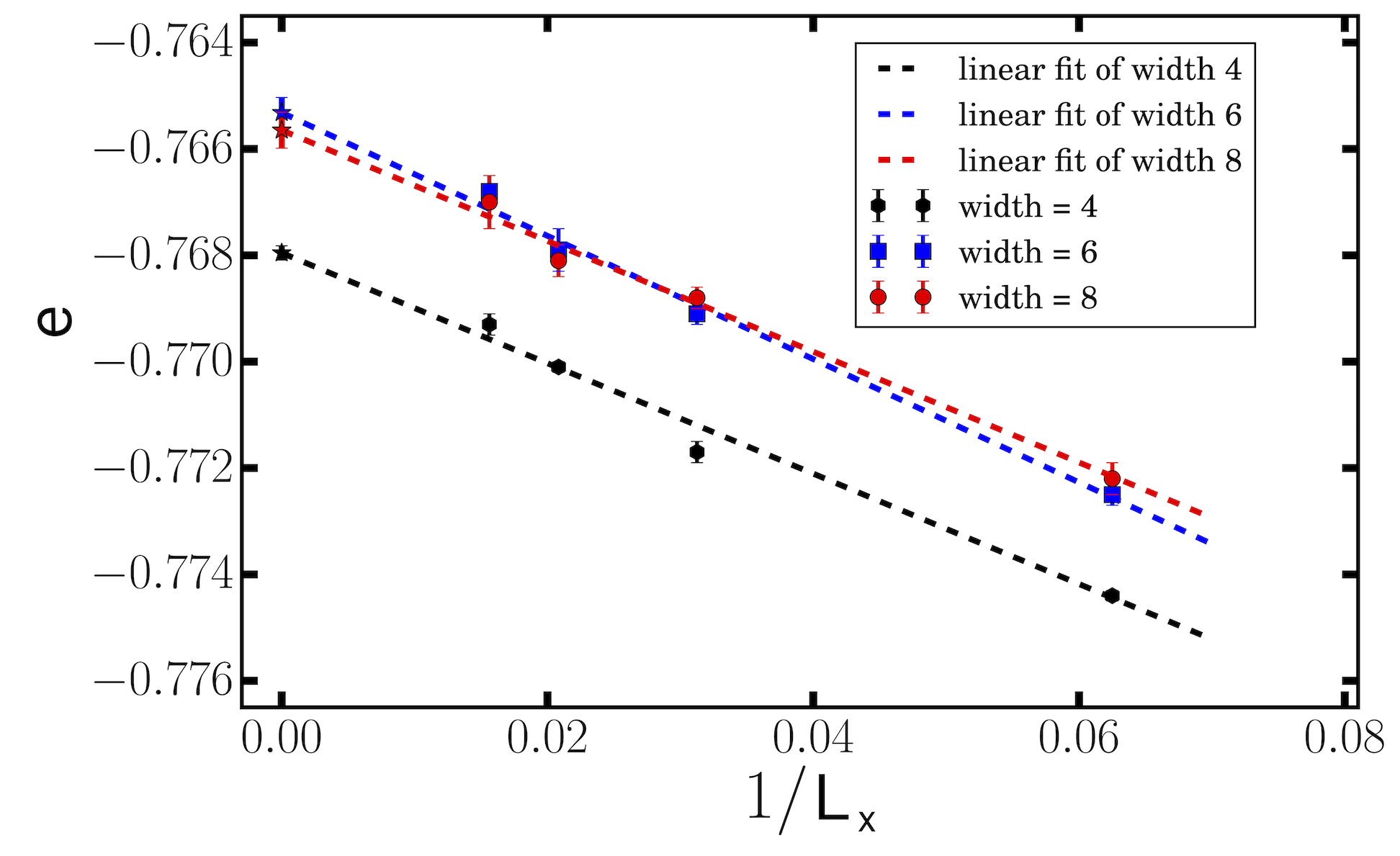}
	\caption{Energies per site for wavelength-$8$ stripes in various cylinder sizes for 
		$U/t = 8$. Pinning fields are applied on the open boundaries. Linear fits of
		$1/L_x$ are shown. The infinite length values and error bars from extrapolation
	are marked as stars in the plot. }
	\label{fig:stripe:afqmc_width}
\end{figure}

Fig.~\ref{fig:stripe:gs_energy} shows the best energy estimate for the
ground state from the different methods. The two different DMRG
formulations (real space and hybrid basis) agree perfectly, providing
a strong independent check of the calculations, and in subsequent figures
we report only the single consistent result. Note that the error bars
for AFQMC, DMRG, and DMET do not reflect the uncontrolled systematic
errors in the methods. However, as described above, the systematic
errors can be estimated by cross checks between the methods.
For example, DMRG and AFQMC calculations on finite clusters with identical
boundary conditions provide an estimate of the small constrained path bias
(see Table~\ref{tab:stripe:energy_cp} and Ref.~\cite{qin2016coupling}) consistent with
the difference in the DMRG and AFQMC energies in Fig.~\ref{fig:stripe:gs_energy}
(around $0.0035t$); similarly AFQMC extrapolations to the thermodynamic limit
indicate that the DMRG energies (using cylinders up to width 7) are essentially
converged with respect to cylinder width (Fig.~\ref{fig:stripe:afqmc_width}).

\begin{table}[h]
	\centering
	\caption{Comparison of energies per site from AFQMC and DMRG for various inhomogeneous states
	on cylinders with pinning fields. The CP-error denotes the energy difference between AFQMC and DRMG
results, as a result of the constrained path error. Note the wavelength-5 stripes are meta-stable.}
	\label{tab:stripe:energy_cp}
	\begin{tabular}{p{0.15\textwidth}p{0.3\textwidth}p{0.15\textwidth}p{0.15\textwidth}p{0.15\textwidth}}
		\toprule
		\textbf{Size} & \textbf{Stripe wavelength} & \textbf{DMRG} & \textbf{AFQMC} & \textbf{CP-Error}\\
		$4\times16$ & 8 & -0.77127(2) & -0.7744(1) &	-0.0031	\\
		$6\times16$ & 5 &-0.7682(3) & -0.7692(1)   &	-0.0010	\\
		$6\times16$ & 8 & -0.7691(5) & -0.7725(2)  &	-0.0034	\\
		$4\times24$ & 8 &-0.76939(3) & -0.7727(2)  &	-0.0033	\\
		$6\times32$ & 5 &-0.7648(3) & -0.7663(1)  &	-0.0015	\\
		$6\times32$ & 8 &-0.7658(7) &-0.7691(2)&	-0.0033	 \\
	\bottomrule
	\end{tabular}
\end{table}

The ground-state energy has close agreement between all the methods, and all energies
lie in the range $-0.767 \pm 0.004t$. If, for a typical HTSC material, we estimate
$t \sim 3000K$, then this corresponds to a range of about $\pm 10K$ per site, or
$\pm 100K$ per hole. For a numerical comparison, this is also more than an order
of magnitude lower than the temperatures accessible in finite temperature,
thermodynamic limit, simulations in this part of the phase diagram, indicating
that we are potentially accessing different physics
from what haven been seen before~\cite{LeBlanc2015,wu2016controlling}.
Shown in the inset are the corresponding best estimates at half filling
from the same methods, where the spread in energies is less than $0.001t$.
This illustrates the significantly greater numerical
challenge encountered in the underdoped region.  Nonetheless, the accuracy and agreement
reached here represents a ten-fold improvement over recent comparisons of numerical
methods at this point in the phase diagram~\cite{LeBlanc2015}.

\noindent\textbf{Ground state stripe order}. For all the methods employed, the lowest energies shown in Fig.~\ref{fig:stripe:gs_energy}
correspond to a {\it vertical striped state}. This corresponds to a co-directional charge
and spin-density wave, with the region of maximum hole density coinciding with a domain wall
and $\pi$-phase shift in the antiferromagnetism. As mentioned, unidirectional stripes of
various kinds are a long-standing candidate order in the doped Hubbard and related models.
Hartree-Fock calculations give filled stripes (i.e. one hole per charge unit cell)
in both vertical and diagonal orientations, while one of the first applications of the DMRG
to 2D systems found strong evidence for half-filled stripes in the $t$-$J$
model~\cite{white1998density}. Finally, one of the earliest examples of inhomogeneity
in doped HTSC's were the static half-filled stripes in LaSrCuO at 1/8 doping
~\cite{tranquada1995evidence}.

The convergence to the same inhomogeneous order in the ground state in the current study,
from multiple methods with very different approximations, strongly suggests that stripes
indeed represent the true ground state order of the Hubbard model in the underdoped regime,
and further highlights the accuracy we achieve with different techniques.
However, the stripe order we observe has some unusual characteristics. We return to the details
of the stripe order, its associated physics, and its relationship with experimentally observed
stripes further below. First, however, we examine the possibility of other competing meta-stable
states.

\begin{figure}[htpb]
  \centering
  \includegraphics[width=\columnwidth]{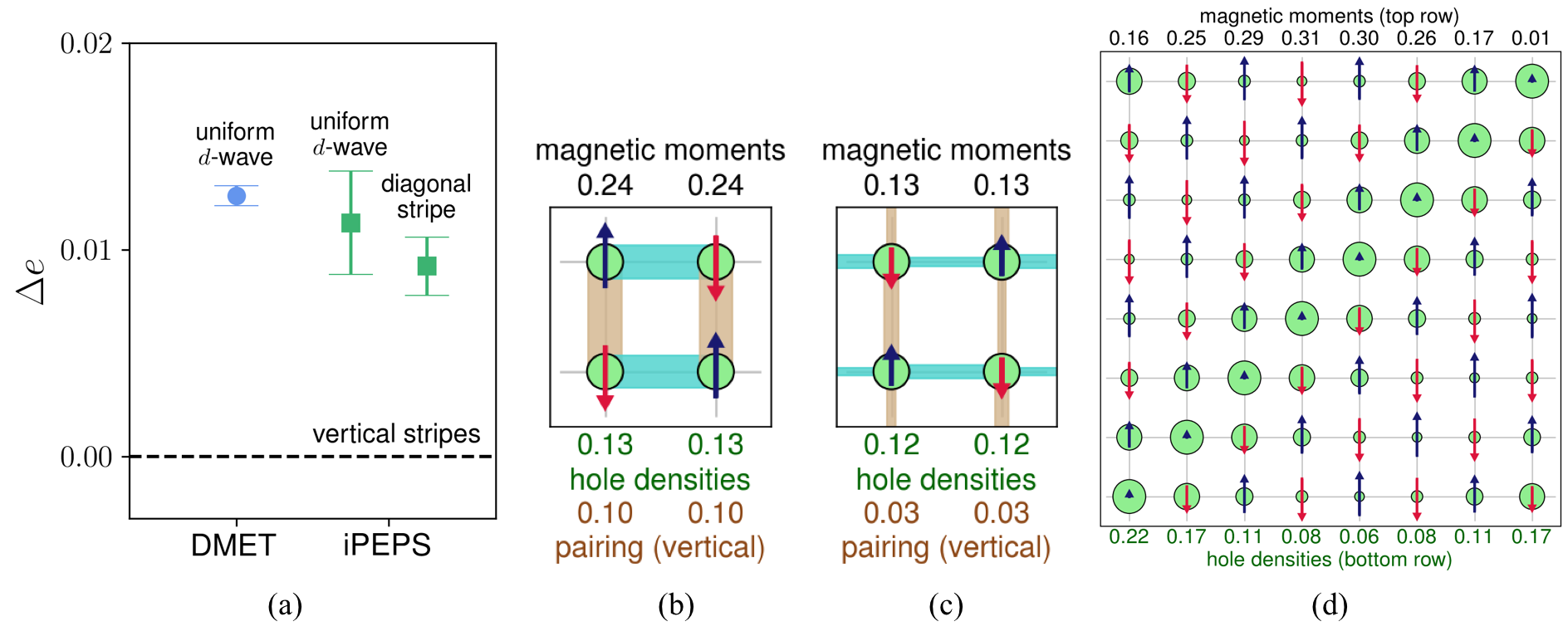}
  \caption{Important candidate states relative to the striped ground state from
	  DMET and iPEPS and the orders.
	  (a) Relative energy of competing states compared to the vertical striped state.
	  Charge, spin and pairing orders of the uniform $d$-wave state from (b) DMET and (c) iPEPS.
	  (d) Charge and spin orders of the diagonal striped state from iPEPS. Note that
  the spins are flipped in the neighboring supercells. (Circle radius is proportional to hole density, arrow height is proportional to spin density, bond width is proportional to pairing density).}
  \label{fig:stripe:competing}
\end{figure}

\noindent {\bf Competing states: uniform $d$-wave state}.
Recent work using iPEPS and DMET on the $t$-$J$ and Hubbard models suggested
close competition between a uniform $d$-wave superconducting ground state and
a striped order~\cite{Corboz2014,Zheng2016}. With an improve energy resolution
in this work, uniform states did not spontaneously appear in any of our calculations
which indicates that they lie higher in energy than the striped order.
We found that we could stabilize a uniform $d$-wave state
in the DMET calculations by constraining the fragment to a $2\times 2$ or
$2\sqrt{2} \times \sqrt{2}$ geometry [Fig.~\ref{fig:stripe:competing}(b)] and 
in the iPEPS calculations by using a $2\times2$ unit cell [Fig.~\ref{fig:stripe:competing}(c)].
DMET calculations on similarly shaped larger clusters (such as a $4\times 4$ cluster)
spontaneously broke symmetry to create a non-commensurate non-uniform state. From these
calculations we estimate that the uniform state lies $\sim 0.01t$ above the lowest energy
state [Fig.~\ref{fig:stripe:competing}(a)], and is not competitive at the energy resolution
we can now achieve.

\noindent {\bf Competing states: diagonal stripes}. While we find the ground state order to be
a vertical stripe type order, other studies of stripes indicate that different orientations
can form~\cite{Kato1990}. On short length scales, the relevant question is whether diagonal
stripes [with a $(\pi, \pi)$ wave vector] are competitive with vertical stripes [with a
$(0, \pi)$ wavevector]. With the boundary conditions used in this work, diagonal stripes would
be frustrated in the DMRG and AFQMC calculations, and did not spontaneously appear.
To stabilize diagonal stripes in the DMET and iPEPS calculations, we used tilted
$n\sqrt 2 \times \sqrt 2$ fragments ($n=2,5)$ for DMET, and a $16 \times 16$ simulation cell
with 16 independent tensors in iPEPS. The $16 \times 16$ iPEPS cell gave a diagonal stripe
[Fig.~\ref{fig:stripe:competing}(d)] that was significantly higher in energy than the vertical
stripe, by $0.009t$. The DMET cluster gave rise to a frustrated diagonal order that we also
estimate to be higher in energy by $\sim 0.005t$. While it is likely that the orientation of
the stripe will depend on doping and coupling, vertical stripes appear to be significantly
preferred at this point in the phase diagram.

\noindent {\bf Competing states: other short-range orders}. While  other types of order have been proposed in the underdoped region,
such as spiral magnetic phases~\cite{Chubukov1995,yamase2016coexistence} and checkerboard order~\cite{Vojta2002},
we find no evidence for other kinds of short-range orders at this point in the phase diagram from any of our methods. The lack  of 
checkerboard order, which would easily fit within the large clusters
in our simulations (e.g. up to $64 \times 6$ in the DMRG calculations)
appears to rule them out as low energy states, in agreement with earlier DMRG simulations on the $t$-$J$ model~\cite{white2004checkerboard}.
While we cannot rule out incommensurate orders, we have found that the variation of energy with unit cell wavelength (see below) is 
quite smooth, thus we do not expect a dramatic energy gain from incommensurability. We note that studies finding
incommensurate magnetic orders have focused on smaller values of $U$~\cite{yamase2016coexistence}.

\begin{figure}[htpb]
  \centering
  \includegraphics[width=\columnwidth]{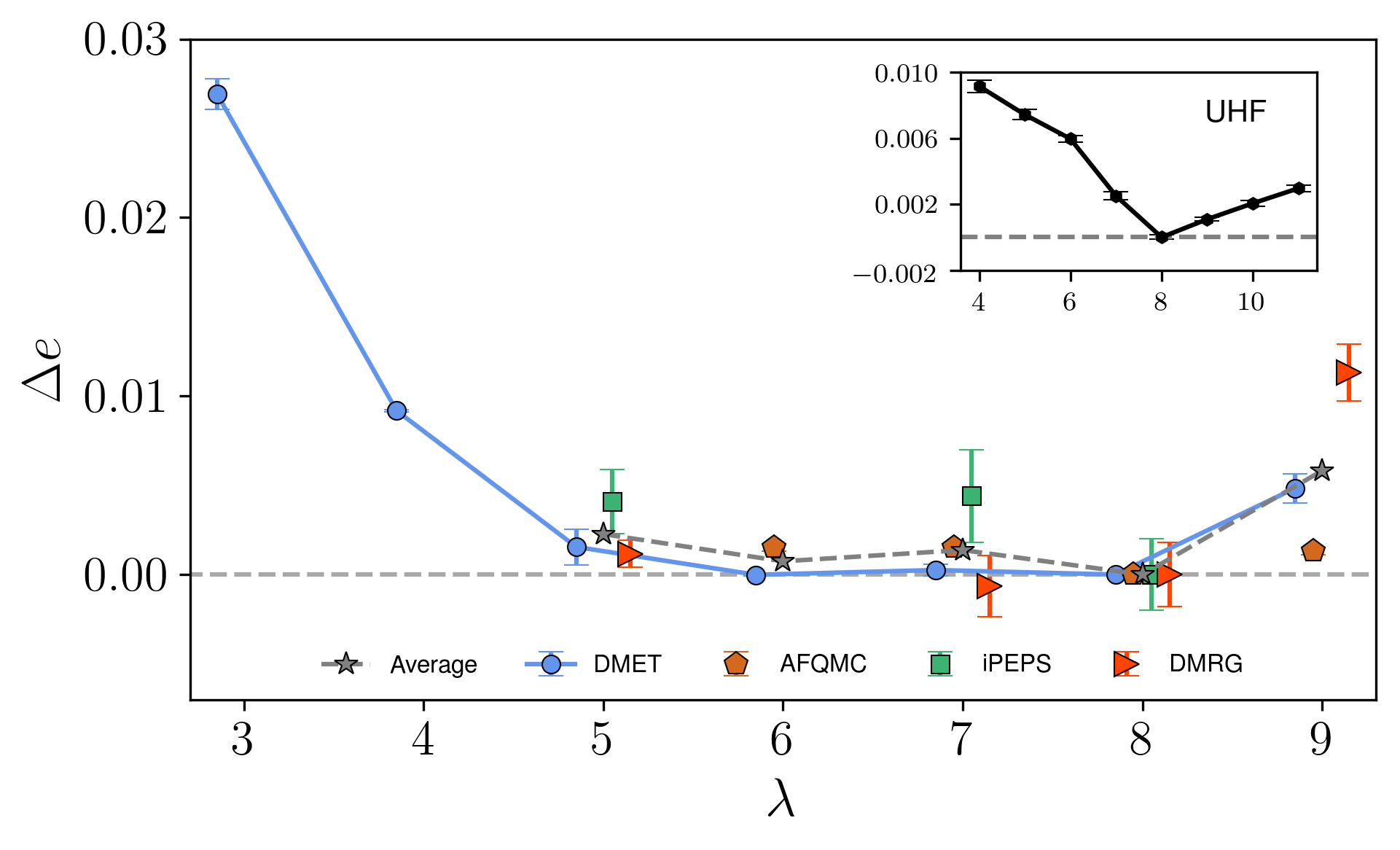}
  \caption{Energies per site for stripes with different wavelengths relative to
  that of the wavelength-8 stripe from DMET, AFQMC, iPEPS and DMRG. To aid
  readability, the data points are shifted horizontally. Inset: Relative energies
  of stripes with different wavelengths from UHF, with an effective coupling $U/t=2.7$ obtained from self-consistent AFQMC simulations.}
  \label{fig:stripe:energy_wavelength}
\end{figure}

\noindent {\bf Ground state stripes: detailed analysis}. We now return to a more
detailed discussion of the vertical stripe order observed in the ground state. Within
the family of vertical stripes, we can consider questions of wavelength (charge and
spin periodicity), type and strength of charge and spin modulation (e.g. bond- versus
site-centered), and coexistence with pairing order.

We first discuss the wavelength $\lambda$. At $1/8$ doping, the filling of the stripe
is related to the wavelength by $\lambda/8$. As described, we can access different
wavelength meta-stable stripes and their relative energetics by carefully choosing
different total cluster dimensions and boundary conditions (in the DMRG and AFQMC
calculations) or unit cell/fragment sizes (in the iPEPS and DMET calculations).
Fig.~\ref{fig:stripe:energy_wavelength} shows the energy per site of the stripe
versus its wavelength $\lambda$ from the multiple methods. Earlier DMRG calculations
on the Hubbard model had focused on $\lambda=4$ (half-filled stripes) which are seen
in HTSC's~\cite{white1998density,white2003stripes}, but we now observe that these are
relatively high in energy. A striking feature is that for $\lambda= 5-8$ the energies
are nearly degenerate. This is clearly seen in the DMET data where stripes of all
wavelengths can be stabilized, as well as the plot of the averaged energy of the methods
between $\lambda=5-8$.

The energy difference between the $\lambda=5$ and $\lambda=8$ stripe in the different methods is estimated to be between $0.0005t$ (DMRG)--$0.0041t$ (iPEPS) with an
average of $0.0022t$.
This suggests that the magnetic domain walls can fluctuate freely, consistent with
proposals for fluctuating stripes~\cite{parker2010fluctuating}.
In particular, the stripes may be distorted at a small cost over long length scales.

Although the different wavelengths are nearly degenerate, there appears
to be a slight  minimum near wavelength $\lambda=8$ (a filled stripe) in all
the methods.
Very recently, similar filled stripes have been reported as the ground state
in part of the frustrated $t$-$J$ model phase diagram~\cite{dodaro2017intertwined}.
 $\lambda=9$ appears significantly higher in energy in both  DMET and DMRG.
In the DMRG calculations, the $\lambda=9$ state was not even meta-stable as boundary 
conditions and initial states were varied, so the high-energy state shown was forced with a static potential.
The AFQMC results show a much weaker dependence on wavelength for longer wavelengths, 
for example the $\lambda=8$ and $\lambda=10$
stripe energies per site appear to be within $0.0015t$. 
However, when a mixture of the $\lambda=8$ and $\lambda=10$ stripe states
is set up on a length 40 cluster that is commensurate with both,
the state that survives is the $\lambda=8$ stripe, suggesting
a preference for this wavelength.
The increase in energy at wavelengths $\lambda > 8$ coincides with
unfavorable double occupancy of the stripe. This simple interpretation
is supported by a  mean-field (unrestricted Hartree-Fock (UHF)) calculation with an effective interaction $U/t=2.7$ chosen within
the self-consistent AFQMC procedure. The mean-field result shows a clear minimum at
a wavelength-$8$ vertical stripe. (Note that this requires the use of an effective $U/t$; at the bare $U/t=8$, mean-field theory
would produce a diagonal stripe~\cite{Jie2013jpcm}.)
The correspondence between the energies and densities in the effective mean-field 
and correlated calculations suggests that the mean-field theory with a renormalized interaction may be surprisingly 
good at describing the energetics of stripes. 
However, mean-field theory appears to somewhat underestimate the degeneracy
of the stripes as a function of wavelength, particularly at shorter wavelengths.

\begin{figure}[htpb]
  \centering
  \subfigure[AFQMC]{
	  \includegraphics[width=0.95\textwidth]{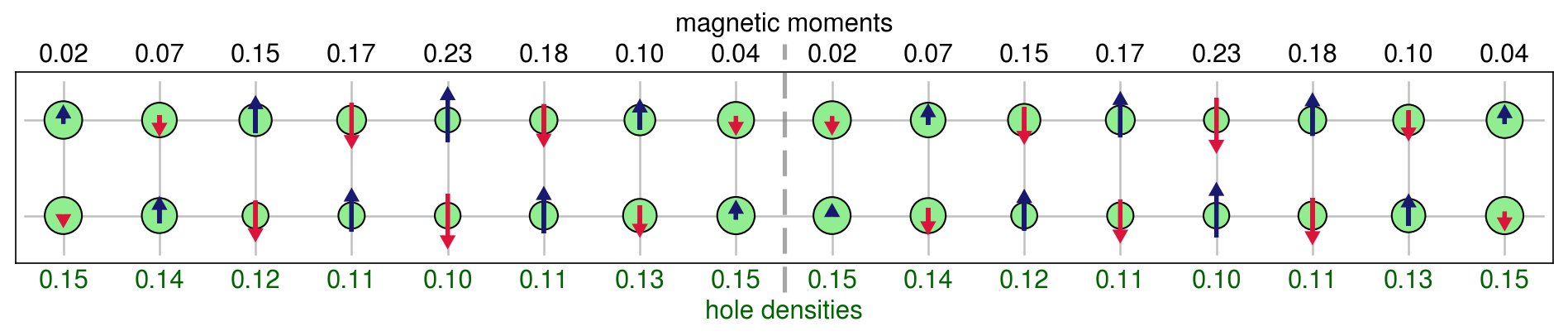}
   }
  \subfigure[DMRG]{
    \includegraphics[width=0.95\textwidth]{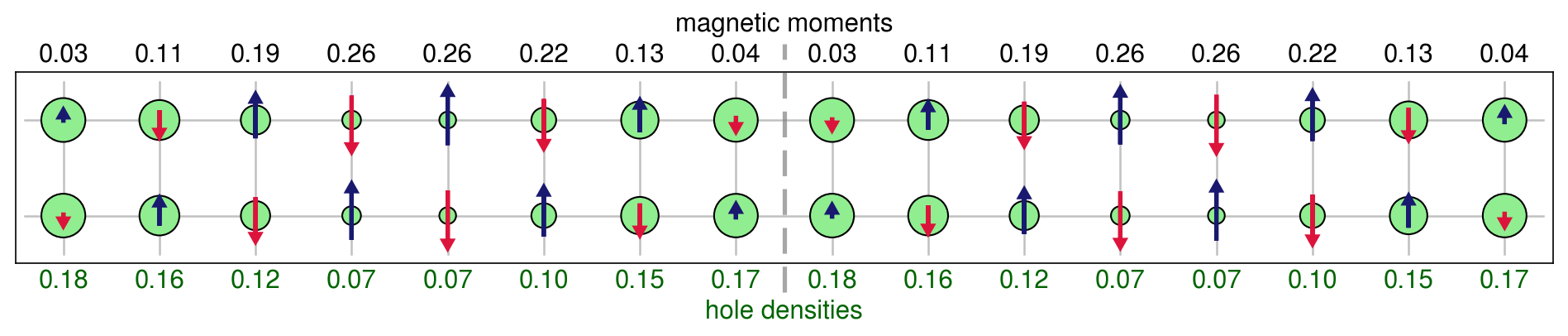}
    }
  \subfigure[DMET]{
    \includegraphics[width=0.95\textwidth]{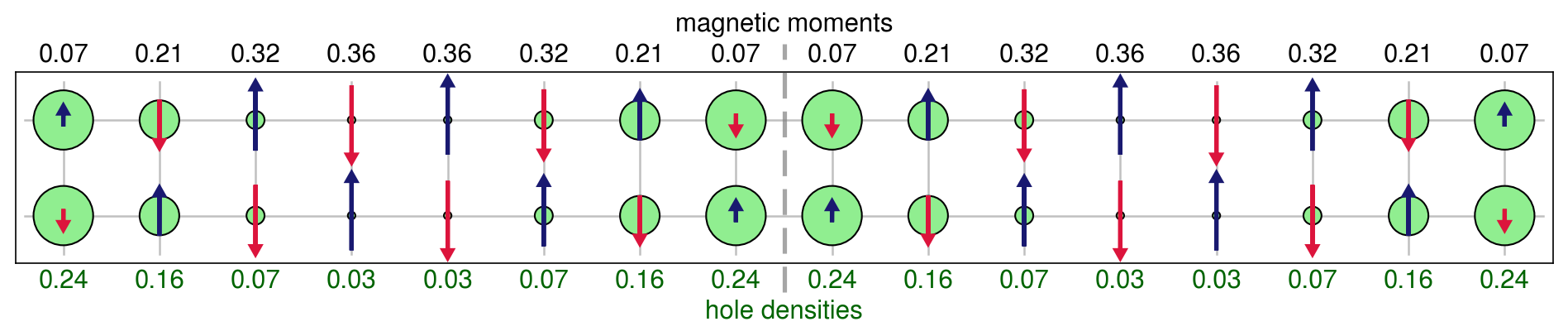}
    }
  \subfigure[iPEPS]{
    \includegraphics[width=0.95\textwidth]{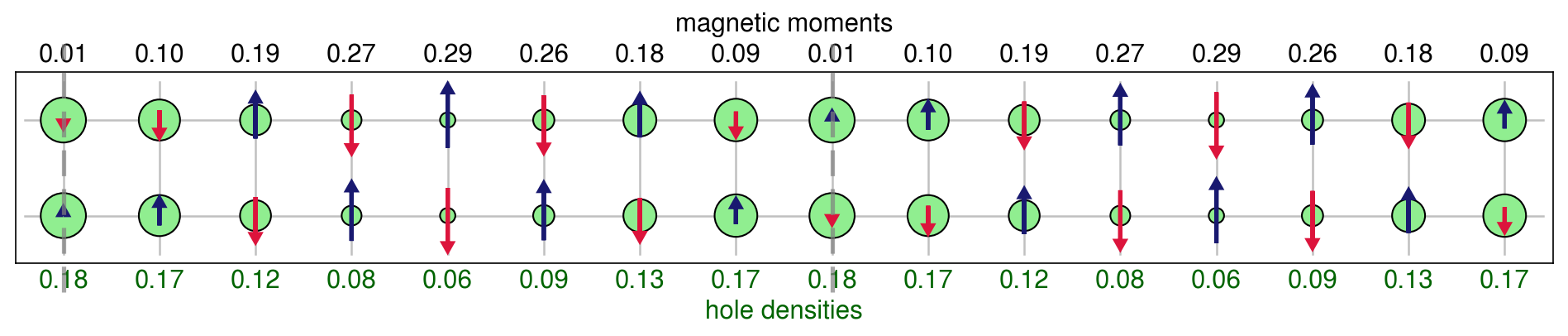}
    }
  \caption{Charge and spin orders in the wavelength-8 stripes from
  AFQMC, DMRG, DMET and iPEPS. The local magnetic moments and hole
  densities are shown above and below the order plots, respectively.
  (Circle radius is proportional to hole density, arrow height is
  proportional to spin density). The gray dashed lines represent the
  positions of maximum hole density.}
  \label{fig:stripe:w8stripes}
\end{figure}

The stripe order for the $\lambda= 8$ stripe from the different methods
is depicted in Fig.~\ref{fig:stripe:w8stripes}.
We show the full period (16) for the spin modulation. The stripe is 
bond-centered in the AFQMC, DMRG, and DMET calculations.
In the iPEPS calculation, the stripe is nominally site-centered.
In all the calculations, the width of the hole domain wall spans several sites,
blurring the distinction between bond- and site-centered stripes, and we
conclude that the energy difference between the two is very small. 
There is substantial agreement in the order observed by the different
numerical techniques, with only some differences in the magnetitude of
modulation of the hole and spin-densities. 

Note that for even-wavelength stripes, the spin wavelength must be twice
that of the charge modulation in order to accommodate the stripe as well
as the antiferromagnetic order. At odd wavelengths,  site-centered stripes
appear in all the calculations, and here charge and spin order can have
the same wavelength. (This odd-even alternation does not affect the peaks
of the structure factor near $(\pi,\pi)$, see Appendix~\ref{sec:stripe:additional}.)

\begin{figure}[htpb]
  \centering
  \subfigure[iPEPS $\lambda=5$]{
	  \includegraphics[width=0.39\textwidth]{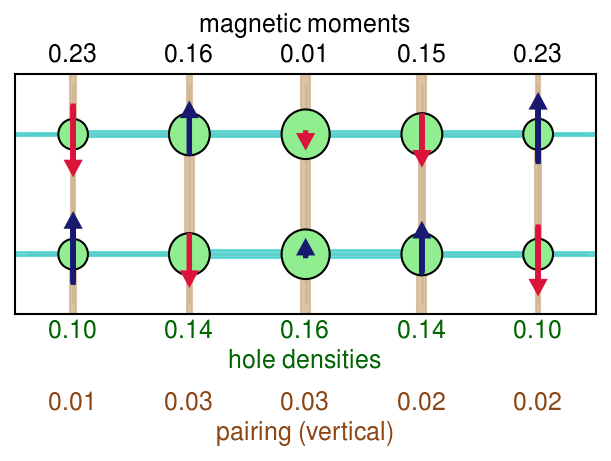}
  }
  \subfigure[iPEPS $\lambda=7$]{
	  \includegraphics[width=0.54\textwidth]{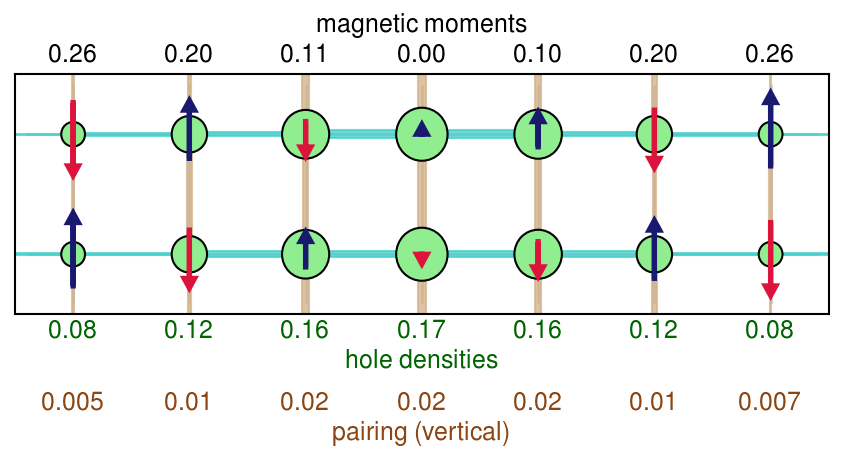}
  }
  \subfigure[DMET meta-stable $\lambda=5$]{
	  \includegraphics[width=0.39\textwidth]{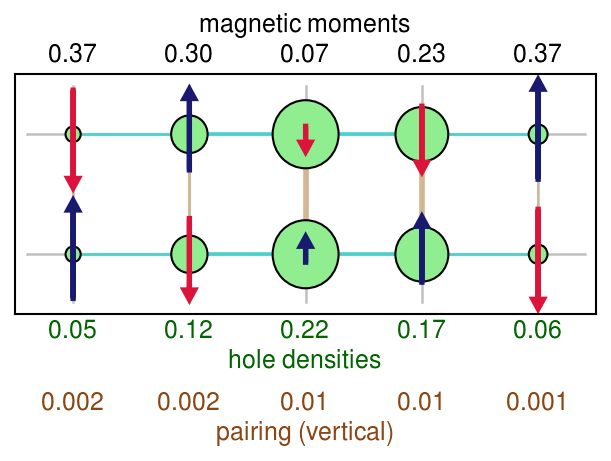}
  }
  \subfigure[DMRG pairing order parameters]{
	  \includegraphics[width=0.55\textwidth]{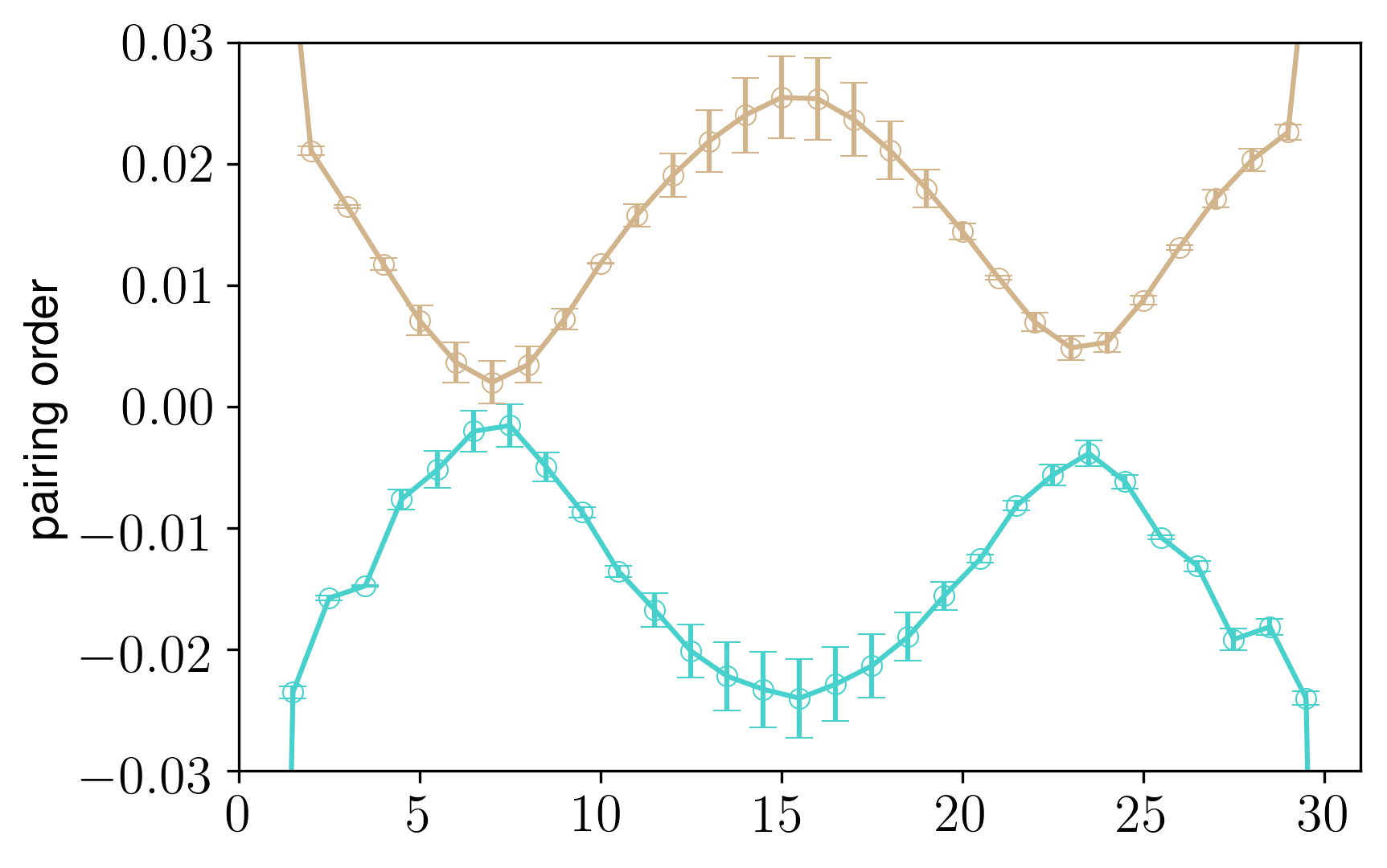}
  }
  \caption{Meta-stable stripe states with $d$-wave pairing from iPEPS,
	  DMET, and DMRG. (a)(b) iPEPS stripes with $\lambda=5$ and $\lambda=7$.
  (c) DMET meta-stable $\lambda=5$ stripe with pairing. (Circle radius is
	  proportional to hole density, arrow height is proportional to spin density,
	bond width is proportional to pairing density).
	(d) DMRG pairing order parameters on a 32$\times$4 cylinder.
    The positive values are from the vertical bonds and the negative
    values from the horizontal bonds. }
  \label{fig:stripe:sc_order}
\end{figure}

\noindent {\bf Pairing order, fluctuations, and superconductivity}. A key question is whether pairing order coexists with stripe order.
Previous work on the $t$-$J$ model with iPEPS found coexisting $d$-wave order for partially filled ($\lambda < 8$) stripes.
No $d$-wave order is found in the Hubbard $\lambda=8$ stripe with any technique.
However, $d$-wave order can be found at other wavelengths. For example,
for $\lambda=5$ and $\lambda=7$ stripes, iPEPS produces $d$-wave order along the
bonds (see Fig.~\ref{fig:stripe:sc_order}) with a maximum $d$-wave expectation
value of $0.026$ and $0.021$, respectively. DMRG calculations with pinning pairing
fields on the boundary for  a $32 \times 4$ cylinder also find $d$-wave order,
with an extrapolated maximum $d$-wave order of $0.025$, consistent with the
iPEPS results. In the DMET calculations, though the lowest energy $\lambda=5$ stripe
has no $d$-wave order, however, at slightly higher energy ($\sim 0.003t$) a
$\lambda=5$ state similar to the iPEPS stripe can be found with coexisting
$d$-wave order, but with a substantially smaller maximum order parameter of
$0.01$. Overall our results support the coexistence of modulated $d$-wave
order with the striped state, although the strength of pairing is intertwined
with the stripe wavelength and filling. The pairing modulation we find
(Fig.~\ref{fig:stripe:sc_order}) is in phase between cells. Other kinds
of pairing inhomogeneities, such as pair density waves, have also been
discussed in the literature~\cite{Fradkin2015}.

It has long been argued that fluctuating stripes could promote
superconductivity~\cite{emery1997spin,kivelson1998electronic,zaanen2001}.
Our work provides some support for this picture, as there is a low energy
scale associated with the deformation of stripe wavelength while we also
find coupling between the wavelength and the pairing channel.
We can imagine fluctuations in wavelength both at low temperatures, as well
as in the ground state. In the latter case, this could lead to a stripe liquid
ground state rather than a stripe crystal. Our calculations are consistent
with both possibilities.

\begin{figure}[htpb]
  \centering
  \subfigure[$U/t=6$]{
  	  \includegraphics[width=0.45\columnwidth]{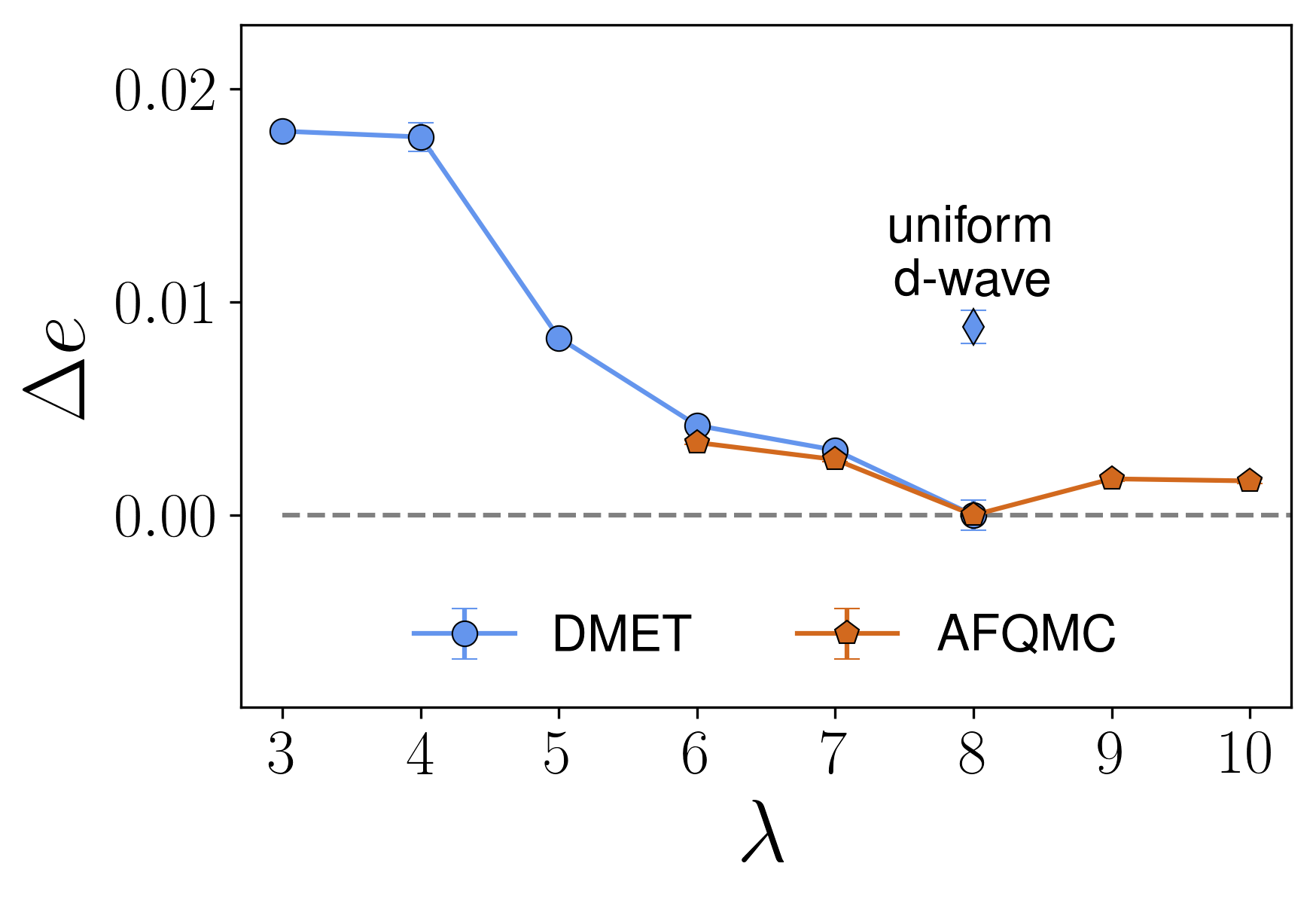}
	  \label{fig:stripe:U6}
  }
  \subfigure[$U/t=12$]{
	  \includegraphics[width=0.45\columnwidth]{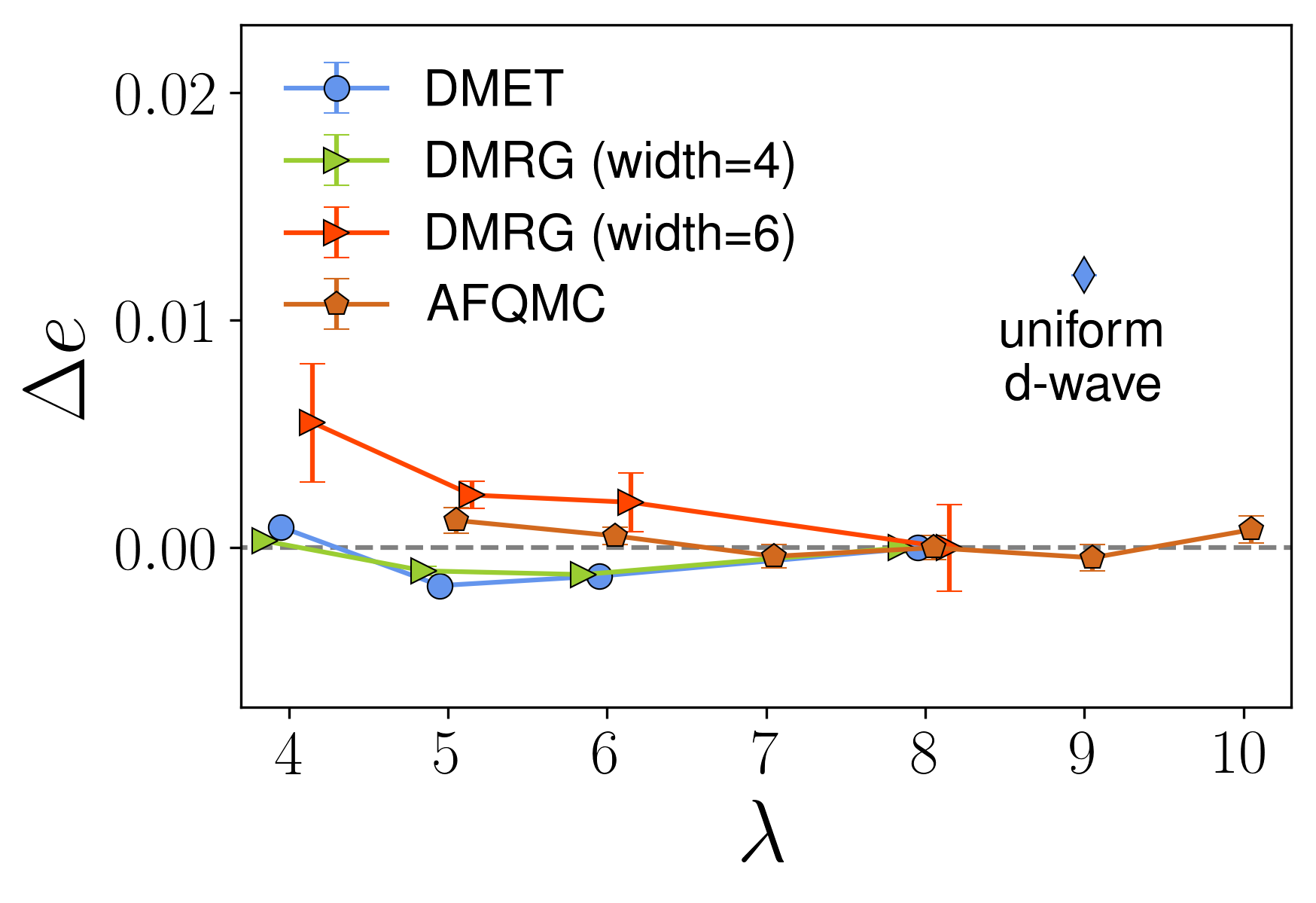}
	  \label{fig:stripe:12}
  }
  \caption{Relative energies of stripe states (with respect to wavelength) and the
  uniform $d$-wave state with $1/8$ doping at weaker and stronger couplings.}
  \label{fig:stripe:coupling}
\end{figure}

\noindent {\bf Changing the coupling}. We may also ask whether the $U/t=8$,
$1/8$ doping point is an anomalous point in the Hubbard phase diagram, and,
if, for example, moving away from this point would alleviate the unusual
stripe compressibility (with respect to wavelength at fixed doping) that we observe.
In Fig.~\ref{fig:stripe:coupling} we show the energies of various striped
states and the uniform state at $U/t=6$ and $U/t=12$, $1/8$ doping, computed
using AFQMC, DMET and DMRG.
At both couplings, the stripes around wavelength 8 are nearly degenerate,
with the degeneracy increasing as the coupling increases. At $U/t=6$, we find
the ground state is a filled stripe state with wavelength $\lambda = 8$, with
a larger energy stabilization than at $U/t=8$. The trend is consistent with
the state observed at $U/t=4$ with a more sinusoidal spin-density wave,
more delocalized holes, and a more pronounced minimum wavelength
~\cite{chang2010spin}.
In particular, at $U/t=6$, at $\lambda=4$, the stripe and spin-density wave state
(distinguished by whether or not the state has the $\pi$-phase shift) are essentially degenerate
in DMET [$e=-8589(6)$ vs. $e=-0.85890(4)t$], compared to an energy difference
of $\sim 0.009t$ at $U/t=8$.
At $U/t=12$, we find a filled stripe with AFQMC and DMRG (width 6), but DMET
and DMRG on a narrower cylinder (width 4) find $\lambda= 5-6$. The similarity
of the DMET and DMRG (width 4) data suggests that the shorter wavelength is
associated with a finite width effect.
We notice that $2/3$-filled stripes consistent with $\lambda= 5-6$ were also seen
in earlier DMRG studies on width-6 cylinders~\cite{hager2005stripe}, but a more
detailed analysis shows that the filled stripe becomes favored when extrapolated
to infinite bond dimension. In fact, similar to Ref.~\cite{hager2005stripe},
at finite $M$,  we always have the $2/3$-filled stripes as the ground state
because of their lower entanglement than the filled stripes. We are only
able to stabilize the filled stripes, which turns out to be the true ground state
when having eliminated the density matrix truncation error, by initializing the DMRG
calculations with favorable initial guess, based on our understanding of the
model and with the help of information from other methods. This example again
illustrates the benefits of combining information from different numerical
methods.
Thus, we conclude that the ground state at $U/t=12$ is also the $\lambda=8$ stripe,
although stripes of other wavelengths become even more competitive than at $U/t=8$.
Overall, the similarity in the physics over a wide range of $U/t$ indicates that
striped orders with low energy fluctuations of domain walls remain a robust feature
in the moderate to strongly coupled underdoped region.

\noindent {\bf Connection to stripe order in HTSC's}. 
In HTSC's the accepted stripe wavelength at 1/8 doping (e.g. in LaSrCuO) is $\lambda \approx 4.3$
(close to half-filled)~\cite{tranquada1995evidence}.
However, we find that the $\lambda=4$ stripe is unfavored in the 2D Hubbard model for the
coupling range ($U/t=6-12$) normally considered most relevant to cuprate physics.
This implies that the detailed charge-ordering of real materials arises from even stronger
coupling or, more likely, quantitative corrections beyond the Hubbard model.
With respect to the latter, one possibility is long-range hopping, which has been seen
to change the preferred stripe wavelength in the frustrated $t$-$J$ model~\cite{dodaro2017intertwined}.

\begin{figure}[htpb]
  \centering
  \includegraphics[width=0.7\textwidth]{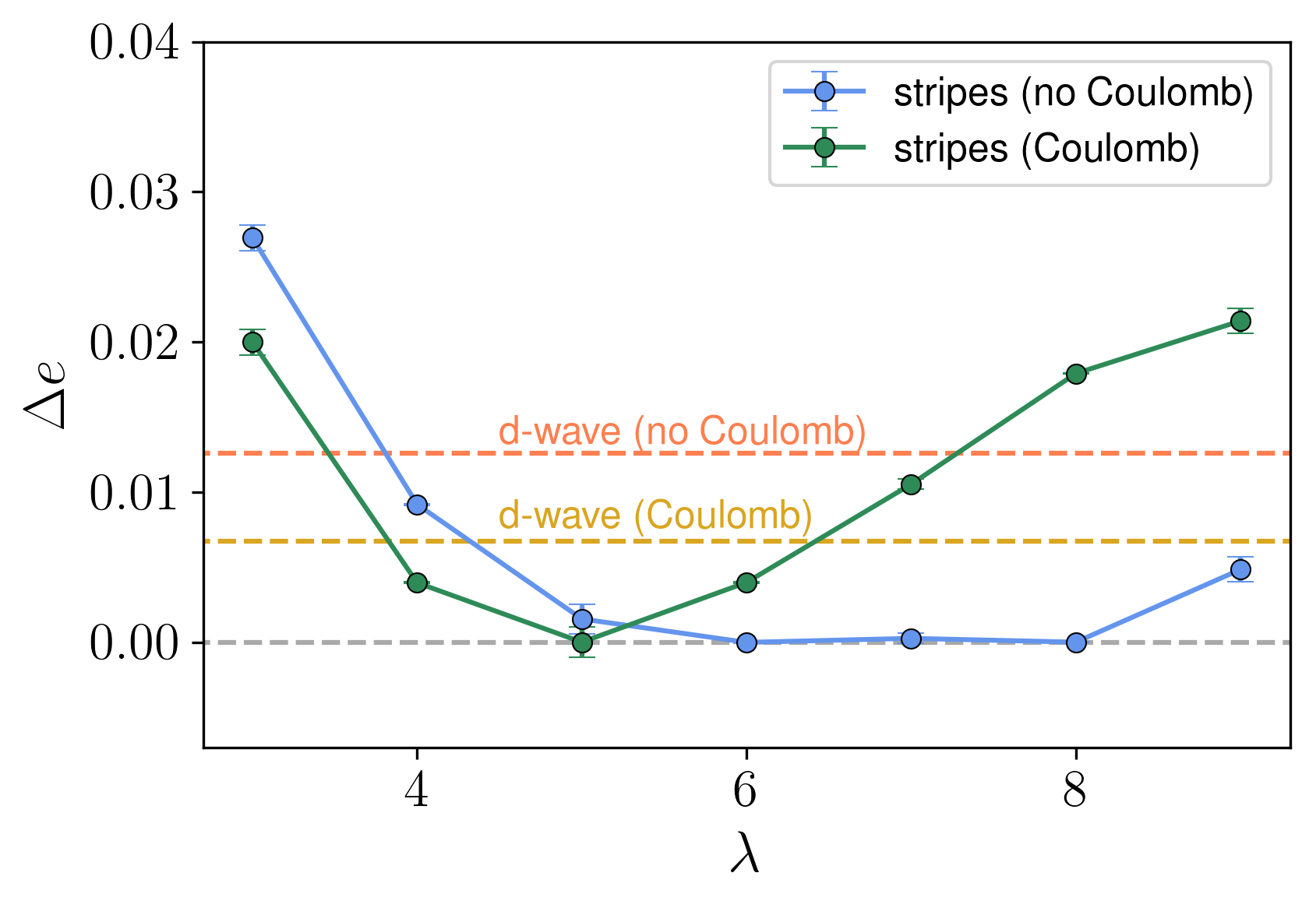}
  \caption{Energy landscape before and after adding the estimated long-range
  Coulomb interaction for vertical stripes of different wavelength. The charge
  distributions are from DMET calculations.}
  \label{fig:stripe:coulomb}
\end{figure}

Another possibility is the long-range Coulomb repulsion. Long-range repulsion can play a dual role,
in both driving charge inhomogeneity, as well as smoothing it out. In the Hubbard model, where stripes
naturally form, the latter property can help drive the ground state towards shorter stripe wavelengths.
In Fig.~\ref{fig:stripe:coulomb} we show the estimated effect of the long-range interactions on
the stripe energetics by computing the Coulomb
energy of the charge distributions of stripes with wavelengths $\lambda=3$--$9$. We use a
dielectric constant of 15.5 (in the range proposed for the cuprate plane~\cite{schuttler2001screening}).
As shown in Fig.~\ref{fig:stripe:coulomb}, the long-range Coulomb interaction favors shorter
wavelength stripes and the homogeneous $d$-wave state, lifting the near-degeneracy of the stripes,
shifting the ground state to $\sim2/3$-filled stripe and making the uniform $d$-wave state also more competitive.
Including the long-range Coulomb interaction favors the shorter wavelength stripes that is $\sim O(0.01t)$
per site for the $\lambda=4$ versus $\lambda=8$ stripe. Although this is only an order of magnitude estimate,
it is on the same energy scale as the stripe energetics in Fig.~\ref{fig:stripe:competing}, and thus provides
a plausible competing mechanism for detailed stripe physics in real materials.

\section{Conclusions} \label{sec:stripe:conclusion}
In summary, in this work we have employed state-of-the-art numerical methods
to determine the ground state of the iconic $1/8$ doping point of the 2D Hubbard model at
moderate to strong coupling. Through careful convergence of all the methods, and exhaustive
cross checks and validations, we are able to eliminate several of the competing orders that
have been proposed for the underdoped region in favor of a vertical striped order with
wavelength near $\lambda \approx 8$. The striped order displays a remarkably low energy
scale associated with its wavelength, which implies strong fluctuations either at low temperature
or in the ground-state itself. This low energy scale can roughly be described at the mean-field
level with a strongly renormalized $U$. We find coexisting pairing order with a strength
intertwined with the stripe wavelength, indicating a coupling of stripe fluctuations to
superconductivity. The stripe degeneracy is robust as the coupling strength is changed.

It has long been a goal of numerical simulations to provide definitive solutions of microscopic
models. Our work demonstrates that even in one of the most difficult condensed matter models,
such unambiguous simulations are now possible. In so far as the 2D Hubbard model is a realistic
model of high-temperature superconductivity, the stripe physics observed here provides a firm
basis for understanding the  diversity of inhomogeneous orders seen in the materials, as well as
a numerical foundation for the theory of fluctuations and its connections to superconductivity.
However, our work also enables us to see limitations of the Hubbard model in understanding real
HTSC's. Unlike the stripes at this doping point in real materials, we find filled stripes rather
than near half-filled stripes.
Given the very small energy scales involved, terms beyond the Hubbard model, such as long-range
Coulomb interactions, will likely play a role in the detailed energetics of stripe fillings.
The work we have presented provides an optimistic perspective that achieving a comprehensive
numerical characterization of more detailed models of the HTSC's will also be within reach.

\begin{subappendices}

\section{Appendix}
\subsection{Additional Information for the Figures and Discussion}\label{sec:stripe:additional}
\noindent \textbf{Figure ~\ref{fig:stripe:gs_energy}}.
The plotted energies (units of $t$) correspond to the following specific calculations.
\begin{itemize}
\item AFQMC: $-0.766\pm 0.001$ from extrapolation to $\infty$ (in both length and width directions) clusters with pinning fields.
\item DMRG: $-0.7627 \pm 0.0005$ from extrapolation to $\infty \times 6$ clusters with pinning fields using the hybrid momentum/real-space representation.
\item DMET: $-0.77063 \pm 0.00001$ from $8\times 2$ clusters with spin-inversion boundary conditions.
\item iPEPS: $-0.7673 \pm 0.002$ from $16 \times 2$ supercells with extrapolation to zero truncation error.
\end{itemize}

\noindent \textbf{Wavelengths of stripes}.  A key feature of the stripes
is that each stripe acts as an antiferromagnetic domain wall. As a well-known
consequence, at $1/8$ doping for half-filled stripes, the wavelength associated
with the AF periodicity ($8$) is twice that of the charge periodicity ($4$).
As an oversimplified but useful characterization of this periodicity, we describe
it by labeling the spin  pattern along a fixed row, assuming the stripe is width $1$:
$\ldots \cdot \uparrow\downarrow\uparrow\cdot\downarrow\uparrow\downarrow\cdot \uparrow\downarrow\uparrow\cdot \ldots$.
Here the $\cdot$'s indicate the positions of the localized hole, and the patterns
$\uparrow \cdot \downarrow$ or $\downarrow \cdot \uparrow$ signify the domain wall
nature of the stripe. Consider a charge wavelength which is an odd
integer, e.g. $5$:  $\ldots \cdot \uparrow\downarrow\uparrow\downarrow\cdot \uparrow\downarrow\uparrow\downarrow\ldots$
We see that the ratio of AF and charge wavelengths is one in this case, not two!
This odd-even alternation is potentially confusing, particularly if one has non-integer
charge periodicity.

However, experimentally, one looks at structure factors, noting peaks near 
$(\pi,\pi)$.  The locations of the peaks nearest $(\pi,\pi)$ do not show any odd/even
alternation. To see this note that the shift of the $k$-space origin to
$(\pi,\pi)$, for one particular row, is equivalent to an alternating sign chain $-1^x$
in the AF pattern, e.g. for charge wavelength $4$,
\[
\ldots \cdot\uparrow\downarrow\uparrow\cdot\downarrow\uparrow\downarrow\cdot\uparrow\downarrow\uparrow\cdot\ldots \to \ldots \cdot\uparrow\uparrow\uparrow\cdot\downarrow\downarrow\downarrow\cdot\uparrow\uparrow\uparrow\cdot\ldots
\]
and for charge wavelength $5$ 
\[
\ldots \cdot\uparrow\downarrow\uparrow\downarrow\cdot\uparrow\downarrow\uparrow\downarrow\ldots \to \ldots \cdot\uparrow\uparrow\uparrow\uparrow\cdot\downarrow\downarrow\downarrow\downarrow\ldots
\]
In both the even and odd cases, the distance of peaks from $(\pi,\pi)$ corresponds to
an AF ``wavelength'' of twice the charge wavelength.

\newpage
\subsection{Summary of Stripe Energy Results}\label{sec:stripe:energies}

\begin{longtable}{p{0.2\textwidth}p{0.2\textwidth}p{0.2\textwidth}p{0.15\textwidth}p{0.15\textwidth}}
  \caption{Best estimates of energies per site for stripes and and other low-energy state
	  for $U/t=8$. For the AFQMC calculations (PBC) denotes periodic boundary
  	 conditions used on both the short- and long-axes of the cylinder.
	 For the DMRG (real-space) calculations, periodic boundary conditions
  	were used along the short axis, open boundary conditions on the long axis.
	For the hybrid basis DMRG (h-DMRG) calculations, periodic or anti-periodic
	boundary conditions were used on the short axis, denoted PBC or APBC.
  	SF denotes that the DMET correlation potential in the spin-channel is flipped,
	doubling the spin wavelength. (Thus the $8\times 2$ (SF) pattern
  	in DMET has a charge wavelength of 8 but a spin wavelength of 16.)}
  \label{tab:stripe:U8}\\
\toprule
\textbf{Method} & \textbf{Size} & \textbf{Wavelength} & \textbf{Energy}  ($t$) &
\textbf{Error}  ($t$) \\
\midrule
\endfirsthead
\toprule
\textbf{Method} & \textbf{Size} & \textbf{Wavelength} & \textbf{Energy}  ($t$) &
\textbf{Error}  ($t$) \\
\midrule
\endhead
\bottomrule
\endfoot
AFQMC & $12 \times 8$ (PBC) & $6$ & $-0.7653$ &  0.0002\\
AFQMC & $14 \times 8$ (PBC) & $7$ & $-0.7653$ & 0.0002\\
AFQMC & $16 \times 8$ (PBC) & $8$ & $-0.7668$ & 0.0002\\
AFQMC & $18 \times 8$ (PBC) & $9$ & $-0.7655$ & 0.0002\\
AFQMC & $20 \times 8$ (PBC) & $10$ & $-0.7653$ & 0.0002\\
AFQMC & $\infty\times 4$ & $8$ & $-0.7680$ & 0.0001\\
AFQMC & $\infty\times 6$ & $8$ & $-0.7653$ & 0.0003\\
AFQMC & $\infty\times 8$ & $8$ & $-0.7656$ & 0.0004\\
\midrule
DMRG & $\infty\times 4$ & $8$ & $-0.76598$ & 0.00003\\
DMRG & $\infty\times 6$ & $5$ & $-0.7615$ & 0.0004\\
DMRG & $\infty\times 6$ & $8$ & $-0.762$ & 0.001\\
DMRG & $\infty\times 7$ & $7$ & $-0.762$ & 0.001\\
DMRG & $\infty\times 6$ & $9$ & $-0.751$ & 0.0016\\
\midrule
h-DMRG & $\infty \times 6$ (PBC) & $5$ & $-0.76210$ & 0.00005\\
h-DMRG & $\infty \times 4$ (APBC) & $8$ &$-0.76057$ & 0.00007\\
h-DMRG & $\infty \times 4$ (PBC) & $8$ &$-0.7657$ & 0.0003\\
h-DMRG & $\infty \times 4$ (av.)\footnote{Average of APBC and PBC results.} & $8$ &$-0.7631$ & 0.0003\\
h-DMRG & $\infty \times 6$ (PBC) & $8$ &$-0.7627$ & 0.0005\\
\midrule
DMET & $2\times2$& $d$-wave &$-0.7580$ &$0.0005$\\
DMET & $3\times2$&  $3$ & $-0.7437$ & $0.0009$\\
DMET & $4\times2$ (SF) & $4$ &$-0.7614$ & $0.00005$\\
DMET & $5\times2$ & $5$ &$-0.7691$ &$0.001$\\
DMET & $6\times2$ (SF) & $6$ & $-0.7706$ & $0.00007$ \\
DMET & $7\times2$&$7$& $-0.7704$ &$0.0003$\\
DMET & $8\times2$ (SF) & $8$ & $-0.7706$ & $0.00001$\\
DMET & $9\times2$& $9$ & $-0.7658$ &$0.0008$\\
DMET & $2\sqrt{2}\times\sqrt{2}$& $d$-wave & $-0.7620$ & $0.00001$\\
DMET & $5\sqrt{2}\times\sqrt{2}$& frustrated\footnote{No clear pattern, order appears to be frustrated.} & $-0.7689$ & $0.0008$\\
\midrule
iPEPS & $2 \times 2$\footnote{Using 2 independent tensors.} & $2$ & $-0.7560$ & $0.0025$ \\ 
iPEPS & $5 \times 2$ & $5$ & $-0.7632$ & $0.0018$ \\
iPEPS & $7\times 2$ & $7$ & $-0.7629$ & $0.0026$ \\
iPEPS & $16 \times 2$ & $8$ & $-0.7673$ & $0.002$ \\
iPEPS & $16 \times 16$\footnote{Using 16 independent tensors.} &  diag. $4 \sqrt{2}$ & $-0.7581$ & $0.0014$  \\
\end{longtable}

\begin{table}[pht]
  \caption{Best estimates of energies per site for stripes and other low-energy states for $U/t=6$.}
  \label{tab:stripe:U6}
  \centering
\begin{tabular}{p{0.2\textwidth}p{0.2\textwidth}p{0.2\textwidth}p{0.15\textwidth}p{0.15\textwidth}}
\toprule
\textbf{Method} & \textbf{Size} & \textbf{Wavelength} & \textbf{Energy}  ($t$) &
\textbf{Error}  ($t$) \\
\midrule
AFQMC & $12 \times 8$ (PBC) & $6$ & $-0.8684$ & 0.0001\\
AFQMC & $14 \times 8$ (PBC) & $7$ & $-0.8692$ & 0.0001\\
AFQMC & $16 \times 8$ (PBC) & $8$ & $-0.8718$ & 0.0001\\
AFQMC & $18 \times 8$ (PBC) & $9$ & $-0.8701$ & 0.0001\\
AFQMC & $20 \times 8$ (PBC) & $10$ &$-0.8702$ & 0.0001\\
\midrule
DMET & $2\times2$& $d$-wave &$-0.8679$ &$0.0007$\\
DMET & $3\times2$&  $3$ & $-0.85867$ & $0.00004$\\
DMET & $4\times2$&  $4$ &$-0.85890$ & $0.00004$\\
DMET & $5\times2$ & $5$ &$-0.86836$ &$0.00001$\\
DMET & $6\times2$ (SF) & $6$ & $-0.87247$ & $0.00001$ \\
DMET & $7\times2$& $7$ & $-0.87363$ &$0.00002$\\
DMET & $8\times2$ (SF) & $8$ & $-0.87667$ & $0.0007$\\
\bottomrule
\end{tabular}
\end{table}

\begin{table}[pht]
  \caption{Best estimates of energies per site for stripes and other low-energy states for $U/t=12$.}
  \label{tab:stripe:U12}
  \centering
\begin{tabular}{p{0.2\textwidth}p{0.2\textwidth}p{0.2\textwidth}p{0.15\textwidth}p{0.15\textwidth}}
\toprule
\textbf{Method} & \textbf{Size} & \textbf{Wavelength} & \textbf{Energy}  ($t$) &
\textbf{Error}  ($t$) \\
\midrule
AFQMC & $10 \times 8$ (TABC) & $5$ & $-0.6446$ & 0.0006\\
AFQMC & $12 \times 8$ (TABC) & $6$ & $-0.6452$ & 0.0004\\
AFQMC & $14 \times 8$ (TABC) & $7$ & $-0.6461$ & 0.0006\\
AFQMC & $16 \times 8$ (TABC) & $8$ & $-0.6458$ & 0.0006\\
AFQMC & $18 \times 8$ (TABC) & $9$ & $-0.6462$ & 0.0006\\
AFQMC & $20 \times 8$ (TABC) & $10$ &$-0.6450$ & 0.0006\\
\midrule
DMRG & $\infty\times4$ & 4 & -0.641379 & 0.000052 \\
DMRG & $\infty\times4$ & 5 & -0.64269 & 0.00019 \\
DMRG & $\infty\times4$ & 6 & -0.64285 & 0.00021 \\
DMRG & $\infty\times4$ & 8 & −0.64168 & 0.00023 \\
DMRG & $\infty\times6$ & 4 & -0.6383 & 0.0026 \\
DMRG & $\infty\times6$ & 5 & -0.64148 & 0.00059 \\
DMRG & $\infty\times6$ & 6 & -0.6418 & 0.0013 \\
DMRG & $\infty\times6$ & 8 & -0.6438 & 0.0019 \\
\midrule
DMET & $2\times2$      & $d$-wave &$-0.63940$  & $0.00001$\\
DMET & $4\times2$ (SF) & $4$ 	  &$-0.6505$   & $0.0001$\\
DMET & $5\times2$      & $5$ 	  &$-0.6531$   & $0.0001$\\
DMET & $6\times2$ (SF) & $6$ 	  &$-0.6526$   & $0.0002$ \\
DMET & $8\times2$ (SF) & $8$ 	  &$-0.6514$   & $0.0001$\\
\bottomrule
\end{tabular}
\end{table}

\end{subappendices}

\chapter[More Realistic Models of High-T$_c$ Superconductors]{More Realistic Models of High-T$_c$ Superconductors~\footnote{This chapter presents work that has not been published before.}} \label{chpt:threeband}
\section{Introduction} \label{sec:threeband:intro}
As discussed in Chapter~\ref{chpt:diagram} and~\ref{chpt:stripe}, numerical studies of the one-band
Hubbard model have achieved substantial progress regarding the ground state energies and orders,
as well as various low-lying competing states even, even in the most difficult part of the phase diagram.
This is possible only through combing various state-of-the-art approximate numerical methods to
explore many candidate states proposed by theorists or discovered by previous computational
and experimental studies.
A great success as it is, the question arises that how relevant the results, or the one-band
Hubbard model itself, are to the cuprates physics. As we discussed, the fact that many of the
candidate states are in a small energy range in the 2D Hubbard model reflects an artificial
degeneracy in this model. Any of these states could become the true ground state if the parameters
are slightly varied to favor that state. Thus, to answer questions in experiments and to understand
the factors that determine the superconductivity transition temperature, one has to go beyond the
one-band Hubbard model and consider more details of the real materials.

In this chapter, we describe the applications of DMET to two more sophisticated systems: the three-band
Hubbard model and a downfolded \textit{ab initio} cuprate Hamiltonian. These models introduce
multi-orbital and dynamical correlation effects, which could be important factors in real materials.
Other interactions, such as long-range Coulomb interactions and interlayer coupling are not studied here,
but could also have significant impact on the ground state and low-temperature behaviors
of cuprate superconductors.

In Sec.~\ref{sec:threeband:three}, we discuss the background and results of the three-band Hubbard
model. Sec.~\ref{sec:threeband:realistic} describes the downfolded cuprate
Hamiltonian and the DMET calculations on the system. We present the main conclusions
in Sec.~\ref{sec:threeband:conclusion}.

\section{The Three-Band Hubbard Model} \label{sec:threeband:three}

\begin{figure}[htpb]
	\centering
\subfigure[]{
	\includegraphics[width=0.48\textwidth]{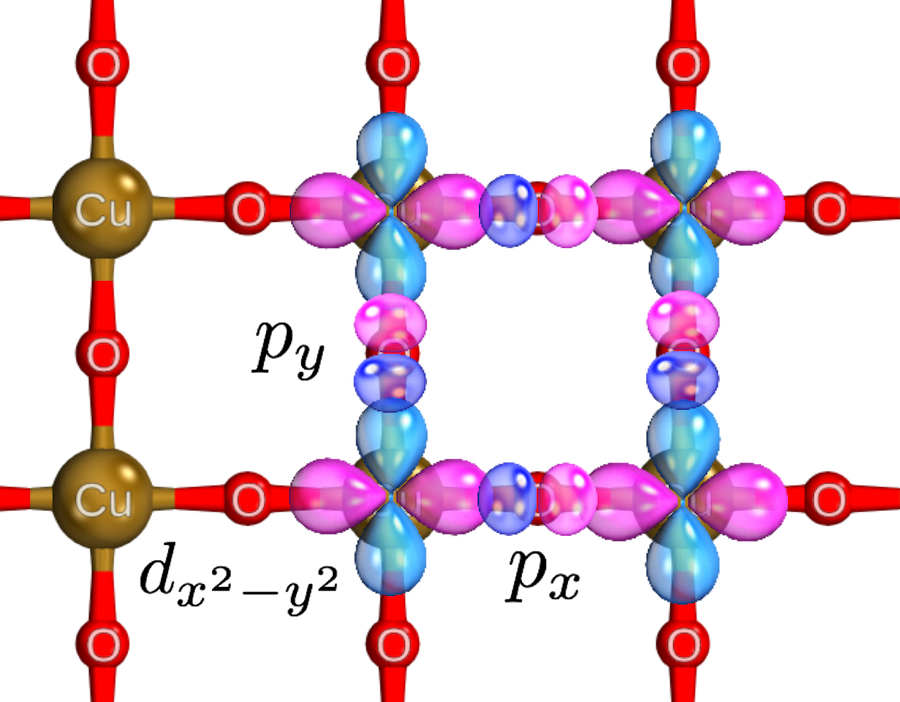}
	\label{fig:threeband:cuo2}
}
\subfigure[]{
	\includegraphics[width=0.45\textwidth]{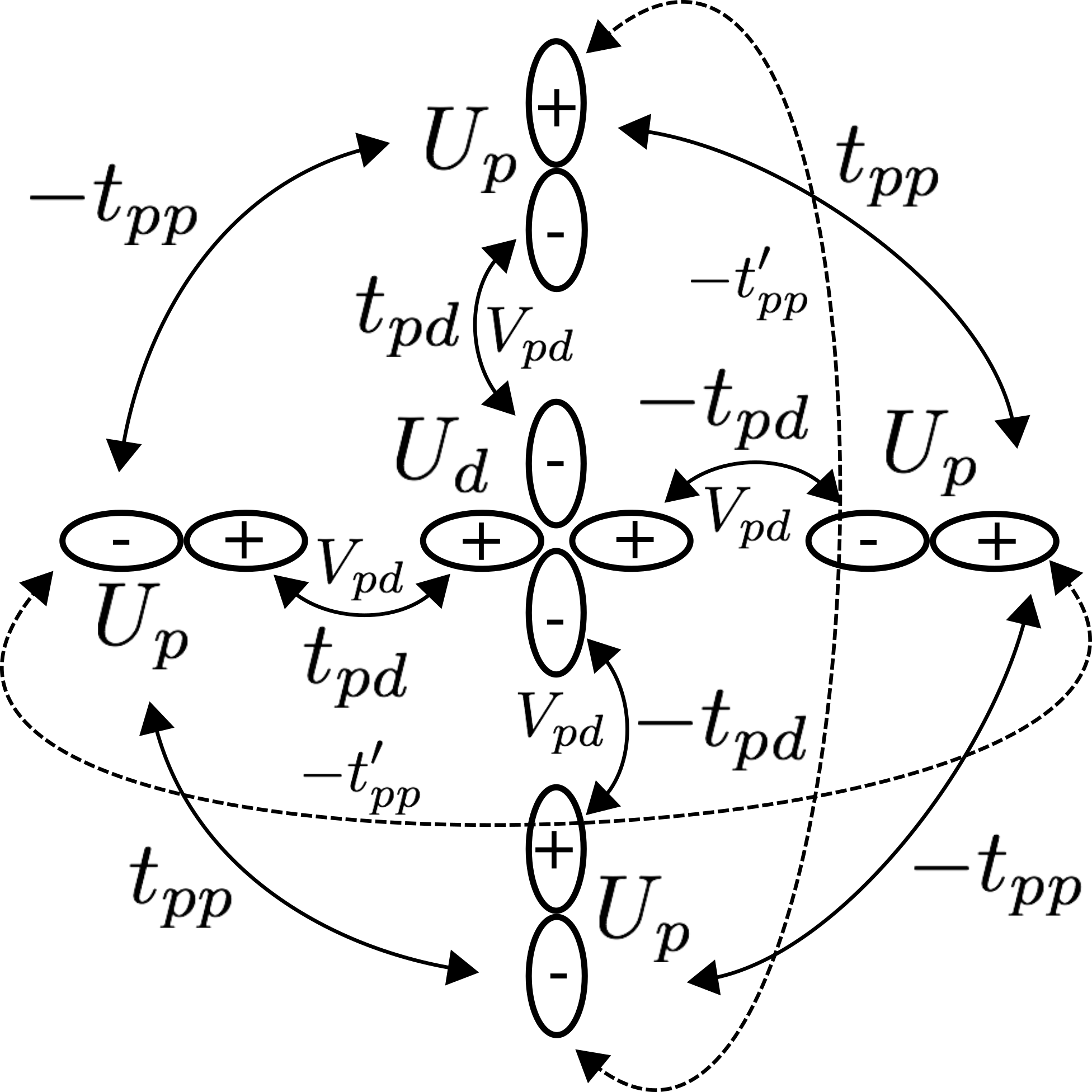}
	\label{fig:threeband:parameter}
}
	\caption{(a) The cuprate CuO$_2$ plane and the orbital arrangement of the three-band
	Hubbard model. (b) The common parameters in the three-band Hubbard model.}
	\label{fig:threeband:threeband}
\end{figure}

\noindent\textbf{The model}. The three-band Hubbard model simulates the physics of the
CuO$_2$ plane of cuprates with the valence orbitals: the copper $d_{x^2-y^2}$ orbital and the oxygen
bridge $p_{x}$ or $p_{y}$ orbitals [Fig.~\ref{fig:threeband:cuo2}]. When undoped, there
are 5 electrons in each unit cell, formally filling up the $p$ sites while the $d$ site
is half filled. Physically, however, it forms an antiferromagnetic charge-transfer insulator,
with occupation higher than one on the $d$ sites. Unlike the one-band Hubbard model, the
three-band model has significant particle-hole asymmetry as the doped holes go primarily
on the $p$ sites while the doped electrons reside on the Cu $d$ sites.

Previous numerical studies of the three-band model
ground state using Lanczos, quantum Monte Carlo, dynamical mean-field theory, variational cluster
approximation and diagrammatic perturbation show even richer
charge, spin and pairing orders than in the one-band model~\cite{scalettar1991antiferromagnetic,
kuroki1996quantum,guerrero1998quantum,de2009correlation,arrigoni2009phase,hanke20103,bulut2014charge,
go2015spatial,maier2014pairing}. Yet, because the model is numerically more difficult, fewer
studies are carried out compared to the
one-band Hubbard model. The numerical studies are far from reaching the thermodynamic limit (TDL)
or having enough energy resolution to determine the ground state, especially in the dope model.
DMET is thus a unique method that has the potential to carry out definitive numerical simulations
on the three-band model with an affordable computational cost.

\noindent \textbf{Computational details}. The full parameterization of the three-band model usually
includes the hopping between the nearest
$d$ and $p$ orbitals $t_{pd}$, and between the nearest and next-nearest $p$-orbitals $t_{pp}$ and
$t_{pp}^\prime$, onsite interactions $U_d$ and $U_p$, nonlocal Coulomb interaction $V_{pd}$ and the
energy difference $\Delta=\varepsilon_d - \varepsilon_p$ [Fig.~\ref{fig:threeband:parameter}]. There
are different approaches to map cuprates to the three-band model, thus a wide range of parameters
are published and used in numerical studies, which do not always agree with each other
~\cite{mcmahan1988calculated,mila1988parameters,hybertsen1989calculation,eskes1989effective,
mcmahan1990cuprate,martin1996electronic,arrigoni2009phase}. In our DMET calculations, we choose to use
only the minimal set of parameters $U_d$, $t_{pd}$ and $\Delta$, from the common parameter sets
summarized in Table~\ref{tab:threeband:parameter}. Note that we use the electron representation, which
is different from literature conventions, where the parameters are written down for the holes.

\begin{table}[htpb]
	\centering
	\caption{Model parameters used in this study, in unit of eV.}
	\label{tab:threeband:parameter}
	\begin{tabular}{p{0.3\textwidth}p{0.2\textwidth}p{0.2\textwidth}p{0.2\textwidth}}
		\toprule
		\textbf{Source}&$\mathbf{U_d}$&$\mathbf{t_{pd}}$&$\mathbf{\Delta}$\\
		\midrule
		Hybertsen (1989)~\cite{hybertsen1989calculation} &10.5&1.3&-6.9\\
		Martin (1996)~\cite{martin1996electronic} &16.5&1.8&-11.1\\
		Hanke (2009)~\cite{arrigoni2009phase} &12&1.5&-7.5\\
		\bottomrule
	\end{tabular}
\end{table}

\begin{figure}[htpb]
	\centering
	\includegraphics[width=0.55\textwidth]{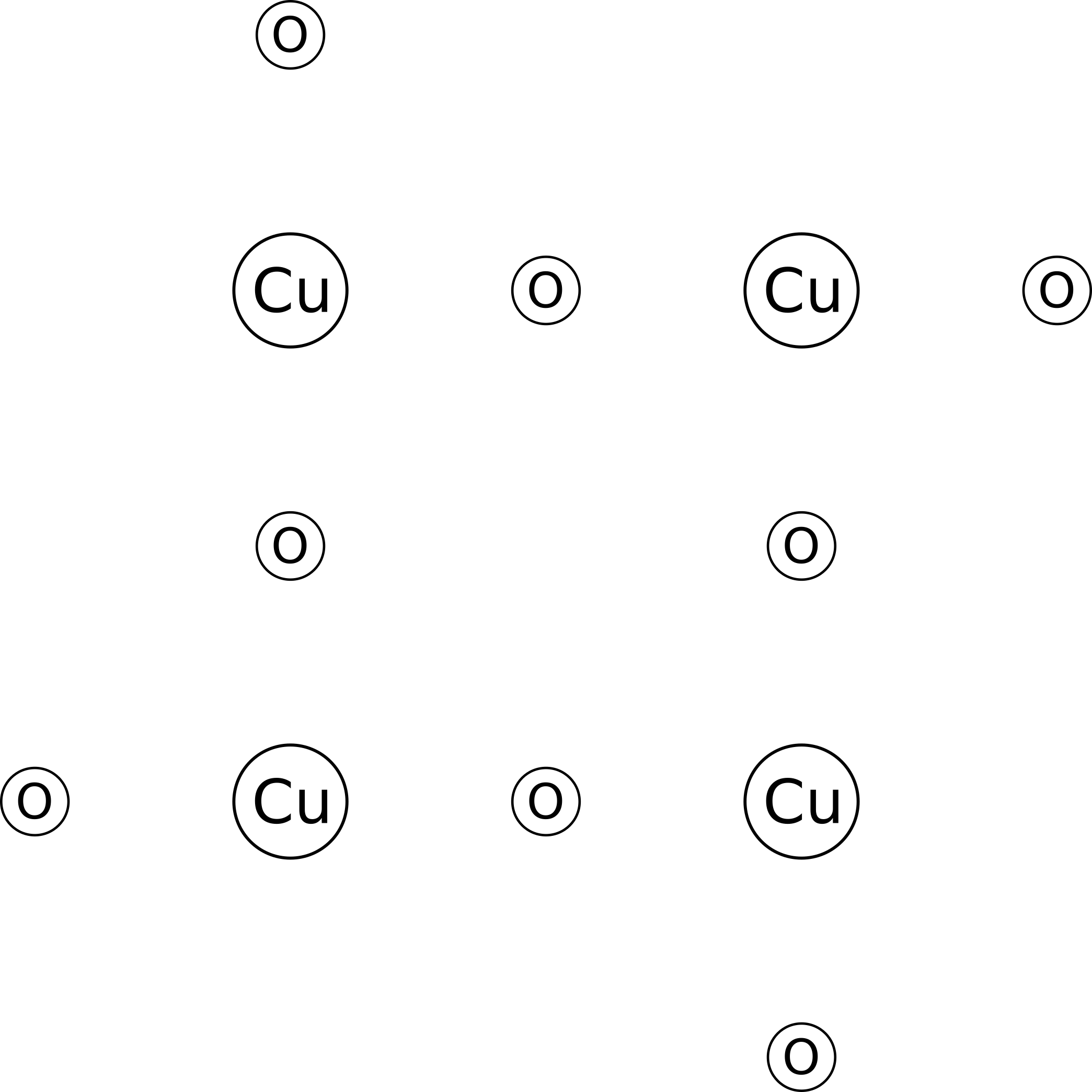}
	\caption{The 12-site fragment used in DMET studies of the three-band Hubbard model.}
	\label{fig:threeband:cell}
\end{figure}

We carried out broken particle-number symmetry DMET calculations
(see Appendix Sec.~\ref{sec:algo:bcs}) on the three-band model using fragments of
12 sites, on an auxiliary lattice of $20\times20$ unit cells. In the three-band model, the conventional
$2\times2$ supercell breaks inversion symmetry and, since the Cu and O atoms become inequivalent in
the impurity model, the symmetry would also break in the correlated wavefunction, leading to unfavored
artifacts. To reduce this effect as much as possible, we follow Ref.~\cite{hanke20103} and use a modified
$2\times2$ cell which restores the inversion symmetry (Fig.~\ref{fig:threeband:cell}). We varied occupations
in the model and computed the energies, magnetic and pairing order parameters. Since it is not clear
how to decompose the pairing order in the symmetry-broken three-band fragments, we use the total norm of
local pairing order parameters as a proxy for the pairing order strength, which reflects the combined
pairing strength in all symmetry sectors.

\noindent\textbf{Results -- undoped}. The results for the undoped three-band model
are summarized in Table~\ref{tab:threeband:undoped}.
We find the charge transfer between the oxygen $p$-orbitals and
the copper $d$ orbitals, and antiferromagnetic order with staggered magnetization $m\sim0.37$, localized
on $d$ orbitals. These results are consistent among all sets of parameters, and with previous studies
~\cite{scalettar1991antiferromagnetic}. We also notice the more recent parameters from Martin (1996)
and Hanke (2009) are more consistent with each other than Hybertsen (1989), indicating increasing consensus on the parameters of the model.

\begin{table}[htpb]
	\centering
	\caption{Charge and staggered magnetization from 12-site DMET calculations on the three-band model.}
	\label{tab:threeband:undoped}
	\begin{tabular}{p{0.3\textwidth}p{0.15\textwidth}p{0.15\textwidth}p{0.15\textwidth}p{0.15\textwidth}}
		\toprule
		\textbf{Model}&$\mathbf{n_d}$&$\mathbf{n_p}$&$\mathbf{m_d}$&$\mathbf{m_p}$\\
		\midrule
		Hybertsen & 1.237& 1.882& 0.363& 0.000\\
		Martin &1.218& 1.891& 0.377& 0.000\\
		Hanke &1.218& 1.891& 0.374& 0.001\\
		\bottomrule
	\end{tabular}
\end{table}

\begin{figure}[htpb]
	\centering
\subfigure[hole energy]{
	\includegraphics[width=0.45\textwidth]{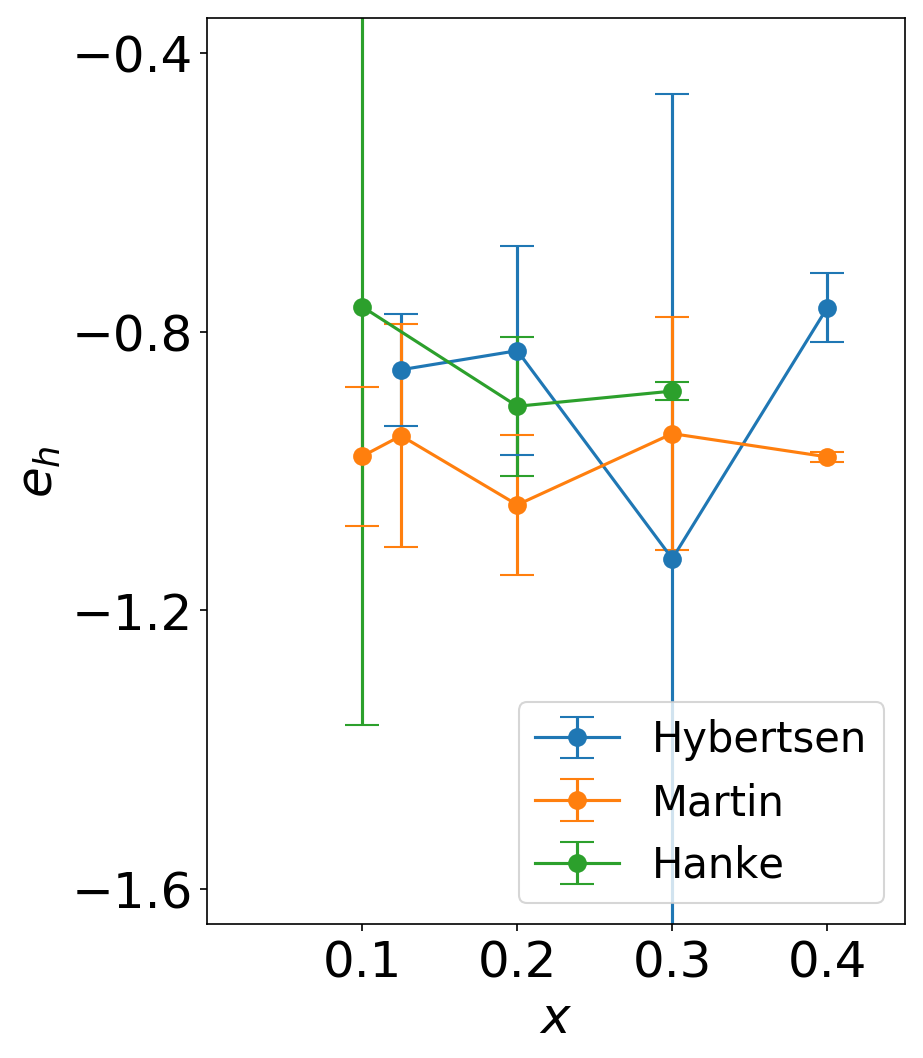}
	\label{fig:threeband:e_h}
}
\subfigure[electron energy]{
	\includegraphics[width=0.45\textwidth]{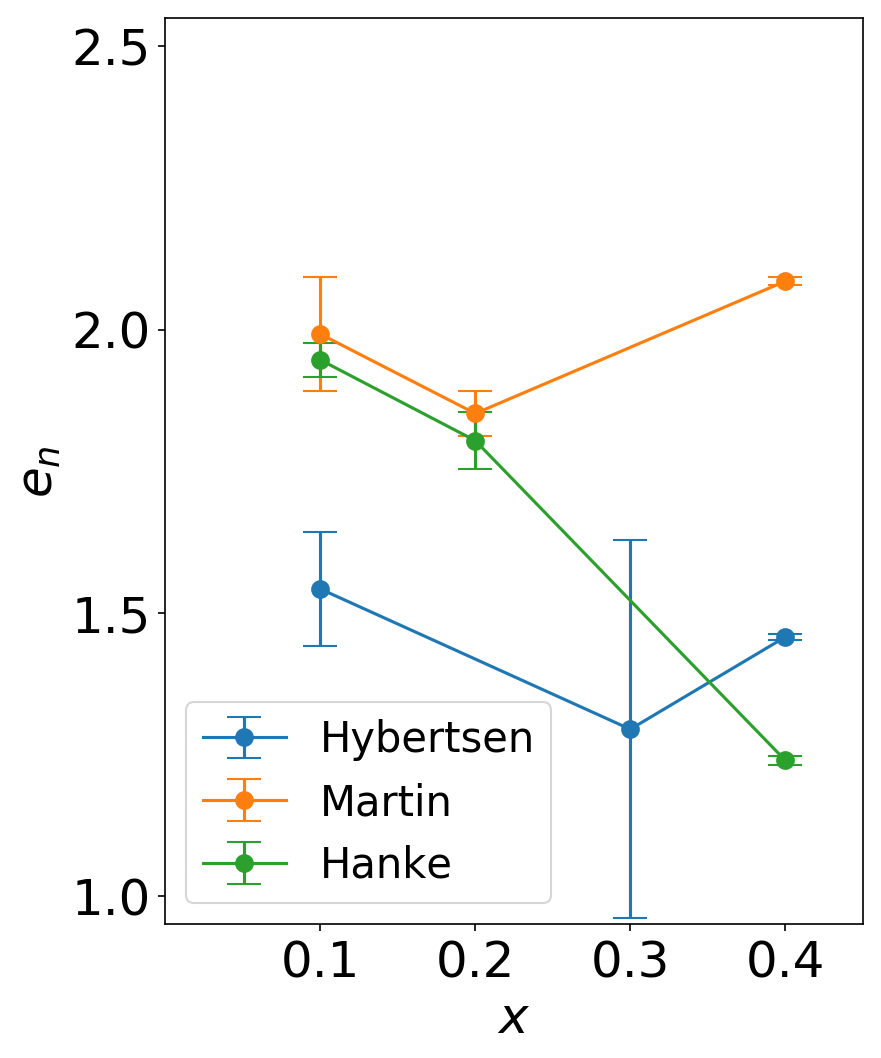}
	\label{fig:threeband:e_n}
}
\caption{Energies per hole (electron) in hole (electron) doped three-band model for all three sets of
parameters, in the unit of eV.}
	\label{fig:threeband:energies}
\end{figure}

\begin{figure}[htpb]
	\centering
\subfigure[Hybertsen]{
	\includegraphics[width=0.305\textwidth]{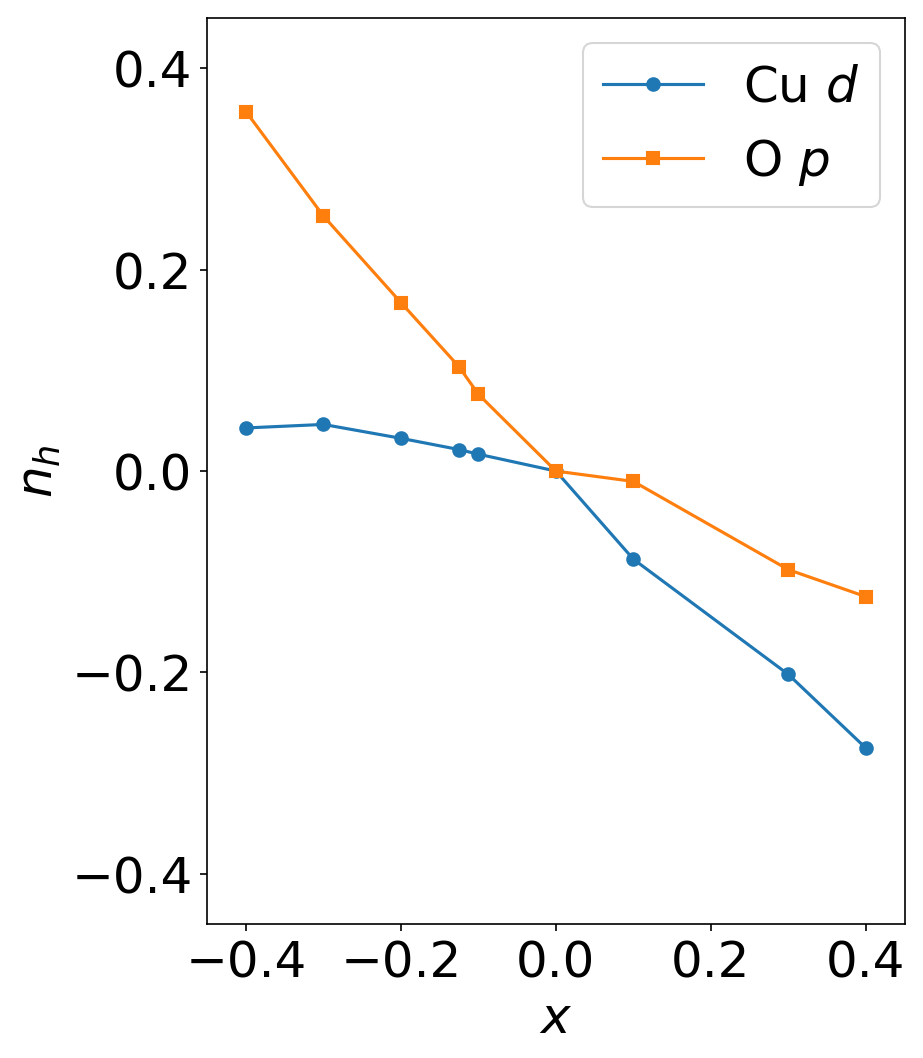}
	\label{fig:threeband:c_Hyb}
}
\subfigure[Martin]{
	\includegraphics[width=0.305\textwidth]{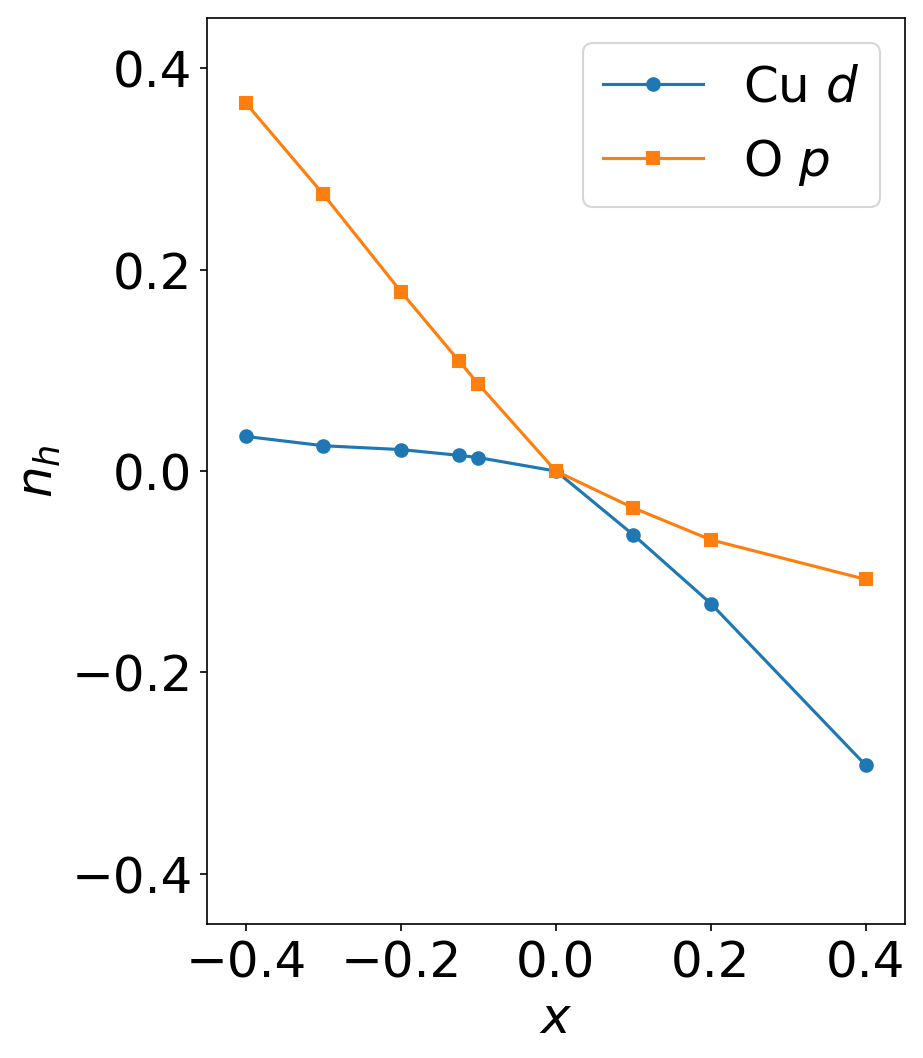}
	\label{fig:threeband:c_Mat}
}
\subfigure[Hanke]{
	\includegraphics[width=0.305\textwidth]{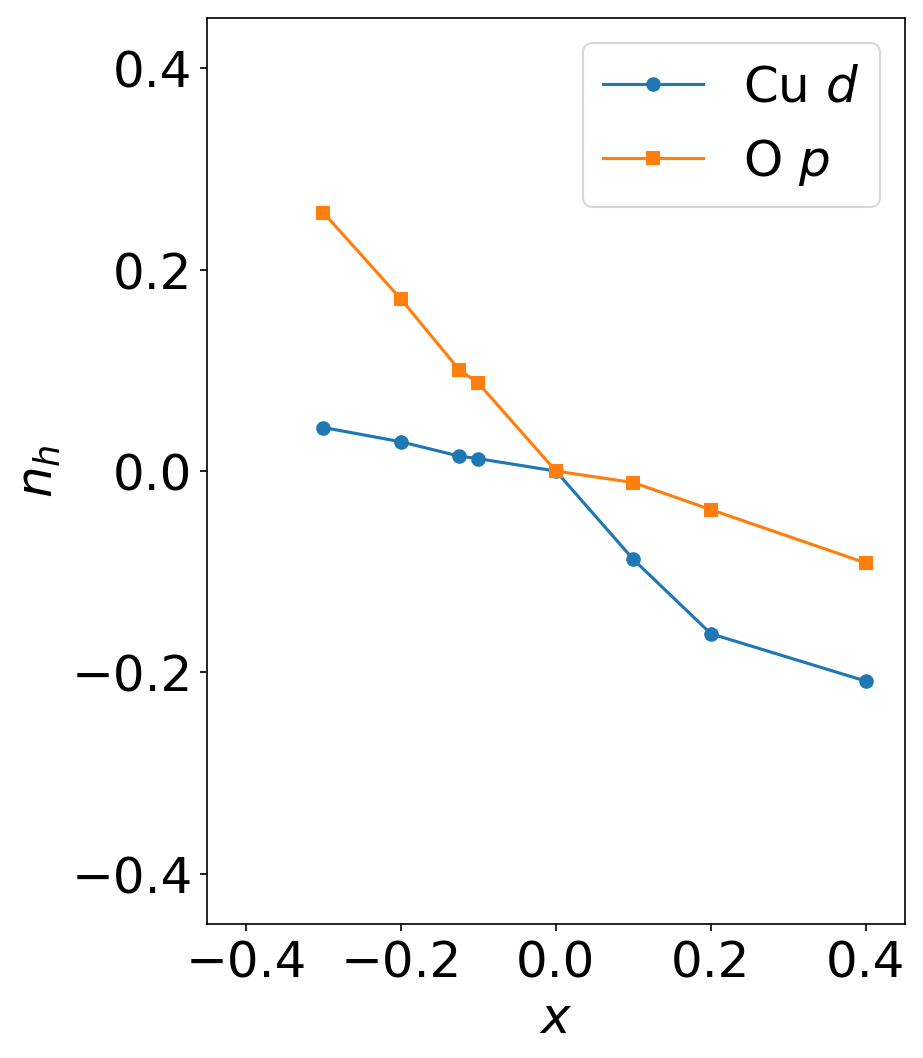}
	\label{fig:threeband:c_Han}
}
\caption{Hole distribution between $d$ and $p$ orbitals in the doped three-band model. The horizontal axis
is doping where negative numbers indicate hole doping while positive numbers are electron doping. The
hole distribution for $p$-orbitals is the sum of the two $p$-orbitals in a unit cell.}
	\label{fig:threeband:charge}
\end{figure}

\noindent\textbf{Results -- energy and charge distribution}. We plot the energies per hole and energies
per electron for the doped three-band model in Fig.~\ref{fig:threeband:energies}. Since we do not perform
fragment size extrapolation, the error bars only show the DMET convergence errors. We notice on both hole-
and electron-doped sides, there are points with large error bars. These reflect either DMET has
experienced
convergence issues, or the energy becomes sensitive to small changes in the correlation potential,
usually because the correlation potential has redundant or unnecessary degrees of freedom. Thus, it
is necessary to explore better parameterization of $u$ to avoid this problem.

Nevertheless, Fig.~\ref{fig:threeband:energies} is still informative. For hole-doped cuprates, the energies
per hole are almost constant (between $-1.2$eV and $-0.8$eV) when doping is varied in all the
parameters, suggesting the phase separation or inhomogeneous orders are likely to appear for a wide range
of doping. On the electron doped side, the energies per (doped) electron have different behaviors for
the parameters we use: Martin and Hybertsen are more or less constant around $2.0$eV and $1.5$eV, respectively,
while Hanke  gives decreasing energy per electron, leading to strong phase separation effect. These
unusual and inconsistent behaviors reflect that these models do not correctly describe the electron-doped
cuprates, simply because they are fitted to the hole-carrier superconductors, such as La$_{2-x}$Sr$_x$CuO$_4$.

In Fig.~\ref{fig:threeband:charge}, we summarize the charge distribution when the model is doped for
all three sets of parameters. The numbers are averaged over local charge distributions. As expected,
when the system is hole-doped, the holes go primarily to the oxygen $p$ sites, while the doped electrons
tend to stay on the copper $d$ sites. This leads to the particle-hole asymmetry in cuprate
superconductors, and is correctly described by the three-band model using DMET.

\begin{figure}[htpb]
	\centering
\subfigure[Hybertsen]{
	\includegraphics[width=0.305\textwidth]{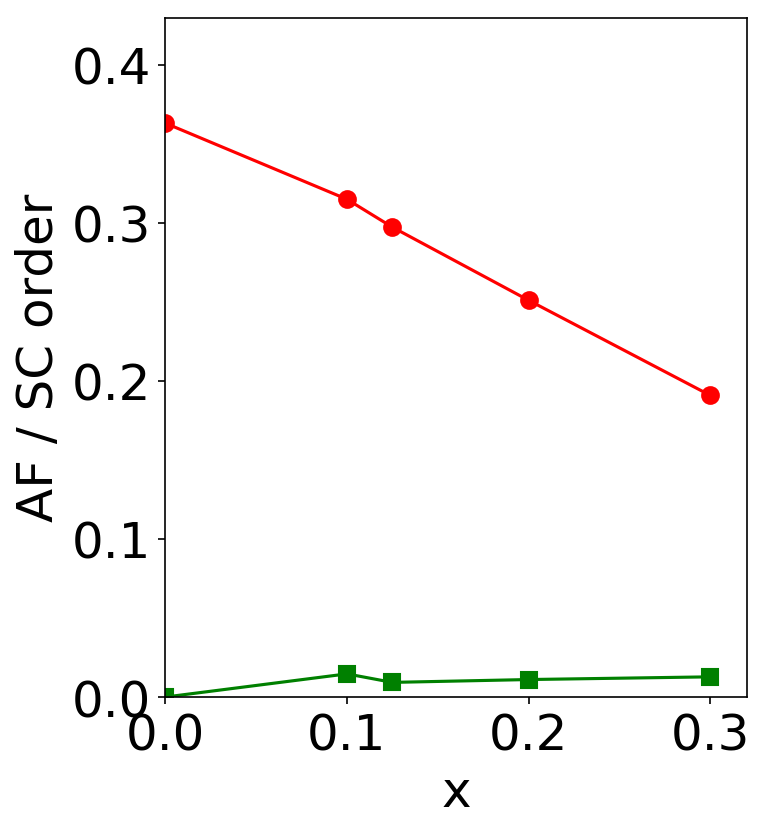}
	\label{fig:threeband:order_Hyb}
}
\subfigure[Martin]{
	\includegraphics[width=0.305\textwidth]{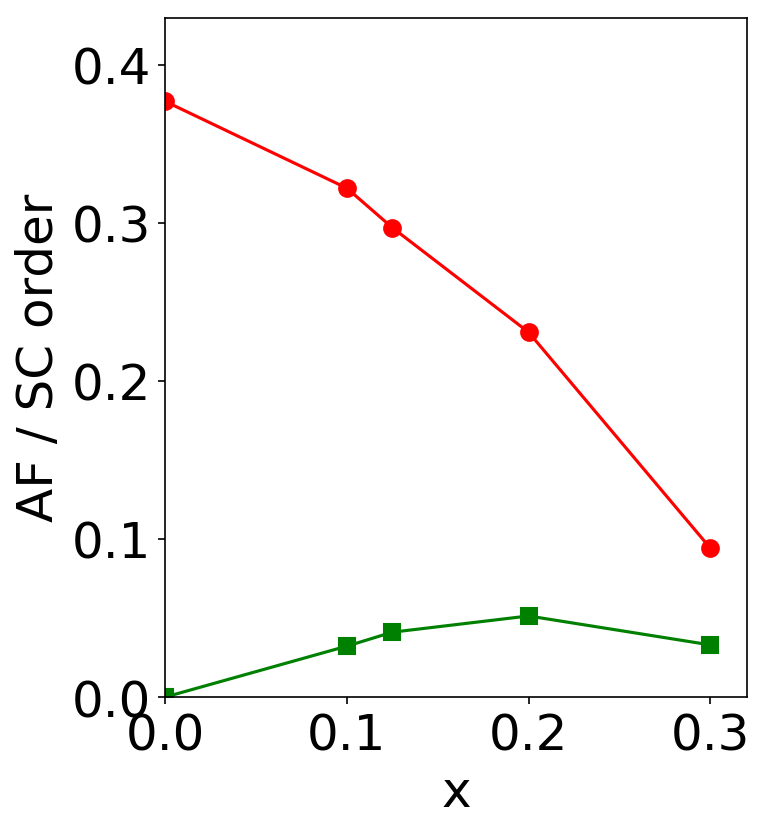}
	\label{fig:threeband:order_Mat}
}
\subfigure[Hanke]{
	\includegraphics[width=0.305\textwidth]{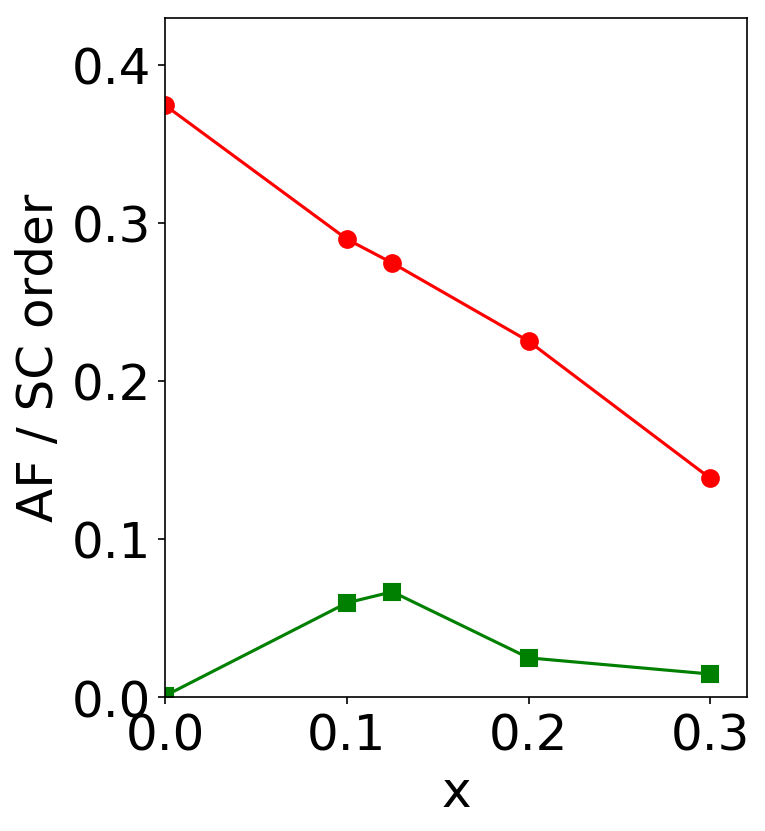}
	\label{fig:threeband:order_Han}
}
\caption{Averaged antiferromagnetic (red circle) and total pairing (green square) order parameters
	in the hole-doped three-band model for all sets of parameters from DMET.}
	\label{fig:threeband:order}
\end{figure}

\noindent\textbf{Results -- spin and pairing orders}. In Fig.~\ref{fig:threeband:order}, we show the
antiferromagnetic (AF) and pairing order parameters in the hole-doped three-band model. We do not display
the electron-doped side because from what we learned before, the parameters we use do not seem to describe
the correct physics on the electron-doped side. The magnetic order parameters are averaged over the
copper $d$ sites with the $(\pi,\pi)$ structure factor, as there is virtually no spin on the $p$ orbitals
when the model is doped. Overall, the shapes of the order parameter curves and sizes are similar to what
we obtained from one-band model calculations.

The AF order decreases with increasing doping, consistent with what we find in the one-band Hubbard model
and cuprates. The order remains robust at $x=0.3$ doping, with the value $\sim0.1$--$0.2$, which is probably
overestimated. It is a known issue in DMET to overestimate magnetic order parameters in small
fragments, which can be eliminated by increasing the fragment size and extrapolating to the TDL.
Both results
from Martin and Hanke show a superconducting dome (in terms of the magnitudes of the order parameters,
rather than transition temperature), with maximum pairing at $x=0.2$ and $x=0.125$, respectively. This is
plausible behavior one would expect from the experience. The Hybertsen parameters, however, give virtually
no pairing order in the whole parameter range.

\begin{figure}[htpb]
	\centering
	\subfigure[]{
	\includegraphics[width=0.3\textwidth]{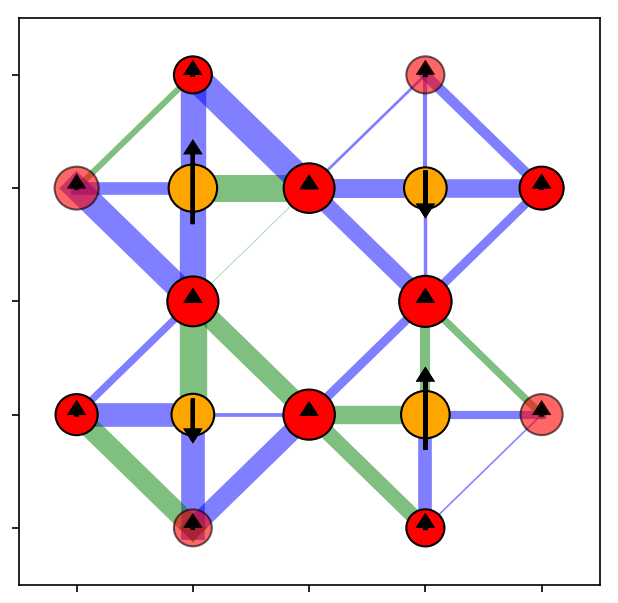}
	\label{fig:threeband:Martin_pattern_1}
}
\subfigure[]{
	\includegraphics[width=0.3\textwidth]{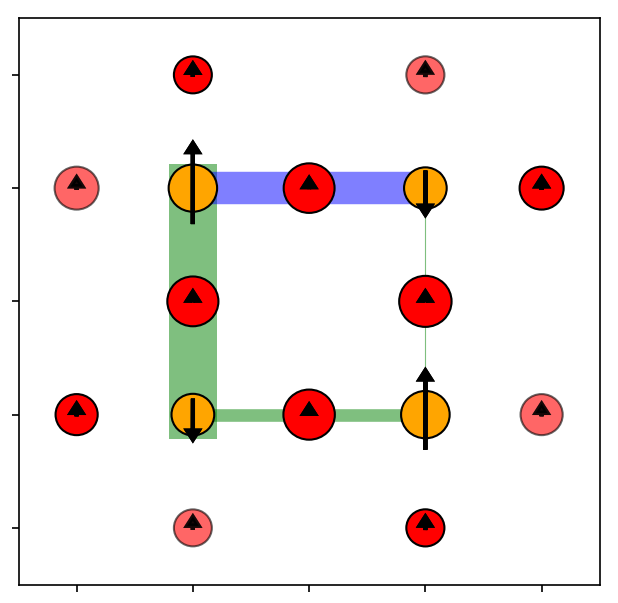}
	\label{fig:threeband:Martin_pattern_2}
}
\subfigure[]{
	\includegraphics[width=0.3\textwidth]{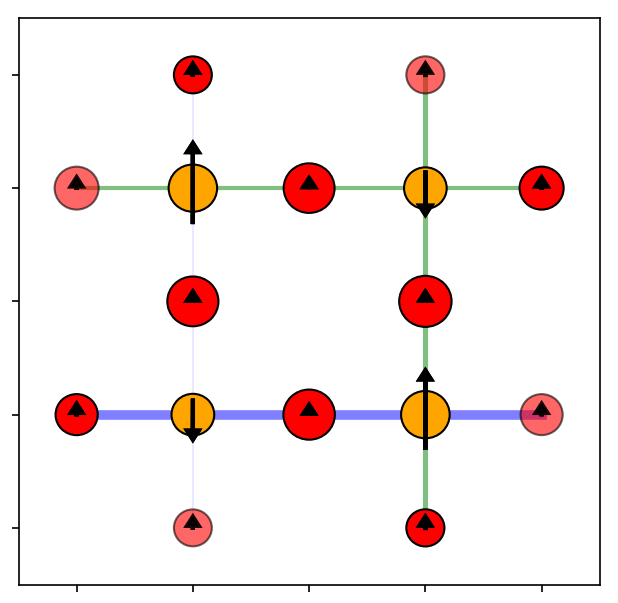}
	\label{fig:threeband:Martin_pattern_3}
}
\subfigure[]{
	\includegraphics[width=0.3\textwidth]{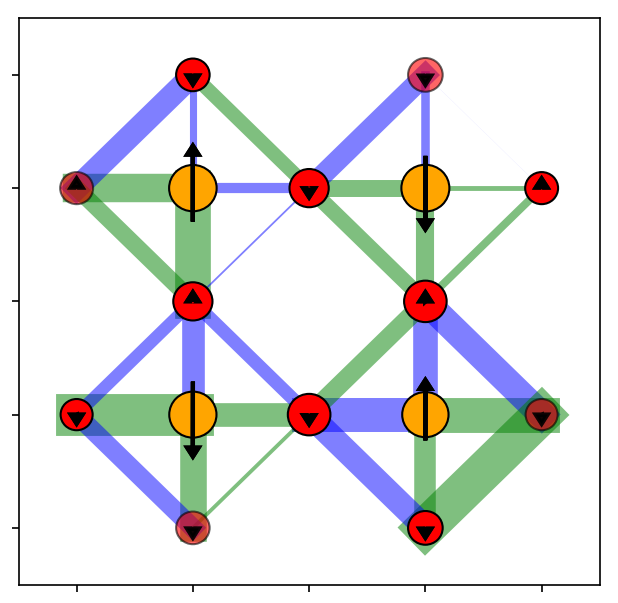}
	\label{fig:threeband:Hanke_pattern_1}
}
\subfigure[]{
	\includegraphics[width=0.3\textwidth]{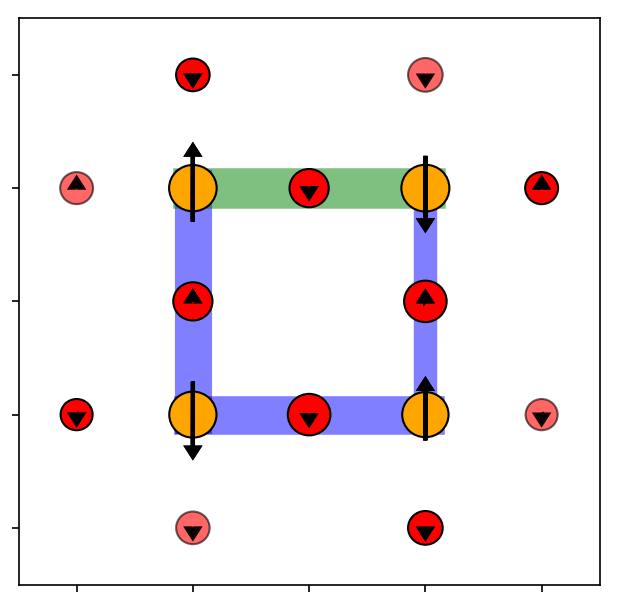}
	\label{fig:threeband:Hanke_pattern_2}
}
\subfigure[]{
	\includegraphics[width=0.3\textwidth]{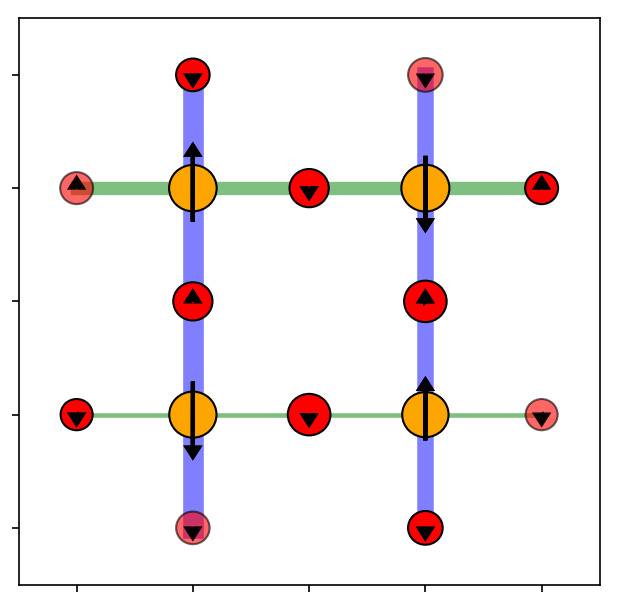}
	\label{fig:threeband:Hanke_pattern_3}
}
\caption{Charge, spin and pairing patterns of the three-band model. The diameters of the red circles are
proportional to the hole densities, the sizes of the arrows are proportional to staggered magnetization, and
the width of the ribbons are proportional to pairing strength. (a) to (c) show parameters from Martin, with
$x=0.2$ doping. (d) to (f) show parameters from Hanke, with $x=0.125$ doping. (a) and (d) show nearest neighbor
pairings; (b) and (e) show pairings between Cu $d$-orbitals; (c) and (f) show pairings between
next-nearest-neighbor O $p$-orbitals.}
	\label{fig:threeband:pattern}
\end{figure}

Fig.~\ref{fig:threeband:pattern} shows the local orders at the optimal doping points using parameters
from Martin and Hanke. The charge and spin orders do not show much inhomogeneity, while the pairing order
is much more complicated than the simple $d$-wave symmetry we saw in the one-band model. Overall, the pairing
strength is strongest between neighboring sites (including nearest $p$ orbitals) and between $d$ orbitals.
Besides the most common $d_{x^2-y^2}$ symmetry, we can also recognize extended $s$-wave
(eg., between nearest $p$ sites), $p$-wave and $d_{xy}$-wave pairing. However, it is not clear enough
how to decompose the real space pairings into angular momentum sectors (coupled with spatial inhomogeneity)
for the DMET results. It is unlikely that the pairing irregularity is an artifact, because the spin and
charge orders seem reasonable, and the trends for the total pairing strength are also as expected.

\noindent\textbf{Conclusions}. We conducted an exploratory investigation of the three-band model using DMET using
a 12-site symmetrized fragment. Using the conventional DMET algorithms designed for the one-band model,
we were not able to control the energy accuracy to similar magnitudes, mainly due to the
difficulties in converging the DMET energies.
Nevertheless, we obtain reasonable charge and spin distributions for the undoped and hole-doped
models with all the parameters we examed, while the local pairing order parameters show complicated
spatial and angular momentum distribution, which requires further investigations with higher energy accuracy
to confirm or reject. Other future directions include employing impurity solvers which are more scalable
such as CASSCF solvers, to study larger fragments on inhomogeneity and to extrapolate to the TDL.

\section{Realistic Cuprates from Hamiltonian Downfolding} \label{sec:threeband:realistic}
\noindent\textbf{The Hamiltonian and computational strategy}.
The ultimate goal in material numerical studies is being able to compute properties of strongly correlated
materials from first principles. To do this, one has to go beyond model systems and form a streamlined
approach to process real materials similar to DFT for normal materials. DMFT+DFT has been
a very successful practice and applied to many important problems so far (See, eg.,
Refs.~\cite{Kotliar2006,yin2011magnetism,park2012site,dang2014covalency}). However, despite the
huge computational cost,
DMFT also suffers from the so-called \textit{double counting} problem~\cite{karolak2010double},
where the correlation energy is included in both DMFT and DFT calculations. To fix the double counting problem,
one usually has to introduce system-dependent terms derived from experience or experimental data.

In addition to exploring a parameter-free treatment of cuprates, we would also like to find
an alternative path to study general strongly correlated materials from first principles using DMET.
We propose the following algorithm to carry out the calculations:
\begin{enumerate}
\item Compute the band structure using DFT or Hartree-Fock, and use the information
	to construct, in a local basis, the Fock matrix of the lattice system,
	the local one-body density matrix and the local two-body integrals.
\item Project the local one-body density matrix to each atom and form the atomic natural orbitals (ANO).
	Choose relevant ANOs to form the basis of the correlated calculations. Remove the Coulomb and exchange
	contributions from the correlated ANOs in the lattice Fock matrix.
\item Perform DMET calculations based on the effective hopping (lattice Fock matrix minus contributions from
	correlated ANOs) and local two-body integrals (in the ANO basis).
\item Once DMET is converged, add the correlation potential $u$ to the DFT or Hartree-Fock calculations in step
1, and perform the calculations in step 1 to 3 again, until convergence.
\end{enumerate}

\begin{figure}[htpb]
	\centering
\subfigure[La$_2$CuO$_4$]{
	\includegraphics[width=0.3\textwidth]{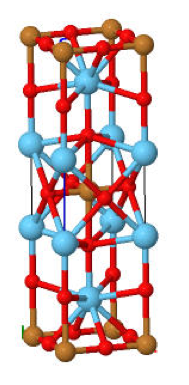}
	\label{fig:threeband:La2CuO4}
}
\subfigure[Ca$_2$CuO$_2$Cl$_2$]{
	\includegraphics[width=0.26\textwidth]{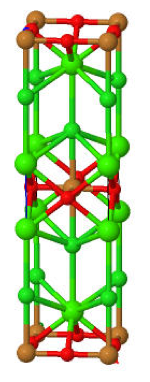}
	\label{fig:threeband:Ca2CuO2Cl2}
}
\caption{Conventional cells of cuprate superconductors La$_2$CuO$_4$ and Ca$_2$CuO$_2$Cl$_2$. For La$_2$CuO$_4$,
we use a cubic structure without John-Teller distortion.}
	\label{fig:threeband:structure}
\end{figure}

\begin{figure}[htpb]
	\centering
\subfigure[]{
	\includegraphics[width=0.42\textwidth]{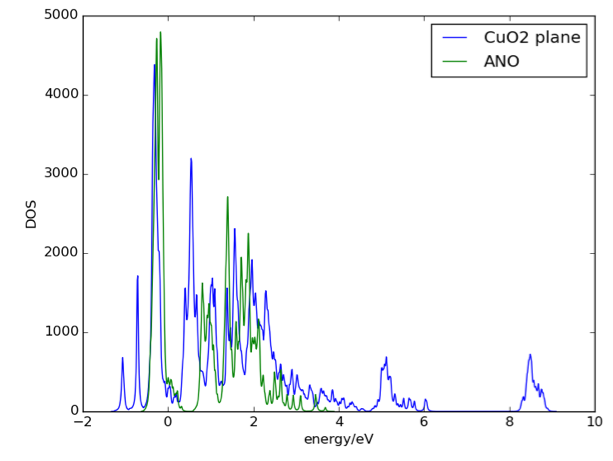}
	\label{fig:threeband:ANO_La}
}
\subfigure[]{
	\includegraphics[width=0.42\textwidth]{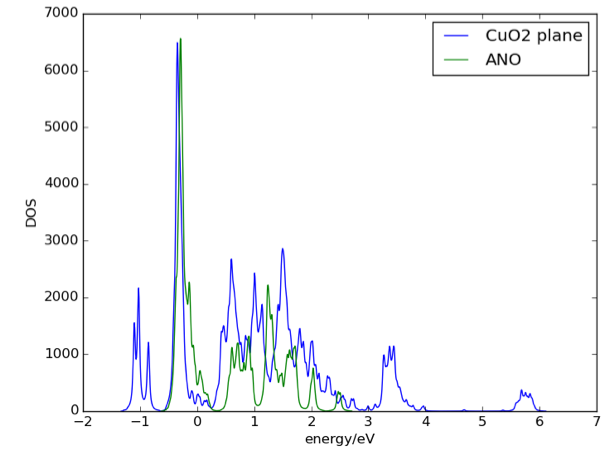}
	\label{fig:threeband:ANO_Ca}
}
\caption{Projected density of states (DOS) for the CuO$_2$ plane from original DFT band structure and projected
	ANO one-body density matrices. (a) La$_2$CuO$_4$ with Cu $3d,4s,4p,4d,5s$ and O $2p,3p$ orbitals (27
	orbitals per unit cell);
(b) Ca$_2$CuO$_2$Cl$_2$ with Cu $3d,4s,4p,4d$ and O $2p,3p$ orbitals (26 orbitals per unit cell).}
	\label{fig:threeband:ANO}
\end{figure}

In this work, we carried out small proof-of-principle calculations on La$_2$CuO$_4$ and Ca$_2$CuO$_2$Cl$_2$,
whose structures are shown in Fig.~\ref{fig:threeband:structure}. For step 1, we performed the spin-restricted
DFT calculations with the periodic Gaussian double-zeta basis fitted as linear
combinations of plane wave functions, detailed in Ref.~\cite{booth2016plane}.
Lately, new periodic DFT / Hartree-Fock codes are available
and would presumably be more efficient and robust to use~\cite{mcclain2017gaussian}.
For step 2, we project the
one-body density matrix to atomic basis by taking the corresponding blocks of the density matrix, and diagonalize
them to obtain the ANOs. Alternatively, we could take the corresponding rows and run SVD to obtain a slightly
different set of ANOs. The ANOs we obtained show clear characteristics of orbital shape and occupation. In
Fig.~\ref{fig:threeband:ANO} we plot the projected density of states to the full CuO$_2$ plane and to the ANO
space. By including $\sim26$ orbitals per unit cell, it is already sufficient to reproduce the PDOS
near the Fermi level (up to $\sim3$eV above), and serves as a realistic correlation model for the real material.
Thus, if we choose the $2\times2$ supercell as the fragment, the total number of orbitals is
around 100 per fragment, or 200 per impurity model, most of which are weakly correlated. Thus, it is possible
to use CASSCF impurity solvers coupled with DMRG to treat the impurity models.

To illustrate the principle, however, we further freeze most of the ANOs but 5 per unit cell: Cu $3d_{x^2-y^2}, 4s, 4d_{x^2-y^2}$
and O $2p_{x/y}$ orbitals. Although the impurity model is now small enough to be solved with DMRG alone, we instead use the
DMRG/CASSCF solver with 12 active orbitals. The calculations are carried out without broken particle-number symmetry and we
did not perform the macro self-consitency between the DFT and DMET calculations (step 4).

\noindent\textbf{Results}. In the undoped material, the average occupations on Cu atoms (the three orbitals) are 1.832 and 2.012
for La$_2$CuO$_4$ and Ca$_2$CuO$_2$Cl$_2$, respectively.
This seems unreasonable but may merely reflect the renormalization between $p$-orbitals and virtual orbitals on Cu. Besides,
since we include two virtual orbitals on each Cu atom but no virtual orbitals on O atoms, there may be an imbalance between
the two species that drives the artificial charge transfer. If we only look at the Cu $3d_{x^2-y^2}$ orbital, the occupations
are 1.382 and 1.410.

\begin{figure}[htpb]
	\centering
\subfigure[La$_2$CuO$_4$]{
	\includegraphics[width=0.42\textwidth]{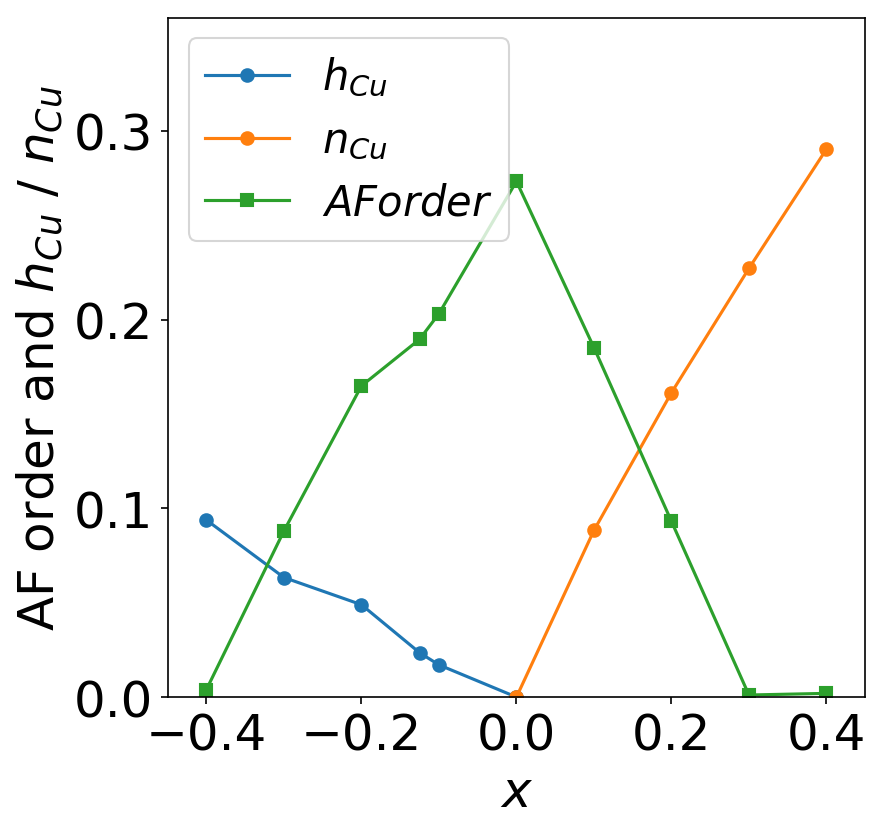}
	\label{fig:threeband:order_La}
}
\subfigure[Ca$_2$CuO$_2$Cl$_2$]{
	\includegraphics[width=0.42\textwidth]{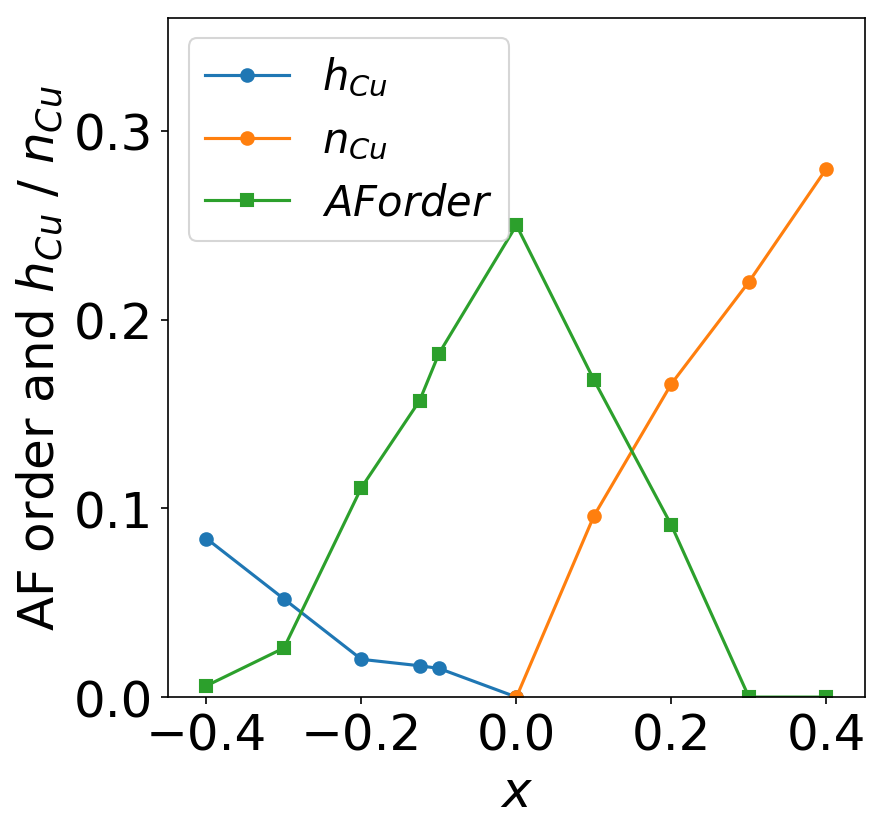}
	\label{fig:threeband:order_Ca}
}
\caption{Averaged antiferromagnetic order parameters and hole/electron densities on Cu atoms for various doping levels in cuprates,
from 5-band DMET calculations. Positive $x$ means electron doping, while negative $x$ means hole doping.}
	\label{fig:threeband:cuprate_order}
\end{figure}

Fig.~\ref{fig:threeband:cuprate_order} shows the hole/electron densities on copper and the AF order, versus doping. When
$x=0$, i.e., undoped, the AF order is slightly higher in La$_2$CuO$_4$ ($m=0.273$) than in Ca$_2$CuO$_2$Cl$_2$ ($m=0.250$).
These numbers are smaller than experimental values, but the order is consistent: the N{\'e}el temperature for La$_2$CuO$_4$
is around 300K, compared to $\sim$250K for Ca$_2$CuO$_2$Cl$_2$~\cite{saylor1989neel,vaknin1997stacking}. The quantitative
disagreement may be due to the spin-restricted Fock matrix from DFT calculations, which suppresses the magnetic order. This
could be fixed by introducing the self-consistency between DFT and DMET. When
the cuprates are doped, the holes go primarily to oxygen $p$ orbitals while electrons mostly go to copper $d$ orbitals, consistent
with our results on the three-band model. Thus, the AF order is destroyed faster on the electron-doped side compared to
hole-doped side. This is opposite to what happens in the generic cuprate phase diagram, and not what one would expect.
However, by inspecting the results more carefully, we find signs of broken spatial symmetry and development of inhomogeneous orders
in the underdoped region of the hole-doped cuprates, which is much weaker for electron-doped cuprates. Thus, once we
go to larger fragments, the inhomogeneous orders may be stabilized and destroy the AF order in the hole-doped cuprates; while the
AF order survives for electron doping. This conjecture is consistent with what we saw in the 1D Hubbard model calculations, as
well as the results here, and seems a plausible explanation for the particle-hole asymmetry in the generic cuprate phase diagram.
Confirming this conjecture may require larger fragments similar to those used in the one-band Hubbard model studies.

\noindent\textbf{Conclusions}. We demonstrate a DFT/HF-DMET approach to study strongly correlated materials from first principles
without any artificial parameters. The approximations involved in these calculations are by no means converged or controlled.
However, we achieve qualitatively correct results in all respects we have access to. With an improved treatment in all
the steps involved, it is promising to finally perform accurate calculations of cuprates from first principles.

\section{Conclusions} \label{sec:threeband:conclusion}
We present two exploratory applications of DMET to study the cuprate physics beyond the one-band Hubbard model. General qualitative
features, such as the interplay of antiferromagnetic and pairing orders, instabilities to inhomogeneous phases, remain in these
systems, while to reach the quantitative details requires more effort to improve the DMET algorithm. With
further developments in the method, impurity solvers and computational resources, \textit{ab initio}, predictive calculations of
high-T$_c$ superconductivity in cuprates is an achievable goal.

\chapter[Finite Temperature Density Matrix Embedding Theory: A Superoperator Approach]{Finite Temperature Density Matrix Embedding Theory: A Superoperator Approach~\footnote{This chapter presents work that has not been published before.}} \label{chpt:finiteT}
\section{Introduction} \label{sec:finiteT:intro}
While ground state wavefunctions contain rich information about the materials, it is still
desirable to access finite-temperature properties, partly because most experimental
measurements happen in finite temperature and it is easy to obtain the temperature
dependence of physical observables.

Thus, extending DMET to finite temperature would significantly expand the scope of the method,
which becomes potentially useful for studying chemical dynamics, spectral properties and
directly observe superconducting phase transition in cuprates.

A simple ``hack'' to enable finite temperature calculations is to use the thermal one-body
density matrix to replace the ground-state density matrix in DMET, and finite-temperature
impurity solvers to replace ground state solvers. A problem of this approach is that the one-body
density matrix is no longer idempotent, and the number of bath orbitals is not bounded anymore;
although we can always truncate the bath orbitals, it becomes increasingly hard to do so
at high temperature. The more serious issue is that, this approach does not have a solid
theoretical framework to guarantee its correctness. Nevertheless, this approach is of low
cost and minimal implementation effort and thus is worth exploring for low-temperature
applications.

In this chapter, however, we discuss another approach based on the singular value decomposition
(SVD) of the density operator, analogous to the SVD of the wavefunction in the ground-state formulation.
The central problem here, is how to define a Hilbert space for the mean-field density operators,
which enables SVD in the single-particle basis, similar to the Schmidt decomposition
of mean-field wavefunctions.
It turns out that a mathematical tool called superoperators can be applied
to solve this problem. In Sec.~\ref{sec:finiteT:theory}, we introduce the theoretical
framework in detail. Then we present the conclusions in Sec.~\ref{sec:finiteT:conclusion}.

\section{Theory} \label{sec:finiteT:theory}

\subsection{The Superoperator Space}\label{sec:finiteT:theory:superop}
The superoperator formalism provides a way to encode thermal density operators as
vectors in a fermion Fock space, thus one can perform operations such as
creation, annihilating, inner product, etc. similar to wavefunctions.

To derive the formulation, we use the spinless notation for simplicity. We start by
writing the normal fermion creation (annihilation) operators as Majorana fermions (we
reserve $i$ as the complex number unit)
\begin{equation}
	w_{2j-1}=a_j+a_j^{\dagger}\ \ w_{2j}=i(a_j-a_j^{\dagger}).
  \label{eq:finiteT:Maj}
\end{equation}

The Majorana fermions are the antiparticles of themselves, thus $w_kw_k=1$ (this can also be
derived rigorously using Eq.~\ref{eq:finiteT:Maj}).
For any operator $\hat{O}$ as finite or infinite polynomial expansions
of $\{a_j^{(\dagger)}\}$, the Majorana
transformation leads to a linear combination of Majorana fermion strings
\begin{displaymath}
  w_1^{\alpha_1}w_2^{\alpha_2}\dots w_{2n}^{\alpha_{2n}}
\end{displaymath}
where $\alpha_i=0$ or $1$. Thus, each of the $2^n$ string in the Majorana fermion
string space can be written as a string of $\{\alpha_i\}$, and
\begin{equation}
	\hat{O}
	= \sum_{\alpha_1,\ldots,\alpha_{2n}}C_{\alpha_1\ldots\alpha_{2n}}
	w_1^{\alpha_1}w_2^{\alpha_2}\dots w_{2n}^{\alpha_{2n}}
	=\sum_{\alpha_1,\ldots,\alpha_{2n}}C_{\alpha_1\ldots\alpha_{2n}}
|\alpha_1\cdots \alpha_{2n}).
\label{eq:finiteT:op_expansion}
\end{equation}
We can define creation and annihilation operators in the superoperator space
\begin{equation}
  f_k|O) = \alpha_k|w_kO)\ \ \ \ \ f_k^{\dagger}|O) = (1-\alpha_k)|w_kO)
  \label{eq:finiteT:fermi_op}
\end{equation}
where $|w_k,\alpha_1,\cdots,\alpha_k,\cdots,\alpha_{2n})
=(-1)^{\alpha_1+\ldots+\alpha_{k-1}}|\alpha_1,\cdots,1-\alpha_k,\cdots,\alpha_{2n})$.
We can easily compute the anticommutation relations
\begin{equation}
	\begin{split}
	\{f_k, f_l\}|O)=&\alpha_k\alpha_l[|w_kw_lO)+|w_lw_kO)]\delta_{k\neq l}
	\\&+2(1-\alpha_k)\alpha_k|O)\delta_{k=l}= 0\\
	\{f_k^{\dagger}, f_l^{\dagger}\}|O)
	=&(1-\alpha_k)(1-\alpha_l)[|w_kw_lO)+|w_lw_kO)]\delta_{k\neq l}
	\\&+2(1-\alpha_k)\alpha_k|O)\delta_{k=l}= 0\\
	\{f_k^{\dagger}, f_l\}|O)=&(1-\alpha_k)\alpha_l[|w_kw_lO)+|w_lw_kO)]\delta_{k\neq l}\\
& +[(1-\alpha_k)^2+\alpha_k^2]|O)\delta_{k=l} = \delta_{kl}.
	\end{split}
\end{equation}
We notice the anticommutation relations are the same as fermions. Thus, the operators
$\{f_k^{(\dagger)}\}$ are (super) fermion operators, and the superoperator space
$\{|\alpha_1\cdots \alpha_{2n})\}$ has the same algebra as the fermion Fock space.

Now we look at a few common notations and operations in the superoperator space.
The vacuum state of the space is the identity operator, denoted as $|E)$.

A common operation between operators is the operator product. Given $\hat{O}_1(w_k)$ and
$\hat{O}_2(w_k)$, the product is written as the concatenation of the Majorana
fermion strings. Notice in Eq.~\ref{eq:finiteT:fermi_op}, $|w_k O)=(f_k+f_k^{\dagger})|w_k O)$,
the product can be expressed in the superoperator space as
\begin{equation}
|O_1O_2)=\hat{O}_1(f_k+f_k^{\dagger})|O_2).
	\label{eq:finiteT:op_product}
\end{equation}
Another important operation is the trace. We notice that
\begin{equation}
	\begin{split}
	\text{Tr}(AB)=&\sum_{\vec{n}}\langle\vec{n}|
	\prod_k\sum_{\{\alpha_k^A\}, \{\alpha_k^B\}}
	C_{\{\alpha_k^A\}}C_{\{\alpha_k^B\}}w_k^{\alpha_k^A+\alpha_k^B}|\vec{n}\rangle\\
	=&\sum_{\{\alpha_k^A\}, \{\alpha_k^B\}}C_{\{\alpha_k^A\}}C_{\{\alpha_k^B\}}\delta_{\{\alpha_k^A\}, \{\alpha_k^B\}}=(A|B).
	\end{split}
\end{equation}
Thus the trace of the product of two operators is the overlap of the their supervectors,
and the trace for one operator is $\text{Tr}A=(E|A)$.
\subsection{Mean-Field Density Operator}\label{sec:finiteT:theory:mf}

Consider the quadratic Hamiltonian
\begin{equation}
	\hat{h}=\sum_{i,j}h_{ij}a_i^{\dagger}a_j=\sum_k \varepsilon_k c_k^{\dagger}c_k + h_0
  \label{eq:finiteT:mf_Ham}
\end{equation}
where $a_i^{(\dagger)}$ annihilates (creates) an electron at site $i$, and
$\{c_k^{\dagger}\}$ is the basis that diagonalizes the mean-field Hamiltonian.
In addition to the one-body part of the full Hamiltonian or the Fock matrix, we absorb
the correlation potential and the shift resulted from chemical potential into $\hat{h}$.
Thus, the mean-field many-body density operator at $\beta=1/kT$ is
\begin{equation}
  \begin{split}
	  \hat{\rho}&=\frac{e^{-\beta \hat{h}}}{\text{Tr}e^{-\beta \hat{h}}}
    =\frac{1}{Z}e^{-\beta\sum_i\varepsilon_ic_i^{\dagger}c_i}
    =\frac{1}{Z}\prod_i[1-c_i^{\dagger}c_i+\sum_{n=0}^{\infty}(-\beta\varepsilon_i)^nc_i^{\dagger}c_i]\\
    &=\frac{1}{Z}\prod_i(c_ic_i^{\dagger}+e^{-\beta\varepsilon_i}c_i^{\dagger}c_i)
  \end{split}
  \label{eq:finiteT:mf_dm}
\end{equation}
where the partition function
\begin{equation}
  \begin{split}
Z&=\text{Tr}\prod_i(c_ic_i^{\dagger}+e^{-\beta\varepsilon_i}c_i^{\dagger}c_i)
    =\sum_{n_1,n_2,\dots ,n_N }\langle n_1n_2\dots n_N|\prod_i(c_ic_i^{\dagger}+e^{-\beta\varepsilon_i}c_i^{\dagger}c_i)|n_1n_2\dots n_N\rangle\\
    &=\prod_i \sum_{n_i=0,1}\langle n_i|c_ic_i^{\dagger}+e^{-\beta\varepsilon_i}c_i^{\dagger}c_i|n_i\rangle
    =\prod_i (1+e^{-\beta\varepsilon_i}).
  \end{split}
  \label{eq:finiteT:Z}
\end{equation}

The density operator is actually separable $\hat{\rho}=\prod_i\hat{\rho}_i$, where
\begin{equation}
\hat{\rho}_i=\frac{c_ic_i^{\dagger}+e^{-\beta\varepsilon_i}c_i^{\dagger}c_i}
{1+e^{-\beta\varepsilon_i}}.
\end{equation}
which is equivalent to considering each canonical orbitals as an independent subsystems.

At $T=0$, the expression reduces to
\begin{equation}
	\hat{\rho}=\prod_{i\in occ}c_i^{\dagger}c_i\prod_{a\in virt}c_ac_a^{\dagger},
  \label{eq:dm_0K}
\end{equation}
whose eigenvector is the ground state Slater determinant.

The mean-field Hamiltonian in the canonical basis (Eq.~\ref{eq:finiteT:mf_Ham}) can be further
written as
\begin{equation}
  \hat{h}=-\frac{i}{2}\sum_j\varepsilon_jw_{2j-1}w_{2j}+\frac{1}{2}\sum_j \varepsilon_j+h_0
  \label{eq:finiteT:H_maj}
\end{equation}
where $w_{j}$ are now the Majorana fermions corresponding to canonical orbitals $c_i^{\dagger}$,
and $h_0=$.
And the density operator is
\begin{equation}
  |\rho)=|e^{-\beta \hat{h}}/Tr(e^{-\beta \hat{h}}))=e^{-\beta\hat{\chi}}|E)
  /(E|e^{-\beta\hat{\chi}}|E)
\end{equation}
where, according to Eq.~\ref{eq:finiteT:op_product}
\begin{equation}
\hat{\chi}=-\frac{i}{2}
\sum_{j}\varepsilon_{j}(f_{2j-1}^{\dagger}+f_{2j-1})(f_{2j}^{\dagger}+f_{2j})
+\frac{1}{2}\sum_j \varepsilon_j + h_0
  \label{eq:finiteT:chi}
\end{equation}
is the mean-field Hamiltonian operator in the superoperator space.
Forget about the normalization factor for the moment, the density operator becomes
\begin{equation}
  \begin{split}
    |\rho)&=e^{-\beta\hat{\chi}}|E)\\
    &=e^{\frac{i}{2}\sum_j\beta\varepsilon_i(f_{2j-1}^{\dagger}+f_{2j-1})(f_{2j}^{\dagger}+f_{2j})}|E)\\
    &=\prod_j(\cosh\frac{\beta\varepsilon_j}{2}+
    i\sinh\frac{\beta\varepsilon_j}{2}f_{2j-1}^{\dagger}f_{2j}^{\dagger})|E).
  \end{split}
  \label{eq:finiteT:rho_superoperator}
\end{equation}
The same result can be obtained by replacing the fermion operators in
Eq.~\ref{eq:finiteT:mf_dm} with Majorana fermions and directly write down the
superoperator space representation. The normalization factor for
Eq.~\ref{eq:finiteT:rho_superoperator} is thus
$(E|e^{-\beta\hat\chi}|E)=(E|\rho)=\prod_j\cosh\frac{\beta\varepsilon_j}{2}$.

\subsection{Finite Temperature DMET Embedding}\label{sec:finiteT:theory:embedding}

The resulted superoperator representation of the mean-field density operator
is a BCS state. To show this, we perform the Bogoliubov transformation
\begin{equation}
	\alpha_{2i-1}^{\dagger}=u_i{f}_{2i-1}^{\dagger}-v_i{f}_{2i},\thinspace
  \alpha_{2i}^{\dagger}=u_i{f}_{2i}^{\dagger}+v_i{f}_{2i-1}
  \label{eq:finiteT:bogoliubov}
\end{equation}
with the inverse transformation
\begin{equation}
	f_{2i-1}^{\dagger}=u_i^*\alpha_{2i-1}^{\dagger}+v_i\alpha_{2i},\thinspace f_{2i}^{\dagger}=u_i^*\alpha_{2i}^{\dagger}-v_i\alpha_{2i-1}.
  \label{eq:finiteT:bogoliubov_inverse}
\end{equation}
The transformation is analogous to the spin-singlet Bogoliubov transformations for normal
fermions, if we view particle indices $2i-1$ as spin-up and indices $2i$ as spin-down.
Using the condition $\alpha_i|\rho)=0$, we obtain
\begin{equation}
  \begin{split}
    u_i&=\frac{1}{\sqrt{1+\tanh^2(\frac{\beta\varepsilon_i}{2})}}=\sqrt{\frac{1}{2}(1+\frac{1}{\cosh(\beta\varepsilon_i)})}\\
    v_i&=-i\tanh(\frac{\beta\varepsilon_i}{2})u_i.
  \end{split}
  \label{eq:finiteT:quasi_particle}
\end{equation}
Thus, the problem becomes the Schmidt decomposition of a BCS wavefunction. In fact,
we can also compute the generalized one-body density matrix of superoperator state $|\rho)$,
in terms of operators $f_{j}^{\dagger}$, where the only non-zero elements are
\begin{equation}
  \begin{split}
    &\langle f_{2j-1}^{\dagger}f_{2j-1}\rangle = \langle f_{2j}^{\dagger}{f}_{2j}\rangle =\frac{1}{2}(1-\frac{1}{\cosh(\beta\varepsilon_j)})\\
    &\langle {f}_{2j-1}{f}_{2j}\rangle = -\langle {f}_{2j}{f}_{2j-1}\rangle = -\frac{i}{2}\tanh(\beta\varepsilon_j).
  \end{split}
  \label{eq:finiteT:grdm}
\end{equation}
Note that the sparsity in the Bogoliubov transformation and the generalized density matrix
is simple because we work in the canonical basis of the mean-field Hamiltonian. Transforming
to the local basis will yield dense representations of both.

The superoperator approach doubles the number of orbitals in the DMET fragment as well as the
size of the impurity model, compared to their ground state counterparts.

\subsection{Impurity Solver}\label{sec:finiteT:theory:solver}
After constructing the impurity model, the mean-field density operator
can be expressed as
\begin{equation}
|\rho)=|\rho_{\text{emb}})\otimes|\rho_{\text{core}}).
\end{equation}
In the impurity solver, we essentially replace the mean-field
$|\rho_{\text{emb}})$ with correlated $|P_{\text{emb}})$. However,
obtaining $|P_{\text{emb}})$ is not as simple as it seems.
We have to be aware that
$|\rho_{\text{emb}})\neq e^{-\beta \hat\chi_{\text{emb}}}|E_{\text{emb}})$,
where $\hat\chi_\text{emb}=P\chi P$, the mean-field Hamiltonian projected to the
impurity model, because the imaginary time evolution also involves the coupling
term $\hat\chi_\text{emb-core}$. Thus, to obtain $|P_{\text{emb}})$, we
replace only the pure impurity model part of the Hamiltonian with the interacting
version, which gives~\footnote{Strictly speaking, the calculation of $|P_{\text{emb}})$ should
also start with $|E_{\text{emb}})$ and use operator $\hat X_\text{emb}+\hat\chi_\text{emb-core}$
to evolve the impurity model. However, under the constraint of fixed $|\rho_{\text{core}})$,
the impurity model solution can be approximated with Eq~\ref{eq:finiteT:solver}.}
\begin{equation}
	|P_{\text{emb}})=e^{-\beta(\hat X_\text{emb}-\hat \chi_\text{emb})}|\rho_{\text{emb}})
	\label{eq:finiteT:solver}
\end{equation}
where $\hat X_\text{emb}=P\hat X P$, and $X$ is the superoperator representation of the
interacting Hamiltonian with two-body terms (non-interacting or interacting bath).
The imaginary time evolution in Eq.~\ref{eq:finiteT:solver} is slightly different
from what we usually encounter, as the operator acts on a pure state (precisely a BCS wavefunction)
in the superoperator space. Thus, the calculation can be carried out
using quantum Monte Carlo simulations such as AFQMC with walkers in the BCS wavefunction
space~\cite{shi2017many}, or with time-dependent matrix product state algorithms
~\cite{vidal2003efficient,vidal2004efficient,white2004real}.

\section{Conclusions} \label{sec:finiteT:conclusion}
In this chapter, we present a superoperator space algorithm for finite-temperature density
matrix embedding.
The theory relies on transforming the mean-field density operator to a BCS state in
the superoperator space, which can form the DMET bath states using the single-particle
based decomposition we introduced in Sec.~\ref{sec:dmet:bcs}. As a result of switching
to the superoperator space, the size of the fragment, and thus the bath, is doubled
compared to the ground-state formulation. It is consistent with the doubled degrees of
freedom in the density operator language of mixed states. The algorithm works best
for higher temperatures. At low temperature, where $\beta$ becomes large, however, the algorithm
may suffer from accumulated errors from the approximations in Eq.~\ref{eq:finiteT:solver},
as well as errors from the finite-temperature impurity solvers.

\appendix
\chapter{Mathematical Derivations and Formula} \label{chpt:formula}
\section{Impurity Model Hamiltonian for Broken Particle-Number Symmetry DMET} \label{sec:formula:bcs_integral}
The impurity model for broken particle-number symmetry DMET is defined with
the reference vacuum $|\Psi_c\rangle$ and a set of allowed excitations (fragment
and bath quasiparticles) defined in Eq.~\ref{eq:dmet:bcs:bath}.

To obtain the impurity Hamiltonian, we project the non-interacting bath Hamiltonian to
the impurity model Hilbert space in the grand canonical ensemble
\begin{equation}
	\begin{split}
	H_{\text{NI}}-\mu\sum_i n_i=&\sum_{ij\sigma}h_{ij,\sigma}a_{i\sigma}^{\dagger}a_{j\sigma}
	+\frac{1}{2}\sum_{ijkl,\mu\nu}(ik||jl)a_{i\mu}^{\dagger}a_{j\nu}^{\dagger}a_{l\nu}a_{k\mu}
	\\
	&+\sum_{ij}\Delta_{ij}a_{i\alpha}^{\dagger}a_{j\beta}^{\dagger}+h.c.
	\end{split}
	\label{eq:formula:H_lattice}
\end{equation}
where the chemical potential is absorbed into the one-body term.
Eq.~\ref{eq:formula:H_lattice} is transformed using the inverse transformation,
using only the fragment and bath orbitals
\begin{equation}
  Pa_\alpha^{\dagger}P=
  \begin{bmatrix}
  c_\alpha^{\dagger}&c_\beta
  \end{bmatrix}
  \begin{bmatrix}
  V_\alpha^{\dagger}\\
    U_\alpha^{\dagger}
  \end{bmatrix},\thinspace
  Pa_\beta^{\dagger}P=\begin{bmatrix}
    c_\beta^{\dagger}&c_\alpha
    \end{bmatrix}\begin{bmatrix}
    V_\beta^{\dagger}\\
    U_\beta^{\dagger}
    \end{bmatrix}
    \label{eq:formula:inverse}
\end{equation}
where we ignore the bath index in Eq.~\ref{eq:dmet:bcs:bath} and use $c$ instead of $d$.
In the impurity model basis, we have
\begin{equation}
	\begin{split}
	H_{\text{imp}}=&E_0+H_1+H_2\\
	H_1=&\sum_{pq\sigma}\bar{h}_{pq,\sigma}c_{p\sigma}^{\dagger}c_{q\sigma}
	+\sum_{pq}\bar\Delta_{pq}c_{p\alpha}^{\dagger}c_{q\beta}^{\dagger}+c.c.\\
    H_2=&\frac{1}{4}\sum_{pqsr}x_{pqsr}c_{p\alpha}^{\dagger}c_{q\alpha}^{\dagger}c_{s\beta}^{\dagger}c_{r\beta}^{\dagger}
    +\frac{1}{2}\sum_{pqsr,\sigma}y_{pqsr,\sigma}c_{p\sigma}^{\dagger}c_{q\sigma}^{\dagger}c_{s\bar\sigma}^{\dagger}c_{r\sigma} + h.c.\\
    &+\frac{1}{2}\sum_{pqsr,\sigma\mu}w_{pqsr,\sigma\mu}c_{p\sigma}^{\dagger}c_{q\mu}^{\dagger}c_{s\mu}c_{r\sigma}.
	\end{split}
	\label{eq:formula:H_imp}
\end{equation}
Note that $H_2$ has two additional terms which break particle-number symmetry. They connect to
the $N\pm2,N\pm4$ number sectors.

\subsection{One-Body Terms}
We can work out the one-body terms using the matrix representation in Eq.~\ref{eq:formula:inverse}.
\begin{equation}
  \begin{split}
	  P\hat{h}P&=P\sum_\sigma a_\sigma^{\dagger}h_\sigma a_\sigma P\\
    &=\sum_\sigma
    \begin{bmatrix}
    c_\sigma^{\dagger}&c_{\bar\sigma}
    \end{bmatrix}
    \begin{bmatrix}
    V_\sigma^{\dagger}\\
    U_\sigma^{\dagger}
    \end{bmatrix}
  h_\sigma
    \begin{bmatrix}
    V_\sigma&U_\sigma
    \end{bmatrix}
    \begin{bmatrix}
    c_\sigma\\
    c_{\bar\sigma}^{\dagger}
    \end{bmatrix}
  =\sum_\sigma
    \begin{bmatrix}
    c_\sigma^{\dagger}&c_{\bar\sigma}
    \end{bmatrix}
    \begin{bmatrix}
    V_\sigma^{\dagger}h_\sigma V_\sigma&V_\sigma^{\dagger}h_\sigma U_\sigma\\
    U_\sigma^{\dagger}h_\sigma V_\sigma&U_\sigma^{\dagger}h_\sigma U_\sigma
    \end{bmatrix}
    \begin{bmatrix}
    c_\sigma\\
    c_{\bar\sigma}^{\dagger}
    \end{bmatrix}\\
  &=\sum_\sigma c_\sigma^{\dagger}(V_\sigma^{\dagger}h_\sigma V_\sigma-U_{\bar\sigma}^{\dagger}h_{\bar\sigma}U_{\bar\sigma})c_\sigma
  +c_\alpha^{\dagger}(V_\alpha^{\dagger}h_\alpha U_\alpha-U_\beta^{\dagger} h_\beta V_\beta)c_\beta^{\dagger} + c.c.
  +\sum_\sigma\text{Tr}(U_\sigma^{\dagger}h_\sigma U_\sigma),
  \end{split}
\end{equation}
and similarly the pairing channel
\begin{equation}
  \begin{split}
  P\hat\Delta P=&a_\alpha^{\dagger}\Delta a_\beta^{\dagger}+h.c.\\
  =&\begin{bmatrix}
    c_\alpha^{\dagger}&c_\beta
	\end{bmatrix}
	\begin{bmatrix}
    V_\alpha^{\dagger}\\
    U_\alpha^{\dagger}
  \end{bmatrix}\Delta
  \begin{bmatrix}
    V_\beta&U_\beta
  \end{bmatrix}
  \begin{bmatrix}    c_\beta^{\dagger}\\
    c_\alpha
  \end{bmatrix}+h.c.\\
  =&c_\alpha^{\dagger}V_\alpha^{\dagger}\Delta V_\beta c_\beta^{\dagger}+c_\alpha^{\dagger}V_\alpha^{\dagger}\Delta U_\beta c_\alpha
  +c_\beta U_\alpha^{\dagger}\Delta V_\beta c_\beta^{\dagger}+c_\beta U_\alpha^{\dagger}\Delta U_\beta c_\alpha+h.c.\\
  =&c_\alpha^{\dagger}(V_\alpha^{\dagger}\Delta U_\beta+U_\beta^{\dagger}\Delta^{\dagger}V_\alpha)c_\alpha
  -c_\beta^{\dagger}(V_\beta^{\dagger}\Delta^{\dagger}U_\alpha+U_\alpha^{\dagger}\Delta V_\beta)c_\beta
  \\&+c_\alpha^{\dagger}(V_\alpha^{\dagger}\Delta V_\beta+U_\beta^{\dagger}\Delta^{\dagger}U_\alpha)c_\beta^{\dagger}+h.c.
  +\text{Tr}(U_\alpha^{\dagger}\Delta V_\beta+V_\beta^{\dagger}\Delta^{\dagger}U_\alpha).
  \end{split}
\end{equation}
Therefore, the contributions from the one-body Hamiltonian to the impurity model are
\begin{equation}
	\begin{split}
		E_0^1=&\sum_\sigma\text{Tr}(U_\sigma^{\dagger}h_\sigma U_\sigma)+
		\text{Tr}(U_\alpha^{\dagger}\Delta V_\beta+V_\beta^{\dagger}\Delta^{\dagger}U_\alpha)\\
		\bar{h}_\alpha^1=&V_\alpha^{\dagger}h_\alpha V_\alpha-U_\beta^{\dagger}h_\beta U_\beta+
		V_\alpha^{\dagger}\Delta U_\beta+U_\beta^{\dagger}\Delta^{\dagger}V_\alpha\\
		\bar{h}_\beta^1=&V_\beta^{\dagger}h_\beta V_\beta-U_{\alpha}^{\dagger}h_{\alpha}U_{\alpha}-V_\beta^{\dagger}\Delta^{\dagger}U_\alpha-U_\alpha^{\dagger}\Delta V_\beta\\
		\bar{\Delta}^1=&V_\alpha^{\dagger}h_\alpha U_\alpha-U_\beta^{\dagger} h_\beta V_\beta+
		V_\alpha^{\dagger}\Delta V_\beta+U_\beta^{\dagger}\Delta^{\dagger}U_\alpha.
	\end{split}
\end{equation}

\subsection{Two-Body Terms}
The contributions from two-body terms are much more complicated. It is evaluated by expanding
\begin{equation}
	\begin{split}
	P\hat{H}_2P=&\frac{1}{2}\sum_{pqrs,ijkl,\mu\nu}(ik||jl)
	(v_{\mu,ip}^*c_{p\mu}^{\dagger}+u_{\mu,ip}^*c_{p\bar\mu})
	(v_{\nu,jq}^*c_{q\nu}^{\dagger}+u_{\nu,ip}^*c_{p\bar\nu})
	\\&\times (v_{\nu,ls}c_{s\nu}+u_{\nu,ls}^*c_{s\bar\nu}^{\dagger})
	(v_{\mu,kr}c_{s\mu}+u_{\mu,ls}^*c_{r\bar\mu}^{\dagger})
	\label{eq:formula:2body}
	\end{split}
\end{equation}
which contains 64 terms requiring normal ordering. We evaluated this expression using an
automatic fermion algebra and code
generator (Sec.~\ref{sec:algo:generator}).

Note that besides the two-body terms $x, v, w$, Eq.~\ref{eq:formula:2body} results in
one-body contributions $\bar{h}_\sigma^2, \Delta^2$ and a constant $E_0^2$ as well.
For the convenience of calculation, we enforce the following symmetries
\begin{equation}
	\begin{split}
	x_{pqsr}&=-x_{pqrs}=-x_{qpsr}=x_{qprs}\\
	y_{pqsr,\sigma}&=-y_{qpsr,\sigma}
	\end{split}
\end{equation}
while for $w_{pqsr,\mu\nu}$, the 8-fold symmetry still applies for like spins, and for
opposite spins, only the 2-fold symmetry $w_{pqsr,\mu\bar\mu}=w_{rsqp,\mu\bar\mu}^*$ applies.
(Of course, we still have $w_{pqsr,\alpha\beta}=w_{qprs,\beta\alpha}$). In the spin-restricted
case, additional symmetries $x_{pqsr}=x_{rsqp}$, $v_{\alpha}=-v_{\beta}$ and
$w_{\alpha\alpha}=w_{\beta\beta}$ exist, but $w_{\alpha\alpha}\neq w_{\alpha\beta}$!

One additional note is that, when we want to evaluate any physical quantities in the impurity
model (or the fragment), we write down the expression in the lattice basis and project to
the impurity model in the same ways as projecting the Hamiltonian. For instance, the chemical
potential term $\mu N$, will no longer be a diagonal one-body matrix in the impurity model
representation. Instead, it will contain normal and pairing one-body terms as well as a constant
contribution.

\section{DMRG Impurity Solver with Broken Particle-Number Symmetry} \label{sec:formula:dmrg_bcs}
Efficient DMRG implementations use quantum numbers to introduce block sparsity~\cite{Chan2011}.
To see how this works, one can simply
associate them with the left and right basis in Eq.~\ref{eq:dmet:solver:dmrg_lrbasis} and prevent mixing
them when performing SVD on the site wavefunction (or equivalently diagonalizing the density matrix).

Normal DMRG calculations usually keep quantum numbers $n$ (particle number) and $S_z$ (spin component
on the $z$-axis).
Some implementations
also keep $S^2$ (total spin)~\cite{sharma2012spin}. When solving the impurity model Hamiltonian Eq.~\ref{eq:formula:H_imp}, we still keep the
quantum numbers $n$ and $S_z$, since the Hamiltonian is still sparse in terms of $n$; Instead of being block
diagonal in the basis with definitive particles, it now becomes a block band matrix, connecting
to $n$ sectors with $n\pm2,n\pm4$ only. The computational cost is thus still lower than completely
discarding the information of particle number.

In realistic DMRG implementations, the Schr\"odinger equation (Eq.~\ref{eq:dmet:solver:dmrg_e}) is
solved using the Davidson algorithm (where one computes $HC$)~\cite{davidson1975iterative}, where the Hamiltonian matrix is
represented as the sum over a number of direct product terms
\begin{equation}
	(H^{(k)})_{i_{k-1},i_k;i_{k-1}^\prime,i_k^\prime}^{n_k,n_{k+1};n_k^\prime,n_{k+1}^\prime}
	=\sum_s [H^L_s]_{i_{k-1}; i_{k-1}^\prime}^{n_k; n_k^\prime} \otimes [H^R_s]_{i_{k}; i_{k}^\prime, }^{n_{k+1}; n_{k+1}^\prime}
\end{equation}
where 
\begin{equation}
	\begin{split}
	[H^L_s]_{i_{k-1}; i_{k-1}^\prime}^{n_k; n_k^\prime}=&
		\langle L_{i_{k-1}}, n_k|\hat{H}^L_s|L_{i_{k-1}^\prime}, n_k^\prime\rangle \\
		[H^R_s]_{i_{k}; i_{k}^\prime, }^{n_{k+1}; n_{k+1}^\prime}=&
		\langle R_{i_{k}}, n_{k+1}|\hat{H}^R_s|R_{i_{k}^\prime}, n_{k+1}^\prime\rangle.
	\end{split}
\end{equation}

\begin{table}[htpb]
	\centering
	\caption{Normal and complimentary operators for normal quantum chemistry Hamiltonian. Numerical
		factors are omitted. See text for detail.}
	\label{tab:formula:dmrg_comp}
	\begin{tabular}{p{0.25\textwidth}p{0.65\textwidth}}
		\toprule
		\textbf{Normal operator}&\textbf{Complimentary operator}\\
		\midrule
		$a_i^{\dagger}$ & $A_i=\sum_j h_{ij} a_j + \sum_{jkl} (ik||jl) a_j^{\dagger}a_k a_l$\\
		$a_i^{\dagger}a_j^{\dagger}$ & $B_{ij}=\sum_{kl}(ik||jl) a_l a_k$\\
		$a_i^{\dagger}a_j$ & $C_{ij}=\sum_{kl}[(ij||kl)-(il||kj)] a_k^{\dagger} a_l$\\
		\bottomrule
	\end{tabular}
\end{table}

\begin{table}[htpb]
	\centering
	\caption{Normal and complimentary operators for BCS impurity model Hamiltonian. Numerical
		factors are omitted. See text for detail.}
	\label{tab:formula:dmrg_comp_bcs}
	\begin{tabular}{p{0.25\textwidth}p{0.65\textwidth}}
		\toprule
		\textbf{Normal operator}&\textbf{Complimentary operator}\\
		\midrule
		$a_i^{\dagger}$ & $A_i=\sum_j h_{ij} a_j + \Delta_{ij}a_j^{\dagger} +
		\sum_{jkl} (ik||jl) a_j^{\dagger}a_k a_l+\sum_{jkl} (y_{ijkl}-y_{jikl}+y_{jkil})
		a_j^{\dagger}a_k^{\dagger}a_l +\sum_{jkl} (x_{ijkl}-x_{jikl}+x_{jkil}-x_{jkli})a_j^{\dagger}
		a_k^{\dagger}a_l^{\dagger}$\\
		$a_i^{\dagger}a_j^{\dagger}$ & $B_{ij}=\sum_{kl}(ik||jl) a_l a_k+\sum_{kl}(y_{ijkl}-y_{ikjl}+y_{kijl})a_k^{\dagger}a_l+\sum_{kl}(x_{ijkl}-x_{ikjl}+x_{kijl}-x_{kilj}+x_{klij}+x_{iklj})a_k^{\dagger}a_l^{\dagger}$\\
		$a_i^{\dagger}a_j$ & $C_{ij}=\sum_{kl}[(ij||kl)-(il||kj)] a_k^{\dagger} a_l+\sum_{kl}(y_{klij}-y_{lkij}-y_{kilj}+y_{likj}+y_{iklj}-y_{ilkj})a_k^{\dagger}a_l^{\dagger}+\sum_{kl}(y_{lkji}^*-y_{klji}^*-y_{ljki}^*+y_{kjli}^*+y_{jlki}^*-y_{jkli}^*)a_l^{\dagger}a_k^{\dagger}$\\
		\bottomrule
	\end{tabular}
\end{table}

For normal state DMET, Table~\ref{tab:formula:dmrg_comp} show the decomposition of the Hamiltonian~\cite{Chan2011}
. Assuming the number of sites on the left is smaller than that on the right, the Hamiltonian
decomposition $H^L_s, H^R_s$ is
\begin{equation}
	H=H_L\otimes I_R+I_L\otimes H_R+(\sum_{i\in L} a_i^{\dagger}A_i + \sum_{i\in R} a_i^{\dagger}A_i
+\sum_{ij\in L}a_i^{\dagger}a_j^{\dagger}B_{ij}+\sum_{ij\in L}a_i^{\dagger}a_j^{\dagger}C_{ij}+h.c.)
	\label{eq:formula:dmrg_comp}
\end{equation}
where the pre-contracted complimentary operators $A, B, C$ are defined in Table.~\ref{tab:formula:dmrg_comp}.
The contractions are performed in the left (right) basis if the normal operator is in the right (left) basis.
$I_L$ and $I_R$ are identity operators on the left and right basis, respectively. Similar decomposition can
be carried out for the impurity model Hamiltonian in Eq.~\ref{eq:formula:H_imp}, with the much more
complicated complimentary operators defined in Table~\ref{tab:formula:dmrg_comp_bcs}.
Note that although the number of operators is still the same, the complimentary operators now have more
than one possible quantum numbers.

Note that in Table~\ref{tab:formula:dmrg_comp} and ~\ref{tab:formula:dmrg_comp_bcs}, we did not make the
numerical factors correct, since they are coupled with the details of the implementation and thus not
very useful when taken out of the context.

\section{CASSCF Formulation with Broken Particle-Number Symmetry} \label{sec:formula:casscf_bcs}
Conceptually, active space calculations based on a BCS wavefunction means using the BCS
wavefunction as a reference state, and allow some of the quasiparticle excitations while
freezing others. The allowed excitations form the active space, while the core quasiparticles
can be thought of as the occupied quasiparticle modes on top of the all-spin-down product state
we mentioned before.

\subsection{Active Space Hamiltonian}
We would like to obtain an effective Hamiltonian with the core contracted. We can
write the original Hamiltonian as
\begin{equation}
	\hat{H}=H_{\text{active}}+H_{\text{core}}+H_{\text{core-active}}.
\end{equation}
The pure core contribution is a constant energy, while $H_{\text{core-active}}$ gives effective
one-body terms similar to the Fock matrix. The pure active part is transformed as described
in Sec.~\ref{sec:formula:bcs_integral}.

To see how $H_{\text{core-active}}$ is transformed, we illustrate the transformation of the
normal two-body integral
\begin{equation}
  \begin{split}
  \frac{1}{2}&(ik||jl)\langle \Psi_c|\langle \Psi|a_{i\mu}^{\dagger}a_{j\nu}^{\dagger}a_{l\nu}a_{k\mu}|\Psi\rangle| \Psi_c\rangle\\
  =&(ik||jl)[(\rho_{ik,\alpha}^\text{core}+\rho_{ik,\beta}^\text{core})(\rho_{jl,\alpha}^\text{active}+\rho_{jl,\beta}^\text{active})
    -\rho_{il,\alpha}^\text{core}\rho_{jk,\alpha}^\text{emb}-\rho_{il,\beta}^\text{core}\rho_{jk,\beta}^\text{emb}
  \\&+\frac{1}{2}(\kappa_{ij}^{\text{core}*}\kappa_{kl}^\text{emb}+\kappa_{ji}^{\text{core}*}k_{lk}^\text{emb})+h.c.]\\
  =&[2J_{ij}^\text{core}-K_{ij,\sigma}^\text{core}]\rho_{ij,\sigma}^\text{emb}+L_{ij}^\text{core}\kappa_{ij}^{\text{emb}*}+h.c.
  \end{split}
\end{equation}
where $J_{ij}^\text{core}=\frac{1}{2}(ij||kl)(\gamma_{kl,\alpha}^\text{core}+\gamma_{kl,\beta}^\text{core})$, $K_{ij,\sigma}^\text{core}=(il||kj)\gamma_{kl,\sigma}^\text{core}$, $L_{ij}^\text{core}=(ik||jl)\kappa_{kl}^\text{core}$. These operators are closely related to the Coulomb, exchange and pairing contributions in Fock matrix. Therefore, the effective one-body terms from the normal two-body integrals are
\begin{equation}
	\bar{H}_1^\text{active}=
	\sum_{ij,\sigma}[2J_{ij}^\text{core}-K_{ij,\sigma}^\text{core}]a_{i\sigma}^{\dagger}a_{j\sigma}-\sum_{ij}L_{ij}^\text{core}a_{i\alpha}^{\dagger}a_{j\beta}^{\dagger}+h.c..
\end{equation}
The derivation for all the two-body terms are more complicated and more terms are involved.
We thus use an automatic fermion algebra and code generator to derive and implement these
terms (Sec.~\ref{sec:algo:generator}).

\subsection{Orbital Rotation}
We derive the equations for orbital gradient and Hessian for BCS-CASSCF. They
are then inserted to the CASSCF routine of
PySCF~\footnote{https://github.com/sunqm/pyscf}~\cite{sun2017general} to perform orbital rotations.

Orbital rotation in the Bogoliubov transformation group is defined as
\begin{equation}
  \begin{bmatrix}
    b_\alpha\\
    b_\beta\\
    b_\alpha^{\dagger}\\
    b_\beta^{\dagger}
  \end{bmatrix}=
  \begin{bmatrix}
    v_\alpha^*&&&u_\beta^*\\
    &v_\beta^*&u_\alpha^*&\\
    &u_\beta&v_\alpha&\\
    u_\alpha&&&v_\beta
  \end{bmatrix}
  \begin{bmatrix}
    a_\alpha\\
    a_\beta\\
    a_\alpha^{\dagger}\\
    a_\beta^{\dagger}
  \end{bmatrix}=U
\begin{bmatrix}
    a_\alpha\\
    a_\beta\\
    a_\alpha^{\dagger}\\
    a_\beta^{\dagger}
  \end{bmatrix}
  \label{eq:formula:rotate}
\end{equation}
where $U$ is a unitary matrix. The orbital rotation is a unitary canonical
transformation~\cite{blaizot1986quantum}, and there exists a unitary operator
$S=e^{i\alpha^{\dagger}K\alpha/2}$, where
\begin{equation}
	\alpha=\begin{bmatrix}
    a_\alpha\\
    a_\beta\\
    a_\alpha^{\dagger}\\
    a_\beta^{\dagger}
  \end{bmatrix}
\end{equation}
and $K$ is Hermitian, such that
\begin{equation}
	b_i^{(\dagger)}=Sa_i^{(\dagger)}S^{-1}.
\end{equation}
The relationship between $U$ and $K$ is
\begin{equation}
	U=e^{-iK}.
\end{equation}
We would like to solve for contraints on $K$. We define a unitary matrix
\begin{equation}
  P=
  \begin{bmatrix}
    &&&1\\
    1&&&\\
    &1&&\\
    &&1&
  \end{bmatrix}
\end{equation}
which block-diagonalizes $U$, such that
\begin{equation}
  P^{\dagger}UP=
  \begin{bmatrix}
    v_\beta^*&u_\alpha^*&&\\
    u_\beta&v_\alpha&&\\
    &&v_\beta&u_\alpha\\
    &&u_\beta^*&v_\alpha^*
  \end{bmatrix}=
  \begin{bmatrix}
    X^*&\\
    &X
  \end{bmatrix}
\end{equation}
where $X$ is also unitary.
We thus have
\begin{equation}
  \begin{bmatrix}
    X^*&\\
    &X
  \end{bmatrix}=P^{\dagger}e^{-iK}P=\sum_{n=0}^\infty P^{\dagger}\frac{(-iK)^n}{n!}P=\sum_{n=0}^\infty \frac{(-iP^{\dagger}KP)^n}{n!}=e^{-iP^{\dagger}KP}.
\end{equation}
Therefore, let $X=e^{-i\Gamma}$, and $X^*=e^{i\Gamma^*}$, then
\begin{equation}
  K=P
  \begin{bmatrix}
    -\Gamma^*&\\
    &\Gamma
  \end{bmatrix}P^{\dagger}.
\end{equation}
If we are dealing with Bogoliubov transformations with real numbers, all the entries of $K$ can be set to imaginary numbers, as well as $\Gamma$, therefore $\Gamma^*=-\Gamma$. We can parameterize $-iK$ as
\begin{equation}
  iK=P
  \begin{bmatrix}
    i\Gamma&\\
    &i\Gamma
  \end{bmatrix}P^{\dagger}=
  \begin{bmatrix}
    \gamma_a&&&\Delta\\
    &\gamma_b&-\Delta^T&\\
    &\Delta&\gamma_a&\\
    -\Delta^T&&&\gamma_b
  \end{bmatrix}
\end{equation}
where $\gamma_a=-\gamma_a^T$, $\gamma_b=-\gamma_b^T$.
We therefore have
\begin{equation}
  \begin{split}
    S=e^{\frac{1}{2}\alpha^{\dagger}(iK)\alpha}=&\exp\{\frac{1}{2}[
    \gamma^a_{ij}(a_{i\alpha}^{\dagger}a_{j\alpha}+a_{i\alpha}a_{j\alpha}^{\dagger})
    +\gamma^b_{ij}(a_{i\beta}^{\dagger}a_{j\beta}+a_{i\beta}a_{j\beta}^{\dagger})\\
    &-\Delta_{ji}(a_{i\beta}a_{j\alpha}+a_{i\beta}^{\dagger}a_{j\alpha}^{\dagger})
    +\Delta_{ij}(a_{i\alpha}a_{j\beta}+a_{i\alpha}^{\dagger}a_{j\beta}^{\dagger})
    ]\}\\
    =&\exp[\gamma^a_{ij}a_{i\alpha}^{\dagger}a_{j\alpha}+\gamma^b_{ij}a_{i\beta}^{\dagger}a_{j\beta}
    +\Delta_{ij}(a_{i\alpha}a_{j\beta}+a_{i\alpha}^{\dagger}a_{j\beta}^{\dagger})].
  \end{split}
\end{equation}
Since $\text{Tr}\gamma_a=\text{Tr}\gamma_b=0$. Given this transformation, for any Fock state
\begin{equation}
	|\Psi\rangle=\Psi_{n_1,\ldots,n_k}(C_1^{\dagger})^{n_1}\cdots(C_k^{\dagger})^{n_k}|0\rangle.
\end{equation}
The Bogoliubov orbital rotation gives
\begin{equation}
  |\Psi^\prime\rangle=\Psi_{n_1,\ldots,n_k}(C_1^{\prime\dagger})^{n_1}\cdots(C_k^{\prime\dagger})^{n_k}|0^\prime\rangle
  =\Psi_{n_1,\ldots,n_k}S(C_1^{\dagger})^{n_1}S^{-1}\cdots S(C_k^{\prime\dagger})^{n_k}S^{-1}S|0^\prime\rangle=S|\Psi\rangle
\end{equation}
and the energy expression, under the rotation of orbitals defined in
Eq.~\ref{eq:formula:rotate}, is
\begin{equation}
  E[S]=\langle\Psi| S^{-1}HS|\Psi\rangle.
\end{equation}
Therefore, the energy gradient is
\begin{equation}
  \begin{split}
    \frac{\partial E}{\partial \gamma^\sigma_{ij}}|_{S=1}&=\langle [H, a_{i\sigma}^{\dagger}a_{j\sigma}-a_{j\sigma}^{\dagger}a_{i\sigma}]\rangle
    =2\Re\langle [H, a_{i\sigma}^{\dagger}a_{j\sigma}]\rangle\\
    \frac{\partial E}{\partial \Delta_{ij}}|_{S=1}&=\langle [H, a_{i\alpha}a_{j\beta}+a_{i\alpha}^{\dagger}a_{j\beta}^{\dagger}]\rangle
    =2\Re\langle [H, a_{i\alpha}a_{j\beta}]\rangle =2\Re\langle [H, a_{i\alpha}^{\dagger}a_{j\beta}^{\dagger}]\rangle.
  \end{split}
  \label{eq:formula:casscf_g}
\end{equation}
And the Hessian is
\begin{equation}
  \begin{split}
    \frac{\partial^2 E}{\partial \gamma^\mu_{ij} \partial\gamma^\nu_{kl}}|_{S=1}&=
    \frac{1}{2}(\langle [[H, a_{i\mu}^{\dagger}a_{j\mu}-a_{j\mu}^{\dagger}a_{i\mu}], a_{k\nu}^{\dagger}a_{l\nu}-a_{l\nu}^{\dagger}a_{k\nu}]\rangle + \\&
    \langle[[H, a_{k\nu}^{\dagger}a_{l\nu}-a_{l\nu}^{\dagger}a_{k\nu}], a_{i\mu}^{\dagger}a_{j\mu}-a_{j\mu}^{\dagger}a_{i\mu}]\rangle)\\
    \frac{\partial^2 E}{\partial \gamma^\sigma_{ij} \partial\Delta_{kl}}|_{S=1}&=
    \frac{1}{2}(\langle [[H, a_{i\mu}^{\dagger}a_{j\mu}-a_{j\mu}^{\dagger}a_{i\mu}], a_{k\alpha}a_{l\beta}+a_{k\alpha}^{\dagger}a_{l\beta}^{\dagger}]\rangle + \\&
    \langle[[H, a_{k\alpha}a_{l\beta}+a_{k\alpha}^{\dagger}a_{l\beta}^{\dagger}], a_{i\mu}^{\dagger}a_{j\mu}-a_{j\mu}^{\dagger}a_{i\mu}]\rangle)\\
    \frac{\partial^2 E}{\partial \Delta_{ij} \partial\Delta_{kl}}|_{S=1}&=
    \frac{1}{2}(\langle [[H, a_{i\alpha}a_{j\beta}+a_{i\alpha}^{\dagger}a_{j\beta}^{\dagger}], a_{k\alpha}a_{l\beta}+a_{k\alpha}^{\dagger}a_{l\beta}^{\dagger}]\rangle + \\&
    \langle[[H, a_{k\alpha}a_{l\beta}+a_{k\alpha}^{\dagger}a_{l\beta}^{\dagger}], a_{i\alpha}a_{j\beta}+a_{i\alpha}^{\dagger}a_{j\beta}^{\dagger}]\rangle).
  \end{split}
  \label{eq:formula:casscf_h}
\end{equation}
If we restrict everything in the real domain, $\Re$ can be ignored. The gradient and Hessian
expressions are analogous to those in normal state CASSCF, with the only additional contribution
from Bogoliubov rotations.
Eq.~\ref{eq:formula:casscf_g} and Eq.~\ref{eq:formula:casscf_h} are thus evaluated using the automatic fermion algebra and code generator, described in Sec.~\ref{sec:algo:generator}.

\section{Constraints for Sign-Problem-Free Correlation Potentials in DMET in the
Half-Filled Hubbard Model} \label{sec:formula:sign}
We first motivate our derivation by recalling how AFQMC becomes sign-problem free in the half-filled Hubbard model
on a bipartite lattice. Given the repulsive Hubbard model with chemical potential $\mu=U/2$
\begin{equation}
	H-\mu n=-t\sum_{\langle ij\rangle\sigma} a_{i\sigma}^{\dagger}a_{j\sigma} + U\sum_{i}[n_{i\alpha}n_{i\beta}-\frac{1}{2}(n_{i\alpha}+n_{i\beta})],
\end{equation}
we perform the partial particle-hole transformation on \textit{only} the spin-up electrons
\begin{equation}
  \hat{P}: a_{i\alpha}^{\dagger}\rightarrow (-)^ia_{i\alpha}, a_{i\alpha}\rightarrow (-)^ia_{i\alpha}^{\dagger}
  \label{eq:ph_transform}
\end{equation}
where the parity term $(-)^i$ is $1$ for sublattice $A$, and $-1$ for the other sublattice, $B$. The transformation results in the attractive Hubbard model
\begin{equation}
  \hat{P}H\hat{P}^{-1}=-t\sum_{\langle ij\rangle,\sigma}a_{i\sigma}^{\dagger}a_{j\sigma} - U\sum_{i}[n_{i\alpha}n_{i\beta}-\frac{1}{2}(n_{i\alpha}+n_{i\beta}-1)]
\end{equation}
which is well-known to be sign-problem free at any occupation. This is seen by performing the Hubbard-Stratonovich transformation,
where the Trotter propagator becomes~\cite{blankenbecler1981monte}
\begin{equation}
  e^{- \tau\hat{P}H\hat{P}^{-1}} =
  \exp(\tau t\sum_{ij\sigma} a_{i\sigma}^{\dagger}a_{j\sigma})\prod_i\sum_{x_i=\pm1}\frac{1}{2}e^{\gamma x_i(n_{i\alpha}+n_{i\beta}-1)}
  \label{eq:projector_Hubbard}
\end{equation}
with $\gamma=\cosh^{-1}e^{\tau U/2}$. Notice that  Eq.~\ref{eq:projector_Hubbard} is spin-symmetric, thus as long as the trial wavefunction $|\Phi_t\rangle$
is  spin-symmetric, the walkers $|\Phi_w\rangle$ are also spin-symmetric. The overlap
\begin{equation}
  \langle \Phi_t|\Phi_w\rangle = \langle \Phi_{t\alpha}|\Phi_{w\alpha}\rangle \langle \Phi_{t\beta}|\Phi_{w\beta}\rangle
  = |\langle\Phi_{t\alpha}|\Phi_{w\alpha}\rangle|^2 \ge 0
\end{equation}
then eliminates the sign problem. From this argument, we also see why the repulsive Hubbard model is sign problem free only at half-filling,
since we require the same number of spin-up holes and spin-down particles in the wavefunction.

In DMET calculations, it is easy to show that if the partial particle-hole symmetry is preserved in the lattice Hamiltonian, the resulting impurity
problem remains sign-problem free.
Consider the partial particle-hole transformation,
Eq.~(\ref{eq:ph_transform}), acting on the non-interacting lattice Hamiltonian in Eq.~\ref{eq:dmet:theory:corr_pot}, with chemical potential $\mu=U/2$
\begin{equation}
  \begin{split}
  &P(h-\mu n)P \\
  &=\hat{P}[h_0+u-\sum_i\frac{U}{2}(n_{i\alpha}+n_{i\beta})]\hat{P}^{-1}\\
  &=h_0+N_c(\sum_{i\in C}u_{ii,\alpha}-UN_{\text{frag}}/2)+\\
  &\sum_C\sum_{i,j\in C}\{[\frac{U}{2}\delta_{ij}-(-)^{i+j}u_{ij,\alpha}]a_{i\alpha}^{\dagger}a_{j\alpha}+(u_{ij,\beta}
  -\frac{U}{2}\delta_{ij})a_{i\beta}^{\dagger}a_{j\beta}\}.
  \end{split}
  \label{eq:h_mf_transformed}
\end{equation}
To impose spin symmetry, we have
\begin{equation}
	\frac{U}{2}\delta_{ij}-(-)^{i+j}u_{ij,\alpha} = u_{ij,\beta}-\frac{U}{2}\delta_{ij}
\end{equation}
which leads to Eq.~\ref{eq:dmet:solver:afqmc_phsymm}. When this condition is satisfied, the ground state of the transformed lattice Hamiltonian $P(h-\mu n)P$
is a spin-symmetric Slater determinant and thus the bath orbitals obey $B^\alpha=B^\beta$. The impurity model Hamiltonian is thus sign-problem free, as the one-body part is clearly spin-symmetric and the fragment interaction $U$ transforms to an attractive Hubbard interaction.

Note that our argument applies to both CDMET and DCA-DMET, since the DCA transformation preserves the partial particle-hole symmetry, which is
the only structure assumed of $h$ in the above derivation.

\section{Translational Symmetries in the DCA-DMET Correlation Potential with AF Order} \label{sec:formula:dca_pot}

We here consider the translational symmetry in the correlation potential in the presence of antiferromagnetic order.
Instead of the normal translational operators, the lattice
Hamiltonian is invariant under the spin-coupled translational operators
\begin{equation}
  T_x: a_{i\sigma}^{(\dagger)} \rightarrow 
  \begin{cases}
    a_{i+x,\sigma}^{(\dagger)},& \text{if } x \text{ is even}\\
    a_{i+x,\bar\sigma}^{(\dagger)},& \text{if } x \text{ is odd}
  \end{cases}
\end{equation}
where the parity of $x$ represents whether a translation brings a site to the same or different sublattice. The Hubbard Hamiltonian is
invariant under $T_x$ operations, because it has both translational and time-reversal symmetry.
Transforming the correlation potential with the spin-coupled translational operators yields
  \begin{equation}
  \begin{split}
  &\text{for even } x,
  T_x u T_x^{-1}
  =\sum_C\sum_{i,j\in C}\sum_\sigma u_{ij\sigma}a_{i+x\sigma}^{\dagger}a_{j+x\sigma}
  =\sum_C\sum_{i,j\in C}\sum_\sigma u_{i-x,j-x,\sigma}a_{i\sigma}^{\dagger}a_{j\sigma} = u\\
  &\text{for odd } x,
  T_x u T_x^{-1}
  =\sum_C\sum_{i,j\in C}\sum_\sigma u_{ij\sigma}a_{i+x\bar\sigma}^{\dagger}a_{j+x\bar\sigma}
  =\sum_C\sum_{i,j\in C}\sum_\sigma u_{i-x,j-x,\bar\sigma}a_{i\sigma}^{\dagger}a_{j\sigma} = u
  \end{split}
\end{equation}
leading to the constraint
\begin{equation}
  u_{ij\sigma}=
  \begin{cases}
    u_{0,j-i,\sigma}, &\text{if } i \text{ is even}\\
    u_{0,j-i,\bar\sigma}, &\text{if } i \text{ is odd}
  \end{cases}
  .
\end{equation}
This constraint, as one can easily verify, is also compatible with the partial particle-hole symmetry required for sign-free AFQMC simulations
in the Hubbard model.

\chapter{Algorithms} \label{chpt:algo}
\section{Normal State DMET Algorithm} \label{sec:algo:normal}
The algorithm for normal state DMET calculations is described below.
\begin{enumerate}
	\item Initial guess for the correlation potential $u = u_0$.
	\item Solve the mean-field wavefunction $|\psi\rangle$, determine the initial value of chemical potential $\mu=\frac{1}{2}(e_{\text{HOMO}}+e_{\text{LUMO}})$.
	\item Calculate bath orbitals and construct impurity model Hamiltonian $H_{\text{emb}}$.
	\item Solve the impurity model while adjusting the chemical potential (Sec.~\ref{sec:algo:chem_fit}).
	\item Compute the one-body matrix $\rho_\Psi$ and fragment energy $E$.
	\item Optimize correlation potential $u^*$.
	\item If $||\Delta u|| < \varepsilon$ and $||\Delta \mu|| < \varepsilon$,
		complete the calculation; otherwise continue.
	\item If $i>I_{\text{DIIS}}$, the starting cycle for DIIS, use DIIS to obtain the new
		correlation potential $u = u_{i+1}=u_{\text{DIIS}}$, otherwise $u_{i+1}=u^*$.
	\item Diagonalize the mean-field Hamiltonian, if the $\mu$ is out of the range of
		the HOMO and LUMO energies, adjust $\mu=\frac{1}{2}(e_{\text{HOMO}}+e_{\text{LUMO}})$.
		Go to Step 3.
\end{enumerate}

\section{Broken Partical-Number Symmetry DMET Algorithm} \label{sec:algo:bcs}
The algorithm for broken particle-number symmetry DMET calculations is described below.
\begin{enumerate}
	\item Initial guess for the correlation potential $u_i = u_0$.
	\item Solve for the BCS wavefunction $|\psi\rangle$, while determining initial value of
		chemical potential $\mu$ by imposing the correct number of electrons to $|\psi\rangle$.
	\item Calculate bath orbitals and construct impurity model Hamiltonian $H_{\text{emb}}$.
	\item Solve the impurity model while adjusting the chemical potential (Sec.~\ref{sec:algo:chem_fit}).
	\item Compute the generalized one-body matrix $G_\Psi$ and fragment energy $E$.
	\item Optimize correlation potential $u^*$.
	\item If $||\Delta u|| < \varepsilon$ and $||\Delta \mu|| < \varepsilon$,
		complete the calculation; otherwise continue.
	\item If $i>I_{\text{trace}}$, the starting cycle for zero-tracing, shift the
		diagonal of $u^*$ to have $\text{Tr}(\Delta u)=0$.
	\item If $i>I_{\text{DIIS}}$, the starting cycle for DIIS, use DIIS to obtain the new
		correlation potential and chemical potential
		$u = u_{i+1}=u_{\text{DIIS}}$, $\mu=\mu_{\text{DIIS}}$; otherwise $u_{i+1}=u^*$.
	\item Go to Step 2.
\end{enumerate}

\section{Correlation Potential Optimization} \label{sec:algo:corr_fit}
In the correlation potential optimization step, we need to minimize the (modified)
cost function
\begin{equation}
	f(u)=||G_\Psi^*-G_\psi(u)||^2
\end{equation}
where $\Psi$ and $\psi$ are correlated and mean-field wavefunctions, respectively.
The density matrix $G_\psi(u)$ is projected to the impurity model. Because of broken
particle-number symmetry, generalized density matrix $G$ is used instead of $\rho$.

To evaluate the cost function, each time when we update $u$, we need to diagonalize
the lattice mean-field Hamiltonian compute the density matrix and project to the
impurity model. This can be costly especially when we compute the gradient.
An approximate approach with good performance is to first project the mean-field Hamiltonian
to the impurity model and when $u$ is updated, use the same linear transformation to project
$u$ to the impurity model, and diagonalize the mean-field Hamiltonian in the impurity model
space. This saves the cost of diagonalizing the giant lattice Hamiltonian (even if k-space symmetry
is used), as well as constructing and transforming the giant density matrix
(instead, the correlation potential is much smaller). The approximate algorithm is called the
fragment fitting, in contrast to the lattice fitting.

We derive the gradient for the broken particle-number symmetry DMET in fragment fitting, while for normal
state it is similar.
There are two components in the gradient to be evaluated
\begin{enumerate}
	\item $\partial f / \partial H$, where $H$ is projected mean-field Hamiltonian.
	\item $\partial H / \partial u$, the derivative of the projected Hamiltonian with
		respect to $u$.
\end{enumerate}

For the first component, we have
\begin{equation}
	s_{ij}=\frac{\partial f}{\partial H_{ij}}=\sum_{kl}\frac{\partial(G^\Psi_{kl}-G^\psi_{kl})^2}{\partial H_{ij}}=\sum_{kl}2(G^\psi_{kl}-G^\Psi_{kl})\frac{\partial G^\psi_{kl}}{\partial H_{ij}}
\end{equation}
where $H_{ij}$ includes the matrix elements in both the normal and the pairing
channel of the BdG equation.

Now we use $G^\psi = CC^{\dagger}$, where $C$ is the solution to the BdG equation (coefficient
matrix). Thus,
\begin{equation}
  \frac{\partial G_{kl}}{\partial H_{ij}}=\sum_{m\in occ}\frac{\partial(c_{km}c_{lm}^*)}{\partial H_{ij}}
  =\sum_{m\in occ}c_{km}\frac{\partial c_{lm}^*}{\partial H_{ij}}+\frac{\partial c_{km}}{\partial H_{ij}}c_{lm}^*.
  \label{eq:algo:partialG}
\end{equation}
By \textit{occ}, we mean the eigenvectors with eigenvalues negative eigenvalues. We use the perturbation theory to evaluate $\partial C/\partial H$. First assume no degeneracy, the first-order perturbation for wavefunction is
\begin{equation*}
	\delta C_n=\sum_{m\neq n}\frac{|m^{(0)}\rangle\langle m^{(0)}|\delta H|n^{(0)}\rangle }{\varepsilon^n-\varepsilon^m}.
\end{equation*}
Translate into matrix language, it becomes
\begin{equation}
  dc_{kl}=\sum_{m\neq l}\sum_{ij}\frac{c_{km}c_{im}^*c_{jl}}{\varepsilon^l-\varepsilon^m}dh_{ij}\Longrightarrow
  \frac{\partial c_{kl}}{\partial h_{ij}}= \sum_{m\neq l}\frac{c_{km}c_{im}^*c_{jl}}{\varepsilon^l-\varepsilon^m}.
\end{equation}
Now insert to Eq.~\ref{eq:algo:partialG}, we have
\begin{equation*}
  \frac{\partial G_{kl}}{\partial H_{pq}}
  =\sum_{m\in occ,n\in virt}\frac{c_{km}c_{ln}^*(c_{pn}c_{qm}^*+c_{pm}^*c_{qn})+c_{kn}c_{lm}^*(c_{pn}^*c_{qm}+c_{pm}c_{qn}^*)}{2(\varepsilon^m-\varepsilon^n)}
\end{equation*}
where we have perform symmetrization as $H_{pq}=H_{qp}$ to eliminate \textit{occ-occ} contributions. Thus we require non-zero HOMO-LUMO gap.
We can define
\begin{equation}
	B_{pqkl}=\frac{\partial G_{kl}}{\partial h_{pq}^{emb}}=\sum_{m\in occ,n\in virt}\frac{c_{km}c_{ln}^*c_{pn}c_{qm}^*+c_{kn}c_{lm}^*c_{pn}^*c_{qm}}{\varepsilon^m-\varepsilon^n}.
\end{equation}

We now look at the second component. Since (up to a constant)
\begin{equation}
	\begin{split}
	PH_{\text{mf}}P=&
	P\begin{bmatrix}
		a_\alpha^{\dagger}& a_\beta
	\end{bmatrix}
	\begin{bmatrix}
		h_\alpha& \Delta\\
		\Delta^{\dagger}& -h_\beta
	\end{bmatrix}
	\begin{bmatrix}
		a_\alpha\\
		a_\beta^{\dagger}
	\end{bmatrix}P\\
	=&\begin{bmatrix}
		c_\alpha^{\dagger}& c_\beta
	\end{bmatrix}
	\begin{bmatrix}
		V_\alpha^{\dagger}&U_\beta^{\dagger}\\
		U_\alpha^{\dagger}&V_\beta^{\dagger}
	\end{bmatrix}
	\begin{bmatrix}
		h_\alpha& \Delta\\
		\Delta^{\dagger}& -h_\beta
	\end{bmatrix}
	\begin{bmatrix}
		V_\alpha&U_\alpha\\
		U_\beta&V_\beta
	\end{bmatrix}
	\begin{bmatrix}
		c_\alpha\\
		c_\beta^{\dagger}
	\end{bmatrix}\\
	=&\begin{bmatrix}
		c_\alpha^{\dagger}& c_\beta
	\end{bmatrix}
	\begin{bmatrix}
		\bar{h}_\alpha& \bar\Delta\\
		\bar\Delta^{\dagger}& -\bar{h}_\beta
	\end{bmatrix}
	\begin{bmatrix}
		c_\alpha\\
		c_\beta^{\dagger}
	\end{bmatrix},
	\end{split}
\end{equation}
when $u$ is arranged correctly, the projection to the impurity model becomes
\begin{equation}
	\delta H = W^{\dagger}\delta u W,\thinspace W=	\begin{bmatrix}
		V_\alpha&U_\alpha\\
		U_\beta&V_\beta.
	\end{bmatrix}
\end{equation}
Note that $W$ is not a square matrix. We thus define
\begin{equation}
	A_{ijpq}=\frac{\partial H_{pq}}{\partial u_{ij}}=W_{ip}^*W_{jq}.
\end{equation}

Therefore, the gradient of the cost function is
\begin{equation}
	g_{ij}=\frac{\partial f}{\partial u_{ij}}=2(G^\psi_{kl}-G^\Psi_{kl})A_{ijpq}B_{pqkl}.
\end{equation}
In practice, tensor $A$ is only computed once, while $B$ is updated in each optimization step.
The procedure to fit the correlation potential is
\begin{enumerate}
  \item Use the independent elements of $u$ as primary variables. Compute tensor $A$.
  \item Compute impurity model BdG matrix $H$ and eigenvectors $C$, eigenvalues $\varepsilon$,
	  as well as generalized density matrix $G^\psi$.
  \item Compute tensor $B$.
  \item Compute gradient $g_{ij}$.
  \item Use conjugate gradient to get optimization direction.
  \item Linear search.
  \item Update $u$ and compute $\Delta G^\psi$. If $||\Delta G^\psi||<\delta$, the threshold of convergence, exit; otherwise return to step 2.
\end{enumerate}

\section{Adaptive Chemical Potential Optimization} \label{sec:algo:chem_fit}
Because impurity model calculations are the most expensive part of DMET, we try to
minimize the number of such calculations. An effective way to do so is to minimize
the number of cycles in chemical potential optimization. In each DMET macro-iteration,
we allow at most three impurity model calculations, with the following algorithm
\begin{enumerate}
	\item Solve the impurity model with initial $\mu_1$ and obtain the number of electron
		$n_1$. If $||n_1 - n^*|| < \varepsilon_n$ ($n^*$ is the target number of
		electrons), complete the optimization; otherwise let $\mu=\mu_2$. We will
		describe the algorithm to determine $\mu_2$ later.
	\item  Solve the impurity model with $\mu=\mu_2$ and obtain the number of electron
		$n_2$. If $||n_1 - n^*|| < \varepsilon_n$ ($n^*$ is the target number of
		electrons), complete the optimization; otherwise let $\mu=\mu_3$. We will
		describe the algorithm to determine $\mu_3$ later.
	\item  Solve the impurity model with $\mu=\mu_3$ and obtain the number of electron
		$n_3$. Return the results.
\end{enumerate}

In this algorithm, $\mu_2$ is determined as a weighted average using predictions from all
previous chemical potential optimization runs, while $\mu_3$, if necessary, is determined
by the linear extrapolation using results from $\mu_1$ and $\mu_2$.

To compute $\mu_2$ at DMET iteration $i$, for a given previous iteration $s$, the weight and
prediction value is determined as follows.
\begin{itemize}
	\item If the algorithm stops at step 1, $w_s=0$.
	\item If the algorithm stops at step 2, we use the two results to compute
		a slope, and use the slope with the current $\mu_1$ and $n_1$ to obtain
		$\mu_s^{\text{pred}}$. The weight is computed as
		\begin{equation}
			w_s=\exp\{-\min[||(n_1,n^*)-(n_1^{(s)}, n_2^{(s)})||^2, ||(n_1,n^*)
			-(n_2^{(s)}, n_1^{(s)})||^2]/2\sigma_2 - (i-s)\}
			\label{eq:algo:chem_weight}
		\end{equation}
	where $n_k^{(s)}$ is the number of electrons in the $k$'th trial of iteration $s$. The
	weight factor depends on both the similarity in the number of electrons and the length
	of th history.
\item If the algorithm stops at step 3, we  we use the three results to fit a parabola,
	and determine the position of the target number of electrons and the current number
	of electrons on the parabola. The difference then becomes the change of $\mu$, i.e.
	$\mu_s^{\text{pred}}=\mu_1+\Delta\mu^{\text{pred}}$. If this does not work, we switch
	to linear regression and use the slope to find $\mu_s^{\text{pred}}$.
	The weight $w_s$ is similar to Eq.~\ref{eq:algo:chem_weight} but we go over all 6
	ordered pairs of $n_k^{(s)}$ to find the minimum, and another parameter $\sigma_3$ is used
	to replace $\sigma_2$.
\end{itemize}
We run this procedure for every previous DMET iterations, and finally determine
$\mu_2=\sum_s w_s \mu_s^{\text{pred}} / \sum_s w_s$.
The parameters $\sigma_2$ and $\sigma_3$ are determined experimentally. For the 2D Hubbard model,
a reasonable set of parameters is $\sigma_2 = 0.00025, \sigma_3 = 0.0005$.
Using this scheme, it usually takes less than 4 DMET iterations to make the number of electrons
sufficiently close to the target value; then it stays close to the target number in further iterations.

\section{Davidson Algorithm} \label{sec:algo:davidson}
The Davidson algorithm~\cite{davidson1975iterative} is an efficient way to find the lowest/highest
eigenvectors of a Hermitian
matrix. It essentially spans a subspace in which the matrix is diagonalized, similar to the
power method. It has better numerical stability and faster convergence because a preconditioner
is used to scale the vectors and the vectors are orthogonalized. Given the Hermitian matrix $A$,
the algorithm is as follows
\begin{enumerate}
	\item Select a guess vector $b_1$.
	\item Compute the subspace representation for $A$ on $\{b_i\} (i\le n)$, $G_{ij}=b_i^T A b_j$.
	\item Diagonalize $G_{ij}$ and obtain the lowest subspace eigenvalues and eigenvector $Gx=\lambda x$. Take the lowest one if we are interested in the ground state only. The current best approximation
		for the lowest eigenvector is thus $c=\sum_{i=1}^{n}x_ib_i$.
	\item Compute the residual vector $r=(A-\lambda)c$. If $||r||<\varepsilon$ the convergence
		threshold, complete the calculation.
		Otherwise, compute the
		rescaled correction vector $\delta_i = (\lambda-A_{ii})^{-1}r_i$.
	\item Orthogonalize the correction vector $\delta$ against  $\{b_i\} (i\le n)$ and normalize it.
		Add to the set of basis as $b_{n+1}$.
	\item Discard the earliest vectors when the subspace becomes too large.
\end{enumerate}

\section{BitGen: An Automatic Fermion Algebra and Code Generator} \label{sec:algo:generator}
The main features of the fermion algebra tool include
\begin{itemize}
	\item Fermion normal ordering;
	\item Operator rotation and transformation;
	\item Evaluating common expectation values, such as density matrices;
	\item Combing like terms;
	\item Conscious about special tensor symmetries;
	\item Transforming derived formula to Python code, using numpy tensor operations;
	\item Recognizing intermediate results to prevent recomputing.
\end{itemize}
The tool is a standalone module in the libDMET package
~\footnote{https://bitbucket.org/zhengbx/libdmet}.

\singlespacing
\bibliographystyle{unsrt}

\cleardoublepage
\ifdefined\phantomsection
  \phantomsection  
\else
\fi
\addcontentsline{toc}{chapter}{Bibliography}

\bibliography{thesis}

\end{document}